\documentclass[acmsmall,nonacm]{acmart}
\makeatletter
\def\mdseries@tt{m}
\makeatother
\hypersetup{breaklinks=true}
\usepackage{breakurl}

\usepackage{pervasives}
\usepackage{oxide}

\renewcommand{\oxnamegrammar}{
  \begin{displaymath}
    \begin{array}{lrclrclr}
      \bnflabel{Variables} & \oxid &&
      \bnflabel{Functions} & \oxfnname &&
      \bnflabel{Type Vars.} & \oxtvar \\
      \bnflabel{Frame Vars.} & \oxenvvar &&
      \bnflabel{Concrete Regions} & \oxcprov &&
      \bnflabel{Abstract Regions} & \oxabsprov \\
    \end{array}
  \end{displaymath}
}

\renewcommand{\bnfdef}{\!\Coloneqq\!}
\renewcommand{\bnfalt}{\; | \;}

\renewcommand{\oxexpressiongrammar}{
  \bnflabel{Expressions} & \oxexpr & \bnfdef &
  \oxprim \bnfalt \oxplaceexpr \bnfalt
  \oxref{\oxcprov}{\oxmuta}{\oxplaceexpr} \bnfalt
  \oxref{\oxcprov}{\oxmuta}{\oxindex{\oxplaceexpr}{\oxexpr}} \bnfalt
  \oxref{\oxcprov}{\oxmuta}{\oxslice{\oxplaceexpr}{\oxexpr_1}{\oxexpr_2}}
  \\ && \bnfalt &
  \oxseq{\oxexpr_1}{\oxexpr_2} \bnfalt
  \oxassign{\oxplaceexpr}{\oxexpr} \bnfalt
  \oxprod{\oxexpr_1 \oxdotsc \oxexpr_n} \bnfalt
  \oxarr{\oxexpr_1 \oxdotsc \oxexpr_n}
  \\ && \bnfalt &
  \oxletrgn{\oxcprov}{\oxexpr} \bnfalt
  \oxlet{\oxid}{\oxsitype}{\oxexpr_1}{\oxexpr_2}
  \\ && \bnfalt &
  \oxclosure{
    \oxascribe{\oxid_1}{\oxsitype_1} \oxdotsc
    \oxascribe{\oxid_n}{\oxsitype_n}
  }{\oxsitype_r}{
    \oxexpr
  }
  \\ && \bnfalt &
  \oxapp{\oxexpr_f}{
    \overline{\oxenv} \oxcomma \overline{\oxprov} \oxcomma \overline{\oxsitype}
  }{
    \oxexpr_1 \oxdotsc \oxexpr_n
  }
  \\ && \bnfalt &
  \oxbranch{\oxexpr_1}{\oxexpr_2}{\oxexpr_3} \bnfalt
  \oxindex{\oxplaceexpr}{\oxexpr} \bnfalt
  \oxabort{\oxstring}
  \\ && \bnfalt &
  \oxfor{\oxid}{\oxexpr_1}{\oxexpr_2} \bnfalt
  \oxwhile{\oxexpr_1}{\oxexpr_2}
  \\ && \bnfalt &
  \oxinl{\oxsitype_1}{\oxsitype_2}{\oxexpr} \bnfalt \oxinr{\oxsitype_1}{\oxsitype_2}{\oxexpr}
  \\ && \bnfalt &
  \oxmatch{\oxexpr}{\oxid_1}{\oxexpr_1}{\oxid_2}{\oxexpr_2}
}

\renewcommand{\oxptrexprgrammar}{
  \bnflabel{Expressions} & \oxexpr & \bnfdef &
  \oxdots \bnfalt \oxframed{\oxexpr} \bnfalt \oxshift{\oxexpr}
  \\ && \bnfalt &
  \oxsliceval{\oxvalue_1 \oxdotsc \oxvalue_n} \bnfalt \oxuninit
  \oxptrextension
}

\renewcommand{\oxvaluegrammar}{
  \bnflabel{Values} & \oxvalue & \bnfdef &
  \oxprim \bnfalt \oxfnname \bnfalt \oxuninit \bnfalt
  \oxprod{\oxvalue_1 \oxdotsc \oxvalue_n} \bnfalt
  \oxarr{\oxvalue_1 \oxdotsc \oxvalue_n} \bnfalt
  \oxsliceval{\oxvalue_1 \oxdotsc \oxvalue_n}
}

\renewcommand{\oxvaluectxgrammar}{
  \bnflabel{Value Contexts} & \oxvaluectx & \bnfdef &
  \oxhole \bnfalt
  \oxprod{\oxvalue_1 \oxdotsc \oxvaluectx \oxdotsc \oxvalue_n} \bnfalt
  \oxarr{\oxvalue_1 \oxdotsc \oxvaluectx_1 \oxdotsc \oxvaluectx_m \oxdotsc \oxvalue_n}
}

\renewcommand{\oxsitypegrammar}{
  \bnflabel{Sized Types} & \oxsitype & \bnfdef &
  \oxbasetype \bnfalt \oxtvar \bnfalt \oxtref{\oxprov}{\oxmuta}{\oxxitype}
  \\ && \bnfalt &
  \oxtprod{\oxsitype_1 \oxdotsc \oxsitype_n} \bnfalt
  \oxtarr{\oxsitype}{\oxnum} \bnfalt
  \oxtsum{\oxsitype_1}{\oxsitype_2}
  \\ && \bnfalt &
  \oxtfunext{
    \overline{\oxenvvar} \oxcomma \overline{\oxabsprov} \oxcomma \overline{\oxtvar}
  }{
    \oxsitype_1 \oxdotsc \oxsitype_n
  }{\oxsitype_r}{\oxenv}{\overline{\oxabsprov_1: \oxabsprov_2}}
}

\renewcommand{\oxevalctxgrammar}{
  \bnflabel{Eval. Contexts} & \oxevalctx & \bnfdef &
  \oxhole \bnfalt
  \oxref{\oxprov}{\oxmuta}{\oxindex{\oxplaceexpr}{\oxevalctx}} \bnfalt
  \oxref{\oxprov}{\oxmuta}{\oxslice{\oxplaceexpr}{\oxevalctx}{\oxnoseqexpr}} \bnfalt
  \oxref{\oxprov}{\oxmuta}{\oxslice{\oxplaceexpr}{\oxvalue}{\oxevalctx}}
  \\ && \bnfalt &
  \oxlet{\oxid}{\oxsitype}{\oxevalctx}{\oxexpr} \bnfalt
  \oxletrgn{\oxcprov}{\oxevalctx}
  \\ && \bnfalt &
  \oxassign{\oxplaceexpr}{\oxevalctx} \bnfalt
  \oxseq{\oxevalctx}{\oxexpr} \bnfalt
  \oxframed{\oxevalctx} \bnfalt
  \oxshift{\oxevalctx} \bnfalt
  \oxshiftprov{\oxevalctx}
  \\ && \bnfalt &
  \oxapp{\oxevalctx}{\overline{\oxenv} \oxcomma \overline{\oxprov} \oxcomma  \overline{\oxsitype}}{
    \oxnoseqexpr_1 \oxdotsc \oxnoseqexpr_n
  }
  \\ && \bnfalt &
  \oxapp{\oxvalue}{\overline{\oxenv} \oxcomma \overline{\oxprov} \oxcomma \overline{\oxsitype}}{
    \oxvalue_1 \oxdotsc \oxvalue_m \oxcomma \oxevalctx \oxcomma
    \oxnoseqexpr_1 \oxdotsc \oxnoseqexpr_n
  }
  \\ && \bnfalt &
  \oxindex{\oxplaceexpr}{\oxevalctx} \bnfalt
  \oxbranch{\oxevalctx}{\oxexpr_1}{\oxexpr_2} \bnfalt
  \oxwhile{\oxexpr_1}{\oxexpr_2}
  \\ && \bnfalt &
  \oxprod{
    \oxvalue_1 \oxdotsc \oxvalue_m \oxcomma \oxevalctx \oxcomma
    \oxnoseqexpr_1 \oxdotsc \oxnoseqexpr_n
  }
  \\ && \bnfalt &
  \oxinl{\oxsitype_1}{\oxsitype_2}{\oxevalctx} \bnfalt
  \oxinr{\oxsitype_1}{\oxsitype_2}{\oxevalctx}
  \\ && \bnfalt &
  \oxmatch{\oxevalctx}{\oxid_1}{\oxexpr_1}{\oxid_2}{\oxexpr_2}
}

\acmJournal{PACMPL}
\acmVolume{1}
\acmNumber{CONF} 
\acmArticle{1}
\acmYear{2018}
\acmMonth{1}
\acmDOI{} 
\startPage{1}

\setcopyright{none}

\begin{document}
\sloppy 

\title{Oxide: The Essence of Rust}


\author{Aaron Weiss}
\affiliation{
  \position{}
  \institution{Northeastern University}
  \city{Boston}
  \state{MA}
  \postcode{02115}
  \country{USA}
}
\email{weiss@ccs.neu.edu}

\author{Olek Gierczak}
\affiliation{
  \position{}
  \institution{Northeastern University}
  \city{Boston}
  \state{MA}
  \postcode{02115}
  \country{USA}
}
\email{gierczak.o@northeastern.edu}

\author{Daniel Patterson}
\affiliation{
  \position{}
  \institution{Northeastern University}
  \city{Boston}
  \state{MA}
  \postcode{02115}
  \country{USA}
}
\email{dbp@dbpmail.net}

\author{Amal Ahmed}
\affiliation{
  \institution{Northeastern University}
  \city{Boston}
  \state{MA}
  \postcode{02115}
  \country{USA}
}
\email{amal@ccs.neu.edu}

\begin{abstract}
  Rust claims to advance industrial programming by bridging the gap between
\emph{low-level} systems programming and \emph{high-level} application
programming, enabling programmers to build more reliable and efficient software.
At the heart of this achievement is the \emph{borrow checker} --- a novel approach
to \emph{ownership} that aims to balance type system expressivity with
usability. And yet, to date there is no type system that fully captures Rust's
notion of ownership and borrowing, and hence no proper foundation for research
on Rust.

We capture the essence of this model of ownership by developing a type systems
account of Rust's borrow checker. We present \lang, a formalized programming
language close to \emph{source-level} Rust (but with fully-annotated types).
\lang takes a new view of \emph{lifetimes} as sets of locations called
\emph{regions} which approximate the origins of references. Our type system is
able to automatically compute this information through a control-flow-based
substructural typing judgment. In doing so, we develop a novel type system for
\emph{region-based alias management}. Significantly, \lang is the first type
system for core Rust that provides a tested semantics and leverages conventional
tools for the formalization and metatheory: it is not built on top of a
separation logic and is proved sound using progress and preservation. As such,
it offers a self-contained model of borrow checking --- including features such
as \emph{non-lexical lifetimes} --- that provides a basis for future research on
Rust.


\end{abstract}



\maketitle

\section{Introduction}
\label{sec:intro}
The Rust programming language exists at the intersection of low-level
``systems'' programming and high-level ``applications'' programming, providing
both fine-grained control over memory and performance \emph{and} high-level
abstractions that make software more reliable and quicker to produce. To
accomplish this, Rust integrates decades of programming languages research into
a production system. Most notably, this includes ideas from linear and ownership
types~\cite{girard87:linear-logic, lafont88:linear-am, clarke98:ownership-types,
  noble98:ownership} and region-based memory
management~\cite{grossman02:cyclone, fluet06:linrgn}. Yet, Rust goes beyond
prior art in developing a particular typing discipline that aims to balance both
\emph{expressivity} and \emph{usability}. As such, Rust has something
interesting to teach us \mbox{about making ownership \emph{practical} for
  programming.}

But without a formal semantics to build upon, it is difficult for researchers to
learn, understand, and investigate this new discipline. This is not a new
problem; the novelty of new languages has often encouraged their formal study to
learn \emph{precisely} what they offer. As \citet{guha10:essence-js} did for
JavaScript, we endeavor to do for Rust --- capturing the essential pieces of
Rust, namely the borrow checker, and providing a foundation for research
with our new formally-defined language, \lang.

While there are existing formalizations of Rust~\cite{reed15:patina,
  benitez16:rusty-types,jung18:rustbelt,jung19:stacked-borrows, pearce21:fr},
none properly convey the essence of Rust's type system. We will discuss all of
them in more detail in \Sec{sec:related}, but for now, we will focus on
RustBelt~\cite{jung18:rustbelt} which represents the most significant effort to
date, and the strongest point of comparison to \lang. RustBelt defines a
calculus called \lambdarust and takes a semantic approach to type
soundness~\cite{milner78:inference,ahmed04:thesis,ahmed10:fpcc-jrnl} to verify
that major parts of Rust's standard library APIs (written using \rusti{unsafe}
code) do not violate its safety guarantees. Yet, \lambdarust's
continuation-passing style and low-level nature --- closer to Rust's Mid-level
Intermediate Representation (MIR) --- make it difficult to use for
\emph{source-level} reasoning. Further, the \lambdarust semantics rely on a
lifetime logic embedded in Iris~\cite{jung18:iris}. While this logic and
embedding is useful for \emph{verifying} the implementation of standard library
APIs, the need to understand the lifetime logic and Iris poses a considerable
cost to other researchers interested in, for instance, investigating new type
features for Rust. Follow-on work by \citet{jung19:stacked-borrows} provides an
operational model called \emph{Stacked Borrows} for the comparatively untyped
``raw pointers'' (usable only in \rusti{unsafe}), which is largely orthogonal to
our efforts as we focus predominantly on the \emph{static} semantics of Rust.

\subsection{Why do another formalism of Rust then?}
RustBelt and other prior work formalize a semantics for Rust based on the notion
of \emph{lifetimes} as the centerpiece of their borrow checking analysis in some
way, and indeed, in context, this was a perfectly sensible decision. After all,
Rust's initial versions of borrow checking relied on lifetimes tied to lexical
scope (i.e., to a first approximation, an object in memory was considered to
live until the end of its lexical scope). However, the work that extended the
language to \emph{non-lexical} lifetimes fundamentally complicated reasoning
about references tied to lifetimes in this way. Using continuation-passing
style~\cite{sussman75:scheme} as \lambdarust does addresses some of the added
complexity of non-lexical lifetimes by providing a natural way for non-lexical
lifetimes to be made contiguous. However, we believe it is necessary to model
how the \emph{source} program works and how to think about borrow checking with
non-lexical lifetimes in that light. To that end, we employ a novel use of
\emph{regions} to track \emph{aliasing} in the static semantics of the program
in \lang.

As we will see, \lang is a higher-level language, with syntax close to that of
surface Rust and a semantics that works with an \emph{abstract} notion of memory
that does not require us to make concrete memory layout decisions for each type.
This is significant because it allows us to focus on the essence of how safe
Rust deals with memory and aliasing, while avoiding a need to address details
caught up in discussions about memory layout and validity guarantees that are
ongoing in the unsafe code guidelines working group~\cite{rust:unsafe-code}. We
also focus our efforts by requiring type annotations on let bindings in \lang to
avoid the orthogonal complexities of type inference, and omitting the trait
system which is largely described in the literature on typeclasses. We also do
not include operations for concurrency, as we believe borrow checking can be
understood clearly \emph{without} it.

\subsection{Our Contributions}
Our efforts to develop \lang have led us to five main contributions: (1)~We
present \lang as the first formal account close to safe, surface Rust. (2)~Most
significantly, we note that while Rust's borrow-checking implementation relies
on constraint generation and an algorithmic constraint solver, we provide
\emph{an inductive definition of borrow checking in terms of conventional
  inference rules}. This definition builds on a view of lifetimes as sets of
locations called \emph{regions} approximating the provenances of references,
rather than abstractions of the lines of code where the referenced memory is
live. (3)~This design represents a novel treatment of regions, leveraging them
to manage \emph{aliasing} rather than memory itself, that we call
\emph{region-based alias management}. (4)~We provide \emph{the first syntactic
  type safety}~\cite{wright92:progress-preservation} result for Rust, which is
challenging because we must maintain the well-typedness of values on
the stack. Ordinarily, this is straightforward, but since our values include
suspended computations which can themselves introduce aliasing, we must show
that the requirements for safe aliasing in that computation are maintained
throughout the program's execution. (5)~\lang features \emph{a tested semantics}
which has been validated in its faithfulness to \rustc borrow checking on the
subset of features supported by \lang using tests from Rust's official borrow
checker and non-lexical lifetimes test suites. Thus, we posit that \lang serves
as an explainable \emph{essence of Rust}, and a solid foundation for research on
and leveraging Rust.

The rest of the paper is organized as follows: \Sec{sec:main} describes the
essence of Rust and \lang at an intuitive level. \Sec{sec:formalization}
presents the formal details of \lang including the syntax (\Sec{sec:terms},
\Sec{sec:types}, and \Sec{sec:regions}), type system (\Sec{sec:typechecking}),
operational semantics (\Sec{sec:dynamics}), and metatheory
(\Sec{sec:metatheory}). \Sec{sec:tested-semantics} provides evidence that \lang
faithfully models Rust, via discussion of our compiler \compiler from Rust to
\lang and a type checker \typechecker used to validate that \lang typechecking
matches Rust on a subset of Rust's official test suite. We discuss related work
in \Sec{sec:related} and some higher-level points about \lang in
\Sec{sec:discussion}. \ifanonymous{
  The anonymous supplementary material includes a technical appendix featuring
  all typing rules, definitions, and proofs, as well as a snapshot of the
  tooling and test suites for our tested semantics.
}{
  The technical appendices include complete definitions (\Sec{sec:full-syntax},
  \Sec{sec:full-statics}, \Sec{sec:metafunctions}, \Sec{sec:full-dynamics}),
  typing rules (\Sec{sec:typing}), and proofs (\Sec{sec:metatheory-full}). Our
  implementation and test suite for our tested semantics are available on
  \href{https://github.com/aatxe/oxide}{GitHub}.
}


\epigraph{Nothing is yours. It is to use. It is to share. If you will not share
  it, you cannot use it.}{\textit{The Dispossessed} \\ \textsc{Ursula K. Le Guin}}
\section{Data They Can Call Their Own}
\label{sec:main}
The essence of Rust lies in its novel approach to \emph{ownership} and
\emph{borrowing}, which account for the most interesting parts of the language's
static semantics and the justification for its claims to \emph{memory safety}
and \emph{data race freedom}. In this section, we gradually introduce
\lang by exploring example programs that illustrate key pieces of how ownership
and borrowing function. At the same time, we'll explain the syntax in each
example for readers unfamiliar with Rust.

\subsection{Ownership as Use-Once Variables}
\label{sec:ownership}

Rust's notion of ownership rests atop a long lineage of work, harkening back to
the early days of linear logic~\cite{girard87:linear-logic}, and especially
efforts by \citet{wadler91:linearity} and \citet{baker92:linear-lisp} to develop
systems for functional programming \emph{without} garbage collection. However,
as noted by \citet{wakeling91:linearity}, Wadler's effort relied greatly on
pervasive copying. This reliance on copying and the associated performance
penalty would not suffice for real world systems programming efforts, and thus,
Rust's ownership model is best understood as instead building off of Baker's
work on Linear Lisp where linearity enabled efficient reuse of objects in
memory~\cite{baker92:linear-lisp, baker94:anchored-ptrs, baker94:linear-stacks,
  baker95:linear-variables}. The resemblance is especially strong between Rust
\emph{without} borrowing and Baker's 'use-once'
variables~\cite{baker95:linear-variables}. We illustrate these ideas at work in
\lang with the following example:

\begin{rustbk}
  struct Point(u32, u32);
  let pt = Point(6, 9);
  let x = pt;
  let y = pt; // ERROR: pt was already moved
\end{rustbk}

In this example, we first declare a type \rusti{Point} that consists of a pair
of unsigned 32-bit integers (\rusti{u32}). Then, on line 2, we create a new
\rusti{Point} from \rusti{(6, 9)} and name it \rusti{pt}. We say that this new
value is \emph{owned} by the identifier \rusti{pt}. Then, on line 3, we
transfer this ownership by \emph{moving} the value from \rusti{pt} to \rusti{x}.
Moving the value out of \rusti{pt}, invalidates this old name.
Subsequently, when we attempt to use it again on line 4, we encounter an error
because \rusti{pt} was already moved in the previous line. If we instead used
\rusti{x} instead of \rusti{pt} on line 4, we would not error as each variable
is used once.

\subsection{Borrowing with Loans}
\label{sec:borrowing}

Rust's main departure from techniques like 'use-once'
variables~\cite{baker95:linear-variables} is a softening of a rather stringent
requirement: namely, that \emph{everything} must be managed uniquely. Instead,
Rust allows the programmer to locally make a decision to use unique
references~\cite{minsky96:unique-ptrs} with unguarded mutation \emph{or} to use
shared references without such mutation.\footnote{The use of ``such'' here is
  intentional as dynamically guarded mutation, e.g. using a
  \rusti{Mutex}, is still allowed through a shared reference. Indeed, this is
  precisely what makes such guards \emph{useful} when programming.} This
flexibility in choosing arises at the point where the programmer creates a new
reference, and draws inspiration from work on ownership types and flexible alias
protection~\cite{noble98:ownership, clarke98:ownership-types}. We again
illustrate its use in \lang with an example:

\begin{rustbk}
  letrgn<'x, 'y> {
    let pt = Point(6, 9);
    let x = &'x shrd pt;
    let y = &'y shrd pt; // OK: sharing is fine!
  }
\end{rustbk}

In the above example, we replaced the \emph{move} expressions in each binding
with \emph{borrow} expressions (written using \texttt{\&}) that each create a
shared reference to \rusti{pt}. Here, we also see our first syntactic
differences from Rust. Namely, in \lang, borrow expressions include an
annotation for their \emph{region} (roughly an analogue of Rust's
\emph{lifetimes}) which are bound earlier using $\oxkey{letrgn}$ on line 2. As
noted in the comment, this program no longer produces an error because the
references allow precisely this kind of sharing, but one should note that this
sharing would be \emph{disallowed} by a standard linear or affine type system.
As a consequence of allowing this sharing, the type systems of both \lang and
Rust prevent mutation through these references. To mutate through a reference,
you must instead have a unique reference (i.e. it is the only usable name for
the underlying data). Our next example replaces our shared references with
unique ones:

\begin{rustbk}
  letrgn<'x, 'y> {
    let pt = Point(6, 9);
    let x = &'x uniq pt;
    let y = &'y uniq pt; // ERROR: cannot borrow pt uniquely twice
    ... // additional code that uses x
  }
\end{rustbk}

We've now chosen to create unique, rather than shared, references to \rusti{pt}.
However, since our program attempts to do so twice, we encounter an error
similar to the one we had in our first example when we tried to move \rusti{pt}
twice. The astute reader might notice that another important change happened ---
we added some additional code afterward that somehow makes use of \rusti{x}.
This is important because of a feature in Rust known as \emph{non-lexical
  lifetimes} (or NLL for short)~\cite{turon17:nll, matsakis16:nll}. With
non-lexical lifetimes and no uses of \rusti{x} in the ensuing code, the compiler
would figure out that this apparent violation of the uniqueness of unique
references would be not be realized since \rusti{x} is never used, and thus may
as well not exist. As such, the program would be accepted by the borrow checker.
With the additional use of \rusti{x}, the violation \emph{is} realized and so
the program is rejected.

Similar to the last example, the borrow checker also prevents us from mixing
unique references with shared ones, as in the following example:

\begin{rustbk}
  letrgn<'x, 'y> {
    let pt = Point(6, 9);
    let x = &'x uniq pt;
    let y = &'y shrd pt; // ERROR: cannot borrow pt while
                         // a unique loan is live
      ... // additional code that uses x
    }
\end{rustbk}

In this case, we've changed the borrow expression on line 4 to create a shared,
rather than unique, reference. This again produces an error because Rust forbids
the creation of a shared reference while a unique \emph{loan} exists. Here, we
use the word loan to refer to the state introduced in the borrow checker (which
records the loan's uniqueness and its origin) by the creation of a reference.
Regions\footnote{Historically, Rust has used the term \emph{lifetime}, rather
  than region. Recent efforts on a borrow checker rewrite called Polonius have
  transitioned to using the term origin~\cite{matsakis18:polonius,
    matsakis20:polonius} for a similar concept to our regions. We discuss
  Polonius further in \Sec{sec:discussion}.} in \lang (denoted \rusti{'a},
\rusti{'b}, etc.) can be understood as collections of these loans which together
form a static approximation of the origins of references annotated with that
region. In this way, we can think of our regions as a sort of \emph{static}
grouping of distinct objects in memory, and by associating a reference with a
region, we identify that the objects in memory that reference could point to
must be from the collection of loans we've associated with it. The borrow
checker leverages this information about the origins of references to determine
whether or not a reference is safe to create or use at a given program point.

We use the term \emph{region} here because the term carries a sort of grouping
association, and past uses of region in the literature also use the term to
represent a grouping of objects in memory. However, it's important to note the
difference in how they are used! In the literature, regions are used to manage
memory~\cite{grossman02:cyclone, fluet06:linrgn,
  tofte94:region-inference, tofte97:regions} with each region representing a
contiguous chunk of memory in which references are managed. By contrast, \lang's
regions correspond to an abstract and purely static grouping of objects in
memory, and doesn't have any influence over where allocation happens. Instead,
their use by the borrow checker during typechecking rules out bad aliasing
patterns (as we have seen so far), leading us to refer to this approach to
regions as \emph{region-based alias management}.

In \lang, we write these loans as a pair of a place and an ownership qualifier
($\oxmut$ or $\oximm$), e.g. $\oxloanpkg{\oxmut}{\oxkey{pt}}$ would be the loan
corresponding to the borrow on line 4. During typechecking, we associate each
of the regions bound with $\oxkey{letrgn}$ (e.g. \rusti{'x} and \rusti{'y})
with sets of these loans. Specifically, after line 4, \rusti{'x} will map to the
loan set $\oxset{\oxloanpkg{\oxmut}{\oxkey{pt}}}$ and, after line 5, \rusti{'y}
will map to the loan set $\oxset{\oxloanpkg{\oximm}{\oxkey{pt}}}$. Although
these examples only have single-element loan sets, more complex programs using
branching will merge loan sets making them approximate. When typechecking a
borrow expression, \lang looks at these loan sets in the environment to
determine whether or not the borrow should be permitted.

While we were unable to create a second reference to the same place as an
existing unique reference in our past examples, \lang and Rust both allow the
programmer to create two unique references to disjoint paths within the same
object, as in the following example:

\begin{rustbk}
  letrgn<'x, 'y> {
    let mut pt = Point(6, 9);
    let x = &'x uniq pt.0;
    let y = &'y uniq pt.1;
    // no error, our loans don't overlap!
    ...
  }
\end{rustbk}

In this example, we're borrowing from specific paths within \rusti{pt} (namely,
the first and second projections respectively). Since these paths give a name to
the places being referenced, we refer to them as \emph{places}. Here, we see
that our notion of ownership is fine-grained, allowing unique loans against
non-overlapping places within \emph{aggregate structures} (like structs,
enumerations, and tuples). Intuitively, this is safe because the parts of memory
referred to by each individual place (in this case, \rusti{pt.0} and
\rusti{pt.1}) do not overlap, and thus they represent portions that can each be
uniquely owned and borrowed.

Rust supports an additional pattern that weakens conventional notions of
flexible alias protection. In particular, it allows the programmer to create a
unique reference by borrowing from one they already have. However, the program
is unable to use the old reference until the \emph{reborrowed} one is destroyed.
We produce the same behavior in \lang, and we can see it at work in the
following example:

\begin{rustbk}
  letrgn<'x, 'y> {
    let mut pt = Point(6, 9);
    let x = &'x uniq pt.0;
    let y = &'y uniq *x;
    // we can use y, cannot use x until we drop y
    ...
  }
\end{rustbk}

In this example, we borrow the first projection of \rusti{pt} (\rusti{pt.0}) and
then reborrow it by creating a borrow to \rusti{*x}. We then can use \rusti{y}
in the continuation, but won't be able to use \rusti{x} until \rusti{y} is
dropped. This particular pattern of \emph{reborrowing} is perhaps one of the
most unique things about Rust's design.

The combination of features discussed above, namely moves, shared and unique
borrows, the ability to create unique references to disjoint paths, reborrowing,
and non-lexical lifetimes, makes borrow checking extremely subtle, even when we
focus on just the safe subset of Rust. 
In the rest of the paper, we present Oxide and discuss how our formalism deals
with this mix of features. 


\section{Oxide, Formally}
\label{sec:formalization}
In this section, we present \lang's formal semantics. We first discuss the terms
in the language (\Sec{sec:terms}), then the types (\Sec{sec:types}) and
\emph{regions} (\Sec{sec:regions}), and our environments and the
mechanics of typechecking (\Sec{sec:typechecking}). Finally, we move on to
discussion of our metatheory (\Sec{sec:metatheory}) and tested semantics
(\Sec{sec:tested-semantics}).

\subsection{The Syntax of \lang}
\label{sec:terms}

\begin{figure}
  \figuresize
  \oxnamegrammar

  \begin{minipage}{.43\textwidth}
    \begin{bnf}
      \oxpathgrammar \\
      \oxplacegrammar \\
      \oxplaceexprgrammar \\
      \oxplaceexprctxgrammar \\[1em]

      \oxprovgrammar \\
      \oxmutagrammar \\[1em]
    \end{bnf}
  \end{minipage}
  \begin{minipage}{.56\textwidth}
    \begin{bnf}
      \oxprimitivegrammar \\
      \oxexpressiongrammar \\
    \end{bnf}
  \end{minipage}

  \vspace{-1em}
  \caption{Term Syntax of \lang}
  \label{fig:term-syntax}
  \vspace{-1em}
\end{figure}

\Fig{fig:term-syntax} presents the syntax of \lang terms, in four broad
groupings: (1)~metavariables for the various kinds of names that exist,
(2)~places, which act as names for abstract memory locations, 
(3)~annotations for references, and (4)~the actual terms of the language. The
first group is fairly conventional, but we'll discuss special names like
frame variables as they come up.

\paragraph{Places and Place Expressions} As we saw in \Sec{sec:borrowing},
places $\oxplace$ and place expressions $\oxplaceexpr$ are names for paths from
a particular variable to a particular part of the object stored there, whether
that be a projection of a tuple, or a field of a struct (where a struct is
really precisely just a named tuple or record type). One might think of place
expressions as a sort of syntactic generalization of variables. They are
analogous to what are called \texttt{lvalues} in C. Places $\oxplace$ are a
subset of place expressions that do not include dereferences. They can
intuitively be thought of as an abstract name of a memory location since when
bound, they will always correspond to one particular value on the stack. Place
expression contexts $\oxplaceexprctx$ are used in various parts of the formalism
to decompose place expressions $\oxplaceexpr$ into an innermost dereferenced
place, $\oxderef{\oxplace}$, and an outer context $\oxplaceexprctx$.

\paragraph{Annotations for References} In \lang, we have two annotations that we
provide for every borrowing expression. First, we annotate references with
ownership qualifiers $\oxmuta$, indicating whether the reference is shared
($\oximm$) or unique ($\oxmut$). We use these rather than their equivalents in
Rust (no annotation and $\rusti{mut}$ respectively) because the terms more
accurately reflect the semantic focus on \emph{aliasing}, rather than
\emph{mutation}. Indeed, in Rust, a value of the type \rusti{&&mut u32}
\emph{cannot} be mutated (because we have a shared reference to a unique
reference), and a value of the type \rusti{&Cell<u32>}\footnote{\rusti{Cell<T>}
  is a Rust standard library type that provides a ``mutable memory location''
  that allows mutation in its API.} \emph{can} be mutated through the method
\rusti{Cell::set}.

Second, we annotate references with \emph{regions}. Regions $\oxprov$ have two
forms: abstract regions $\oxabsprov$ (pronounced \emph{var-rho}) and concrete
regions $\oxcprov$. Abstract regions correspond to lifetime variables
\rusti{'a}, \rusti{'b}, etc.\ in Rust, and are used polymorphically in function
types to indicate that the function is agnostic to the particular regions of
reference-type parameters. Concrete regions, by contrast, carry concrete
information in the environment where they correspond to a set of loans. A loan
$\oxloanpkg{\oxmuta}{\oxplaceexpr}$ indicates a possible origin
($\oxplaceexpr$), qualified by whether the loan is unique or shared ($\oxmuta$).
Intuitively, each loan tells us a single possible origin for a reference, while
a concrete region maps to \emph{all} possible origins of a reference. As we will
see in \Sec{sec:regions}, regions are essential to enabling our type system to
guarantee the correct use of unique and shared references.

\paragraph{Expressions} Expressions $\oxexpr$ in \lang are numerous, but largely
standard. For example, constants $\oxprim$ consist of the unit value $\oxunit$,
unsigned 32-bit integers $\oxnum$, and boolean values $\oxtrue$ and $\oxfalse$.
The most interesting expressions in \lang are the ones we've already seen by
example: place expression usage (written simply $\oxplaceexpr$) and borrowing
(with several forms that we explain shortly). The former may be thought of as
variables that behave linearly for non-copyable data (removing the place from
the environment after use), and traditionally for copyable data. (As a first
approximation, one can think of all data that is not a unique pointer as safely
copyable.)

There are three borrowing forms, and all work in fundamentally the same way:
they are each used as introduction forms for references. The simplest case,
written $\oxref{\oxcprov}{\oxmuta}{\oxplaceexpr}$, introduces an
$\oxmuta$-reference (with region $\oxcprov$) to the location that the place
expression $\oxplaceexpr$ evaluates to. The next form borrows from
$\oxindex{\oxplaceexpr}{\oxexpr}$ instead of simply $\oxplaceexpr$, and is used
to borrow an element out of an array or slice $\oxplaceexpr$ at the index given
by $\oxexpr$. The final form borrows from
$\oxslice{\oxplaceexpr}{\oxexpr_1}{\oxexpr_2}$, and is used to borrow a slice of
$\oxplaceexpr$ using the range given by $\oxexpr_1$ and $\oxexpr_2$. A
\emph{slice} is Rust terminology for a dynamically-sized subsection of an array.

In these last two cases, one might wonder ``why are indexing and slicing not
places themselves?'' The answer comes in two parts: (1)~indexing and slicing
take arbitrary expressions, while places are entirely static, and (2)~unlike
tuple projections which have a fine-grained notion of ownership, indexing and
slicing affect the ownership of the array or slice overall. This second part
means that while you can create two unique references to different projections
of the same tuple, you cannot create two unique references to different indices
of an array.

The remainder of our expressions are standard or discussed already. These
include sequencing, assignment, and creation of tuples and arrays. Our closure
syntax follows the syntax of Rust, and thus uses vertical bars to denote the
closure's parameters. As in Rust, closures are not polymorphic; only global
functions (shown in \Fig{fig:envs}) may be polymorphic and specify where-bounds
on regions.\footnote{In Rust, where-bounds in functions are used to constrain one
  lifetime to outlive another, meaning that a reference with the larger lifetime
  must be valid at least as long as a reference with the shorter lifetime.} We
use function application when applying closures as well as global functions.
Hence, function application additionally includes polymorphic instantiation
written using Rust's turbofish syntax (\rusti{::<>}). An $\oxabort{\oxstring}$
indicates irrecoverable failure; it terminates the program with the given string
as a diagnostic message. Finally, \lang includes tagged sums, which are
introduced using the \texttt{Left} and \texttt{Right} forms and eliminated using
\texttt{match}.\footnote{Rust, of course, supports more general n-ary tagged sums
  with user-definable tags (calling the whole system enumerations), but binary
  sums suffice to get at the essence of Rust without requiring a complicated
  formalization for pattern matching}

\subsection{Types in \lang}
\label{sec:types}

\begin{figure}
  \figuresize
  \begin{bnf}
    \oxkindgrammar \\
    \oxbasetypegrammar \\
  \end{bnf}
  \begin{bnf}
    \oxsitypegrammar \\
  \end{bnf}
  \begin{bnf}
    \oxxitypegrammar \\
    \oxsdtypegrammar \\
    \oxsxtypegrammar \\
    \oxtypegrammar \\
  \end{bnf}

  \vspace{-1em}
  \caption{Type Syntax of \lang}
  \label{fig:type-syntax}
  \vspace{-1em}
\end{figure}

In \lang, we have five distinct categories of types (based on two features we
need to distinguish: sized vs. unsized, and initialized vs. dead), and a kind
system to track the three kinds of polymorphism in the language. While these
distinctions may seem complex, they greatly simplify the well-formedness
conditions required on types during typechecking. The grammars are all present
in \Fig{fig:type-syntax}, and are explained in detail in the rest of the
section.

\paragraph{Sized and Unsized Types} We need to distinguish between types based
on sizedness, which is a direct consequence of Rust itself. All bindings in Rust
(and in \lang) must be able to fit on the stack which requires that they have a
statically-known size. In Rust, this is dealt with using a special
automatically-derived \emph{marker trait} called \rusti{Sized} which serves as a
tag during typechecking to indicate that a type has a statically-knowable size.
For pragmatic reasons (since one typically works with sized types), Rust decided
on using \rusti{Sized?} to indicate that a type is ``possibly unsized'' (and
thus could only be part of a type for a let binding if it is behind a
reference). In \lang, we have a comparable syntactic distinction between
\emph{sized} types $\oxsitype$ and \emph{maybe unsized} types $\oxxitype$. Sized
types characterize all the types with statically-known sizes and maybe unsized
types $\oxxitype$ include all such types \emph{and} the slice type
$\oxtslice{\oxsitype}$ which corresponds to a dynamically-sized portion of an
array.

\paragraph{Initialized and Dead Types} We also need to distinguish between types
based on initialization, which we use to model the 'use-once' linearity of
variables referring to non-copyable data. To that end, we introduce two
categories. First, \emph{dead} types $\oxsdtype$ which is either a sized and
initialized type with a dagger on it (indicating that it is dead) or a product
of dead types. These correspond to totally moved types. Second, we have
\emph{maybe dead} types $\oxsxtype$ which can be either initialized, dead, or a
product of maybe dead types, corresponding to types where some of their
components have been moved. Though not supported directly in our formalism,
these dead and maybe dead types also can be used directly to support
uninitialized and partially-initialized variable bindings.

\paragraph{Kinds and Polymorphism} \lang has three kinds $\oxkind$: the kind of
ordinary types $\oxktype$, the kind of regions $\oxkprov$, and the kind of
\emph{frame typings} $\oxkenv$. (Frame typings are relevant for closures, as
we'll see below.) We abstract over variables of each kind in \lang and, to aid
the reader, we have separate syntax for each: $\oxtvar$, $\oxabsprov$, and
$\oxenvvar$, respectively. For simplicity, \lang restricts type variables
$\oxtvar$ to being instantiated only with sized and initialized types, but this
limitation could be addressed by enriching kinds further with a unique kind for
each sort of type.

\paragraph{The Types Themselves} The majority of types in \lang are sized \&
initialized types, including base types $\oxbasetype$, type variables $\oxtvar$,
tuples $\oxtprod{\oxsitype_1 \oxdotsc \oxsitype_n}$, arrays of length $n$
$\oxtarr{\oxsitype}{\oxnum}$, binary sums $\oxtsum{\oxsitype_1}{\oxsitype_2}$,
references $\oxtref{\oxprov}{\oxmuta}{\oxxitype}$, and function types.
With the exception of references, any types that occur within these types are
themselves required to be both sized and initialized. For reference types
$\oxtref{\oxprov}{\oxmuta}{\oxxitype}$, we include both the region $\oxprov$ and
ownership qualifier $\oxmuta$ in the type which allow us to understand
statically both a reference's origin as well as its aliasing requirements. We
allow potentially unsized types under references since the reference itself will
always have a fixed size regardless of what it points to (e.g. 64-bit on a
64-bit machine).

Function types have three notable features. First, each function type can
possibly include a frame expression $\oxenv$ (syntax in \Fig{fig:envs}) over the
arrow indicating what bindings, if any, were caught up in the closure
environment (when nothing is captured, we put nothing over the arrow). Next,
functions are polymorphic in type and region variables, as well as in frame
variables $\oxenvvar$ to enable the use of higher-order functions. Finally,
functions can relate types with abstract regions using outlives bounds:
\texttt{where} $\oxabsprov_1: \oxabsprov_2$ means $\oxabsprov_1$ outlives
$\oxabsprov_2$. These \texttt{where} bounds come directly from Rust, and are
useful in making functions that, e.g., reborrow from one of several
reference-typed parameters.

\begin{figure}
  \figuresize
  \begin{bnf}
    \oxglobalctxgrammar \\
    \oxglobalentrygrammar \\
  \end{bnf}
  \begin{bnf}
    \oxtvarctxgrammar \\
    \oxkontctxgrammar \\
    \oxvarctxgrammar \\
    \oxframegrammar \\
    \oxenvgrammar \\
  \end{bnf}

  \vspace{-1em}
  \caption{Environments in \lang}
  \label{fig:envs}
  \vspace{-1em}
\end{figure}

\subsection{Environments for Typechecking}
\label{sec:envs}

With the syntax of terms and types in hand, we can look more closely at some
example Oxide programs to understand the environments we'll be using for
typechecking. We'll start with a simple example using reborrowing, much like our
last example in \Sec{sec:main}.

\begin{rustbk}
  struct Obj(u32);
  letrgn<'y, 'z> { // 'y -> {}, 'z -> {}
    let mut x = Obj(5); // x : Obj
    let y = &'y uniq x /* 'y -> { $\oxloanpkg{\oxmut}{x}$ } */;
    let z = &'z uniq *y /* 'z -> { $\oxloanpkg{\oxmut}{x}$, $\oxloanpkg{\oxmut}{\oxderef{\! y}}$ } */;
  }
\end{rustbk}

Here, we create an object named \rusti{x} and in the comment, we see how our
stack typing (written $\oxvarctx$) will record the new binding and its type.
Then, on line 4, we produce a unique reference to \rusti{x}. In the comment, we
see the metadata produced by this borrow associating the region \rusti{'y} with
the set of loans $\oxset{\oxloanpkg{\oxmut}{x}}$. This metadata means that a
reference with the region \rusti{'y} must necessarily point to \rusti{x}. Note
that this metadata is produced immediately after the borrow expression, and
doesn't depend on the binding \rusti{y} being introduced for it. Then, on line
5, we reborrow from \rusti{y} as \rusti{z}, and again, we can see the
corresponding metadata produced associating the region \rusti{'z} with the set
of loans
$\oxset{ \oxloanpkg{\oxmut}{x} \oxcomma \oxloanpkg{\oxmut}{\oxderef{\! y}} }$.
Note that borrowing $\oxderef{y}$ here means reborrowing from \rusti{y}. This
tells us two things: (1)~that a reference with the region \rusti{'z} points to
\rusti{x}, and (2)~that a reference with the region \rusti{'z} was created by
reborrowing from \rusti{y}. That latter means that \rusti{y} ought to be
rendered unusable as long as our reference \rusti{z} (with region \rusti{'z})
exists. The two pieces of information we've seen here, in-scope bindings with
their types and borrowing metadata, are precisely what's necessary for us to
track in our stack typing $\oxvarctx$, and each entry follows the syntax seen
here. In this next example, we'll see a bit more complexity by defining and
using a closure.

\begin{rustbk}
  let x = Obj(5); // x : Obj
  let y = Obj(9); // y : Obj
  letrgn<'z> { // z -> {}
    let f = |obj: &'z uniq Obj| -> Obj { Obj((*obj).0 + y.0) }; // $\oxvarctxentry{y}{\oxtuninit{Obj}}$
    let z = &'z uniq x /* 'z -> { $\oxloanpkg{\oxmut}{x}$ } */;
    f(z) // $\oxvarctxentry{z}{\oxtuninit{(\oxtref{'z}{\oxmut}{Obj})}}$
  }
\end{rustbk}

We create two objects named \rusti{x} and \rusti{y} respectively. Then, on lines
4-6, we define a closure named \rusti{f} that moves \rusti{y} from the context.
This movement is described by the annotation on line 6 that shows the new entry
for \rusti{y} in our stack typing $\oxvarctx$. This new entry differs in that
the type associated with \rusti{y} is marked with a dagger indicating that it is
now dead. On line 7, we create a reference \rusti{z} that we then pass on line 8
as an argument for \rusti{f}. On line 7, we also see the annotation for the
region indicating that a reference with region \rusti{'z} must point to
\rusti{x}, much like we saw in the previous example. In the comment on line 8,
we see that the use of \rusti{z} \emph{moved} it into the function call as well,
thus we mark its whole type with the dagger.

\begin{figure}
  \figuresize
  \begin{flushleft}
    \oxmusafetyform where $\oxmusafetyshape$ means
    $\oxmusafetyinner{\oxtvarctx}{\oxkontctx}{\oxvarctx}{\oxmuta}{\oxemptyctx}{\oxplaceexpr}{ \oxset{\overline{\oxloanpkg{\oxmuta}{\oxplaceexpr^\prime}}} }$.
  \end{flushleft}

  \begin{mathpar}
    \OSafePlace \and \ODeref \and \ODerefAbs
  \end{mathpar}

  \vspace{-1em}
  \caption{Ownership Safety in \lang}
  \label{fig:safety}
  \vspace{-1em}
\end{figure}

One question that might arise here is ``given we marked it dead, how are we
still allowed to use \rusti{y} \emph{inside} the closure?'' This is an important
point that leads to a key aspect of the structure of our stack typing
$\oxvarctx$. The low-level nature of \lang means we need to actually statically
record the frame-based structure of the stack at runtime. So, when we construct
this closure, the moved objects (in this case, solely \rusti{y}) are recorded in
a new frame $\oxframe$ that is tracked in the type of the closure,
$\oxtfun{}{\oxtref{'z}{\;\oxmut\;}{\texttt{Obj}}}{\texttt{Obj}}{\oxframe}$. This
frame provides the type information necessary to typecheck the body in the
future, and the closure at runtime has a corresponding stack frame as part of
its value form. When we typecheck the body of the closure, they are appended to
the current stack typing with $\oxframesep$. So, our stack typing is really a
collection of frames $\oxframe$ separated by $\oxframesep$ and our frames
$\oxframe$ contain both in-scope bindings with their types and in-scope regions
with their associated loan sets. As the language of stacks and frames may imply,
both stack typings $\oxvarctx$ and frames $\oxframe$ are \emph{ordered} which
makes it easier to state invariants like outlives (which we will see in
\Sec{sec:typechecking}) and resolves many of the typical issues that arise
formally with variable binding.

This is all formally-defined in \Fig{fig:envs} which features a grammar for all
of the environments used in the static semantics. As suggested by the grammars,
there are two other environments that are non-conventional. First, we have a
global environment $\oxglobalctx$ which consists of a series of top-level
function definitions. We define the well-formedness of these function
definitions by saying that their bodies must be well-typed assuming that all
other functions (including itself) are well-typed at their annotated type, which
enables a simple treatment of mutually recursive function definitions. Second,
we have a temporary typing $\oxkontctx$ which consists of a sequence of types
for parts of objects that will be in the continuation of the term being
typechecked. The need for this is subtle, but we will explain it using the
following example:

\begin{rustbk}
  letrgn<'a, 'b> {
    let x = Obj(5);
    let y = Obj(9);
    let tup = (&'a uniq x, (); &'b uniq x);
  }
\end{rustbk}

In this example, we create two objects \rusti{x} and \rusti{y} and then attempt
to create a pair named \rusti{tup} that consists of two unique references.
Exactly as written, we first uniquely borrow \rusti{x} with region \rusti{'a}
and then in the second component, we sequence a unit value with a unique borrow
of \rusti{x} with region \rusti{'b}. Of course, the very idea of having a
product of unique references to the same data sounds like a contradiction and so
we would hope to reject this program! However, to capture the expressivity of
Rust's borrow-checking in \lang, we also have to clear loan sets associated with
unused regions at sequencing points in programs. This leads to a dilemma: by
reading the program, we know that the first reference with region \rusti{'a} is
still used when we go to define the second one, but the naive definition of use
would correspond to ``is there currently a reference with that region in the
stack typing?'' The continuation typing comes in to resolve this. We don't have
a name (yet) for the already-typechecked portions of a product, but we can
record their types in the continuation typing to record that they are around
since they may then be let-bound and used further in the program. We'll see
formally how this interaction happens during typechecking in the
\oxname{T-Tuple} rule in \Sec{sec:typechecking}.

Now, we can look at the shape of our typing judgment, which we will return to
define in \Sec{sec:typechecking}. The shape of our typing judgement is
$\oxtypjudgeshape$. This is read: with global environment $\oxglobalctx$, type
environment $\oxtvarctx$, temporary typing $\oxkontctx$, and stack typing
$\oxvarctx$, $\oxexpr$ is a well-typed expression of type $\oxtype$ with an
updated stack typing $\oxvarctx^\prime$ for use in typechecking the continuation
of $\oxexpr$.

\subsection{Region-Based Alias Management}
\label{sec:regions}

As discussed in \Sec{sec:terms}, borrow expressions in \lang all have
\emph{region} annotations which are used to associate references statically with
information about their possible referents. This information is essential to our
formulation of borrow checking since we leverage it to determine if new
references would be safe to create. This is done formally with a judgement
called \emph{ownership safety} which has the form $\oxmusafetyshape$. We can
read it as saying ``in the environments $\oxtvarctx$ and $\oxvarctx$, it is safe
to use the place expression $\oxplaceexpr$ $\oxmuta$-ly and producing all of the
loans in $\oxset{\overline{\oxloanpkg{\oxmuta}{\oxplaceexpr^\prime}}}$ (called
the borrow chain).,'' where $\oxmuta$ is either $\oxmut$ or $\oximm$. That is,
if we have a derivation where $\oxmuta$ is $\oxmut$, we know that we can use the
place expression $\oxplaceexpr$ uniquely because we have a proof that there are
no live loans against the section(s) of memory that $\oxplaceexpr$ represents.
Further, when we have a derivation where $\oxmuta$ is $\oximm$, we know that we
can use the place expression $\oxplaceexpr$ sharedly because we have a proof
that there are no live \emph{unique} loans against the section(s) of memory that
$\oxplaceexpr$ represents. The produced borrow chain is used to create the
borrowing metadata to store in the environment when the ownership safety check
is guarding a borrow expression, and in the simplest case, is just precisely
$\oxplaceexpr$.

Since it is precisely this ownership safety judgment that captures the essence
of Rust's ownership semantics, we understand Rust's borrow checking system as
ultimately being a system for statically building a proof that data in memory is
either \emph{uniquely owned} (and thus able to allow unguarded mutation) or
\emph{collectively shared}, but not both. To do so, intuitively, ownership
safety looks at all of the concrete regions in $\oxvarctx$, and ensures that all
the loans they map to are not in conflict with the place expression
$\oxplaceexpr$ we are attempting to use. For a $\oxmut$ borrow, a conflict
occurs if \emph{any} loan maps to an overlapping place, but for a $\oximm$
borrow, a conflict occurs only when a $\oxmut$ loan maps to an overlapping
place. Since places are abstract memory locations, we can consider two places as
overlapping if they are equal or one is a prefix of the other (meaning that the
the longer place corresponds to a piece of the larger object in memory that the
shorter place refers to).

Unfortunately, reborrowing complicates matters. To support reborrowing,
ownership safety uses an expanded inner form written $\oxmusafetyinnershape$,
which says that $\oxplaceexpr$ is $\oxmuta$-safe under $\oxtvarctx$ and
$\oxvarctx$, with \emph{reborrow exclusion list} $\overline{\oxplace}$, and
produces the loans
$\oxset{\overline{\oxloanpkg{\oxmuta}{\oxplaceexpr^\prime}}}$. Intuitively, we
use this reborrow exclusion list $\overline{\oxplace}$ to rule out precisely the
loan conflicts that arise from reborrowing as a programming pattern --- namely,
a reference conficting with loans from a reference it was reborrowed from.
Reborrowing is also what causes the borrow chain to contain more than just
$\oxplaceexpr$ for two reasons: (1)~we may be reborrowing from a reference whose
region has already lost precision, i.e.\ contains multiple loans, and thus we
cannot have perfect information about the reborrowed reference either, and
(2)~the reborrowed borrow chain will include an additional loan that records
that it was reborrowed (e.g. a unique loan for a reborrow from \rusti{x} would
appear as $\oxloanpkg{\oxmut}{\oxderef{\! x}}$). This second point is precisely
what produces the \emph{chain} aspect of the borrow chain since we can use this
information in each loan set to follow a series of consecutive reborrows.

Formally, the first rule, \oxname{O-SafePlace}, checks if a place $\oxplace$ is
$\oxmuta$-safe by looking at each loan in every region $\oxcprov^\prime$ in
$\oxvarctx$ and either (1) making sure that if either that loan or $\oxmuta$ is
$\oxmut$ then $\oxplace$ does not overlap with the loan; or (2) checking that
all references in $\oxvarctx$ with region $\oxcprov^{\prime}$ are in the
reborrow exclusion list (meaning we need not check if there is overlap with
$\oxplace$).

The next two rules check if a place expression $\oxplaceexpr$ is $\oxmuta$-safe,
decomposing the place expression into a place expression context
$\oxplaceexprctx$ (see \Fig{fig:term-syntax}) with $\oxderef{\oxplace}$ in the
hole. The last two lines of premises for both essentially ensure that either (1)
or (2) holds, but each one adds to the incoming reborrow exclusion list when
checking (2) by also including the place itself $\oxplace$ along with all of the
places $\oxplace_j$ that $\oxplace$ was borrowed from according to the loan set
$\oxcprov$ associated with its type. Both rules also check
$\oxmuta \oxsubtype \oxmuta_{\oxplace}$ (defined as the reflexive closure of
$\oximm \oxsubtype \oxmut$) in order to ensure that the reference has sufficient
permission to be used, preventing a dereference of a $\oxmut$ reference in a
$\oximm$ context.

Unlike \oxname{O-DerefAbs}, \oxname{O-Deref} is dereferencing a reference
$\oxplace$ with a concrete region $\oxcprov$. As such, we can look at the loans
present for $\oxcprov$ in the stack typing. These loans consist of both direct
loans to places $\oxplace_j$ which correspond to a possible origin for the
reference, and indirect loans to place expressions $\oxplaceexpr_j$ which
capture how this reference was reborrowed from other references. As such, when
we recursively check for the safety of these regions, we append the reborrow
origins (the $\oxplace_j$ prefixes of these $\oxplaceexpr_j$) to the reborrow
exclusion list. This means that they will not be considered as possible
conflicts in the rest of ownership safety. At the end, we union together the
borrow chains from all the possible origins to determine our final borrow chain.
We also include an additional loan
$\oxloanpkg{\oxmuta}{ \oxplaceexprctx[\oxderef{\oxplace}] }$ to indicate that
this use was reborrowed from $\oxderef{\oxplace}$.

\subsection{Typechecking \lang Programs}
\label{sec:typechecking}

\begin{figure*}
  \figuresize{ \vspace{-0.4em}
    \begin{flushleft}\oxtypjudgeform\end{flushleft}
    \begin{mathpar}
      \TuThreeTwo \and \TMove \and \TCopy \and \TBorrow \and \TLetProv \and
      \TBranch \qquad \TSeq \and \TLet \and \TDrop \and \TAssignDeref \and
      \TAssign \and \TClosure \and \TTuple \and \TWhile \and \TAbort
    \end{mathpar}
  }

  \caption{Selected \lang Typing Rules}
  \label{fig:typing}
\end{figure*}

\Fig{fig:typing} presents a selection of \lang typing rules. In every rule, we
highlight the expression being typechecked with a $\fbox{framebox}$. The shape
of our typing judgement is $\oxtypjudgeshape$: we typecheck $\oxexpr$ in a
global environment $\oxglobalctx$, type environment $\oxtvarctx$, temporary
typing $\oxkontctx$, and stack typing $\oxvarctx$, producing an updated stack
typing $\oxvarctx^\prime$ for typing the continuation of $\oxexpr$. These rules
rely on the region rewriting and outlives judgments (\Fig{fig:unification}),
which we'll discuss below, and the ownership safety judgment (\Fig{fig:safety}),
which we discussed in \Sec{sec:regions}. We elide the various well-formedness
judgments (for types, stack typings, etc.); \ifanonymous{see supplementary
  material.}{ see the appendix (\Sec{sec:well-formedness}). }

To best understand our typing judgment as a whole, it is useful to first know a
bit about what lies ahead in the metatheory (\Sec{sec:metatheory}). In our type
preservation proof, we need to maintain the well-typedness of values stored on
the stack in \lang. Since our values include closures which themselves may
introduce more aliasing, we then need to maintain our ownership safety judgment.
To make this possible, there are a number of restrictions that arise throughout
the type system on how regions annotate the program (and discussed further in
\Sec{sec:discussion}). We urge readers to keep this in mind.

\paragraph{Moving} In Oxide, as in Rust, owned values are moved or copied out of
a place $\oxplace$ when used, just as we saw in our first example in
\Sec{sec:ownership} where \rusti{pt} was moved to \rusti{x} and thus could not
be bound again to \rusti{y}. In order to move $\oxplace$, three conditions must
hold: (1)~$\oxplace$ must be able to be used $\oxmut$-ly (checked using the
ownership safety judgment in \Fig{fig:safety} from \Sec{sec:regions});
(2)~$\oxplace$ must have a sized and initialized (not dead) type $\oxsitype$ in
$\oxvarctx$; and (3)~this type $\oxsitype$ must be
$\oxkey{noncopyable}$.\footnote{We've elided definitions of $\oxkey{copyable}$
  and $\oxkey{noncopyable}$, but they're straightforward. Intuitively, a type is
  safe to copy if none of its constituent parts are unique. Thus, all types that
  don't contain a unique reference are copyable. Generic types are always
  non-copyable. In Rust, \oxkey{copyable} is actually the \rusti{Copy} trait,
  but \oxkey{copyable} can be thought of as special casing it.} If these
requirements hold, then we use a dagger to mark the place $\oxplace$ dead in the
continuation stack typing, preventing further expressions from reusing it.
Requirement~(1) is needed to ensure that we do not invalidate any existing
references to $\oxplace$ by moving it and requirement~(2) ensures that there is
currently data owned by $\oxplace$. If requirement~(3) does not hold, we'll copy
rather than move which permits more programs to typecheck since \oxname{T-Copy}
does not mark the copied place dead in the continuation stack typing as
\oxname{T-Move} does. Further, unlike moves (which are disallowed through
dereferences, leading to the restriction to places $\oxplace$ rather than place
expressions $\oxplaceexpr$ in \oxname{T-Move}), the copying variant
\oxname{T-Copy} can happen through a dereference and thus handles place
expressions generally.

\paragraph{Borrowing} As with moving, borrowing
$\oxref{\oxcprov}{\oxmuta}{\oxplaceexpr}$ relies on our ownership safety
judgment (\Fig{fig:safety}) with the $\oxmut$ and $\oximm$ modalities
corresponding precisely to the invariants of unique and shared pointers. Namely,
when $\oxmuta$ is $\oxmut$, we require that the place expression $\oxplaceexpr$
have no extant loans in $\oxvarctx$ and when $\oxmuta$ is $\oximm$, we require
no extant \emph{unique} loans. To actually know the type of the reference as a
whole, we also have to know the type of the place expression itself and we rely
on an auxillary judgment
$\oxcomputetynoprov{\oxtvarctx}{\oxvarctx}{\oxmuta}{\oxplaceexpr}{\oxxitype}$
(defined in the appendix) to compute the type $\oxxitype$ by starting with the
type of its root identifier and following the sequence of projections and
dereferences from $\oxplaceexpr$ through the type. For example, if we had the
place expression \rusti{*(x.0)} where x is a product of references with the type
\rusti{&'a uniq u32}, this judgment would produce the type \rusti{u32}. Much
like how \oxname{T-Move} updates the continuation stack typing to prevent
further uses of the moved place, \oxname{T-Borrow} updates the continuation
stack typing to associate the region $\oxcprov$ used for the borrow with the
loans that represent where the new reference may point (i.e.\ its provenance).
In many simple cases (such as borrowing a binding \rusti{x} with type
\rusti{u32}), there will be only one loan (corresponding to precise knowledge of
where it points), but as we will see with branching, these loan sets may grow
larger to account for information loss inherent to static analysis of reference
provenance.

\paragraph{Region Rewriting and Outlives} We examine region rewriting next
(\Fig{fig:unification}) since some of the typing rules discussed below require
it. The region rewriting judgment
$\oxtunify{\oxtvarctx}{\oxkontctx}{\oxvarctx}{\oxtype_1}{\oxtype_2}{\oxvarctx^\prime}$
says under $\oxtvarctx$, $\oxkontctx$, and $\oxvarctx$, we can rewrite the
regions from $\oxtype_1$ into the corresponding regions in $\oxtype_2$ which
will then be interpreted under $\oxvarctx^\prime$. We produce an output
$\oxvarctx^\prime$ with updated regions to be used when typing the continuation
after an appeal to region rewriting. The need for this judgment arises from the
need for values to be given the same type, in spite of the fact that they can be
annotated differently originally. Consider, for example, a branching term such
as \rusti{if cond { &'a uniq x } else { &'b uniq y }}. In this case, the type of
each side of the branch will be something like \rusti{&'a uniq u32} and
\rusti{&'b uniq u32} respectively, but we need to give an overall type for the
term. To deal with this, we pick one of the two types and use rewriting to write
the other one into the chosen type. For safety reasons, we pick the region with
the \emph{shortest} scope.

The rewriting judgment itself is fairly straightforward: it is reflexive and
transitive. Each rule for larger types recursively rewrites in any smaller
types, threading the output environment much like our typing judgment. The
actual rewriting portion arises, as one might expect, in the rule for references
(\oxname{RR-Reference}) which appeals to a judgment on regions which says that
the region from $\oxtype_1$ \emph{outlives} the region from $\oxtype_2$ while
performing the work required to enable the rewriting.

The outlives judgment (\Fig{fig:unification})
$\oxrunify{\oxtvarctx}{\oxvarctx}{\oxprov_1}{\oxprov_2}{\oxvarctx^\prime}$ says
under $\oxtvarctx$ and $\oxvarctx$, $\oxprov_1$ outlives $\oxprov_2$ rewriting
the latter in $\oxvarctx^\prime$ according to the mode $\oxrewritemode$. Every
region outlives itself (reflexivity). An abstract region outlives another if
there's a corresponding outlives relation in $\oxtvarctx$
(\oxname{OL-BothAbstract}) or if we can transitively put together outlives
relations from $\oxtvarctx$ (\oxname{OL-Trans}).\footnote{We do not need
  transitivity for concrete regions beyond what we can already conclude from the
  remaining \oxname{OL} rules.} \oxname{OL-CombineConcrete} (in $\oxcombine$
mode) says that $\oxcprov_1$ outlives $\oxcprov_2$ if it occurs earlier than
$\oxcprov_2$ in $\oxvarctx$. It also requires that there not exist any
references with either region which have been reborrowed
($\oxnotreborrowed{\oxvarctx}{\oxcprov_1}$ and
$\oxnotreborrowed{\oxvarctx}{\oxcprov_2}$, where \oxkey{rnrb} is an abbreviation
for region-not-reborrowed). This invariant ensures that value typing is
preserved under region rewriting. When this is the case, the output typing
$\oxvarctx^\prime$ is updated to associate $\oxcprov_2$ with the union of the
loans from both $\oxcprov_1$ and $\oxcprov_2$. The variant
\oxname{OL-CheckConcrete} (in $\oxnoop$ mode) has the same obligations, but
instead leaves the output unchanged. The behavioral difference between these
rules is the reason for the rewriting modes.

The last two rules say when a concrete region outlives an abstract one and vice
versa. In essence, a concrete region $\oxcprov$ can only outlive an abstract
region $\oxabsprov$ (\oxname{OL-ConcreteAbstract}) if $\oxcprov$ was
\emph{reborrowed}. The first two premises check for reborrowing: $\oxcprov$'s
loan set must be non-empty (otherwise there is no reborrow), and must consist
solely of place expressions $\overline{\oxplaceexpr}$ (since place expressions,
unlike places, contain dereferences, which identifies this as a reborrow instead
of a borrow). The third premise collects all the regions $\overline{\oxprov_i}$
that annotate any references dereferenced in each place expression
$\oxplaceexpr_i$ (see the type-computation judgment $\oxcomputetyshape$ in the
\ifanonymous{supplementary material}{appendix
  (\Sec{sec:additional-judgments})}), while the last premise ensures that all of
these outlive $\oxabsprov$. The final rule, \oxname{OL-AbstractConcrete}, says
that an abstract region \emph{always} outlives a concrete region. This is subtle
but makes sense because any abstract region $\oxabsprov$ is bound in a top-level
function (recall that closures don't abstract over regions), while a concrete
region $\oxcprov$ must be bound by $\oxkey{letrgn}$s \emph{inside} the function
body. Ultimately, any concrete region $\oxcprov^\prime$ that gets substituted
for $\oxabsprov$ upon application will already exist before $\oxcprov$ (even for
recursive calls), meaning it outlives $\oxcprov$.

\begin{figure*}
  \figuresize
  \begin{bnf}
    \oxrewritemodegrammar
  \end{bnf}


  \begin{flushleft}\oxtunifyform\end{flushleft}

  \begin{mathpar}
    \URefl \and \UTrans \and \UUninit \and \UTuple \and \USharedRef
  \end{mathpar}

  \vspace{1ex}
  \begin{flushleft}\oxrunifyform\end{flushleft}
  \vspace{-1ex}
  \begin{mathpar}
    \UReflProv \and \UTransProv \and \UAbsProvs \and \UCombineLocalProvs \and
    \UCombineLocalProvsUnrest \and \UCheckLocalProvs \and \OLocalAbsProvs \and
    \OAbsLocalProvs
  \end{mathpar}

  \vspace{-1em}
  \caption{Region Rewriting and Outlives Relations in \lang}
  \label{fig:unification}
  \vspace{-1em}
\end{figure*}

\paragraph{Branching and Sequencing} The next two rules illustrate how stack
typings are threaded through larger programs since the form of our typing
judgment requires each rule to specify its continuation's stack typing.
\oxname{T-Branch} uses the stack typing $\oxvarctx_1$ that we get from typing
the conditional $\oxexpr_1$ when typing each of the two branches. The type
$\oxsitype$ ascribed to the overall expression must be a supertype of the types
$\oxsitype_2$ and $\oxsitype_3$ of each branch and equal to one of them.
Additionally, branching uses a union operation $\oxintersect$ to combine the
output stack typings from each branch to produce the final stack typing
$\oxvarctx^{\prime}$ for the overall expression. $\oxintersect$ requires that
types of bound variables in the two stack typings be equal (which potentially
demands use of \oxname{T-Drop} when typing the branches), and unions the loan
sets for each region $\oxcprov$ from both stack typings (full definition in
technical appendix). Note that we only need to union stack typings with
identical domains --- we typecheck both branches under $\oxvarctx_1$ so they
produce output stack typings with the same domains (since $\oxkey{let}$ and
$\oxkey{letrgn}$ are the only means for introducing variables and regions, but
both are lexically-scoped), and region rewriting does not change the domain of
stack typings between its input and its output.

When typing $\oxseq{\oxexpr_1}{\oxexpr_2}$, we typecheck $\oxexpr_2$ under the
stack typing $\oxvarctx_1$ we got from typechecking $\oxexpr_1$. But,
importantly, we apply a metafunction $\oxgcloans{\oxkontctx}{\cdot}$ to
$\oxvarctx_1$ to empty out the loan sets of regions not used in $\oxvarctx_1$
before typing $\oxexpr_2$ because $\oxexpr_1$ may have been a unique reference
that is thrown away at runtime before moving on to $\oxexpr_2$. Without
\emph{garbage collecting} loans, \lang would reject programs that are safe and
accepted by Rust. Namely, this clearing allows us to handle the sort of ``early
dropping of references'' inherent to non-lexical lifetimes. Specifically,
$\oxgcloans{\oxkontctx}{\oxvarctx}$ empties out the loan set of each $\oxcprov$
that does not appear in any of the types in $\oxvarctx$ or $\oxkontctx$. The
full formal definition is present in the technical appendix.

\paragraph{Binding} In \lang, \oxname{T-Let} is interesting in three ways.
First, we allow rewriting in \oxname{T-Let} to a specified annotated type. This
rewriting allows us to change the regions in the computed type to match the
annotated type by conservatively combining the loans associated with each region
into the output, as described earlier in the section on region rewriting. Then,
similar to sequencing, \oxname{T-Let} uses the metafunction
$\oxgcloans{\oxkontctx}{\cdot}$ to eliminate any loans that might be unnecessary
as a result of $\oxexpr_1$ potentially being promoted to the annotated type
$\oxsitype_a$. Additionally, in the output stack typing from $\oxexpr_2$, we see
that our binding for $\oxid$ must have a dead type $\oxsdtype$ with the whole
binding being dropped in the overall stack typing $\oxvarctx_2$ output from
\oxname{T-Let} (since the scope of $\oxid$ ends at that point). The requirement
that the type be dead means we must have either used \oxname{T-Move} to move out
of that binding or we must have explicitly used \oxname{T-Drop} on $\oxid$ in
the derivation for $\oxexpr_2$, and can be thought of as a formalization of the
``resource acquisition is initialization'' pattern~\cite{stroustrup94:raii}
since we are explicitly requiring a first-in-last-out allocation/deallocation
pattern and require everything to have been used in either a move or a drop
before we can end its scope.

\paragraph{Drop} As alluded to in the previous two paragraphs, \lang has a rule
called \oxname{T-Drop} which is used non-deterministically during typechecking
to mark a particular place $\oxplace$ as being dead. This rule corresponds
roughly to a conventional weakening rule where $\oxplace$ ``doesn't exist'' (in
this case, is dead) in the premise, but exists in the conclusion. The main
difference is that while the data is thought to be deallocated, the name is
still in-scope to be dropped when its scope ends in \oxname{T-Let}.

\paragraph{Assignment} Assignment is interesting in a few ways. First,
assignment is broken up into two rules \oxname{T-Assign} and
\oxname{T-AssignDeref} where the former is able to assign to a place $\oxplace$
that is dead, and the latter is able to assign to a place through a reference
(i.e. by using dereferencing). The basic structure of each rule is the same. For
both rules, we typecheck the new expression to be assigned, look up the type of
the place we're assigning to (a lookup in \oxname{T-Assign} and a type
computation in \oxname{T-AssignDeref}), check compatibility of the new
expression's type with the type of the place we're assigning to, and then
finally check that it's safe to use the place we're assigning to uniquely
according to ownership safety.

The differences between the two play a fundamental role in allowing us to
appropriately model how assignment works in Rust. Notably, the region rewriting
judgment in \oxname{T-Assign} uses the checking mode (denoted $\oxnoop$) to
limit how conservative the borrow checker need be after an assignment. As
discussed earlier, this mode does not change its output environment (thus,
$\oxvarctx^\prime = \oxrsub{\oxvarctx_1}{\oxderef{\oxplace}}$ in
\oxname{T-Assign}). This is okay in context because after the typing rule is
done, the type of $\oxplace$ is updated to $\oxsitype$ (the type of its new
value) in the continuation. A similar update in \oxname{T-AssignDeref} would
entail updating the types of arbitrarily many bindings, and so instead the more
conservative combine mode (denoted $\oxcombine$) is used. Further, in
\oxname{T-Assign}, we employ the operation
$\oxrsub{\oxvarctx_1}{\oxderef{\oxplace}}$ for the input environment to the
region rewriting judgment. This operation is defined to remove any loans
prefixed by $\oxderef{\oxplace}$ from every loan set in $\oxvarctx_1$. The
$\rsub$ operation (informally called the ``kill rules'' in Rust) amounts to
erasing reborrowing relationships that no longer hold as a result of this
assignment.

Concretely, consider an environment with two references, one named \rusti{x}
with type \rusti{&'x uniq u32} and another named \rusti{y} reborrowed from
\rusti{x} with type \rusti{&'y uniq u32}. This means in our environment we would
have loan sets that look something like
$\oxloanctxentry{\rusti{'x}}{\oxset{\overline{\oxloan}}}$ and
$\oxloanctxentry{\rusti{'y}}{ \oxset{\overline{\oxloan} \oxcomma \oxloanpkg{\oxmut}{\oxderef{\!\oxid}}} }$.
If we were to then assign to \rusti{x} some unrelated reference, the kill rules
would delete the loan $\oxderef{\oxid}$ in the loan set of \rusti{'y} since
after the assignment ran, the two would represent distinct and disjoint
references.

The last difference between the two assignment rules is presence of an
additional obligation in \oxname{T-Assign}:
$\oxrgnuniqueto{\oxsxtype}{\oxplace}{\oxvarctx_1}$. This obligation means that
there are not any places in $\oxvarctx_1$ that share the outermost region of
$\oxsxtype$ (i.e. if $\oxsxtype = \oxtref{\oxcprov}{\oxmuta}{\oxxitype}$, then
$\oxcprov$ would be its outermost region). This allows us to guarantee that the
garbage collection discussed for the sequencing rule will always clear out this
outermost region $\oxcprov$ before the subsequent expression is typechecked,
helping us greatly in our proofs. It's important to note that this obligation
pertains to annotations added to the \lang program compared to Rust, and so it
only limits the patterns of region annotation that can be applied, rather than
the space of Rust programs that can be typechecked in \lang.

\begin{figure*}
  \figuresize
  \begin{mathpar}
    \TAppClosure
  \end{mathpar}

  \vspace{-1em}
  \caption{\lang Typing Rule for Application}
  \label{fig:typing-app}
  \vspace{-1em}
\end{figure*}

\paragraph{Closures and Application}
Closures in \lang correspond to \emph{move closures} in Rust which move or copy
their free variables from the outer environment into the
closure.\footnote{Rust's standard closures implicitly introduce borrowed
  temporaries for all the free variables. We can recover this behavior via a
  simple, local transformation to move closures.} As such, \oxname{T-Closure}
must compute the captured frame by looking at the free variables (and free
regions) of the closure's body, and it must mark dead (add daggers to the types
of) any variables in the stack typing with \emph{non-copyable} types. The
captured frame is suspended over the arrow in the function type to keep track of
the fact that the data caught up in the closure is still alive (and thus must be
considered in ownership safety). We elide the rule for top-level function
definitions, which gives a function $\oxfnname$ the type that $\oxfnname$ is
annotated with in $\oxglobalctx$, relying on well-formedness of $\oxglobalctx$
to know that this is okay.

The rule for application (\oxname{T-AppClosure} in \Fig{fig:typing-app}) is
roughly as one would expect: we typecheck the function, and then the arguments,
threading through the environments. However, at each step, we do a region
rewriting using the unrestricted combine mode (denoted $\oxcombineunrest$) for
the computed argument types to the annotated parameter types. This unrestricted
mode allows us to push loans from the surrounding context into the regions used
in the closure's signature, a behavior ruled out by the conventional combine
mode ($\oxcombine$) because they would enable degenerate annotation patterns
that make it difficult to prove soundness. For top-level functions, we have an
additional rule \oxname{T-AppFunction} in the technical appendix that does not
use a rewriting for the types, but (1)~substitutes all frame, type, and region
variables in the types and (2)~checks that any outlives bounds specified on the
function signature hold (via outlives in \Fig{fig:unification}). This does not
appear in \oxname{T-AppClosure} since closures cannot be polymorphic, nor can
they possess where bounds.

\paragraph{Values and Aggregates} The typing rules for base types
(\oxname{T-u32}, \oxname{T-True}, \oxname{T-False}, etc.) are standard, and
leave the type environment unchanged in their output. Aggregate structures like
tuples check the types of their components while threading through the
environments in left-to-right order. This left-to-right ordering for
typechecking corresponds to the ordering implemented by Rust's typechecker and
borrow checker. One subtlety to note (discussed already with the last example in
\Sec{sec:envs}) is that when typechecking a component $e_i$ of the tuple, we add
the types of all earlier tuple components to the temporary typing $\oxkontctx$.
This is needed because $e_i$ might be a let-binding or sequencing expression
that invokes $\oxgcloans{\oxkontctx}{\cdot}$ on its environment during
typechecking. If the types of the earlier tuple components we've just
typechecked aren't present in the environment, to serve as roots for the
regions mentioned in those types, we may end up incorrectly garbage collecting
the loans that these regions map to. This would make programs with subsequent
borrows typecheck even when the borrow should not be allowed. The elided typing
rule for arrays is similar.

The formalism of \lang omits a specific treatment of structs, but we note that
they are essentially the same as tuples, only featuring a tag that must also be
checked. Our implementation which we discuss in \Sec{sec:tested-semantics}
relies on exactly this approach to support structs.

\paragraph{Remaining Rules} The remaining rules in \Fig{fig:typing} are
straightforward or covered earlier. Elided typing rules all concern arrays and
are given in the \ifanonymous{supplementary material}{ technical appendix
  (\Sec{sec:typing}) }.

\subsection{Operational Semantics}
\label{sec:dynamics}

\begin{figure}
  \figuresize
  \begin{bnf}
    \oxreferentgrammar \\[0em]
    \oxptrexprgrammar \\[0em]
    \oxevalctxgrammar \\[1em]
    \oxvaluegrammar \oxptrextension \\[0em]
    \oxvaluectxgrammar \\[0em]
    \oxstoregrammar \\[0em]
    \oxstackframegrammar \\
  \end{bnf}

  \vspace{-1em}
  \caption{\lang Syntax Extensions for Dynamics}
  \label{fig:dynamics-syntax}
  \vspace{-1em}
\end{figure}

\begin{figure*}
  \figuresize
  \begin{flushleft}\oxnormform \hspace{2em}
    $\oxnorm{\oxstore}{\oxplaceexprctx[\oxid]}
    {\oxreferent}{\oxvaluectx}{\oxvalue} ~~\defeq~~
    \oxnorminner{\oxstore}{\oxplaceexprctx}{\oxid}
    {\oxreferent}{\oxvaluectx}{\oxvalue}$.
  \end{flushleft}

  \begin{flushleft}\oxnorminnerform \hspace{1em}
    read: ``$\oxreferent$ in a context $\oxplaceexprctx$ computes to
    $\oxreferent^\prime$ which maps to $\oxvalue$ in
    $\oxstore$.''\end{flushleft}
  \begin{mathpar}
    \PId \qquad \PProj \qquad \PDerefPtr
  \end{mathpar}

  \vspace{0.5em}

  \begin{flushleft}\oxreduceform\end{flushleft}

  \begin{mathpar}
    \EMove \and \ECopy \and \EBorrow \and \ESeq \quad \ELetProv \quad \EAssign
    \and \ELet \and \EShift \and \EClosure \and \EApp \and \EFramed \and \EWhile
  \end{mathpar}

  \vspace{-1em}
  \caption{Selected Place Expression Evaluation Rules (top) and Reduction Rules
    (bottom)}
  \label{fig:dynamics}
  \vspace{-1em}
\end{figure*}

For our operational semantics, we extend the syntax of \lang in
\Fig{fig:dynamics-syntax} with terms that only arise at runtime. First, to be
able to specify what ``address'' a pointer points to, we introduce an abstract
form of memory addresses called \emph{referents}. Referents $\oxreferent$
essentially record what the offsets are from a variable on the stack in order to
specify a precise ``memory address,'' (e.g., a particular element of an array or
tuple, or a particular slice of an array). We also include some administrative
forms: (1)~$\oxframed{\oxexpr}$ and $\oxshift{\oxexpr}$ which are used when
evaluating application and let bindings discussed below, (2)~$\oxuninit$ (the
dead value), and (3)~$\oxsliceval{\oxvalue_1 \oxdotsc \oxvalue_n}$ which is a
dynamically-sized slice of an array. Then, we introduce value forms including
pointers to referents, and closures packaged with their environment
$\oxstackframe$. \Fig{fig:dynamics-syntax} also includes stacks $\oxstore$ as
a sequence of stack frames $\oxstackframe$, and value contexts $\oxvaluectx$
which allow array values to be decomposed with multiple holes when dealing with
slices.

In \Fig{fig:dynamics}, we present a selection of our small-step operational
semantics which is defined using Felleisen and Hieb-style left-to-right
evaluation contexts~\cite{felleisen92:eval-contexts} over configurations of the
form $\oxconfigurationshape$. Since our semantics uses referents $\oxreferent$
as an abstract version of memory addresses, some of our rules rely on a notion
of place-expression evaluation,
$\oxnorm{\oxstore}{\oxplaceexpr}{\oxreferent}{\oxvaluectx}{\oxvalue}$
(\Fig{fig:dynamics}, top), which should be read as: $\oxplaceexpr$ evaluates to
$\oxreferent$, which maps to $\oxvalue$ with a surrounding context $\oxvaluectx$
in $\oxstore$.

The evaluation rules are straightforward: \oxname{E-Move} returns a value by
moving it off of the stack $\oxstore$, replacing it with $\oxuninit$.
\oxname{E-Copy} copies the value from the stack. \oxname{E-Borrow} creates a
pointer value to the referent $\oxreferent$. Branching is completely standard,
hence elided. Assignment, similar to \oxname{E-Copy} and \oxname{E-Borrow}, uses
the place-expression evaluation rules, but instead cares specifically about the
value context $\oxvaluectx$, rather than the value $\oxvalue$. Assignment also
decomposes the computed referent $\oxreferent$ to get its root identifier
$\oxid$. Then, it updates the stack by maintaining this value context when it
updates $\oxid$ (mapping it to $\oxvaluectx[\oxvalue]$).

\paragraph{Binding and the Stack} Bindings are interesting in that they
introduce our two administrative forms, $\oxframed{\oxexpr}$ and
$\oxshift{\oxexpr}$. For instance, in \oxname{E-Let}, we step to
$\oxshift{\oxexpr}$ rather than $\oxexpr$ alone in order to ensure that the
binding for $\oxid$ is well-scoped and ends when it should (seen in
\oxname{E-Shift}). In \oxname{E-AppClosure}, we similarly step to
$\oxframed{\oxexpr}$ to ensure that after evaluating the body of the closure we
drop the stack frame from that function call (seen in \oxname{E-Framed}). Both
\oxname{E-Shift} and \oxname{E-Framed} rely crucially on the fact that our stack
$\oxstore$ is ordered --- they must match the most recent entry.

\subsection{Well-typed \lang programs won't go wrong!}
\label{sec:metatheory}

We prove syntactic type safety for \lang using progress and
preservation~\cite{wright92:progress-preservation}.

\begin{oxlemma}{Progress}{lemmap:progress}
  If \oxtypjudge{\oxglobalctx}{\oxemptyctx}{\oxkontctx}{\oxvarctx}{\oxexpr}
  {\oxsitype}{\oxvarctx^\prime} and
  \oxstorevalidity{\oxglobalctx}{\oxvarctx}{\oxstore}, then either $\oxexpr$ is
  a value, $\oxexpr$ is an \oxabort{\,\dots\,}, or
  $\oxexists \oxstore^\prime, \oxexpr^\prime. \;
  \oxreduce{\oxglobalctx}{\oxstore}{\oxexpr}{\oxstore^\prime}{\oxexpr^\prime}$.
\end{oxlemma}

The Progress lemma says that if we can typecheck $\oxexpr$ under a valid global
environment $\oxglobalctx$, temporary typing $\oxkontctx$, and stack typing
$\oxvarctx$, \emph{and} we have a stack $\oxstore$ that satisfies this stack
typing $\oxvarctx$, then either $\oxexpr$ is a value, an \rusti{abort!}
expression, or we can take a step. Our stack typing judgment
$\oxstorevalidity{\oxglobalctx}{\oxvarctx}{\oxstore}$ says that each value in
the stack $\oxstore$ has the corresponding type attributed to it in the typing
$\oxvarctx$. The proof proceeds by induction on the typing derivation for
$\oxexpr$, and relies on a Canonical Forms lemma and
\Lemma{lemmap:place-exprs-reduce} which says that place expressions can be
reduced at runtime to values with their computed types. This lets us apply the
rules for moves, copies, borrowing, and assignment.

\begin{oxlemma}{Place Expressions Reduce}{lemmap:place-exprs-reduce}
  If $\oxcomputetynoprov{\oxtvarctx}{\oxvarctx}{\oxmuta}
  {\oxplaceexpr}{\oxxitype}$ and
  $\oxstorevalidity{\oxglobalctx}{\oxvarctx}{\oxstore}$, then
  $\oxnorm{\oxstore}{\oxplaceexpr}{\oxreferent}{\oxvaluectx}{\oxvalue}$ and
  $\oxtypjudge{\oxglobalctx}{\oxtvarctx}{\oxkontctx}{\oxvarctx}
  {\oxvalue}{\oxxitype}{\oxvarctx}$.
\end{oxlemma}

Our proof of \Lemma{lemmap:place-exprs-reduce} (ifanonymous{Lemma 5.3}{\Lemma{lemma:place-exprs-reduce}} in appendix) relies on
the shared inductive structure of type computation
$\oxcomputetynoprov{\oxtvarctx}{\oxvarctx}{\oxmuta}{\oxplaceexpr}{\oxxitype}$
and place expression evaluation
$\oxnorm{\oxstore}{\oxplaceexpr}{\oxreferent}{\oxvaluectx}{\oxvalue}$.

\begin{oxlemma}{Preservation}{lemmap:preservation}
  If \oxtypjudge{\oxglobalctx}{\oxemptyctx}{\oxkontctx}{\oxvarctx}{\oxexpr}
  {\oxsitype_1}{\oxvarctx_f} and
  \oxstorevalidity{\oxglobalctx}{\oxvarctx}{\oxstore} and
  $\oxwfkontctx{\oxglobalctx}{\oxvarctx}{\overline{\oxvalue}}{\oxkontctx}$ and
  \oxreduce{\oxglobalctx}{\oxstore}{\oxexpr}{\oxstore^\prime}{\oxexpr^\prime}, then there
  exists $\oxvarctx_i$ such that
  \oxstorevalidity{\oxglobalctx}{\oxvarctx_i}{\oxstore^\prime} and
  $\oxwfkontctx{\oxglobalctx}{\oxvarctx_i}{\overline{\oxvalue}}{\oxkontctx}$ and
  \oxtypjudge{\oxglobalctx}{\oxemptyctx}{\oxkontctx}{\oxvarctx_i}{\oxexpr^\prime}
  {\oxsitype_2}{\oxvarctx_f^\prime} and
  $\oxtunify[\oxcombine]{\oxemptyctx}{\oxvarctx_f^\prime}
  {\oxsitype_2}{\oxsitype_1}{\oxvarctx_s}$ and there exists $\oxvarctx_o$ such
  that $\oxvarctx_f = \oxvarctx_s \oxintersect \oxvarctx_o$.
\end{oxlemma}

The Preservation lemma says that if $\oxexpr$ has type $\oxsitype_1$ under a
valid global environment $\oxglobalctx$, temporary typing $\oxkontctx$, and
stack typing $\oxvarctx$, \emph{and} we have a stack $\oxstore$ that satisfies
$\oxvarctx$ and a sequence of values $\overline{\oxvalue}$ that satisfies the
temporary typing $\oxkontctx$, \emph{and} we know that $\oxexpr$ can take a step
under $\oxstore$ to the new configuration
$\oxconfiguration{\oxstore^\prime}{\oxexpr^\prime}$, then the following conditions hold.
Our updated stack $\oxstore^\prime$ satisfies $\oxvarctx_i$, our sequence of
temporary values $\overline{\oxvalue}$ continue to satisfy $\oxkontctx$, and our
new expression $\oxexpr^\prime$ typechecks with the type $\oxsitype_2$. In each of
these judgments, we use an intermediate stack typing $\oxvarctx_i$ that
corresponds to the changes the evaluated portion of code made to the
environment. Rather than constrain the type to be the same as the type of our
original $\oxexpr$, our Preservation lemma allows $\oxsitype_2$ to differ in its
regions by the region rewriting judgment, since evaluation potentially can lead
to a type having more precise regions. Further, the output stack typing from
typechecking $\oxexpr^\prime$ is threaded through the rewriting and then ultimately
said to union with some other stack typing $\oxvarctx_o$ in order to capture the
relationship between the old output environment $\oxvarctx_f$ and the new one
$\oxvarctx_f^\prime$ when we have taken a step into one side or the other of a
branch.

As discussed in \Sec{sec:typechecking}, the most challenging part of proving
preservation is in showing that the various changes to the environment preserve
the well-typedness of values, with closures in particular posing the greatest
issue. Indeed, we believe it's clear that a formalization of Rust without
closures misses a significant piece of the language's essence since closures
interact consistently with all parts of the formalism. To that end, the
proof of preservation uses several families of lemmas that
follow the same pattern: since our preservation theorem has to maintain the
well-typedness of the stack $\oxstore$ and temporary values
$\overline{\oxvalue}$, we must show that various judgments in our system
preserve the well-typedness of values. We will highlight these lemmas here,
noting that each require sublemmas for expressions in closure bodies
remaining well-typed which subsequently requires that region rewriting and
ownership safety judgments are preserved by these judgments.

\begin{oxlemma}{Values are Well-Typed after Region Rewriting}
  {lemmap:value-typing-region-rewriting}
  \ifanonymous{(Lemma 5.12}{(\Lemma{lemma:subtyping-value-typing}} in appendix) \hfill \\
  If \ $\oxtypjudge{\oxglobalctx}{\oxtvarctx}{\oxkontctx}{\oxvarctx}
  {\oxvalue}{\oxtype}{\oxvarctx}$ and
  $\oxtunify{\oxtvarctx}{\oxvarctx}{\oxtype_2}{\oxtype_1}{\oxvarctx^\prime}$
  then $\oxtypjudge{\oxglobalctx}{\oxtvarctx}{\oxkontctx}{\oxvarctx^\prime}
  {\oxvalue}{\oxtype}{\oxvarctx^\prime}$.
\end{oxlemma}

\begin{oxlemma}{Values are Well-Typed after Drop/GC-Loans}
  {lemmap:value-typing-related-envs}
  (\ifanonymous{Lemma 5.25}{\Lemma{lemma:value-typing-related-envs}}) \hfill \\
  If \ $\oxsubtypectx{\oxglobalctx}{\oxemptyctx}{\oxvarctx} {\oxvarctx^\prime}$ and
  $\oxtypjudge{\oxglobalctx}{\oxemptyctx}{\oxemptyctx}{\oxvarctx}
  {\oxvalue}{\oxvarctx(x)}{\oxvarctx}$, then
  $\oxtypjudge{\oxglobalctx}{\oxemptyctx}{\oxemptyctx}{\oxvarctx^\prime} {\oxvalue}{\oxvarctx^\prime(x)}{\oxvarctx^\prime}$.
\end{oxlemma}

\begin{oxlemma}{Values are Well-Typed under Well-Typed Extensions}
  {lemmap:value-typing-extension}
  (\ifanonymous{Lemma 5.34}{\Lemma{lemma:values-well-typed-extension}}) \hfill \\
  If \ $\oxtypevalidity{\oxglobalctx}{\oxemptyctx}{\oxvarctx}{\oxsitype_\oxid}$ and
  $\forall \oxcprov \in \oxfprovs{\oxsitype_\oxid}. \;
  \oxnotreborrowed{\oxvarctx}{\oxcprov}$ and
  $\oxtypjudge{\oxglobalctx}{\oxemptyctx}{\oxkontctx}{\oxvarctx}
  {\oxvalue}{\oxsitype}{\oxvarctx}$, then
  $\oxtypjudge{\oxglobalctx}{\oxemptyctx}{\oxkontctx}{ \oxextendctx{\oxvarctx}{
      \oxvarctxentry{\oxid}{\oxsitype_\oxid} } }{\oxvalue}{\oxsitype}{
    \oxextendctx{\oxvarctx}{ \oxvarctxentry{\oxid}{\oxsitype_\oxid} } }$.
\end{oxlemma}

\begin{oxlemma}{Values are Well-Typed after Assignment}
  {lemmap:values-well-typed-assignment}
  (\ifanonymous{Lemma 5.40}{\Lemma{lemma:values-well-typed-assignment}}) \hfill \\
  If \ $\oxlookup{\oxvarctx}{\oxplace_a}{\oxsxtype}$ $\wedge$
  $\oxtunify[\oxnoop]{\oxtvarctx}{\oxkontctx}{
    \oxrsub{\oxvarctx}{\oxderef{\oxplace_a}}
  }{\oxsitype}{\oxsxtype}{\oxvarctx^\prime}$ $\wedge$
  $\oxmusafety{\oxemptyctx}{\oxkontctx}{\oxvarctx^\prime}{\oxmut}{\oxplace_a}{
    \oxset{\oxloanpkg{\oxmut}{\oxplace_a}}
  }$ $\wedge$
  $\oxtypjudge{\oxglobalctx}{\oxemptyctx}{\oxkontctx}{\oxvarctx}
  {\oxvalue}{\oxtype}{\oxvarctx}$, then
  $\oxtypjudge{\oxglobalctx}{\oxemptyctx}{\oxkontctx}{
    \oxgcloans{\oxkontctx}{\oxtupdate{\oxvarctx^\prime}{\oxplace_a}{\oxsitype}}
  }{
    \oxvalue
  }{\oxtype}{
    \oxgcloans{\oxkontctx}{\oxtupdate{\oxvarctx^\prime}{\oxplace_a}{\oxsitype}}
  }$.
\end{oxlemma}

\begin{oxlemma}{Values are Well-Typed under Safe Loan Updates}
  {lemmap:value-typing-loan-update}
  (\ifanonymous{Lemma 5.62}{\Lemma{lemma:values-well-typed-loan-update}}) \hfill \\
  If \ $\oxmusafety{\oxemptyctx}{\oxkontctx}{\oxvarctx}{\oxmuta}{\oxplaceexpr}{
    \oxset{\overline{\oxloan}} }$ and
  $\oxnotinclosure{\oxkontctx}{\oxvarctx}{\oxcprov}$ and
  $\oxlookup{\oxvarctx}{\oxcprov}{\emptyset}$ and
  $\oxtypjudge{\oxglobalctx}{\oxemptyctx}{\oxkontctx}{\oxvarctx}
  {\oxvalue}{\oxsitype}{\oxvarctx}$, then
  $\oxtypjudge{\oxglobalctx}{\oxemptyctx}{\oxkontctx}{ \oxupdate{\oxvarctx}{
      \oxloanctxentry{\oxcprov}{ \oxset{\overline{\oxloan}} } }
  }{\oxvalue}{\oxsitype}{ \oxupdate{\oxvarctx}{ \oxloanctxentry{\oxcprov}{
        \oxset{\overline{\oxloan}} } } }$.
\end{oxlemma}

This pattern of a value typing lemma demanding a whole family of related lemmas
is consistent throughout the supporting lemmas, and they eventually rely on an
ownership safety lemma where we need to actually consider how the various
judgments each change the environment and argue that the separation provided by
closures --- including for regions --- is sufficient to prevent those changes
from breaking ownership safety derivations in the closure bodies themselves. In
each case, the hardest part of proving each family is ensuring that the
induction hypothesis for the ownership safety lemma was rich enough to address
the changes effects on the loans in the environment and the reborrow exclusion
list discussed in \Sec{sec:regions}.

With \Lemma{lemmap:progress} and \Lemma{lemmap:preservation} in hand, we also
prove a conventional type safety theorem as a corollary. However, we note that
Preservation itself represents a more interesting metatheoretic result because
it requires us to show that all of the type system invariants (namely, the
aliasing requirements) are maintained throughout the execution of well-typed
programs.

\begin{figure*}
  \figuresize
  \begin{tabular}{*{8}{c|}c}
  \multicolumn{2}{c|}{passing} & \multicolumn{7}{c}{disqualified} \\ \hline
  borrowck & nll & heap & out-of-scope library & enums & statics \& consts &
  traits & uninitialized variables & misc. \\ \hline
  89  & 119 & 63 & 40 & 50 & 40 & 93 & 40 & 81 \\
\end{tabular}

  \vspace{-1em}
  \caption{Tested Semantics Results}
  \label{fig:test-results}
  \vspace{-1em}
\end{figure*}

\subsection{Tested Semantics}
\label{sec:tested-semantics}

We set out at the onset to solve a particular problem --- there is no high-level
specification of the Rust programming language and its borrowchecker. If there
were, this would be the point where we might present a proof that every
expression that typechecks in \lang also typechecks in Rust and vice versa.
Since doing that is not possible, we follow \citet{guha10:essence-js} in
developing a \emph{tested semantics} for \lang. We built an implementation of
our \lang typechecking algorithm, \typechecker, alongside a compiler, \compiler,
from a subset of Rust (with a small number of additional annotations) to \lang.
In addition to the features described in \Sec{sec:formalization}, our
implementation supports \rusti{struct}s by treating them as tagged tuples or
records. Together, \compiler and \typechecker allowed us to use tests from the
official borrow checker (\rusti{borrowck}) and non-lexical lifetime
(\rusti{nll}) test suites to validate \lang against Rust's implementation,
\rustc. The results of this testing are summarized in \Fig{fig:test-results}.

For the 208 passing tests, we can compile the test case into \lang with
\compiler and then use \typechecker to either successfully typecheck the
program or to produce a type error. We compare this typechecking result to the
expected behavior according to the \rustc test suite. All 208 tests either type
check when \rustc does so, or produce an error corresponding to the error
produced by \rustc.

The remaining 407 tests were taken out of consideration on the basis of being
out-of-scope for this work. There were 20 categories for exclusion, the majority
of which had fewer than 10 applicable tests. \Fig{fig:test-results} includes the
6 largest categories: (1)~heap allocation, (2)~out-of-scope libraries,
(3)~enumerations, (4)~statics and constants, (5)~traits, and (6)~uninitialized
variables. One specialized category (multithreading) was folded into
out-of-scope libraries in this table, with miscellaneous aggregating the
remaining smaller categories: control flow, casting, first-class constructors,
compiler internals dumping, function mutability, inline assembly, macros, slice
patterns, two-phase borrows, uninitialized variables, universal function-call
syntax, unsafe, and variable mutability.

Combined, heap allocation and out-of-scope libraries (of which the former is a
specialization of the latter) make up for the largest excluded category with 103
tests, and can be extended in future work using the strategy outlined by
\citet{weiss19:oxide}. The next largest category, traits, accounts for 93 tests.
Though the trait system is in some ways novel, the bulk of its design is rooted
in the work on Haskell typeclasses and their extensions. As such, we feel that
they are not an \emph{essential} part of Rust, though exploring the
particularities of their design may be a fruitful avenue for future work on
typeclasses. We are working on extending our implementation with sums to support
enumerations, and they are already present in the formalism. Many of the other
categories describe features (e.g., macros, control flow, casting, statics, and
constants) that are well-studied in the literature, and in which we believe Rust
has made relatively standard design choices.

The last issue to discuss involving the tested semantics is the aforementioned
annotation burden. This burden comes directly out of the syntactic differences
between \lang and Rust, and so are overall rather minor. The most immediately
apparent need is to provide a region annotation on borrow expressions, which we
handle using Rust's compiler annotation support. In our tests, a borrow
expression like $\oxref{\texttt{'a}}{\,\oxmut\,}{\texttt{x}}$ appears as
\rusti{#[lft="a"] &mut x}. However, we reduce the need for this by automatically
generating a fresh local region for borrow expressions without an annotation.
This suffices for the majority of expressions without change. Relatedly, one
might also expect to see the introduction of \oxkey{letrgn} throughout. To
alleviate the need for this, our implementation automatically binds free region
at the beginning of each function body.

The other main change we had to make relates to the use of explicit environment
polymorphism in \lang. In Rust, every closure has a unique type without a syntax
for writing it down. To work with higher-order functions, these closures
implement one of three language-defined traits (\rusti{Fn}, \rusti{FnMut}, and
\rusti{FnOnce}) which can be used as bounds in higher-order functions. We
compile the use of these trait bounds to environment polymorphism in a
straight-forward manner (turning instances of the same \rusti{Fn}-bound
polymorphic type into uses of function types with the same environment
variable), but need to introduce a way of writing down which environment to use
at instantiation. We use a compiler annotation (\rusti{#[envs(c1, ..., cn)]}) on
applications which says to instantiate the environment variables with the
captured environments of the types of these bindings. If the bindings are
unbound or not at a function type, we produce an error indicating as much.

Aside from these two changes, there are a handful of smaller changes that we
made by hand to simplify implementation of \compiler and \typechecker, though
the need for these could be obviated with more work. Our implementation does not
support method call syntax, and so we translate method definitions (which take
\rusti{self}, \rusti{&self}, or \rusti{&mut self} as their first argument) into
function definitions with a named first argument at the method receiver's type.
Relatedly, some of the tests used traits in a trivial way to define methods
polymorphic in their receiver type. Like other methods, we translated these into
function definitions, but used a polymorphic type for the receiver. \rustc also
allows for a number of convenient programming patterns (like borrowing from a
constant, e.g. \rusti{&0}) which are not supported by our implementation. To
deal with these cases, we manually introduced temporaries (a process that \rustc
does automatically). As a simplification for the typechecker, \typechecker only
reports the first error that occurs in the program. To ensure that we find a
correspondence between all errors, we split up test files with multiple errors
into one file per test.

Finally, an earlier version of our implementation required type annotations on
all let bindings, and so currently many tests include fully-annotated types. We
later realized our typing judgment is very-nearly a type \emph{synthesis}
judgment as in bidirectional typechecking, and changed the implementation to
support unannotated bindings by using the type synthesized for the expression
being bound. This works for all expressions except \oxkey{abort!} which can
produce any type and so requires annotation.

\ifanonymous{
  We provide a snapshot of the tooling as anonymized supplementary materials to
  our submission, including the entire \rusti{borrowck} and \rusti{nll} test
  suites and all disqualified tests categorized appropriately. The test suite
  features a full, anonymized \oxkey{git} history that captures the changes
  summarized above.
}{}


\section{Related Work}
\label{sec:related}
\subsection{Semantics for Rust}

\paragraph{Early Work} \citet{reed15:patina} developed \emph{Patina}, a formal
semantics for an early version of Rust (pre-1.0) focused on proving memory
safety for a language with a syntactic version of borrow checking and unique
pointers. Unfortunately, the design of the language was not yet stable, and the
language overall has drifted from their model. Also, unlike \lang, Patina made
concrete decisions about memory layout and validity which is problematic as Rust
itself has not yet made such commitments.

\citet{benitez16:rusty-types} developed \emph{Metal}, a formal calculus that, by
their characterization, has a Rust-like type system using an \emph{algorithmic}
borrow-checking formulation. Their model relies on capabilities as in the
Capability Calculus of \citet{crary99:capabilities}, but manages them indirectly
(compared to the first-class capabilities of \citet{crary99:capabilities} or
\citet{morrisett07:linloc}). Compared to Rust and our work on \lang, Metal is
unable to deal with the proper LIFO ordering for object destruction and their
algorithmic formulation is less expressive than our declarative formulation.

\paragraph{RustBelt} In the RustBelt project, \citet{jung18:rustbelt} developed
a formal semantics called $\lambda_{Rust}$ for a continuation-passing style
intermediate language in the Rust compiler known as MIR. They mechanized this
formal semantics in Iris~\cite{jung18:iris} and used it to verify the extrinsic
safety of important Rust standard library abstractions that make extensive use
of \rusti{unsafe} code. Their goal was distinct from ours in that we instead
wish to reason about how programs work at the source-level, and our goals are
fortunately complementary. As argued by \citet{weiss18:oxide}, we can
incorporate \rusti{unsafe} code in the standard library by adding primitives to
\lang, and the verified specifications from RustBelt provide further
justification for their safety.

\paragraph{Featherweight Rust} Recent work by \citet{pearce21:fr} developed a
calculus called FR that, like us, takes inspiration from the Featherweight Java
of \citet{igarashi01:featherweight-java}. Indeed, they take this inspiration so
seriously that FR is limited solely to let bindings, assignment, moves, and
borrows. Such a simplification misses much of the interesting parts of borrow
checking. Without branching, it is possible to statically maintain total
knowledge of pointer provenance for every reference, trivializing checking for
conflicting borrows. Without aggregate data types like tuples and enumerations,
there's no notions of partial ownership and no need for the infrastructure of
\emph{places} and \emph{place expressions}. Further, without closures, there is
no ability for computation to be suspended with ownership effects caught up in
it. Dealing with closures correctly was an immense part of the effort in
designing \lang, and ruled out many simpler borrow checking schemes we developed
along the way. \citeauthor{pearce21:fr} attempts to address this in their work
by describing extensions for branching, tuples, and top-level functions with
very brief arguments as to why the extension would not break their proofs.
However, the answers there are unsatisfying: the argument for branching, for
instance, is roughly that one could individually consider each straightline
execution path through the program as its own program that then has a precise
environment in their calculus. Perhaps most importantly, they limit their
attention to modeling Rust with ``lexical lifetimes,'' a language that has not
actually existed in \emph{five years} at the time of writing. Like with
closures, \lang required a great deal of careful design work to appropriately
handle the behavior of Rust's \emph{non-lexical lifetimes}.

\paragraph{Polonius} Polonius~\cite{matsakis18:polonius} is a new alias-based
implementation of Rust's borrow checker that uses information from the Rust
compiler as input facts for a logic program that checks the safety of borrows in
a program. Much as we have done with \lang, Polonius shifts the view of
\emph{lifetimes} to a model of \emph{origins} as sets of loans which approximate
the possible provenances of a reference. As described by
\citet{matsakis18:polonius}, a reference is no longer valid when any of the
constituent loans of an origin are invalidated. In \lang, we take an analogous
view: a reference type is valid only when its constituent loans are bound in the
stack typing $\oxvarctx$. Though we have not formally explored the connection,
based on the commonality between both new views on lifetimes, we feel that \lang
corresponds to a sort of type-systems analogue of Polonius' constraint solving
approach.

\subsection{Practical Substructural Programming}

As a practical programming language with substructural typing, Rust does not
exist in a vacuum. There have been numerous efforts in the programming languages
community to produce languages that rely on substructurality. Though different
in their design from Rust, these languages sit in the same broader design space,
finding a balance between usability and expressivity.

\citet{pottier13:mezzo} developed Mezzo, an ML-family language
with a static discipline of duplicable and affine permissions to control
aliasing and ownership. Similar to Rust, Mezzo is able to have types refer
directly to values, rather than always requiring indirection as in work on
ownership types~\cite{noble98:ownership,clarke98:ownership-types}. However,
unlike Rust, Mezzo uses a permissions system that works as a sort of type-system
formulation of separation logic~\cite{reynolds02:seplogic}. By contrast, Rust
relies on a borrow checking analysis to ensure that its guarantees about
aliasing and ownership are maintained. In \lang, we formalized this analysis as
the ownership safety judgment which determines if it is safe to use a place
uniquely or sharedly in a given context.


\citet{munch-maccagnoni18:resource-poly} has recently proposed a
backwards-compatible model of resource management for OCaml. Though not yet a
part of OCaml, the proposal is promising and aims to integrate ideas from Rust
and C++ (like ownership and so-called ``resource acquisition is
initialization''~\cite{stroustrup94:raii}) with a garbage-collected runtime
system for a functional language. \citet{munch-maccagnoni18:resource-poly}
argues that these efforts can learn from Rust, and we hope that \lang provides a
strong footing to do so.

\citet{grossman02:cyclone} developed Cyclone as a safe C alternative. To do so,
they rely on techniques from region-based memory
management~\cite{tofte94:region-inference, tofte97:regions}. For Cyclone,
regions indicate where an object is located in memory (e.g. on the stack or
heap), while in \lang regions are used for managing \emph{aliasing} by
abstracting over a reference's possible origins, regardless of the memory mode
at runtime. Like \lang, \citet{fluet06:linrgn} developed a formal semantics to
demonstrate the essence of Cyclone.


\section{Discussion}
\label{sec:discussion}

\paranegspc
\paragraph{Region Reuse in \lang} Overall, in \lang, we've seen a number of
restrictions related to the concrete region annotations that are added to the
source program relative to Rust. This includes the region-not-reborrowed and
region-not-in-closure judgments in rules such as \oxname{T-Borrow} and
\oxname{T-Let}, as well as in the outlives judgment (\Fig{fig:unification}).
Overall, these restrictions may seem to risk limiting our support for Rust's
diverse borrowing patterns, but we've found with our implementation that this is
not the case. In general, we are able to employ a strategy of always preferring
a new region except when required (to pass multiple distinct references to a
polymorphic function such as \rusti{fn choose_ref<'a>(&'a uniq u32, &'a uniq
u32) -> &'a uniq u32}) and indeed, our \lang implementation can do virtually
all of this work automatically. Polonius~\cite{matsakis18:polonius,
  matsakis20:polonius}, a new borrow-checker for \rustc discussed in
\Sec{sec:discussion}, relies on a similar scheme of generating new origins and
constraining them to be equal only when strictly necessary.

\paranegspc
\paragraph{Substructurality in \lang} Since Rust's release, the folklore has
said that, of course, ``Rust is an affine language.'' As such, one might have
expected to see the explicit removal of the structural rule of contraction in a
formal calculus. However, with behavior like copyable types and implicit drops,
the substructurality story for \lang is a bit more complicated. Like an ordered
type system, \lang does not allow exchange to maintain the ordered end of scopes
for bindings, but its rules for variable use (moving, copying, and borrowing)
all employ judgments that enable out-of-order \emph{use} of variables. Like an
affine type system, \lang has a rule \oxname{T-Drop} which resembles a weakening
rule by allowing a program to typecheck with a binding whenever it is possible
to typecheck with that binding \emph{dead}. Unlike conventional weakening,
however, the binding itself must still be present (with a dead type) because of
the ordering requirement! Finally, \lang even has something resembling
contraction in the form of \oxname{T-Copy} which allows many types to be used
multiple times, lowering the friction of the duplicable of-course types common
in the substructural typing literature.


\section{Conclusion and Future Work}
\label{sec:conclusion}
In this paper, we have presented \lang as a formal model of \emph{the essence of
  Rust} with a novel approach for reasoning about the behavior of source-level
Rust programs with \emph{region-based alias management}. We leveraged syntactic
techniques to prove type safety for \lang (\Sec{sec:metatheory}), and
implemented a prototype typechecker in OCaml along side a compiler from Rust to
\lang which we used to validate our semantics against a suite of over
two-hundred tests from the official \rustc test suite.

With \lang in hand, we believe there is a host of new possibilities for research
involving Rust. For instance, while there are some early efforts to bring formal
verification to Rust~\cite{ullrich16:electrolysis, toman15:crust,
  baranowski18:smack, astrauskas18:rust-viper}, the possibilities are limited
without an appropriate semantics to work from. As one particular example, the
work by \citet{astrauskas18:rust-viper} builds verification support for Rust
into Viper~\cite{muller16:viper}, but uses an ad-hoc subset without support for
shared references. Further, Rust's memory safety guarantees lend themselves well
to security-critical applications. However, the existing compiler toolchain
(leveraging LLVM~\cite{lattner04:llvm}) does not lend itself well to preserving
these kinds of guarantees. As such, another avenue for future work using \lang
would be to build an alternative verified compiler toolchain, perhaps by
compilation to Vellvm~\cite{zhao12:vellvm} or CompCert's
Clight~\cite{blazy09:clight}. Overall, we hope that \lang can serve as a rich
platform for research with Rust even beyond our own imaginations.


\bibliography{awe}

\appendix

\renewcommand{\oxnamegrammar}{
  \begin{displaymath}
    \begin{array}{lrclrclrclr}
      \bnflabel{Variables} & \oxid &&
      \bnflabel{Functions} & \oxfnname &&
      \bnflabel{Type Vars.} & \oxtvar &&
      \bnflabel{Frame Vars.} & \oxenvvar \\
      \bnflabel{Concrete Regions} & \oxcprov &&
      \bnflabel{Abstract Regions} & \oxabsprov &&
      \bnflabel{Strings} & \oxstring &&
      \bnflabel{Naturals} & m, \oxnum, k \\
    \end{array}
  \end{displaymath}
}

\renewcommand{\oxexpressiongrammar}{
  \bnflabel{Expressions} & \oxexpr & \bnfdef &
  \oxprim \bnfalt \oxplaceexpr \bnfalt
  \oxref{\oxcprov}{\oxmuta}{\oxplaceexpr} \bnfalt
  \oxref{\oxcprov}{\oxmuta}{\oxindex{\oxplaceexpr}{\oxexpr}} \bnfalt
  \oxref{\oxcprov}{\oxmuta}{\oxslice{\oxplaceexpr}{\oxexpr_1}{\oxexpr_2}}
  \bnfalt \oxassign{\oxplaceexpr}{\oxexpr}
  \\ && \bnfalt &
  \oxletrgn{\oxcprov}{\oxexpr} \bnfalt
  \oxlet{\oxid}{\oxsitype}{\oxexpr_1}{\oxexpr_2} \bnfalt
  \oxseq{\oxexpr_1}{\oxexpr_2}
  \\ && \bnfalt &
  \oxclosure{
    \oxascribe{\oxid_1}{\oxsitype_1} \oxdotsc
    \oxascribe{\oxid_n}{\oxsitype_n}
  }{\oxsitype_r}{
    \oxexpr
  } \bnfalt
  \oxapp{\oxexpr_f}{
    \overline{\oxenv} \oxcomma \overline{\oxprov} \oxcomma \overline{\oxsitype}
  }{
    \oxexpr_1 \oxdotsc \oxexpr_n
  }
  \\ && \bnfalt &
  \oxbranch{\oxexpr_1}{\oxexpr_2}{\oxexpr_3} \bnfalt
  \oxprod{\oxexpr_1 \oxdotsc \oxexpr_n} \bnfalt
  \oxarr{\oxexpr_1 \oxdotsc \oxexpr_n}
  \\ && \bnfalt &
  \oxindex{\oxplaceexpr}{\oxexpr} \bnfalt
  \oxfor{\oxid}{\oxexpr_1}{\oxexpr_2} \bnfalt
  \oxwhile{\oxexpr_1}{\oxexpr_2} \bnfalt
  \oxabort{\oxstring}
  \\ && \bnfalt &
  \oxinl{\oxsitype_1}{\oxsitype_2}{\oxexpr} \bnfalt \oxinr{\oxsitype_1}{\oxsitype_2}{\oxexpr}
  \\ && \bnfalt &
  \oxmatch{\oxexpr}{\oxid_1}{\oxexpr_1}{\oxid_2}{\oxexpr_2}
}

\renewcommand{\oxptrexprgrammar}{
  \bnflabel{Expressions} & \oxexpr & \bnfdef &
  \oxdots \bnfalt \oxframed{\oxexpr} \bnfalt \oxshift{\oxexpr}
  \bnfalt \oxsliceval{\oxvalue_1 \oxdotsc \oxvalue_n} \bnfalt \oxuninit
  \oxptrextension
}

\renewcommand{\oxvaluegrammar}{
  \bnflabel{Values} & \oxvalue & \bnfdef &
  \oxprim \bnfalt
  \oxprod{\oxvalue_1 \oxdotsc \oxvalue_n} \bnfalt
  \oxarr{\oxvalue_1 \oxdotsc \oxvalue_n} \bnfalt
  \oxsliceval{\oxvalue_1 \oxdotsc \oxvalue_n} \bnfalt \oxfnname \bnfalt
  \oxuninit
}

\renewcommand{\oxvaluectxgrammar}{
  \bnflabel{Value Contexts} & \oxvaluectx & \bnfdef &
  \oxhole \bnfalt
  \oxprod{\oxvalue_1 \oxdotsc \oxvaluectx \oxdotsc \oxvalue_n} \bnfalt
  \oxarr{\oxvalue_1 \oxdotsc \oxvaluectx_1 \oxdotsc \oxvaluectx_m \oxdotsc \oxvalue_n}
}

\renewcommand{\oxsitypegrammar}{
  \bnflabel{Sized Types} & \oxsitype & \bnfdef &
  \oxbasetype \bnfalt \oxtvar \bnfalt \oxtref{\oxprov}{\oxmuta}{\oxxitype}
  \bnfalt \oxtarr{\oxsitype}{\oxnum} \bnfalt \oxtprod{\oxsitype_1 \oxdotsc \oxsitype_n} \bnfalt \oxtsum{\oxsitype_1}{\oxsitype_2}
  \\ && \bnfalt &
  \oxtfunext{
    \overline{\oxenvvar} \oxcomma \overline{\oxabsprov} \oxcomma \overline{\oxtvar}
  }{
    \oxsitype_1 \oxdotsc \oxsitype_n
  }{\oxsitype_r}{\oxenv}{\overline{\oxabsprov_1: \oxabsprov_2}}
}

\renewcommand{\oxevalctxgrammar}{
  \bnflabel{Eval. Contexts} & \oxevalctx & \bnfdef &
  \oxhole
  \\ && \bnfalt &
  \oxref{\oxprov}{\oxmuta}{\oxindex{\oxplaceexpr}{\oxevalctx}} \bnfalt
  \oxref{\oxprov}{\oxmuta}{\oxslice{\oxplaceexpr}{\oxevalctx}{\oxnoseqexpr}} \bnfalt
  \oxref{\oxprov}{\oxmuta}{\oxslice{\oxplaceexpr}{\oxvalue}{\oxevalctx}}
  \\ && \bnfalt &
  \oxlet{\oxid}{\oxsitype}{\oxevalctx}{\oxexpr} \bnfalt
  \oxletrgn{\oxcprov}{\oxevalctx}
  \\ && \bnfalt &
  \oxassign{\oxplaceexpr}{\oxevalctx} \bnfalt
  \oxseq{\oxevalctx}{\oxexpr} \bnfalt
  \oxframed{\oxevalctx}
  \\ && \bnfalt &
  \oxshift{\oxevalctx} \bnfalt
  \oxshiftprov{\oxevalctx}
  \\ && \bnfalt &
  \oxapp{\oxevalctx}{\overline{\oxenv} \oxcomma \overline{\oxprov} \oxcomma  \overline{\oxsitype}}{
    \oxnoseqexpr_1 \oxdotsc \oxnoseqexpr_n
  }
  \\ && \bnfalt &
  \oxapp{\oxvalue}{\overline{\oxenv} \oxcomma \overline{\oxprov} \oxcomma \overline{\oxsitype}}{
    \oxvalue_1 \oxdotsc \oxvalue_m \oxcomma \oxevalctx \oxcomma
    \oxnoseqexpr_1 \oxdotsc \oxnoseqexpr_n
  }
  \\ && \bnfalt &
  \oxindex{\oxplaceexpr}{\oxevalctx} \bnfalt
  \oxbranch{\oxevalctx}{\oxexpr_1}{\oxexpr_2}
  \\ && \bnfalt &
  \oxfor{\oxid}{\oxevalctx}{\oxexpr}
  \\ && \bnfalt &
  \oxprod{
    \oxvalue_1 \oxdotsc \oxvalue_m \oxcomma \oxevalctx \oxcomma
    \oxnoseqexpr_1 \oxdotsc \oxnoseqexpr_n
  }
  \\ && \bnfalt &
  \oxarr{
    \oxvalue_1 \oxdotsc \oxvalue_m \oxcomma \oxevalctx \oxcomma
    \oxnoseqexpr_1 \oxdotsc \oxnoseqexpr_n
  }
  \\ && \bnfalt &
  \oxinl{\oxsitype_1}{\oxsitype_2}{\oxevalctx} \bnfalt
  \oxinr{\oxsitype_1}{\oxsitype_2}{\oxevalctx}
  \\ && \bnfalt &
  \oxmatch{\oxevalctx}{\oxid_1}{\oxexpr_1}{\oxid_2}{\oxexpr_2}
}

\section{\lang Syntax}
\label{sec:full-syntax}

\begin{figure}[H]
  \small
  \oxnamegrammar

  \begin{bnf}
    \oxpathgrammar \\[0em]
    \oxplacegrammar \\[0em]
    \oxplaceexprgrammar \\[0em]
    \oxplaceexprctxgrammar \\[0.75em]

    \oxprovgrammar \\[0em]
    \oxmutagrammar \\[0em]
    \oxrewritemodegrammar \\[0em]
    \oxloangrammar \\[0.75em]
    
    \oxkindgrammar \\[0em]
    \oxbasetypegrammar \\[0em]
    \oxsitypegrammar \\[0em]
    \oxxitypegrammar \\[0em]
    \oxsdtypegrammar \\[0em]
    \oxsxtypegrammar \\[0em]
    \oxtypegrammar \\[0.75em]

    \oxprimitivegrammar \\[0em]
    \oxexpressiongrammar \\[0em]
    \oxenvgrammar \\[0.75em]

    \oxglobalctxgrammar \\[0em]
    \oxglobalentrygrammar \\[0.75em]
    \oxtvarctxgrammar \\[0em]
    \oxframegrammar\\[0em]
    \oxvarctxgrammar \\[0em]
    \oxkontctxgrammar \\[0.75em]
  \end{bnf}
\end{figure}

\section{Statics}
\label{sec:full-statics}

\subsection{Well-Formedness Judgments}
\label{sec:well-formedness}

\begin{figure}[H]
  \figuresize
  \begin{flushleft}\oxglobalctxvalidityform \\
    read: ``$\oxglobalctx$ is well-formed'' \end{flushleft}

  \begin{mathpar}
    \WFGlobalCtx
  \end{mathpar}
\end{figure}

\begin{figure}[H]
  \figuresize
  \begin{flushleft}\oxglobalentryvalidityform  \\
    read: ``$\oxglobalentry$ is a well-formed function definition in $\Sigma$'' \end{flushleft}

  \begin{mathpar}
    \WFFunctionDefn
  \end{mathpar}
\end{figure}

\begin{figure}[H]
  \figuresize
  \begin{flushleft}\oxtvarctxwellformedform  \\
    read: ``$\oxtvarctx$ is well-formed'' \end{flushleft}

  \begin{mathpar}
    \WFTVarEmpty \and \WFTVarExtendEnv \and \WFTVarExtendProv \and \WFTVarExtendType \and
    \WFTVarExtendConstraint
  \end{mathpar}
\end{figure}

\begin{figure}[H]
  \figuresize
  \begin{flushleft}\oxvarctxwellformedform  \\
    read: ``$\oxvarctx$ is well-formed under $\oxglobalctx$ and $\oxtvarctx$'' \end{flushleft}

  \begin{mathpar}
    \WFEmptyVarCtx \and \WFVarCtx
  \end{mathpar}
\end{figure}

\begin{figure}[H]
  \figuresize
  \begin{flushleft}\oxkontctxwellformedform  \\
    read: ``$\oxkontctx$ is well-formed under $\oxglobalctx$, $\oxtvarctx$, and $\oxvarctx$'' \end{flushleft}

  \begin{mathpar}
    \WFEmptyTemporaryTyping \and \WFTemporaryTyping
  \end{mathpar}
\end{figure}

\begin{figure}[H]
  \figuresize
  \begin{flushleft}\oxctxswellformedform  \\
    read: ``$\oxglobalctx$, $\oxtvarctx$, and $\oxvarctx$ are well-formed.'' \end{flushleft}

  \begin{mathpar}
    \WFCtx
  \end{mathpar}
\end{figure}

\begin{figure}[H]
  \figuresize
  \begin{flushleft}\oxenvwellformedform  \\
    read: ``$\oxenv$ is a well-formed captured environment''\end{flushleft}

  \begin{mathpar}
    \WFEnvVar \and \WFEnv
  \end{mathpar}
\end{figure}

\begin{figure}[H]
  \figuresize
  \begin{flushleft}\oxrgnvalidityform  \\
    read: ``$\oxprov$ is a well-formed region''\end{flushleft}

  \begin{mathpar}
    \VLocalProv \and \VAbsProv
  \end{mathpar}
\end{figure}

\begin{figure}[H]
  \figuresize
  \begin{flushleft}
    \oxtypevalidityform  \\
    read: ``$\oxtype$ is a well-formed type under $\oxglobalctx$, $\oxtvarctx$, and $\oxvarctx$''
  \end{flushleft}

  \begin{mathpar}
    \VBaseType \and \VTVar \and \VRef \and \VTuple \and \VFunction
    \and \VUninit \and \VArray \and \VSlice
  \end{mathpar}
\end{figure}

\subsection{Region Rewriting \& Outlives Relations}
\label{sec:subtyping}

\begin{figure}[H]
  \figuresize
  \begin{flushleft}\oxtunifyform \\
    read: ``terms at the type $\oxtype_1$ under $\oxtvarctx$ and $\oxvarctx$ can be
    rewritten according to $\oxrewritemode$ as type $\oxtype_2$ under $\oxvarctx^\prime$''
  \end{flushleft}

  \begin{mathpar}
    \URefl \and \UTrans \and \UArray \and \USlice \and \USharedRef \and \UTuple \and \UUninit
  \end{mathpar}
\end{figure}

\begin{figure}[H]
  \figuresize
  \begin{flushleft}\oxrunifyform \\
    read: ``$\oxprov_1$ outlives $\oxprov_2$ under $\oxtvarctx$ and $\oxvarctx$,
    and can be rewritten according to $\oxrewritemode$ under the environment $\oxvarctx^\prime$''
  \end{flushleft}

  \begin{mathpar}
    \UReflProv \and \UAbsProvs \and \UTransProv \and \UCombineLocalProvs \and
    \UCombineLocalProvsUnrest \and \UCheckLocalProvs \and \OLocalAbsProvs \and \OAbsLocalProvs
  \end{mathpar}
\end{figure}

\begin{figure}[H]
  \figuresize
  \begin{flushleft}\oxrunifymanyform\end{flushleft}

  \begin{mathpar}
    \OBounds
  \end{mathpar}
\end{figure}

\subsection{Ownership Safety}
\label{sec:ownership-safety}

\begin{figure}[H]
  \figuresize
  \begin{flushleft}
    \oxmusafetyform where $\oxmusafetyshape$ means
    $\oxmusafetyinner{\oxtvarctx}{\oxkontctx}{\oxvarctx}{\oxmuta}{ \oxemptyctx }{\oxplaceexpr}{ \oxset{\overline{\oxloanpkg{\oxmuta}{\oxplaceexpr}}}
    }$. \\
    read: ``$\oxplaceexpr$ is $\oxmuta$-safe under $\oxtvarctx$ and $\oxvarctx$, with reborrow
    exclusion list $\overline{\oxplace}$, and may point to any of the loans in
    $\overline{\oxloanpkg{\oxmuta}{\oxplaceexpr}}$''
  \end{flushleft}

  \begin{mathpar}
    \OSafePlace \and \ODeref \and \ODerefAbs
  \end{mathpar}
\end{figure}

\subsection{Typing}
\label{sec:typing}

\begin{figure}[H]
  \figuresize
  \begin{flushleft}
    \oxtypjudgeform where \oxctxswellformed{\oxglobalctx}{\oxtvarctx}{\oxvarctx}{\oxkontctx} and
    \oxtypevalidity{\oxglobalctx}{\oxtvarctx}
    {\oxvarctx^\prime}{\oxtype} \\
    read: ``$\oxexpr$ has type $\oxtype$ under $\oxglobalctx$, $\oxtvarctx$, and $\oxvarctx$,
    producing output context $\oxvarctx$''
  \end{flushleft}

  \begin{mathpar}
    \TMove \and \TCopy \and \TBorrow \and \TBorrowIndex \and \TBorrowSlice \and \TIndexCopy \and
    \TSeq \and \TBranch \and \TLet \and \TLetProv \and \TAssignDeref \and \TAssign
  \end{mathpar}
\end{figure}

\begin{figure}[H]
  \figuresize
  
  \begin{mathpar}
    \TWhile \and \TForArray \and \TForSlice \and \TFunction \and \TClosure \and \TApp \and
    \TAppClosure \and \TAbort \and \TUnit \and \TuThreeTwo \and \TTrue \and \TFalse \and \TTuple
  \end{mathpar}
\end{figure}

\begin{figure}[H]
  \figuresize

  \begin{mathpar}
    \TArray \and \TSlice \and \TDrop \and \TInl \and \TInr \and \TMatch
  \end{mathpar}
\end{figure}

\subsection{Additional Judgments}
\label{sec:additional-judgments}

\begin{figure}[H]
  \figuresize
  \begin{flushleft}\oxmutasubtypeform \\
    read: ``$\oxmuta$ is less than $\oxmuta^\prime$ in the qualifier
    ordering''\end{flushleft}

  \begin{mathpar}
    \SRefl \and \SShrdUniq
  \end{mathpar}
\end{figure}

\begin{figure}[H]
  \figuresize
  \begin{flushleft}\oxsubtypectxform \\
    read: ``$\oxvarctx$ is related to $\oxvarctx^\prime$ under $\oxglobalctx$ and $\oxtvarctx$'' \end{flushleft}

  \begin{mathpar}
    \SEnv
  \end{mathpar}
\end{figure}

\begin{figure}[H]
  \figuresize
  \begin{flushleft}\oxcomputetyform \\
    read: ``$\oxplaceexpr$ in an $\oxmuta$ context has type
    $\oxtype$ under $\oxtvarctx$ and $\oxvarctx$, passing through
    the regions in $\overline{\oxprov}$''\end{flushleft}

  \begin{mathpar}
    \TCVar \and \TCProj \and \TCDeref
  \end{mathpar}

  \begin{flushleft}\oxcomputetynoprovform \\
  read: ``$\oxplaceexpr$ in an $\oxmuta$ context has type
  $\oxtype$ under $\oxtvarctx$ and $\oxvarctx$''\end{flushleft}

  $\oxcomputetynoprov{\oxtvarctx}{\oxvarctx}{\oxmuta}{\oxplaceexpr}{\oxtype} = \oxcomputety{\oxtvarctx}{\oxvarctx}{\oxmuta}{\oxplaceexpr}{\oxtype}{\_}$ \\[0.5em]
\end{figure}



\begin{figure}[H]
  \figuresize
  \begin{flushleft}\oxwfkontctxform \\
    read: ``the given values $\overline{\oxvalue}$ satisfy $\oxkontctx$ under $\oxglobalctx$ and $\oxvarctx$''\end{flushleft}

  \begin{mathpar}
    \WFKontCtx
  \end{mathpar}

\end{figure}

\begin{figure}[H]
  \figuresize
  \begin{flushleft}\oxnotreborrowedform \\
    read: ``the region $\oxcprov$ is not reborrowed in $\oxvarctx$'' \end{flushleft}

  \begin{mathpar}
    \NRBOrigin
  \end{mathpar}
\end{figure}

\begin{figure}[H]
  \figuresize
  \begin{flushleft}\oxclosrestrictionform \\
    read: ``the regions $\overline{\oxcprov}$ follow the closure restriction in $\oxkontctx$ or
    $\oxvarctx$''
  \end{flushleft}

  \begin{mathpar}
    \CLSRestriction
  \end{mathpar}

  \begin{flushleft}\oxnotinclosureform \\
    read: ``the region $\oxcprov$ is not in a closure's signature in $\oxkontctx$ or $\oxvarctx$''
  \end{flushleft}

  \begin{mathpar}
    \CLSRegionNotIn
  \end{mathpar}
\end{figure}
   
\section{Metafunctions}
\label{sec:metafunctions}

\begin{figuresize}
  \begin{flushleft}
    \DefFvars 
    \DefRelevant\\
    \DefPRelevant\\[1em]

    \oxintersectform \\[0.5em]
    \DefIntersect

    \oxframeintersectform \\[0.5em]
    \DefFrameIntersect

    \vspace{0.5em}

    \oxplacesform
    
  \begin{lstlisting}
  $\texttt{places} \oxspace (\oxemptyctx) \oxspace = \oxspace \emptyset$
  $\texttt{places} \oxspace (\oxextendctx{\oxvarctx}{
    \oxloanctxentry{\oxcprov}{\oxset{\overline{\oxloanpkg{\oxmuta}{\oxplaceexpr}}}}
  }) \oxspace = \oxspace \oxset{\oxplace \; | \; \oxloanpkg{\oxmuta_i}{\oxplaceexpr_i} \in
    \oxset{\overline{\oxloanpkg{\oxmuta}{\oxplaceexpr}}} \wedge (\oxplaceexpr_i
    = \oxplace \vee \oxplaceexpr_i = \oxplaceexprctx[\oxderef{\oxplace}]) } \oxspace \cup \oxspace \texttt{places}
  \oxspace \oxvarctx$
  $\texttt{places} \oxspace (\oxextendctx{\oxvarctx}{\oxvarctxentry{\oxid}{\oxtype}}) =
   \texttt{places} \oxspace (\oxnewframe{\oxvarctx}{\oxemptyctx}) = \texttt{places} \oxspace (\oxvarctx)$
\end{lstlisting}

    \oxvaluedecomposeform
    \begin{mathpar}
      \DVEnd \and \DVProj
    \end{mathpar}

    \oxvupdateform
    
  \begin{lstlisting}
  $\oxtupdate{\oxstore}{\oxpathcons{\oxid}{\oxpath}}{\oxvalue} =
  \oxtupdate{\oxstore}{\oxid}{\oxevalctx[\oxvalue]}$
    where $\oxdecompose{\oxstore(\oxid)}{\oxpath}{\oxevalctx}{\_}$
\end{lstlisting}

    \oxvlookupform
    
  \begin{lstlisting}
  $\oxlookup{\oxstore}{\oxpathcons{\oxid}{\oxpath}}{\oxvalue}$
    where $\oxdecompose{\oxstore(\oxid)}{\oxpath}{\_}{\oxvalue}$
\end{lstlisting}

    \oxdecomposeform
    \begin{mathpar}
      \DEnd \and \DProj
    \end{mathpar}

    \oxtupdateform
    
  \begin{lstlisting}
  $\oxtupdate{\oxvarctx}{\oxpathcons{\oxid}{\oxpath}}{\oxtype} =
  \oxtupdate{\oxvarctx}{\oxid}{\oxtypectx[\oxtype]}$
    where $\oxdecompose{\oxvarctx(\oxid)}{\oxpath}{\oxtypectx}{\_}$
\end{lstlisting}

    \pagebreak
    
    \oxtlookupform
    
  \begin{lstlisting}
  $\oxlookup{\oxvarctx}{\oxpathcons{\oxid}{\oxpath}}{\oxtype}$
    where $\oxdecompose{\oxvarctx(\oxid)}{\oxpath}{\_}{\oxtype}$
\end{lstlisting}

    \oxnoncopyableform
    
  \begin{align*}
  \texttt{noncopyable}_{\,\oxglobalctx} \  \oxbasetype &= \bot \\
  \texttt{noncopyable}_{\,\oxglobalctx} \  \alpha &= \top \\
  \texttt{noncopyable}_{\,\oxglobalctx} \  \oxtref{\_}{\;\oxmut\;}{\_} &= \top \\
  \texttt{noncopyable}_{\,\oxglobalctx} \  \oxtref{\_}{\;\oximm\;}{\_} &= \bot \\
  \texttt{noncopyable}_{\,\oxglobalctx} \  \oxtfun{\_}{\_}{\_}{\_} &= \bot \\
  \texttt{noncopyable}_{\,\oxglobalctx} \  \oxtarr{\oxtype}{\_} &= \texttt{noncopyable}_{\,\oxglobalctx} \  \oxtype \\
  \texttt{noncopyable}_{\,\oxglobalctx} \  \oxtslice{\oxtype} &= \texttt{noncopyable}_{\,\oxglobalctx} \  \oxtype \\
  \texttt{noncopyable}_{\,\oxglobalctx} \  \oxtprod{\oxtype, \ldots} &= \texttt{noncopyable}_{\,\oxglobalctx} \  \oxtype \, \vee \,  \ldots \\
\end{align*}

    \oxcopyableform
    
  \begin{align*}
  \texttt{copyable}_{\,\oxglobalctx} \, \oxtype &= \neg \, \texttt{noncopyable}_{\,\oxglobalctx} \, \oxtype
\end{align*}

    \oxexplodeform \\[0.5em]
    \DefExplode \\[0.5em]
    \oxexplodeinnerform \\
    
  \begin{align*}
  \texttt{explode} \  \oxvarctxentry{\oxplace}{\oxbasetype}
  &= \oxset{\oxvarctxentry{\oxplace}{\oxbasetype}} \\
  \texttt{explode} \  \oxvarctxentry{\oxplace}{\oxtvar}
  &= \oxset{\oxvarctxentry{\oxplace}{\oxtvar}} \\
  \texttt{explode} \  \oxvarctxentry{\oxplace}{\oxtref{\oxprov}{\oxmuta}{\oxxitype}}
  &= \oxset{\oxvarctxentry{\oxplace}{\oxtref{\oxprov}{\oxmuta}{\oxxitype}}} \\
  \texttt{explode} \  \oxvarctxentry{\oxplace}{\oxtarr{\oxsitype}{\oxnum}}
  &= \oxset{\oxvarctxentry{\oxplace}{\oxtarr{\oxsitype}{\oxnum}}} \\
  \texttt{explode} \  \oxvarctxentry{\oxplace}{\oxtprod{\oxsxtype_1 \oxdots \oxsxtype_n}}
  &= \underset{i \;\in\; 1 \oxdots n}{\cup} \ \oxexplode{
    \oxvarctxentry{\oxproj{\oxplace}{i}}{\oxsxtype_i}
    } \\
  \texttt{explode} \  \oxvarctxentry{\oxplace}{
    \oxtfunext{
      \overline{\oxenvvar} \oxcomma \overline{\oxabsprov} \oxcomma \overline{\oxtvar}
    }{
      \oxsitype_1 \oxdotsc \oxsitype_n
    }{\oxsitype_r}{\oxenv}{\overline{\oxabsprov_1: \oxabsprov_2}}
  }
  &= \oxset{\oxvarctxentry{\oxplace}{
        \oxtfunext{
          \overline{\oxenvvar} \oxcomma \overline{\oxabsprov} \oxcomma \overline{\oxtvar}
        }{
          \oxsitype_1 \oxdotsc \oxsitype_n
        }{\oxsitype_r}{\oxenv}{\overline{\oxabsprov_1: \oxabsprov_2}}
      }
    } \\
  \texttt{explode} \  \oxvarctxentry{\oxplace}{\oxsidtype}
  &= \oxset{\oxvarctxentry{\oxplace}{\oxsidtype}} \\
\end{align*}

    \oxoccursbeforeinform
    \begin{mathpar}
      \OCOccursBase \and \OCOccursExtendFrame \and \OCOccursNewFrame
    \end{mathpar}

    \oxgcloansform \\[0.5em]
    \Defgcloans \\[0.5em]

    \oxrgnuniquetoform \\[0.5em]
    \DefRgnUniqueTo \\[0.5em]

    \oxrsubform \\[0.5em]
    \Defrsub \\[0.5em]
    \oxrsubmanyform \\[0.5em]
    \Defrsubmany \\[0.5em]

    \oxloanmappingsform\\[0.5em]
    \Defoxloanmappings\\[0.5em]
  \end{flushleft}
\end{figuresize}

\section{Dynamics}
\label{sec:full-dynamics}

\begin{figure}[H]
  \figuresize
  \begin{bnf}
    \oxreferentgrammar \\[0em]
    \oxreferentctxgrammar \\[0em]
    \oxptrexprgrammar \\[1em]

    \oxvaluegrammar \oxptrextension \\[0em]

    \oxevalctxgrammar \\[0em]
    \oxvaluectxgrammar \\[0em]
    \oxstoregrammar \\[0em]
    \oxstackframegrammar \\[0em]
  \end{bnf}
\end{figure}

\begin{figure}[H]
  \figuresize
  \begin{flushleft}\oxreferentvalidityform\end{flushleft}

  \begin{mathpar}
    \WFRefId \and \WFRefProj \and \WFRefIndexArray \and \WFRefIndexSlice \and
    \WFRefSliceArray \and \WFRefSliceSlice
  \end{mathpar}
\end{figure}

\begin{figure}[H]
  \figuresize
  \begin{flushleft}\oxevalreferentform\end{flushleft}

  \begin{mathpar}
    \ERId \and \ERProj \and \ERIndexArray \and \ERIndexSlice \and \ERSliceArray
    \and \ERSliceSlice
  \end{mathpar}
\end{figure}

\begin{figure}[H]
  \figuresize
  \begin{flushleft}\oxnormform \\
    read: ``$\oxplaceexpr$ computes to $\oxreferent$, which maps to
    $\oxvalue$ in $\oxstore$.'' \\[0.5em]
    Let $\oxnorm{\oxstore}{\oxplaceexprctx[\oxid]}{\oxreferent}{\oxvaluectx}{\oxvalue} =
    \oxnorminner{\oxstore}{\oxplaceexprctx}{\oxid}{\oxreferent}{\oxvaluectx}{\oxvalue}$.
  \end{flushleft}
\end{figure}

\begin{figure}[H]
  \figuresize
  \begin{flushleft}
    \oxnorminnerform \\
    read: ``$\oxreferent$ in a context $\oxplaceexprctx$ computes to
    $\oxreferent^\prime$ which maps to $\oxvalue$ in $\oxstore$ with a context of $\oxvaluectx$.''
  \end{flushleft}

  \begin{mathpar}
    \PId \and \PProj \and \PDerefPtr \and \PDerefIndexPtrArray \and
    \PDerefIndexPtrSlice \and \PDerefSlicePtrArray \and \PDerefSlicePtrSlice
  \end{mathpar}
\end{figure}

\begin{figure}[H]
  \figuresize
  \begin{flushleft}\oxstorevalidityform \\
    read: ``$\oxstore$ satisfies $\oxvarctx$ under global context
    $\oxglobalctx$''\end{flushleft}

  \begin{mathpar}
    \VStackEmpty \and \VStack
  \end{mathpar}
\end{figure}

\begin{figure}[H]
  \figuresize
  \begin{flushleft}\oxsubstorevalidityform \\
    read: ``$\oxstackframe$ satisfies $\oxframe_c$ under $\oxglobalctx$ and
    $\oxvarctx$'' \end{flushleft}

  \begin{mathpar}
    \VSubStack
  \end{mathpar}
\end{figure}

\begin{figure}[H]
  \figuresize
  \begin{flushleft}\oxtypjudgeform where
    \oxctxswellformed{\oxglobalctx}{\oxtvarctx}{\oxvarctx}{\oxkontctx} and
    \oxtypevalidity{\oxglobalctx}{\oxtvarctx}
    {\oxvarctx^\prime}{\oxtype} \end{flushleft}

  \begin{mathpar} \oxdots \end{mathpar}
  \begin{mathpar}
    \TShift \and \TFramed \and \TPointer \and \TClosureVal \and \TUninit
  \end{mathpar}
\end{figure}

\begin{figure}[H]
  \figuresize
  \begin{flushleft}\oxreduceform \\
  read: ``$\oxstore$ and $\oxexpr$ step to $\oxstore^\prime$ and $\oxexpr^\prime$ under $\oxglobalctx$'' \end{flushleft}

  \begin{mathpar}
    \EMove \and \ECopy \and \EBorrow \and \EBorrowIndex \and \EBorrowSlice
  \end{mathpar}
\end{figure}

\begin{figure}[H]
  \figuresize

  \begin{mathpar}
    \EBorrowIndexOOB \and \EBorrowSliceOOB \and \EIndexCopy \and \EIndexCopyOOB \and \EFramed \and
    \EShift \and \EIfTrue \and \EIfFalse \and \EMatchLeft \and \EMatchRight \and \ELetProv \and
    \ELet \and \ESeq \and \EAssign \and \EWhile \and \EForArray \and \EForSlice \and \EForEmptyArray
    \and \EForEmptySlice \and \EClosure
  \end{mathpar}
\end{figure}

\begin{figure}[H]
  \figuresize
  \begin{mathpar}
    \EApp \and \EAppFun \and \EEvalCtx \and \EEvalCtxAbort
  \end{mathpar}
\end{figure}

\pagebreak

\section{Metatheory}
\label{sec:metatheory-full}

\subsection{Standard Lemmas}
\label{sec:lemmas}

\begin{oxlemma}{Canonical Forms}{lemma:canonical-forms}
  If
  \oxtypjudge{\oxglobalctx}{\oxtvarctx}{\oxkontctx}{\oxvarctx}{\oxvalue}{\oxtype}{\oxvarctx}
  then

  \begin{enumerate}
  \item if $\oxtype = \oxtbool$, then $\oxvalue = \oxtrue$ or $\oxvalue =
    \oxfalse$.
  \item if $\oxtype = \oxtnum$, then $\oxvalue = \oxnum$.
  \item if $\oxtype = \oxtunit$, then $\oxvalue = \oxunit$.
  \item if $\oxtype = \oxtref{\oxprov}{\oxmuta}{\oxsitype}$, then $
    \oxvalue$ is of the form $\oxptr{\oxreferent}$.
  \item if $\oxtype = \oxtref{\oxprov}{\oxmuta}{\oxtslice{\oxsitype}}$, then $
    \oxvalue$ is of the form $\oxsliceptr{\oxreferent}{i}{j}$.
  \item if $\oxtype = \oxtfunext{
      \overline{\oxenvvar} \oxcomma \overline{\oxabsprov} \oxcomma
      \overline{\oxtvar}
    }{\oxsitype_1 \oxdotsc \oxsitype_n}{\oxsitype_r}{}{
      \overline{\oxabsprov_1 : \oxabsprov_2}
    }$, then $\oxvalue$ is of the form $\oxfnname$.
  \item if $\oxtype = \oxtclosure{
        \oxsitype_1 \oxdotsc \oxsitype_n
      }{\oxsitype_r}{\oxframe}$, then $\oxvalue$ is of the form
      $\oxclosureval{\oxstore}{
        \oxascribe{\oxid_1}{\oxsitype_1} \oxdotsc
        \oxascribe{\oxid_n}{\oxsitype_n}
      }{\oxsitype_r}{\oxexpr}$.
  \item if $\oxtype = \oxtarr{\oxtype^\prime}{\oxnum}$, then $\oxvalue$ is of
    the form $\oxarr{\oxvalue_1 \oxdotsc \oxvalue_n}$.
  \item if $\oxtype = \oxtslice{\oxtype^\prime}$, then $\oxvalue$ is of
    the form $\oxsliceval{\oxvalue_1 \oxdotsc \oxvalue_n}$.
  \item if $\oxtype = \oxtprod{\oxtype_1 \oxdotsc \oxtype_n}$, then $\oxvalue$
    is of the form $\oxprod{\oxvalue_1 \oxdotsc \oxvalue_n}$.
  \item if $\oxtype = \oxtsum{\oxtype_1}{\oxtype_2}$, then $\oxvalue$ is of
    either the form $\oxinl{\oxtype_1}{\oxtype_2}{\oxvalue^\prime}$ or
    $\oxinr{\oxtype_1}{\oxtype_2}{\oxvalue^\prime}$.
  \end{enumerate}
\end{oxlemma}

\begin{proof}
  By inspection of the grammar of values and typing rules.
\end{proof}

\begin{oxlemma}{Preservation of Types under Substitution}{lemma:substitution}
  \hfill
  \begin{enumerate}
  \item If $\oxtypjudge{\oxglobalctx}{ \oxextendctx{\oxtvarctx}{
        \oxtvarctxentry{\oxtvar}{\oxktype} }
    }{\oxkontctx}{\oxvarctx}{\oxexpr}{\oxtype}{\oxvarctx^\prime}$ and
    $\oxtypevalidity{\oxglobalctx}{\oxtvarctx}{\oxvarctx}{\oxtype^\prime}$, then
    $\oxtypjudge{\oxglobalctx}{\oxtvarctx}{\oxkontctx}{\oxvarctx}{
      \oxsubst{\oxexpr}{\oxtype^\prime}{\oxtvar}
    }{\oxsubst{\oxtype}{\oxtype^\prime}{\oxtvar}}
    {\oxsubst{\oxvarctx^\prime}{\oxtype^\prime}{\oxtvar}}$
  \item If $\oxtypjudge{\oxglobalctx}{ \oxextendctx{\oxtvarctx}{
        \oxtvarctxentry{\oxabsprov}{\oxkprov} }
    }{\oxkontctx}{\oxvarctx}{\oxexpr}{\oxtype}{\oxvarctx^\prime}$ and
    $\oxrgnvalidity{\oxtvarctx}{\oxvarctx}{\oxprov}$, then
    $\oxtypjudge{\oxglobalctx}{\oxtvarctx}{\oxkontctx} {\oxvarctx}{
      \oxsubst{\oxexpr}{\oxprov}{\oxabsprov}
    }{\oxsubst{\oxtype}{\oxprov}{\oxabsprov}}
    {\oxsubst{\oxvarctx^\prime}{\oxprov}{\oxabsprov}}$
  \item If $\oxtypjudge{\oxglobalctx}{ \oxextendctx{\oxtvarctx}{
        \oxtvarctxentry{\oxenvvar}{\oxkenv} }
    }{\oxkontctx}{\oxvarctx}{\oxexpr}{\oxtype}{\oxvarctx^\prime}$ and
    $\oxenvwellformed{\oxglobalctx}{\oxtvarctx}{\oxvarctx}{\oxenv}$, then
    $\oxtypjudge{\oxglobalctx}{\oxtvarctx}{\oxkontctx} {\oxvarctx}{
      \oxsubst{\oxexpr}{\oxenv}{\oxenvvar}
    }{\oxsubst{\oxtype}{\oxenv}{\oxenvvar}}
    {\oxsubst{\oxvarctx^\prime}{\oxenv}{\oxenvvar}}$
  \end{enumerate}
\end{oxlemma}

\begin{proof}
  By induction on the typing derivation.
\end{proof}

\subsection{Referent Lemmas}
\label{sec:ref-lemmas}

\begin{oxlemma}{Well-Formed References Evaluate to Well-Typed Values}
  {lemma:wf-refs-evalref}
  If $\oxreferentvalidity{\oxglobalctx}{\oxvarctx}{\oxreferent}{\oxxitype}$ and
  $\oxstorevalidity{\oxglobalctx}{\oxvarctx}{\oxstore}$, then
  $\oxevalreferent{\oxstore}{\oxreferent}{\oxvaluectx}{\oxvalue}$.
\end{oxlemma}

\begin{proof}
  We proceed by induction on $\oxreferentvalidity{\oxglobalctx}{\oxvarctx}
  {\oxreferent}{\oxxitype}$. There are six cases: \oxname{WF-RefId},
  \oxname{WF-RefProj}, \oxname{WF-RefIndexArray}, \oxname{WF-RefIndexSlice},
  \oxname{WF-RefSliceArray}, and \oxname{WF-RefSliceSlice}. Each of these cases
  has a corresponding evaluation rule:

  \vspace{1em}
  \noindent \framebox[\textwidth]{
    \figuresize
    \begin{mathpar}
      \WFRefId \and \ERId
    \end{mathpar}
  }
  \vspace{1em}

  For the base case, we consider the frame of $\oxvarctx$ which contains
  $\oxid$. By inversion of \oxname{WF-StackFrame} for the portion of the
  derivation $\oxstorevalidity{\oxglobalctx}{\oxvarctx}{\oxstore}$ pertaining to
  that frame, we have $\forall \oxid \in \oxdomain{\oxstackframe}. \;
  \oxtypjudge{\oxglobalctx}{\oxemptyctx}{\oxemptyctx}{\oxnewframe{\oxvarctx}{\oxframe}}
  {(\oxnewframe{\oxstore}{\oxstackframe})(\oxid)}{
    (\oxnewframe{\oxvarctx}{\oxframe})(\oxid)
  }{\oxnewframe{\oxvarctx}{\oxframe}}$. Focusing on our particular $\oxid$, we
  have both that $\oxlookup{\oxstore}{\oxid}{\oxvalue}$ and that
  $\oxtypjudge{\oxglobalctx}{\oxemptyctx}{\oxemptyctx}{\oxnewframe{\oxvarctx}{\oxframe}}
  {\oxvalue}{(\oxnewframe{\oxvarctx}{\oxframe})(\oxid)}{
    \oxnewframe{\oxvarctx}{\oxframe}
  }$, finishing the case. The remaining cases follow:

  \vspace{1em}

  \noindent \framebox[\textwidth]{
    \figuresize
    \begin{mathpar}
      \WFRefProj \and \ERProj
    \end{mathpar}
  }

  \vspace{1em}

  \noindent \framebox[\textwidth]{
    \figuresize
    \begin{mathpar}
      \WFRefIndexArray \and \ERIndexArray
    \end{mathpar}
  }

  \vspace{1em}

  \noindent \framebox[\textwidth]{
    \figuresize
    \begin{mathpar}
      \WFRefIndexSlice \and \ERIndexSlice
    \end{mathpar}
  }

  \vspace{1em}

  \noindent \framebox[\textwidth]{
    \figuresize
    \begin{mathpar}
      \WFRefSliceArray \and \ERSliceArray
    \end{mathpar}
  }

  \vspace{1em}

  \noindent \framebox[\textwidth]{
    \figuresize
    \begin{mathpar}
      \WFRefSliceSlice \and \ERSliceSlice
    \end{mathpar}
  }

  The proof for each case is identical: apply the induction hypothesis and then
  \Lemma{lemma:canonical-forms} and then the evaluation rule on the right. For
  the well-typed portion, apply inversion on the typing rule for the appropriate
  value.
\end{proof}

\begin{oxlemma}{Place Expressions Reduce}{lemma:place-exprs-reduce}
  If $\oxcomputetynoprov{\oxtvarctx}{\oxvarctx}{\oxmuta}{\oxplaceexpr}{\oxxitype}$
  and $\oxstorevalidity{\oxglobalctx}{\oxvarctx}{\oxstore}$,
  then $\oxnorm{\oxstore}{\oxplaceexpr}{\oxreferent}{\oxvaluectx}{\oxvalue}$ and
  $\oxtypjudge{\oxglobalctx}{\oxtvarctx}{\oxkontctx}{\oxvarctx}{\oxvalue}{\oxxitype}{\oxvarctx}$.
\end{oxlemma}

\begin{proof} 
  We proceed by induction on
  $\oxcomputetynoprov{\oxtvarctx}{\oxvarctx}{\oxmuta}{\oxplaceexpr}{\oxxitype}$.
  There are three cases: \oxname{TC-Var}, \oxname{TC-Proj}, and \oxname{TC-Deref}.

  \vspace{1em}
  \noindent \framebox[\textwidth]{
    \figuresize
    \begin{mathpar}
      \TCVar \and \PId
    \end{mathpar}
  }
  \vspace{1em}
  
  For \oxname{TC-Var}, we consider the piece of the derivation for
  $\oxstorevalidity{\oxglobalctx}{\oxvarctx}{\oxstore}$ (from our premise)
  for the frame containing $\oxid$. By inversion on \oxname{WF-StackFrame},
  we have $\forall \oxid \in \oxdomain{\oxstackframe}. \;
  \oxtypjudge{\oxglobalctx}{\oxemptyctx}{\oxemptyctx}{\oxnewframe{\oxvarctx}{\oxframe}}
  {(\oxnewframe{\oxstore}{\oxstackframe})(\oxid)}{
    (\oxnewframe{\oxvarctx}{\oxframe})(\oxid)
  }{\oxnewframe{\oxvarctx}{\oxframe}}$. This immediately gives us that
  $\oxlookup{\oxstore}{\oxid}{\oxvalue}$ and that
  $\oxtypjudge{\oxglobalctx}{\oxtvarctx}{\oxkontctx}{\oxvarctx}{\oxvalue}
  {\oxxitype}{\oxvarctx}$. To construct our premise for \oxname{P-Referent},
  we apply \oxname{ER-Id} to $\oxlookup{\oxstore}{\oxid}{\oxvalue}$.

  \vspace{1em}
  \noindent \framebox[\textwidth]{
    \figuresize
    \begin{mathpar}
      \TCProj \and \PProj
    \end{mathpar}
  }
  \vspace{1em}
  
  For \oxname{TC-Proj}, we apply our induction hypothesis to
  $\oxcomputety{\oxtvarctx}{\oxvarctx}{\oxmuta}{
    \oxplaceexpr
  }{
    \oxtprod{\oxsitype_1 \oxdotsc \oxsitype_i \oxdotsc \oxsitype_n}
  }{\oxset{\overline{\oxprov_p}}}$ from the premise of \oxname{TC-Proj} and get
  $\oxnorm{\oxstore}{\oxplaceexpr}{\oxreferent}{\oxvaluectx}{\oxvalue}$ and
  $\oxtypjudge{\oxglobalctx}{\oxtvarctx}{\oxkontctx}{\oxvarctx}{\oxvalue}
  {\oxtprod{\oxsitype_1 \oxdotsc \oxsitype_i \oxdotsc \oxsitype_n}}{\oxvarctx}$.
  Then, by \Lemma{lemma:canonical-forms}, we know that $\oxvalue$ must be of the
  form $\oxprod{\oxvalue_1 \oxdotsc \oxvalue_i \oxdotsc \oxvalue_n}$. We can use
  this and the definition of $\oxnorm{\oxstore}{\oxplaceexpr}{\oxreferent}
  {\oxvaluectx}{\oxvalue}$ to get
  $\oxnorminner{\oxstore}{\oxplaceexprctx}{\oxid}{\oxreferent}{\oxvaluectx}{
    \oxprod{\oxvalue_1 \oxdotsc \oxvalue_i \oxdotsc \oxvalue_n}
  }$ (where $\oxplaceexprctx[\oxid] = \oxplaceexpr$). This is precisely the
  premise of \oxname{P-Proj} and thus we can use that. We also have by inversion
  of \oxname{T-Tuple} for
  $\oxtypjudge{\oxglobalctx}{\oxtvarctx}{\oxkontctx}{\oxvarctx}{\oxvalue}
  {\oxtprod{\oxsitype_1 \oxdotsc \oxsitype_i \oxdotsc \oxsitype_n}}{\oxvarctx}$
  that $\oxtypjudge{\oxglobalctx}{\oxtvarctx}{\oxkontctx}{\oxvarctx}{\oxvalue_i}
  {\oxsitype_i}{\oxvarctx}$.

  \vspace{1em}
  \noindent \framebox[\textwidth]{
    \figuresize
    \begin{mathpar}
      \TCDeref
    \end{mathpar}
  }
  \vspace{1em}

  For \oxname{TC-Deref}, we apply our induction hypothesis to 
  $\oxcomputety{\oxtvarctx}{\oxvarctx}{\oxmuta}{
    \oxplaceexpr
  }{
    \oxtref{\oxprov}{\oxmuta^\prime}{\oxxitype}
  }{\oxset{\overline{\oxprov_p}}}$ to get
  $\oxnorm{\oxstore}{\oxplaceexpr}{\oxreferent}{\oxvaluectx}{\oxvalue}$ and
  $\oxtypjudge{\oxglobalctx}{\oxtvarctx}{\oxkontctx}{\oxvarctx}{\oxvalue}
  {\oxtref{\oxprov}{\oxmuta^\prime}{\oxxitype}}{\oxvarctx}$. Then, by
  \Lemma{lemma:canonical-forms}, we know that $\oxvalue$ must of the form
  $\oxptr{\oxreferent}$. We now have five subcases to consider depending on
  whether $\oxreferent$ is of $\oxplace$, $\oxindex{\oxreferent_3}{i}$, or
  $\oxslice{\oxreferent}{i}{j}$, and for the latter two, whether $\oxxitype$ is
  $\oxtarr{\oxsitype}{\oxnum}$ or $\oxtslice{\oxsitype}$.

  \vspace{1em}
  \noindent \framebox[\textwidth]{
    \figuresize
    \begin{mathpar}
      \PDerefPtr \and \PDerefIndexPtrArray \and \PDerefIndexPtrSlice \and
      \PDerefSlicePtrArray \and \PDerefSlicePtrSlice
    \end{mathpar}
  }
  \vspace{1em}

  In all these cases, we know structurally that $\oxplaceexprctx = \oxhole$
  since \oxname{TC-Deref} has no context outside of the dereference. So, for
  each of them, we need to be able to show $\oxnorm{\oxhole}{\oxreferent}{
    \oxreferent^\prime
  }{\oxvaluectx}{\oxvalue^\prime}$. Inversion on \oxname{T-Pointer} gives us
  $\oxreferentvalidity{\oxglobalctx}{\oxvarctx}{\oxreferent}{\oxxitype}$. We can
  then apply \Lemma{lemma:wf-refs-evalref} to get
  $\oxevalreferent{\oxglobalctx}{\oxvarctx}{\oxvaluectx}{\oxvalue}$. Then, we
  can apply \oxname{P-Referent} to this to produce the derivation we need to
  apply the appropriate rule. For \oxname{P-DerefIndexPtrArray} and
  \oxname{P-DerefSlicePtrArray}, we apply \Lemma{lemma:canonical-forms}
  to get that the value is an array. For \oxname{P-DerefIndexPtrSlice} and
  \oxname{P-DerefSlicePtrSlice}, we apply \Lemma{lemma:canonical-forms} to get
  that the value is a slice value.
\end{proof}

\begin{oxlemma}{Reduced Place Expressions Produce Valid Referents}
  {lemma:norm-for-valid-referents}
  If $\oxstorevalidity{\oxglobalctx}{\oxvarctx}{\oxstore}$
  and $\oxnorm{\oxstore}{\oxplaceexpr}{\oxreferentctx[\oxplace]}
  {\oxvaluectx}{\oxvalue}$,
  then
  $\oxreferentvalidity{\oxglobalctx}{\oxvarctx}{\oxreferentctx[\oxplace]}{\oxxitype}$.
\end{oxlemma}

\begin{proof}
  We start by rewriting 
  $\oxnorm{\oxstore}{\oxplaceexpr}{\oxreferentctx[\oxplace]}
  {\oxvaluectx}{\oxvalue}$ with its definition to get
  $\oxnorminner{\oxstore}{\oxplaceexprctx}{\oxid}{
    \oxreferentctx[\oxplace]
  }{\oxvaluectx}{\oxvalue}$ where $\oxplaceexpr = \oxplaceexprctx[\oxid]$. We
  then proceed by induction by cases (note this means our induction hypothesis
  is really about the rewritten form).

  \vspace{1em}
  \noindent \framebox[\textwidth]{
    \figuresize
    \begin{mathpar}
      \PId \and \WFRefId
    \end{mathpar}
  }
  \vspace{1em}

  \oxname{P-Referent} only applies if the context is $\oxhole$ which is only the
  case if our original place expression was $\oxid$. We can rewrite with this
  knowledge to see that we really have
  $\oxevalreferent{\oxstore}{\oxid}{\_}{\oxvalue}$ in our premise. Inversion on
  \oxname{ER-Id} gives us $\oxlookup{\oxstore}{\oxvalue}$ Then, we consider the
  frame of $\oxvarctx$ which contains $\oxid$. By inversion of
  \oxname{WF-StackFrame} for the portion of the derivation
  $\oxstorevalidity{\oxglobalctx}{\oxvarctx}{\oxstore}$ pertaining to that
  frame, we have $\forall \oxid \in \oxdomain{\oxstackframe}. \;
  \oxtypjudge{\oxglobalctx}{\oxemptyctx}{\oxemptyctx}{\oxnewframe{\oxvarctx}{\oxframe}}
  {(\oxnewframe{\oxstore}{\oxstackframe})(\oxid)}{
    (\oxnewframe{\oxvarctx}{\oxframe})(\oxid)
  }{\oxnewframe{\oxvarctx}{\oxframe}}$. Focusing on our particular $\oxid$, we
  have both that $\oxlookup{\oxvarctx}{\oxid}{\oxvalue}$. We can then apply
  \oxname{WF-RefId}.

  \vspace{1em}
  \noindent \framebox[\textwidth]{
    \figuresize
    \begin{mathpar}
      \PProj \and \WFRefProj
    \end{mathpar}
  }
  \vspace{1em}

  Applying the induction hypothesis to
  $\oxnorminner{\oxstore}{\oxplaceexprctx}{\oxreferent_1}{\oxreferent_2}{\_}{
    \oxprod{\oxvalue_0 \oxdotsc \oxvalue_i \oxdotsc \oxvalue_n}
  }$ gives us $\oxreferentvalidity{\oxglobalctx}{\oxvarctx}{\oxreferent_2}{
    \oxtprod{\oxsitype_0 \oxdotsc \oxsitype_i \oxdotsc \oxsitype_n}
  }$. We can then apply \oxname{WF-RefProjection}.
  
  \vspace{1em}
  \noindent \framebox[\textwidth]{
    \figuresize
    \begin{mathpar}
      \PDerefPtr
    \end{mathpar}
  }
  \vspace{1em}

  Applying the induction hypothesis to 
  $\oxnorminner{\oxstore}{\oxplaceexprctx}{\oxplace}{\oxreferent_2}{\_}{\oxvalue}$
  gives us $\oxreferentvalidity{\oxglobalctx}{\oxvarctx}{\oxreferent_2}{
    \oxxitype
  }$.

  \vspace{1em}
  \noindent \framebox[\textwidth]{
    \figuresize
    \begin{mathpar}
      \PDerefIndexPtrArray \and \WFRefIndexArray
    \end{mathpar}
  }
  \vspace{1em}

  Applying the induction hypothesis to
  $\oxnorminner{\oxstore}{\oxplaceexprctx}{\oxreferent_2}{\oxreferent_3}{\_}{
    \oxarr{\oxvalue_0 \oxdotsc \oxvalue_i \oxdotsc \oxvalue_n}
  }$ gives us $\oxreferentvalidity{\oxglobalctx}{\oxvarctx}{\oxreferent_3}{
    \oxtarr{\oxsitype}{\oxnum}
  }$. Then, we can apply \oxname{WF-RefIndexArray} to get
  $\oxreferentvalidity{\oxglobalctx}{\oxvarctx}{\oxindex{\oxreferent_3}{i}}{
    \oxsitype
  }$.

  \vspace{1em}
  \noindent \framebox[\textwidth]{
    \figuresize
    \begin{mathpar}
      \PDerefIndexPtrSlice \and \WFRefIndexSlice
    \end{mathpar}
  }
  \vspace{1em}

  Applying the induction hypothesis to
  $\oxnorminner{\oxstore}{\oxplaceexprctx}{\oxreferent_2}{\oxreferent_3}{\_}{
    \oxsliceval{\oxvalue_0 \oxdotsc \oxvalue_i \oxdotsc \oxvalue_n}
  }$ gives us $\oxreferentvalidity{\oxglobalctx}{\oxvarctx}{\oxreferent_3}{
    \oxtslice{\oxsitype}
  }$. Then, we can apply \oxname{WF-RefIndexSlice} to get
  $\oxreferentvalidity{\oxglobalctx}{\oxvarctx}{\oxindex{\oxreferent_3}{i}}{
    \oxsitype
  }$.

  \vspace{1em}
  \noindent \framebox[\textwidth]{
    \figuresize
    \begin{mathpar}
      \PDerefSlicePtrArray \and \WFRefSliceArray
    \end{mathpar}
  }
  \vspace{1em}

  Applying the induction hypothesis to
  $\oxnorminner{\oxstore}{\oxplaceexprctx}{\oxreferent_2}{\oxreferent_3}{\_}{
    \oxarr{
      \oxvalue_0 \oxdotsc \oxvalue_i \oxdotsc \oxvalue_j \oxdotsc \oxvalue_n
    }
  }$ gives us $\oxreferentvalidity{\oxglobalctx}{\oxvarctx}{\oxreferent_3}{
    \oxtarr{\oxsitype}{\oxnum}
  }$. Then, we can apply \oxname{WF-RefSliceArray} to get
  $\oxreferentvalidity{\oxglobalctx}{\oxvarctx}{\oxslice{\oxreferent_3}{i}{j}}{
    \oxsitype
  }$.

  \vspace{1em}
  \noindent \framebox[\textwidth]{
    \figuresize
    \begin{mathpar}
      \PDerefSlicePtrSlice \and \WFRefSliceSlice
    \end{mathpar}
  }
  \vspace{1em}

  Applying the induction hypothesis to
  $\oxnorminner{\oxstore}{\oxplaceexprctx}{\oxreferent_2}{\oxreferent_3}{\_}{
    \oxsliceval{
      \oxvalue_0 \oxdotsc \oxvalue_i \oxdotsc \oxvalue_j \oxdotsc \oxvalue_n
    }
  }$ gives us $\oxreferentvalidity{\oxglobalctx}{\oxvarctx}{\oxreferent_3}{
    \oxtslice{\oxsitype}
  }$. Then, we can apply \oxname{WF-RefSliceSlice} to get
  $\oxreferentvalidity{\oxglobalctx}{\oxvarctx}{\oxslice{\oxreferent_3}{i}{j}}{
    \oxsitype
  }$.
\end{proof}

\begin{oxlemma}{Reduced Place Expressions Have Roots in Loan Sets}
  {lemma:norm-place-prefix-in-loans}
  If $\oxstorevalidity{\oxglobalctx}{\oxvarctx}{\oxstore}$,
  $\oxnorm{\oxstore}{\oxplaceexpr}{\oxreferentctx[\oxplace]}
  {\oxvaluectx}{\oxvalue}$, and
  $\oxmusafety{\oxemptyctx}{\oxkontctx}{\oxvarctx}{\oxmuta}{\oxplaceexpr}{
    \oxset{\overline{\oxloan}}
  }$, then $\oxreferent = \oxreferentctx[\oxplace]$ and
  $\oxloanpkg{\oxmuta}{\oxplace} \in \oxset{\overline{\oxloan}}$.
\end{oxlemma}

\begin{proof}
  We proceed by induction on $\oxmusafety{\oxemptyctx}{\oxkontctx}{\oxvarctx}{\oxmuta}{
    \oxplaceexpr
  }{
    \oxset{\overline{\oxloan}}
  }$. There are ordinarily three cases: \oxname{O-SafePlace}, \oxname{O-Deref},
  and \oxname{O-DerefAbs}. However, \oxname{O-DerefAbs} requires the type
  variable context to contain entries, and thus can be immediately discharged by
  contradiction. This leaves us with only \oxname{O-SafePlace} and
  \oxname{O-Deref}.

  \vspace{1em}
  \noindent \framebox[\textwidth]{
    \figuresize
    \begin{mathpar}
      \OSafePlace
    \end{mathpar}
  }
  \vspace{1em}

  \oxname{O-SafePlace} tells us that our $\oxplaceexpr$ is in fact a place
  $\oxplace$ meaning that it does not contain any dereferences. As such, we know
  that $\oxnorm{\oxstore}{\oxplaceexpr}{\oxreferentctx[\oxplace]}
  {\oxvaluectx}{\oxvalue}$ must have been derived using a combination of
  \oxname{P-Referent} and \oxname{P-Proj} corresponding to the structure of
  $\oxplace$. The resulting referent in such a case is precisely $\oxplace$
  (meaning $\oxreferentctx = \oxhole$), which we know is in the output
  immediately from the definition of \oxname{O-SafePlace}.
  
  \vspace{1em}
  \noindent \framebox[\textwidth]{
    \figuresize
    \begin{mathpar}
      \ODeref
    \end{mathpar}
  }
  \vspace{1em}

  In the premise of \oxname{O-Deref}, we have a number of ownership safety
  derivations corresponding to each of the loans for the pointer being
  dereferenced. Since we know we have a dereference, we know that we must have
  derived $\oxnorm{\oxstore}{\oxplaceexpr}{\oxreferentctx[\oxplace]}
  {\oxvaluectx}{\oxvalue}$
  using one of the five dereference rules at the appropriate point
  (\oxname{P-DerefPtr}, \oxname{P-DerefIndexPtrArray},
  \oxname{P-DerefIndexPtrSlice}, \oxname{P-DerefSlicePtrArray}, and
  \oxname{P-DerefSlicePtrSlice}). Each of which share a common premise (at least
  when sufficiently generalized):
  $\oxnorminner{\oxstore}{\oxplaceexprctx}{\oxreferent_2}
  {\oxreferent_3}{\oxvaluectx}{\oxvalue}$. Here, $\oxreferent_2$ corresponds to
  the referent of the pointer we are dereferencing. As such, we know that one of
  the derivations of ownership safety corresponds to that particular referent.
  So, we can apply our induction hypothesis and get that
  $\oxloanpkg{\oxmuta}{\oxplace} \in \oxset{
    \overline{\oxloanpkg{\oxmuta}{\oxplaceexpr^\prime_i}} }$ for the appropriate
  ownership safety derivation numbered i. The final output is the union of all
  of these sets, and thus we can generalize to $\oxloanpkg{\oxmuta}{\oxplace}
  \in \oxset{ \overline{\oxloanpkg{\oxmuta}{\oxplaceexpr^\prime_1}} \oxdotsc
    \overline{\oxloanpkg{\oxmuta}{\oxplaceexpr^\prime_n}} \oxdotsc
    \oxloanpkg{\oxmuta}{\oxplaceexprctx[\oxderef{\oxplace}]}}$.
\end{proof}

\subsection{Preservation under Region Rewriting Lemmas}
\label{sec:rewriting-lemmas}

\begin{oxlemma}{Ownership Safety is Preserved under Region Rewriting}
  {lemma:subtyping-ownership-safety} If $\oxmusafetyinner{\oxemptyctx}{\oxkontctx}{\oxvarctx}
  {\oxmuta}{\overline{\oxplace_e}}{\oxplaceexpr}{\oxset{\overline{\oxloan^\prime}}}$ and
  $\oxtunify{\oxemptyctx}{\oxkontctx}{\oxvarctx}{\oxtype_1}{\oxtype_2}{\oxvarctx^\prime}$ and
  $\oxset{\overline{\oxplace_e}} \subseteq \oxset{\overline{\oxplace_e^\prime}}$ then
  $\oxmusafetyinner{\oxemptyctx}{\oxkontctx}{\oxvarctx^\prime}
  {\oxmuta}{\overline{\oxplace_e}}{\oxplaceexpr}{\oxset{\overline{\oxloan^{\prime\prime}}}}$.
\end{oxlemma}
\begin{proof}
  We proceed by induction on the region rewriting judgement. We note that if
  $\oxrewritemode$ is $=$, then by inspection of the outlives judgement, $\oxvarctx
  = \oxvarctx^\prime$, so the proof follows immediately from the premise. So
  consider when $\oxrewritemode$ is $+$. The only case that doesn't follow
  immediately by induction and application of premises is \oxname{RR-Reference},
  and in this case the only interesting part of the proof is the outlives
  constraint.

  Proceeding by induction on the outlives constraint, the only interesting case
  is \\ \oxname{OL-CombineConcrete}.

  \vspace{1em}
  \noindent \framebox[\textwidth]{ \figuresize
    \begin{mathpar}
      \UCombineLocalProvs
    \end{mathpar}
  } \vspace{1em}

  We want to show that $\oxmusafetyinner{\oxtvarctx}{\oxkontctx}
  {\oxvarctx[\oxcprov_2 \mapsto \oxset{\overline{\oxloan}}]}
  {\oxmuta}{\overline{\oxplace_e}}{\oxplaceexpr}{\oxset{\overline{\oxloan^{\prime\prime}}}}$.
  Proceed by induction on the ownership safety judgement in the premise.

  \vspace{1em}
  \noindent \framebox[\textwidth]{ \figuresize
    \begin{mathpar}
      \OSafePlace
    \end{mathpar}
  } \vspace{1em}

  Let $\oxcprov^\prime$ be an arbitrary region. There are two cases to prove,
  depending which part of the disjunction is true for the premise
  $\oxmusafetyinner{\oxtvarctx}{\oxkontctx}{\oxvarctx}{\oxmuta}{\overline{\oxplace_e}}{\oxplace}
  {\oxset{\overline{\oxloan^\prime}}}$.

  If the first part was true, then we need to show that $\forall
  \oxloanpkg{\oxmuta^\prime}{\oxplaceexprctx[\oxplace^\prime]} \in
  \oxvarctx[\oxcprov_2 \mapsto
    \oxset{\overline{\oxloan^{\prime\prime}}}](\oxcprov^\prime).\ (\oxmuta =
  \oxmut \vee \oxmuta^\prime = \oxmut) \implies
  \oxrelevant{\oxplace^\prime}{\oxplace}$. This is only interesting when
  $\oxcprov^\prime = \oxcprov_2$. Using the fact that $\oxvarctx[\oxcprov_2
    \mapsto \oxset{\oxloans}](\oxcprov_2) = \oxvarctx(\oxcprov_1) \cup
  \oxvarctx(\oxcprov_2)$, we need to show $\forall
  \oxloanpkg{\oxmuta^\prime}{\oxplaceexprctx[\oxplace^\prime]} \in
  \oxvarctx(\oxcprov_1) \cup \oxvarctx(\oxcprov_2).\ (\oxmuta = \oxmut \vee
  \oxmuta^\prime = \oxmut) \implies \oxrelevant{\oxplace^\prime}{\oxplace}$.
  This is immediate if we can show that $\oxcprov_1$ and $\oxcprov_2$ are not
  excluded. This is immediate from the region not reborrowed judgement. For
  $\oxcprov_1$ or $\oxcprov_2$ to be excluded, for each reference
  $\oxplace^{\prime\prime}$ that has $\oxcprov_1$ or $\oxcprov_2$, there would
  have to be a loan of the form $\oxplaceexprctx[\oxderef
    \oxplace^{\prime\prime}]$, but such loans are precisely what the region not
  reborrowed judgement excludes.

  If the second part was true, then we can prove the second part immediately
  from the hypothesis, because the types of references are unchanged,
  $\oxkontctx$ is unchanged, and the exclusion list can only grow.

  \vspace{1em}
  \noindent \framebox[\textwidth]{ \figuresize
    \begin{mathpar}
      \ODeref
    \end{mathpar}
  } \vspace{1em}

  Firstly, note that the exclusion list will be equal if $\oxcprov \neq
  \oxcprov_2$, and will be potentially larger if $\oxcprov = \oxcprov_2$.
  Therefore we can immediately apply our induction hypothesis to get ownership
  safety for $\oxplaceexprctx[\oxplaceexpr_i]$ under
  $\oxvarctx[\oxcprov_2 \mapsto \oxloans]$.

  For the rest of the case, apply identical reasoning to that in the
  \oxname{O-SafePlace} case.

  \vspace{1em}
  \noindent \framebox[\textwidth]{ \figuresize
    \begin{mathpar}
      \ODerefAbs
    \end{mathpar}
  } \vspace{1em}

  Since $\oxtvarctx = \oxemptyctx$, there are no valid reference types that have
  an abstract region, meaning the first hypothesis is a contradiction.

\end{proof}

\begin{oxlemma}{Type Computation is Preserved under Region Rewriting}
  {lemma:subtyping-type-computation} If
  $\oxcomputetynoprov{\oxemptyctx}{\oxnewframe{\oxvarctx}{\oxframe}}{\oxmuta}{\oxplaceexpr}{\oxsitype}$
  and
  $\oxtunify{\oxemptyctx}{\oxkontctx}{\oxvarctx}{\oxtype_1}{\oxtype_2}{\oxvarctx^\prime}$
  then
  $\oxcomputetynoprov{\oxemptyctx}{\oxnewframe{\oxvarctx^\prime}{\oxframe}}{\oxmuta}{\oxplaceexpr}{\oxsitype}$.
\end{oxlemma}

\begin{proof}
  The proof is immediate by inspection of the type computation judgement,
  because the only things considered in the judgement are the types of places in
  $\oxvarctx$, which cannot change through the region rewriting judgement (in
  other words, $\oxdomain{\oxvarctx} = \oxdomain{\oxvarctx^\prime}$).
\end{proof}

\begin{oxlemma}{Outlives is Preserved under Region Rewriting}
  {lemma:subtyping-outlives} If $\oxrunify{\oxemptyctx}{\oxkontctx}{\oxnewframe{\oxvarctx}{\oxframe}}{\oxcprov_1}{\oxcprov_2}{\oxnewframe{\oxvarctx_o}{\oxframe_o}}$ and
  $\oxtunify{\oxemptyctx}{\oxkontctx}{\oxvarctx}{\oxtype_1}{\oxtype_2}{\oxvarctx^\prime}$ then
  $\oxrunify{\oxemptyctx}{\oxkontctx}{\oxnewframe{\oxvarctx^\prime}{\oxframe}}{\oxcprov_1}{\oxcprov_2}{\oxnewframe{\oxvarctx_o^\prime}{\oxframe_o^\prime}}$.
\end{oxlemma}
\begin{proof}
  We proceed by induction on the outlives judgement. The only interesting cases
  are \oxname{OL-CombineConcrete} and \oxname{OL-CheckConcrete}. In both cases,
  the only non immediate premise is the region not reborrowed judgement. Proceed
  by induction over the region rewriting hypothesis, and in the interesting case
  \oxname{RR-Reference}, proceed by induction over the outlives judgement. In
  this case, the only interesting case is when the loan sets in $\oxvarctx$
  potentially change, \oxname{OL-CombineConcrete}. But note that no new loans
  are generated, only loans are copied into other sets. For this reason, the
  region not reborrowed judgements we already have are sufficient, because these
  loans that are now potentially in two loan sets were already found to not
  contain any problematic reborrows.
\end{proof}

\begin{oxlemma}{Region Rewriting is Preserved under Region Rewriting}
  {lemma:subtyping-subtyping} If
  $\oxtunify{\oxemptyctx}{\oxkontctx}{\oxnewframe{\oxvarctx}{\oxframe}}{\oxtype_1}{\oxtype_2}{\oxnewframe{\oxvarctx_o}{\oxframe_o}}$ and
  $\oxtunify{\oxemptyctx}{\oxkontctx}{\oxvarctx}{\oxtype_1^\prime}{\oxtype_2^\prime}{\oxvarctx^\prime}$ then
  $\oxtunify{\oxemptyctx}{\oxkontctx}{\oxnewframe{\oxvarctx^\prime}{\oxframe}}{\oxtype_1}{\oxtype_2}{\oxnewframe{\oxvarctx_o^\prime}{\oxframe_o^\prime}}$.
\end{oxlemma}
\begin{proof}
  Proceed by induction over the the region rewriting judgement. The only
  interesting case is \oxname{RR-Reference}, for which we just apply
  \Lemma{lemma:subtyping-outlives}.
\end{proof}

\begin{oxlemma}{Region Rewriting is Preserved by Garbage Collecting Loans}
  {lemma:subtyping-gc} If
  $\oxtunify{\oxemptyctx}{\oxkontctx}{\oxvarctx}{\oxtype_1^\prime}{\oxtype_2^\prime}{\oxvarctx^\prime}$
  then
  $\oxtunify{\oxemptyctx}{\oxkontctx}{\oxgcloans{\oxkontctx}{\oxvarctx}}{\oxtype_1^\prime}{\oxtype_2^\prime}{\oxvarctx^{\prime\prime}}$.
\end{oxlemma}
\begin{proof}
  We proceed by induction over the region rewriting judgement, in which the only
  interesting case is \oxname{RR-Reference}. We then proceed by induction over
  the outlives relation, in which the only interesting cases are
  \oxname{OL-CombineConcrete} and \oxname{OL-CheckConcrete}. The only
  interesting part of the judgement is the region not reborrowed, and this is
  immediate because garbage collection will only potentially remove some loans.
\end{proof}

\begin{oxlemma}{Closure Body Typing is Preserved under Region Rewriting}
  {lemma:subtyping-closure-typing} If
  $\oxtypjudge{\oxglobalctx}{\oxemptyctx}{\oxkontctx}{\oxnewframe{\oxvarctx}{\oxframe}}
  {\oxexpr}{\oxtype}{\oxnewframe{\oxvarctx_o}{\oxframe_o}}$ and
  $\oxtunify{\oxemptyctx}{\oxkontctx}{\oxvarctx}{\oxtype_1}{\oxtype_2}{\oxvarctx^\prime}$
  then
  $\oxtypjudge{\oxglobalctx}{\oxemptyctx}{\oxkontctx}{\oxnewframe{\oxvarctx^\prime}{\oxframe}}
  {\oxexpr}{\oxtype}{\oxnewframe{\oxvarctx_o^\prime}{\oxframe_o^\prime}}$ and
  $\oxtunify{\oxemptyctx}{\oxkontctx}{\oxvarctx_o^\prime}{\oxtype_1}{\oxtype_2}{\oxvarctx_o^{\prime\prime}}$.
\end{oxlemma}
\begin{proof}
  Proceed by induction over the typing derivation for $\oxexpr$.

  The \oxname{T-Abort}, \oxname{T-Function}, \oxname{T-Unit}, \oxname{T-u32},
  \oxname{T-True}, and \oxname{T-False} cases follow immediately.

  The \oxname{T-LetRegion}, \oxname{T-While}, \oxname{T-Closure},
  \oxname{T-Tuple}, \oxname{T-Array}, \oxname{T-Slice}, \oxname{T-Drop},
  \oxname{T-Left}, and \oxname{T-Right} cases all follow immediately from the
  induction hypothesis.

  The \oxname{T-Seq} case follows from the induction hypothesis and
  \Lemma{lemma:subtyping-gc}.

  The \oxname{T-Branch}, \oxname{T-Let}, and \oxname{T-Match} cases follow from
  the induction hypothesis, \Lemma{lemma:subtyping-subtyping}, and
  \Lemma{lemma:subtyping-gc}. Note the reborrow restriction follows immediately
  from the fact that rewriting can at most union together loan sets, which means
  the overall loans considered for the region not reborrowed judgement are the
  same in the context after rewriting.

  The \oxname{T-Move}, \oxname{T-Copy}, \oxname{T-Borrow},
  \oxname{T-BorrowIndex}, \oxname{T-BorrowSlice}, \oxname{T-IndexCopy},
  \oxname{T-ForArray}, and \oxname{T-ForSlice} cases follow from the induction
  hypothesis, \Lemma{lemma:subtyping-ownership-safety}, and
  \Lemma{lemma:subtyping-type-computation}.

  The \oxname{T-AppFunction} and \oxname{T-AppClosure} cases follow from the
  induction hypothesis, \Lemma{lemma:subtyping-outlives}, and the fact that
  context, region, and type well formedness aren't affected by changes in the
  loan sets.
\end{proof}

\begin{oxlemma}{Value Typing is Preserved under Region Rewriting}
  {lemma:subtyping-value-typing} If
  $\oxtypjudge{\oxglobalctx}{\oxemptyctx}{\oxkontctx}{\oxvarctx}
  {\oxvalue}{\oxtype}{\oxvarctx}$ and
  $\oxtunify{\oxemptyctx}{\oxkontctx}{\oxvarctx}{\oxtype_1}{\oxtype_2}{\oxvarctx^\prime}$
  then $\oxtypjudge{\oxglobalctx}{\oxemptyctx}{\oxkontctx}{\oxvarctx^\prime}
  {\oxvalue}{\oxtype}{\oxvarctx^\prime}$.
\end{oxlemma}

\begin{proof}
  We proceed by induction on the value typing.

  \vspace{1em}
  \noindent \framebox[\textwidth]{ \figuresize
    \begin{mathpar}
      \inferrule[T-Pointer]{
        \oxreferentvalidity{\oxglobalctx}{\oxvarctx}{\oxreferentctx[\oxplace]}{\oxxitype} \\
        \oxloanpkg{\oxmuta}{\oxplace} \in \oxvarctx(\oxcprov) \\
      }{
        \oxtypjudge{\oxglobalctx}{\oxemptyctx}{\oxkontctx}{\oxvarctx}{
          \oxptr{\oxreferentctx[\oxplace]} }{
          \oxtref{\oxcprov}{\oxmuta}{\oxxitype} }{\oxvarctx} }
    \end{mathpar}
  } \vspace{1em}

  The \oxname{T-Pointer} case is immediate, because by inspection of the
  referent well formedness, there is no reliance on loan sets, and the loan is
  preserved since by inspection of the rewriting judgement, the loan sets either
  stay the same or potentially grow.
  
  \vspace{1em}
  \noindent \framebox[\textwidth]{ \figuresize
    \begin{mathpar}
      \TClosureVal
    \end{mathpar}
  } \vspace{1em}

  First, we invert the stack frame typing hypothesis to get that $\forall \oxid
  \in \oxdomain{\oxstackframe}.\;
  \oxtypjudge{\oxglobalctx}{\oxemptyctx}{\oxkontctx}{\oxnewframe{\oxvarctx}{\oxframe_c}}{\oxstackframe(\oxid)}{\oxframe_c(\oxid)}{\oxnewframe{\oxvarctx}{\oxframe_c}}$.
  We can apply the induction hypothesis to each of these statements, and apply
  \oxname{WF-Frame} to get
  $\oxsubstorevalidity{\oxglobalctx}{\oxvarctx^\prime}{\oxstackframe_c}{\oxframe_c}$.

  For the typing of the body, we can apply \Lemma{lemma:subtyping-closure-typing}.
\end{proof}

\begin{oxlemma}{Stack Well-Formedness is Preserved under Region Rewriting}
  {lemma:stack-validity-region-rewriting} If
  $\oxstorevalidity{\oxglobalctx}{\oxvarctx}{\oxstore}$ and
  $\oxtunify{\oxemptyctx}{\oxkontctx}{\oxvarctx}{\oxtype_1}{\oxtype_2}{\oxvarctx^\prime}$
  then $\oxstorevalidity{\oxglobalctx}{\oxvarctx^\prime}{\oxstore}$.
\end{oxlemma}

\begin{proof}
  We proceed by induction on the stack typing derivation.
  
  \vspace{1em}
  \noindent \framebox[\textwidth]{ \figuresize
    \begin{mathpar}
      \VStack \and \VStackEmpty
    \end{mathpar}
  } \vspace{1em}

  The \oxname{WF-StackEmpty} case is immediate. In the \oxname{WF-StackFrame}
  case, we get the well formedness in the premise from our induction hypothesis.
  What's left to show is that for all of the values $\oxvalue$ in the stack
  frame, they remain well typed in $\oxvarctx^\prime$. This follows from
  applying \Lemma{lemma:subtyping-value-typing}.
\end{proof}

\subsection{Preservation under Drops and Garbage Collection Lemmas}
\label{sec:environment-relation-lemmas}

\begin{oxlemma}{Values Change Environments in Limited Ways}{lemma:values-dont-change}
  If \oxtypjudge{\oxglobalctx}{\oxtvarctx}{\oxkontctx}{\oxvarctx}{\oxvalue}
  {\oxtype}{\oxvarctx^\prime}, then
  $\oxsubtypectx{\oxglobalctx}{\oxtvarctx}{\oxvarctx}{\oxvarctx^\prime}$.
\end{oxlemma}

\begin{proof}
  We proceed by induction on the structure of the typing derivation. Since we
  assume that the expression being typed is a value, we need only consider the
  cases that can be used to type a value.

  For many cases, the output environments are precisely the input environments,
  and thus this holds immediately. These cases are \oxname{T-Unit},
  \oxname{T-u32}, \oxname{T-True}, \oxname{T-False}, \oxname{T-Pointer},
  \oxname{T-Function}, \oxname{T-ClosureValue}, and \oxname{T-Dead}.

  For \oxname{T-Tuple}, \oxname{T-Array}, \oxname{T-Left}, and \oxname{T-Right},
  knowing that we have a value means that all of the subterms are themselves
  values, and thus we can apply our induction hypothesis to them in sequence
  (relying on the transitivity of $\oxsubtype$ for stack typings).

  This leaves us with one remaining case: \oxname{T-Drop}.

  \vspace{1em}
  \noindent \framebox[\textwidth]{ \figuresize
    \begin{mathpar}
      \TDrop
    \end{mathpar}
  } \vspace{1em}

  For \oxname{T-Drop}, we apply our induction hypothesis to
  $\oxtypjudge{\oxglobalctx}{\oxtvarctx}{\oxkontctx}{
    \oxtupdate{\oxvarctx}{\oxplace}{\oxsidtype_\oxplace}
  }{\oxexpr}{\oxsxtype}{\oxvarctx_f}$ which tells us that
  $\oxsubtypectx{\oxglobalctx}{\oxtvarctx}{
    \oxtupdate{\oxvarctx}{\oxplace}{\oxsidtype_\oxplace} }{\oxvarctx_f}$. Then,
  by \oxname{R-Env}, we have that
  $\oxsubtypectx{\oxglobalctx}{\oxtvarctx}{\oxvarctx}
  {\oxtupdate{\oxvarctx}{\oxplace}{\oxsidtype_\oxplace}}$. Then, by
  transitivity, we have
  $\oxsubtypectx{\oxglobalctx}{\oxtvarctx}{\oxvarctx}{\oxvarctx_f}$.
\end{proof}

\begin{oxlemma}{Type Computation is Preserved in Related Environments}
  {lemma:tc-related-contexts} If $\oxsubtypectx{\oxglobalctx}{\oxtvarctx}
  {\oxvarctx}{\oxvarctx^\prime}$ and
  $\oxcomputety{\oxtvarctx}{\oxvarctx}{\oxmuta}{\oxplaceexprctx[\oxplace]}{\oxtype}{\oxset{\overline{\oxprov}}}$
  and $\oxvarctx(\oxplace) = \oxvarctx^\prime(\oxplace)$, then
  $\oxcomputety{\oxtvarctx}{\oxvarctx^\prime}{\oxmuta}{\oxplaceexprctx[\oxplace]}{\oxtype}{\oxset{\overline{\oxprov}}}$.
\end{oxlemma}
\begin{proof}
  We proceed by induction on the type computation derivation. \oxname{TC-Var}
  follows immediately by the same type hypothesis, and \oxname{TC-Proj} follows
  from applying the induction hypothesis. All that is left is \oxname{TC-Deref}.

  \vspace{1em}
  \noindent \framebox[\textwidth]{ \figuresize
    \begin{mathpar}
      \TCDeref
    \end{mathpar}
  } \vspace{1em}

  First, we can apply the induction hypothesis to get the type computation for
  $\oxplaceexpr$. Then, all that's left is to show the outlives constraint, but
  this is immediate because $\oxtvarctx$ is unchanged and both $\oxvarctx$ and
  $\oxvarctx^\prime$ have the exact same domains.

\end{proof}

\begin{oxlemma}{Ownership Safety Preserved in Related Environments}
  {lemma:ownership-safety-related-envs} If
  $\oxmusafetyinner{\oxtvarctx}{\oxkontctx}{\oxvarctx}{\oxmuta}{\overline{\oxplace_e}}{\oxplaceexpr}
  {\oxset{\overline{\oxloan}}}$ and
  $\oxsubtypectx{\oxglobalctx}{\oxtvarctx}{\oxvarctx}{\oxvarctx^\prime}$ and
  $\oxcomputetynoprov{\oxtvarctx}{\oxvarctx^\prime}{\oxmuta}{\oxplaceexpr}{\oxxitype}$
  and $\oxplaceexpr = \oxplaceexprctx[\oxplace_\oxplaceexpr]$ and
  $\oxvarctx(\oxplace_\oxplaceexpr) = \oxvarctx^\prime(\oxplace_\oxplaceexpr)$,
  then
  $\oxmusafetyinner{\oxtvarctx}{\oxkontctx}{\oxvarctx^\prime}{\oxmuta}{\overline{\oxplace_e}}{\oxplaceexpr}
  {\oxset{\overline{\oxloan}}}$.
\end{oxlemma}

\begin{proof}
  We proceed by induction on the $\omega$-safety derivation, for which there are
  three cases to consider.

  \vspace{1em}
  \noindent \framebox[\textwidth]{ \figuresize
    \begin{mathpar}
      \OSafePlace
    \end{mathpar}
  } \vspace{1em}

  We'd like to show that \oxname{O-SafePlace} can be applied with context
  $\oxvarctx^\prime$. First, note that for any $\oxcprov^\prime$, if the right
  side of the or is true for $\oxvarctx$ with $\overline{\oxplace}$ then it will
  be true for $\oxvarctx^\prime$ with $\overline{\oxplace}$. That is, if all of
  the pointers with region $\oxcprov^\prime$ in $\oxvarctx$ are in the exclusion
  list $\overline{\oxplace}$, then all of the pointers with region
  $\oxcprov^\prime$ in $\oxvarctx^\prime$ are also in the exclusion list
  $\overline{\oxplace}$. Note that $\oxkontctx$ is unchanged between the two.
  Therefore, the only cases we need to consider are where $\oxcprov^\prime$
  occurs in pointers in $\oxvarctx$ and $\oxvarctx^\prime$ that do not occur in
  $\overline{\oxplace}$.

  Since the only allowed change to loan sets is emptying, and an emptied loan
  set has the left side of the disjunction as vacuously true, and if the loan
  set is the same we have the condition from the ownership safety in the
  premise, we are done.

  \vspace{1em}
  \noindent \framebox[\textwidth]{ \figuresize
    \begin{mathpar}
      \ODeref
    \end{mathpar}
  } \vspace{1em}

  Firstly, we have that $\oxvarctx(\oxplace_i) = \oxvarctx^\prime(\oxplace_i)$,
  because $\oxvarctx^\prime(\oxplace_i)$ must be an initialized type by the type
  computation premise, and the only changes in types between $\oxvarctx$ and
  $\oxvarctx^\prime$ allowed by the environment relation is dropping some types
  to uninitialized.

  Second, note that $\oxvarctx^\prime(\oxcprov) = \oxvarctx(\oxcprov)$ since
  $\oxvarctx^\prime(\oxplace)$ being a reference with region $\oxcprov$ means we
  can't empty the loan set. So we proceed by applying the induction hypothesis
  for all $n$ loans, noting that the type computation requirement follows from
  the well formedness of $\oxvarctx^\prime$.

  Finally, we have to show the statement about no conflicting loans, but here
  the argument is identical to that in the \oxname{O-SafePlace} case. If the
  loan set is empty then we're done, otherwise we just use the ownership safety
  premise.

  \vspace{1em}
  \noindent \framebox[\textwidth]{ \figuresize
    \begin{mathpar}
      \ODerefAbs
    \end{mathpar}
  } \vspace{1em}

  This case proceeds similarly to the \oxname{O-Deref} case, but with an added
  application of \Lemma{lemma:tc-related-contexts} to get the type computation,
  and no application of any induction hypothesis.
\end{proof}

\begin{oxlemma}{Types Are Well Formed in Related Environments}
  {lemma:types-well-formed} If $\oxsubtypectx{\oxglobalctx}{\oxtvarctx}
  {\oxvarctx}{\oxvarctx^\prime}$ and $\oxtypevalidity{\oxglobalctx}{\oxtvarctx}
  {\oxvarctx}{\oxxitype}$ and $\forall \oxcprov$ that occur in $\oxxitype$,
  $\oxvarctx(\oxcprov) = \oxvarctx^\prime(\oxcprov)$, then
  $\oxtypevalidity{\oxglobalctx}{\oxtvarctx} {\oxvarctx^\prime}{\oxxitype}$.
\end{oxlemma}

\begin{proof}
  We proceed by induction on the type well formedness derivation. The only case
  that doesn't follow directly from induction and the fact that $\oxtvarctx$ and
  $\oxkontctx$ are unchanged between the two related environments is
  \oxname{WF-Ref}.

  \vspace{1em}
  \noindent \framebox[\textwidth]{ \figuresize
    \begin{mathpar}
      \VRef
    \end{mathpar}
  } \vspace{1em}

  Firstly we apply our induction hypothesis to get that
  $\oxtypevalidity{\oxglobalctx}{\oxtvarctx}
  {\oxvarctx^\prime}{\oxxitype_\oxplaceexpr}$. What's left to show is the loan
  set condition on $\oxcprov$. If $\oxvarctx^\prime(\oxcprov) = \emptyset$, then
  we're done. Otherwise, we just need that the type computation still holds,
  which we get from \Lemma{lemma:tc-related-contexts}. We know the places in
  these place expressions all have the same type in $\oxvarctx$ and
  $\oxvarctx^\prime$ because between these two contexts the only changes allowed
  that could cause problems here are dropping one of these places, but then
  $\oxvarctx^\prime$ would not be well formed since there would be an invalid
  loan.

\end{proof}

\begin{oxlemma}{Related Environments Remain Well-Formed}
  {lemma:related-contexts-well-formed} If
  $\oxsubtypectx{\oxglobalctx}{\oxtvarctx} {\oxvarctx}{\oxvarctx^\prime}$ and
  $\oxctxswellformed{\oxglobalctx}{\oxtvarctx}
  {\oxnewframe{\oxvarctx}{\oxframe_{c}}}{\oxkontctx}$ then
  $\oxctxswellformed{\oxglobalctx}{\oxtvarctx}
  {\oxnewframe{\oxvarctx^\prime}{\oxframe_{c}}}{\oxkontctx}$.
\end{oxlemma}

\begin{proof}
  From the well formedness of $\oxnewframe{\oxvarctx}{\oxframe_c}$, we know that
  the places and disjointness conditions both hold. We also know that the occurs
  in restriction holds, because we at most have the same alive types. By
  \Lemma{lemma:types-well-formed}, noting that
  $\oxsubtypectx{\oxglobalctx}{\oxtvarctx}{\oxnewframe{\oxvarctx}{\oxframe_c}}{\oxnewframe{\oxvarctx^\prime}{\oxframe_c}}$
  is immediate, we know that the types remain well formed in the environment. We
  also have the well formedness of $\oxvarctx^\prime$ as a premise of the
  related environments judgement. All that's left to show is the loan set
  condition. But for this all we have to show is that each place computes to
  some type, which follows from \Lemma{lemma:tc-related-contexts}. We know the
  types of the places in each place expression remain the same because the only
  allowed changes between $\oxvarctx$ and $\oxvarctx^\prime$ are that places can
  be dropped and loan sets emptied, but if one such place was dropped, then
  $\oxvarctx^\prime$ would have not been well formed.
\end{proof}

\begin{oxlemma}{Related Input Environments Produce Similar Output Environments}
  {lemma:related-input-output-differences} If:
  \begin{itemize}
  \item
    $\oxtypjudge{\oxglobalctx}{\oxtvarctx}{\oxkontctx}{\oxvarctx_1}{\oxexpr_1}{\oxtype_1}{\oxvarctx_2}$
  \item
    $\oxtypjudge{\oxglobalctx}{\oxtvarctx}{\oxkontctx}{\oxvarctx_1}{\oxexpr_2}{\oxtype_2}{\oxvarctx_3}$
  \item
    $\oxsubtypectx{\oxglobalctx}{\oxtvarctx}{\oxvarctx_1}{\oxvarctx_1^\prime}$
  \item
    $\oxtypjudge{\oxglobalctx}{\oxtvarctx}{\oxkontctx}{\oxvarctx_1^\prime}{\oxexpr_1}{\oxtype_1}{\oxvarctx_2^\prime}$
  \item
    $\oxtypjudge{\oxglobalctx}{\oxtvarctx}{\oxkontctx}{\oxvarctx_1^\prime}{\oxexpr_2}{\oxtype_2}{\oxvarctx_3^\prime}$
  \item
    $\oxsubtypectx{\oxglobalctx}{\oxtvarctx}{\oxvarctx_2}{\oxvarctx_2^\prime}$
  \item
    $\oxsubtypectx{\oxglobalctx}{\oxtvarctx}{\oxvarctx_3}{\oxvarctx_3^\prime}$
  \item $\forall \oxid \in \oxdomain{\oxvarctx_2}$, $\oxvarctx_2(\oxid) =
    \oxvarctx_3(\oxid)$ and $\oxvarctx_2^\prime(\oxid) =
    \oxvarctx_3^\prime(\oxid)$
  \item $\forall \oxcprov$ that occur in $\oxexpr_1$ or $\oxexpr_2$ or
    $\oxtype_1$ or $\oxtype_2$, $\oxvarctx_1(\oxcprov) =
    \oxvarctx_1^\prime(\oxcprov)$
  \end{itemize}

  then $\forall \oxcprov \in \oxdomain{\oxvarctx_1}$, if
  $\oxvarctx_2^\prime(\oxcprov) = \emptyset$ and $\oxvarctx_3^\prime(\oxcprov)
  \neq \emptyset$, then $\oxvarctx_2(\oxcprov) = \emptyset$, and if
  $\oxvarctx_3^\prime(\oxcprov) = \emptyset$ and $\oxvarctx_2^\prime(\oxcprov)
  \neq \emptyset$, then $\oxvarctx_3(\oxcprov) = \emptyset$.
\end{oxlemma}
\begin{proof}
  The proofs for both statements in the conclusion follow identically, so
  without loss of generality it suffices to show that if
  $\oxvarctx_2^\prime(\oxcprov) = \emptyset$ and $\oxvarctx_3^\prime(\oxcprov)
  \neq \emptyset$, then $\oxvarctx_2(\oxcprov) = \emptyset$. Note there are two
  cases to consider: that the loan set was empty all along, or that the loan set
  was at some point non empty, but then got garbage collected.

  First, at some point between $\oxvarctx_1^\prime$ and $\oxvarctx_2^\prime$,
  $\oxcprov$ mapped to a non empty set of loans but then was garbage collected.
  In this case, $\oxvarctx_2^\prime$ must not contain any references that
  contain $\oxcprov$, since otherwise it would have been invalid to garbage
  collect $\oxcprov$. But then since $\oxvarctx_2^\prime$ and
  $\oxvarctx_3^\prime$ agree on types, it must be the case that it was also
  garbage collected in $\oxvarctx_3^\prime$, which is a contradiction with the
  fact that $\oxvarctx_3^\prime(\oxcprov)$ is non empty, so this case is
  impossible.

  Second, at each step of the derivation between $\oxvarctx_1^\prime$ and
  $\oxvarctx_2^\prime$, $\oxcprov$ mapped to empty. If $\oxvarctx_1(\oxcprov)$
  also was empty, then this means $\oxvarctx_2(\oxcprov)$ is also empty, and
  we're done. Otherwise, $\oxcprov$ was garbage collected between $\oxvarctx_1$
  and $\oxvarctx_1^\prime$. But then $\oxcprov$ must be free in $\oxexpr_2$ for
  loans to have been added between $\oxvarctx_1^\prime$ and
  $\oxvarctx_3^\prime$, which means the loan set could not have been emptied
  between $\oxvarctx_1$ and $\oxvarctx_1^\prime$, which is a contradiction.
\end{proof}

\begin{oxlemma}{Outlives Preserves Related Environments}
  {lemma:outlives-related} If
  $\oxrunify{\oxtvarctx}{\oxkontctx}{\oxvarctx}{\oxprov_1}{\oxprov_2}{\oxvarctx_o}$, and
  $\oxsubtypectx{\oxglobalctx}{\oxtvarctx}{\oxvarctx}{\oxvarctx^\prime}$ and
  $\oxctxswellformed{\oxglobalctx}{\oxtvarctx}{\oxvarctx_o}{\oxkontctx}$ and
  $\oxvarctx(\oxprov_1) = \oxvarctx^\prime(\oxprov_1)$ and $\oxvarctx(\oxprov_2)
  = \oxvarctx^\prime(\oxprov_2)$, then
  $\oxrunify{\oxtvarctx}{\oxkontctx}{\oxvarctx^\prime}{\oxprov_1}{\oxprov_2}{\oxvarctx_o^\prime}$,
  and
  $\oxsubtypectx{\oxglobalctx}{\oxtvarctx}{\oxvarctx_o}{\oxvarctx_o^\prime}$.
  and $\oxvarctx_o(\oxprov_1) = \oxvarctx_o^\prime(\oxprov_1)$ and
  $\oxvarctx_o(\oxprov_2) = \oxvarctx_o^\prime(\oxprov_2)$
\end{oxlemma}

\begin{proof}
  Proceed by induction on the outlives derivation. \oxname{OL-Refl},
  \oxname{OL-Trans}, \oxname{OL-AbstractConcrete}, and \oxname{OL-BothAbstract}
  are immediate.

  \oxname{OL-ConcreteAbstract} follows from additionally applying
  \Lemma{lemma:tc-related-contexts}. The condition on the place having the same
  type follows from the fact that $\oxplaceexpr$ is a loan and
  $\oxvarctx^\prime(\oxcprov)$ is not emptied, so we could not have dropped the
  place.

  \oxname{OL-CheckConcrete} is immediate, because the occurs before condition is
  unaffected since the domains are equal, and the region not reborrowed
  judgement is unaffected by adding loans that are already in other loan sets.

  This leaves two cases which proceed similarly: \oxname{OL-CombineConcrete} and
  \oxname{OL-CombineConcreteUnrestricted}.
  
  \vspace{1em}
  \noindent \framebox[\textwidth]{ \figuresize
    \begin{mathpar}
      \UCombineLocalProvs
    \end{mathpar}
  } \vspace{1em}

  Since $\oxvarctx^\prime(\oxcprov_1) = \oxvarctx(\oxcprov_1)$ and
  $\oxvarctx^\prime(\oxcprov_2) = \oxvarctx(\oxcprov_2)$,
  $\oxvarctx^\prime(\oxcprov_1) \cup \oxvarctx^\prime(\oxcprov_2) =
  \oxvarctx(\oxcprov_1) \cup \oxvarctx^\prime(\oxcprov_2)$. The region not
  reborrowed judgement is unaffected by adding loans that are already in other
  loan sets, so those conditions are also still true. The rest of the conditions
  are immediate: the equality on $\oxcprov_1$ and $\oxcprov_2$'s loan sets, the
  closure restriction since types are at most the same, and well formedness.
\end{proof}

\begin{oxlemma}{Related Environments Preserved by Region Rewriting}
  {lemma:unified-types-well-formed} If
  $\oxtunify{\oxtvarctx}{\oxkontctx}{\oxvarctx}{\oxsitype_1}{\oxsitype_2}{\oxvarctx_o}$, and
  $\oxsubtypectx{\oxglobalctx}{\oxtvarctx}{\oxvarctx}{\oxvarctx^\prime}$ and
  $\oxctxswellformed{\oxglobalctx}{\oxtvarctx}{\oxvarctx_o}{\oxkontctx}$ and $\forall
  \oxcprov$ that occur in $\oxsitype_1$ or $\oxsitype_2$, $\oxvarctx(\oxcprov) =
  \oxvarctx^\prime(\oxcprov)$, then
  $\oxtunify{\oxtvarctx}{\oxkontctx}{\oxvarctx^\prime}{\oxsitype_1}{\oxsitype_2}{\oxvarctx_o^\prime}$,
  and
  $\oxsubtypectx{\oxglobalctx}{\oxtvarctx}{\oxvarctx_o}{\oxvarctx_o^\prime}$,
  and $\forall \oxcprov$ that occur in $\oxsitype_1$ or $\oxsitype_2$,
  $\oxvarctx_o(\oxcprov) = \oxvarctx_o^\prime(\oxcprov)$.
\end{oxlemma}

\begin{proof}
  Proceed by induction on the region rewriting derivation. The only interesting
  case is \oxname{RR-Reference}, which proceeds by
  \Lemma{lemma:outlives-related} in addition to applying the induction
  hypothesis.
\end{proof}

\begin{oxlemma}{Expression Typing Preserved in Related Environments}
  {lemma:expression-typing-related-envs} Let $\oxexpr$ be a surface expression
  as defined on page 1. If
  $\oxtypjudge{\oxglobalctx}{\oxtvarctx}{\oxkontctx}{\oxnewframe{\oxvarctx}{\oxframe}}
  {\oxexpr}{\oxtype}{\oxnewframe{\oxvarctx_o}{\oxframe_o}}$ and
  $\oxsubtypectx{\oxglobalctx}{\oxtvarctx}{\oxnewframe{\oxvarctx}{\oxframe}}
  {\oxnewframe{\oxvarctx^\prime}{\oxframe}}$ and $\oxfvars{\oxexpr} =
  \overline{\oxid_f} \subseteq \oxdomain{\oxframe}|_\oxid$ and $\forall \oxcprov
  \in \oxfprovs{\oxexpr}.$ $\oxcprov \in \oxdomain{\oxframe}$, and $\forall
  \oxcprov$ that occur a type in $\overline{\oxframe(\oxid_f)}$,
  $\oxvarctx(\oxcprov) = \oxvarctx^\prime(\oxcprov)$ then
  $\oxtypjudge{\oxglobalctx}{\oxtvarctx}{\oxkontctx}{\oxnewframe{\oxvarctx^\prime}{\oxframe}}
  {\oxexpr}{\oxtype}{\oxnewframe{\oxvarctx_o^\prime}{\oxframe_o}}$ and
  $\oxsubtypectx{\oxglobalctx}{\oxtvarctx}{\oxnewframe{\oxvarctx_o}{\oxframe_o}}
  {\oxnewframe{\oxvarctx_o^\prime}{\oxframe_o}}$ and $\forall \oxcprov$ that
  occur a type in $\overline{\oxframe(\oxid_f)}$, $\oxvarctx_o(\oxcprov) =
  \oxvarctx_o^\prime(\oxcprov)$.
\end{oxlemma}
\begin{proof}
  Proceed by induction on the typing derivation for $e$. In the cases of
  \oxname{T-Abort}, \oxname{T-Function}, \oxname{T-Unit}, \oxname{T-u32},
  \oxname{T-True}, and \oxname{T-False}, the results are immediate.

  In the cases of \oxname{T-LetRegion}, \oxname{T-While}, \oxname{T-ForArray},
  \oxname{T-ForSlice}, \oxname{T-Closure}, \oxname{T-Left}, and
  \oxname{T-Right}, they all follow immediately from induction hypotheses.

  For each of the following cases, the convention is that the statement in the
  box is our assumption, and we want to prove the same statement with
  $\oxvarctx^\prime$ replaced for each $\oxvarctx$.

  \vspace{1em}
  \noindent \framebox[\textwidth]{ \inferrule[T-Tuple]{
      \forall i \in \oxset{1 \oxdots n}. \;
      \oxtypjudge{\oxglobalctx}{\oxtvarctx}{
        \oxextendctx{\oxkontctx}{
          \oxkontctxentry{\oxsitype_1} \oxdotsc \oxkontctxentry{\oxsitype_{i-1}}
        }
      }{
        \oxnewframe{\oxvarctx_{i-1}}{\oxframe_{i-1}}
      }{ \oxexpr_i
      }{\oxsitype_i}{\oxnewframe{\oxvarctx_i}{\oxframe_i}} 
    }{
      \oxtypjudge{\oxglobalctx}{\oxtvarctx}{\oxkontctx}{\oxnewframe{\oxvarctx_0}{\oxframe_0}}{
        \oxprod{\oxexpr_1 \oxdotsc \oxexpr_n} }{\oxtprod{\oxsitype_1 \oxdotsc
          \oxsitype_n}}{ \oxnewframe{\oxvarctx_n}{\oxframe_n} } } }

  We have $n$ induction hypotheses, each giving us the properties for input
  context $\oxnewframe{\oxvarctx^\prime_{i-1}}{\oxframe^\prime_{i-1}}$ and output context
  $\oxnewframe{\oxvarctx^\prime_i}{\oxframe^\prime_i}$.

  Given these resulting $n$ typing judgements, we get from applying
  \oxname{T-Tuple} that
  $\oxtypjudge{\oxglobalctx}{\oxtvarctx}{\oxkontctx}{\oxnewframe{\oxvarctx^\prime_0}{\oxframe_0}}{
    \oxprod{\oxexpr_1 \oxdotsc \oxexpr_n} }{\oxtprod{\oxsitype_1 \oxdotsc
      \oxsitype_n}}{ \oxnewframe{\oxvarctx^\prime_o}{\oxframe^\prime_o} }$, as
  well as the related environments judgement
  $\oxsubtypectx{\oxglobalctx}{\oxtvarctx}{\oxnewframe{\oxvarctx_n}{\oxframe_n}}{\oxnewframe{\oxvarctx_n^\prime}{\oxframe_n^\prime}}$.

  The cases for \oxname{T-Array} and \oxname{T-Slice} proceed identically. This
  reasoning is also used in the \oxname{T-App} case.

  \vspace{1em}
  \noindent \framebox[\textwidth]{ \figuresize
    \begin{mathpar}
      \inferrule[T-Branch]{
        \oxtypjudge{\oxglobalctx}{\oxtvarctx}{\oxkontctx}{\oxnewframe{\oxvarctx}{\oxframe}}{
          \oxexpr_1
        }{\oxtbool}{\oxnewframe{\oxvarctx_1}{\oxframe_1}} \\
        \oxtypjudge{\oxglobalctx}{\oxtvarctx}{\oxkontctx}{\oxnewframe{\oxvarctx_1}{\oxframe_1}}{
          \oxexpr_2
        }{\oxsitype_2}{\oxnewframe{\oxvarctx_2}{\oxframe_2}} \\\\
        \oxtypjudge{\oxglobalctx}{\oxtvarctx}{\oxkontctx}{\oxnewframe{\oxvarctx_1}{\oxframe_1}}{
          \oxexpr_3
        }{\oxsitype_3}{\oxnewframe{\oxvarctx_3}{\oxframe_3}} \\
        \oxsitype = \oxsitype_2 \vee \oxsitype = \oxsitype_3 \\\\
        \oxtunify{\oxtvarctx}{\oxkontctx}{\oxnewframe{\oxvarctx_2}{\oxframe_2}}
        {\oxsitype_2}{\oxsitype}{\oxnewframe{\oxvarctx_{2s}}{\oxframe_{2s}}} \\
        \oxtunify{\oxtvarctx}{\oxkontctx}{\oxnewframe{\oxvarctx_3}{\oxframe_3}}
        {\oxsitype_3}{\oxsitype}{\oxnewframe{\oxvarctx_{3s}}{\oxframe_{3s}}} \\
        \oxnewframe{\oxvarctx_{2s}}{\oxframe_{2s}} \oxintersect \oxnewframe{\oxvarctx_{3s}}{\oxframe_{3s}} = \oxnewframe{\oxvarctx_o}{\oxframe_o} \\
      }{
        \oxtypjudge{\oxglobalctx}{\oxtvarctx}{\oxkontctx}{\oxnewframe{\oxvarctx}{\oxframe}}{
          \oxbranch{\oxexpr_1}{\oxexpr_2}{\oxexpr_3}
        }{\oxsitype}{\oxnewframe{\oxvarctx_o}{\oxframe_o}} }
    \end{mathpar}
  } \vspace{1em}

  By our induction hypothesis we get that
  $\oxtypjudge{\oxglobalctx}{\oxtvarctx}{\oxkontctx}{\oxnewframe{\oxvarctx^\prime}{\oxframe}}{\oxexpr_1}{\oxtbool}{\oxnewframe{\oxvarctx_1^\prime}{\oxframe_1}}$,
  and
  $\oxtypjudge{\oxglobalctx}{\oxtvarctx}{\oxkontctx}{\oxnewframe{\oxvarctx_1^\prime}{\oxframe_1}}{\oxexpr_2}{\oxtbool}{\oxnewframe{\oxvarctx_2^\prime}{\oxframe_2}}$,
  and
  $\oxtypjudge{\oxglobalctx}{\oxtvarctx}{\oxkontctx}{\oxnewframe{\oxvarctx_1^\prime}{\oxframe_1}}{\oxexpr_3}{\oxtbool}{\oxnewframe{\oxvarctx_3^\prime}{\oxframe_3}}$
  with
  $\oxsubtypectx{\oxglobalctx}{\oxtvarctx}{\oxnewframe{\oxvarctx_2}{\oxframe_2}}{\oxnewframe{\oxvarctx_2^\prime}{\oxframe_2}}$
  and
  $\oxsubtypectx{\oxglobalctx}{\oxtvarctx}{\oxnewframe{\oxvarctx_3}{\oxframe_3}}{\oxnewframe{\oxvarctx_3^\prime}{\oxframe_3}}$.

  Next we want to show that
  $\oxtunify{\oxtvarctx}{\oxkontctx}{\oxnewframe{\oxvarctx_2^\prime}{\oxframe_2}}{\oxsitype_2}{\oxsitype}{\oxnewframe{\oxvarctx_{2s}^\prime}{\oxframe_{2s}}}$,
  $\oxsubtypectx{\oxglobalctx}{\oxtvarctx}{\oxnewframe{\oxvarctx_{2s}}{\oxframe_{2s}}}{\oxnewframe{\oxvarctx_{2s}^\prime}{\oxframe_{2s}}}$,
  $\oxtunify{\oxtvarctx}{\oxkontctx}{\oxnewframe{\oxvarctx_3^\prime}{\oxframe_3}}{\oxsitype_3}{\oxsitype}{\oxnewframe{\oxvarctx_{3s}^\prime}{\oxframe_{3s}}}$,
  and
  $\oxsubtypectx{\oxglobalctx}{\oxtvarctx}{\oxnewframe{\oxvarctx_{3s}}{\oxframe_{3s}}}{\oxnewframe{\oxvarctx_{3s}^\prime}{\oxframe_{3s}}}$,
  which all follow from applying \Lemma{lemma:unified-types-well-formed}. To do
  this lemma application, we just need to show that for all $\oxcprov$ in
  $\oxsitype_1$, $\oxsitype_2$ and $\oxsitype$, $\oxvarctx(\oxcprov) =
  \oxvarctx^\prime(\oxcprov)$ which follows from the premise.

  Finally, we need to show that
  $\oxsubtypectx{\oxglobalctx}{\oxtvarctx}{\oxnewframe{\oxvarctx_o}{\oxframe_o}}{\oxnewframe{\oxvarctx_o^\prime}{\oxframe_o}}$.
  The well formedness condition on $\oxvarctx_o^\prime$ follows immediately
  since all types are the same as in $\oxvarctx_2^\prime$ and
  $\oxvarctx_3^\prime$ and the loan sets are just unioned, meaning reference
  types remain valid and we can compute types for all loans.

  The equal or empty condition follows from the fact that $\oxvarctx_2^\prime$
  and $\oxvarctx_3^\prime$ both agree on types by
  \Lemma{lemma:related-input-output-differences}, which means they drop exactly
  the same entries. For any regions emptied, either the same regions are
  emptied, or the region was emptied in the corresponding smaller context
  $\oxvarctx_2$ or $\oxvarctx_3$. Otherwise the loan sets are untouched.

  Finally, both of these are preserved when adding on the same frame, so we're
  done.

  \vspace{1em}
  \noindent \framebox[\textwidth]{ \figuresize
    \begin{mathpar}
      \inferrule[T-Match]{
        \oxtypjudge{\oxglobalctx}{\oxtvarctx}{\oxkontctx}{\oxnewframe{\oxvarctx}{\oxframe}}{
          \oxexpr
        }{\oxtsum{\oxsitype_l}{\oxsitype_r}}{\oxnewframe{\oxvarctx_1}{\oxframe_1}} \\
        \forall \oxcprov \in \oxfprovs{\oxtsum{\oxsitype_l}{\oxsitype_r}}. \;
        \oxnotreborrowed{\oxnewframe{\oxvarctx_1}{\oxframe_1}}{\oxcprov} \\\\
        \oxtypjudge{\oxglobalctx}{\oxtvarctx}{\oxkontctx}
        {\oxextendctx{\oxnewframe{\oxvarctx_1}{\oxframe_1}}{
            \oxvarctxentry{\oxid_1}{\oxsitype_l} }}{ \oxexpr_1
        }{\oxsitype_2}{\oxextendctx{\oxnewframe{\oxvarctx_2}{\oxframe_2}}{
            \oxvarctxentry{\oxid_1}{\oxsdtype_l}
          }} \\\\
        \oxtypjudge{\oxglobalctx}{\oxtvarctx}{\oxkontctx}{\oxextendctx{\oxnewframe{\oxvarctx_1}{\oxframe_1}}{
            \oxvarctxentry{\oxid_2}{\oxsitype_r} }}{ \oxexpr_2
        }{\oxsitype_3}{\oxextendctx{\oxnewframe{\oxvarctx_3}{\oxframe_3}}{
            \oxvarctxentry{\oxid_2}{\oxsdtype_r}
          }} \\
        \oxsitype = \oxsitype_2 \vee \oxsitype = \oxsitype_3 \\\\
        \oxtunify{\oxtvarctx}{\oxkontctx}{\oxnewframe{\oxvarctx_2}{\oxframe_2}}
        {\oxsitype_2}{\oxsitype}{\oxnewframe{\oxvarctx_{2s}}{\oxframe_{2s}}} \\
        \oxtunify{\oxtvarctx}{\oxkontctx}{\oxnewframe{\oxvarctx_3}{\oxframe_3}}
        {\oxsitype_3}{\oxsitype}{\oxnewframe{\oxvarctx_{3s}}{\oxframe_{3s}}} \\
        \oxnewframe{\oxvarctx_{2s}}{\oxframe_{2s}} \oxintersect \oxnewframe{\oxvarctx_{3s}}{\oxframe_{3s}} = \oxnewframe{\oxvarctx_o}{\oxframe_o} \\
      }{
        \oxtypjudge{\oxglobalctx}{\oxtvarctx}{\oxkontctx}{\oxnewframe{\oxvarctx}{\oxframe}}{
          \oxmatch{\oxexpr}{\oxid_1}{\oxexpr_1}{\oxid_2}{\oxexpr_2}
        }{\oxsitype}{\oxnewframe{\oxvarctx_o}{\oxframe_o}} }
    \end{mathpar}
  } \vspace{1em}

  This case follows almost identically to the \oxname{T-Branch} case above. The
  only structural difference is that the expression typing judgements for
  $\oxexpr_1$ and $\oxexpr_2$ have $\oxid_1$ and $\oxid_2$ respectively in their
  environments, but we know we can remove $\oxid_1$ and $\oxid_2$ from each side
  and keep the contexts well formed, since nothing that comes before $\oxid_1$
  or $\oxid_2$ can refer to it, and we know that $\oxsitype_1$ and $\oxsitype_2$
  cannot in any way refer to $\oxid_1$ or $\oxid_2$ because we have from the
  region rewriting judgements in the premises that the types are well formed in
  $\oxnewframe{\oxvarctx_2}{\oxframe_2}$ and
  $\oxnewframe{\oxvarctx_3}{\oxframe_3}$ respectively. We also need to show that the region not reborrowed judgement still holds, but this is immediate because at most $\oxvarctx_1^\prime$ has the same types as $\oxvarctx_1$. 
  
  \vspace{1em}
  \noindent \framebox[\textwidth]{ \figuresize
    \begin{mathpar}
      \inferrule[T-Let]{
        \oxtypjudge{\oxglobalctx}{\oxtvarctx}{\oxkontctx}{\oxnewframe{\oxvarctx}{\oxframe}}{
          \oxexpr_1
        }{\oxsitype_1}{\oxnewframe{\oxvarctx_1}{\oxframe_1}} \\
        \oxtunify{\oxtvarctx}{\oxkontctx}{\oxnewframe{\oxvarctx_1}{\oxframe_1}}
                 {\oxsitype_1}{\oxsitype_a}{\oxnewframe{\oxvarctx_{1s}}{\oxframe_{1s}}} \\\\
        \forall \oxcprov \in \oxfprovs{\oxsitype_a}. \;
        \oxnotreborrowed{\oxnewframe{\oxvarctx_{1s}}{\oxframe_{1s}}}{\oxcprov} \\\\
        \oxtypjudge{\oxglobalctx}{\oxtvarctx}{\oxkontctx}{
          \oxgcloans{\oxkontctx}{\oxextendctx{\oxnewframe{\oxvarctx_{1s}}{\oxframe_{1s}}}{\oxvarctxentry{\oxid}{\oxsitype_a}}}
        }{ \oxexpr_2 }{\oxsitype_2}{
          \oxextendctx{\oxnewframe{\oxvarctx_2}{\oxframe_2}}{
            \oxvarctxentry{\oxid}{\oxsdtype} }
        } \\
      }{
        \oxtypjudge{\oxglobalctx}{\oxtvarctx}{\oxkontctx}{\oxnewframe{\oxvarctx}{\oxframe}}{
          \oxlet{\oxid}{\oxsitype_a}{\oxexpr_1}{\oxexpr_2} }{\oxsitype_2}{
          \oxnewframe{\oxvarctx_2}{\oxframe_2} } }
    \end{mathpar}
  } \vspace{1em}

  Firstly, we apply our induction hypothesis to get that $\oxexpr_1$ is well
  typed with input environment $\oxnewframe{\oxvarctx^\prime}{\oxframe}$ and
  output environment $\oxnewframe{\oxvarctx_1^\prime}{\oxframe_1}$ with
  $\oxsubtypectx{\oxglobalctx}{\oxtvarctx}{\oxnewframe{\oxvarctx_1}{\oxframe_1}}{\oxnewframe{\oxvarctx_1^\prime}{\oxframe_1}}$.
  Then, we apply \Lemma{lemma:unified-types-well-formed} to get
  $\oxsubtypectx{\oxglobalctx}{\oxtvarctx}{\oxnewframe{\oxvarctx_{1s}}{\oxframe_{1s}}}{\oxnewframe{\oxvarctx_{1s}^\prime}{\oxframe_{1s}}}$.
  In order to apply this lemma we need to know that for any $\oxcprov$ that
  occur in $\oxsitype_1$ or $\oxsitype_a$,
  $\oxnewframe{\oxvarctx_1}{\oxframe_1}(\oxcprov) =
  \oxnewframe{\oxvarctx_1^\prime}{\oxframe_1}(\oxcprov)$, which we have as a
  conclusion from the previous application of the induction hypothesis.

  The region not reborrowed judgement holds immediately, because at most
  $\oxvarctx_1^\prime$ has the same types as $\oxvarctx_1$.

  To apply our induction hypothesis on $\oxexpr_2$ and continue the case, we need that \\
  $\oxsubtypectx{\oxglobalctx}{\oxtvarctx}{\oxgcloans{\oxkontctx}{\oxnewframe{\oxvarctx_{1s}}{\oxframe_{1s}},
      \oxid :
      \oxsitype_a}}{\oxgcloans{\oxkontctx}{\oxnewframe{\oxvarctx_{1s}^\prime}{\oxframe_{1s}},
      \oxid : \oxsitype_a}}$. But this is immediate by definition since gcloans
  can only empty loan sets for regions for which there are no types that contain
  them, which is allowed by \oxname{R-Env}.

  Our final obligation to apply the induction hypothesis is that for any
  $\oxcprov$ that occurs in a type in $\oxframe_{1s}$ but is not in
  $\oxframe_{1s}$, we need that
  $\oxgcloans{\oxkontctx}{\oxnewframe{\oxvarctx_{1s}}{\oxframe_{1s}}}(\oxcprov)
  =
  \oxgcloans{\oxkontctx}{\oxnewframe{\oxvarctx_{1s}^\prime}{\oxframe_{1s}}}(\oxcprov)$.
  We already have that $\oxnewframe{\oxvarctx_{1s}}{\oxframe_{1s}}(\oxcprov) =
  \oxnewframe{\oxvarctx_{1s}^\prime}{\oxframe_{1s}}(\oxcprov)$, so we just need
  to know that $\exists \oxplace : \oxtype \in \oxvarctx_{1s}$, where $\oxcprov$
  occurs in $\oxtype$, and $\oxnewframe{\oxvarctx_{1s}}{\oxframe_{1s}}(\oxplace)
  = \oxnewframe{\oxvarctx_{1s}}{\oxframe_{1s}}^\prime(\oxplace)$. But we said
  that $\oxcprov$ is contained in a type in $\oxframe_{1s}$, so the place for
  that type is one such place, so we cannot empty the loan set.

  \vspace{1em}
  \noindent \framebox[\textwidth]{ \figuresize
    \begin{mathpar}
      \inferrule[T-Seq]{
        \oxtypjudge{\oxglobalctx}{\oxtvarctx}{\oxkontctx}{\oxnewframe{\oxvarctx}{\oxframe}}{
          \oxexpr_1
        }{\oxsitype_1}{\oxnewframe{\oxvarctx}{\oxframe}_1} \\\\
        \oxtypjudge{\oxglobalctx}{\oxtvarctx}{\oxkontctx}{\oxgcloans{\oxkontctx}{\oxnewframe{\oxvarctx_1}{\oxframe_1}}}{
          \oxexpr_2
        }{\oxsitype_2}{\oxnewframe{\oxvarctx_2}{\oxframe_2}} \\
      }{
        \oxtypjudge{\oxglobalctx}{\oxtvarctx}{\oxkontctx}{\oxnewframe{\oxvarctx}{\oxframe}}{
          \oxseq{\oxexpr_1}{\oxexpr_2}
        }{\oxsitype_2}{\oxnewframe{\oxvarctx_2}{\oxframe_2}} }
    \end{mathpar}
  } \vspace{1em}

  Firstly, we apply our induction hypothesis to get that $\oxexpr_1$ is well
  typed with input environment $\oxnewframe{\oxvarctx^\prime}{\oxframe}$ and
  output environment $\oxnewframe{\oxvarctx_1^\prime}{\oxframe_1}$, with
  $\oxsubtypectx{\oxglobalctx}{\oxtvarctx}{\oxnewframe{\oxvarctx_1}{\oxframe_1}}{\oxnewframe{\oxvarctx_1^\prime}{\oxframe_1}}$.
  We need to know that
  $\oxsubtypectx{\oxglobalctx}{\oxtvarctx}{\oxgcloans{\oxkontctx}{\oxnewframe{\oxvarctx_1}{\oxframe_1}}}{\oxgcloans{\oxkontctx}{\oxnewframe{\oxvarctx_1^\prime}{\oxframe_1}}}$
  before we can apply our induction hypothesis to finish the proof. But this
  fact is trivial by the definitions, since \textrm{gc-loans} can only empty
  regions that are not in initialized types in the context, which is allowed in
  \oxname{R-Env}.

  Our final obligation to apply the induction hypothesis is that for any
  $\oxcprov$ that occurs in a type in $\oxframe_{1}$ but is not in
  $\oxframe_{1}$, we need that
  $\oxgcloans{\oxkontctx}{\oxnewframe{\oxvarctx_{1}}{\oxframe_{1}}}(\oxcprov) =
  \oxgcloans{\oxkontctx}{\oxnewframe{\oxvarctx_{1}^\prime}{\oxframe_{1}}}(\oxcprov)$.
  We already have that $\oxnewframe{\oxvarctx_{1}}{\oxframe_{1}}(\oxcprov) =
  \oxnewframe{\oxvarctx_{1}^\prime}{\oxframe_{1}}(\oxcprov)$, so we just need to
  know that $\exists \oxplace : \oxtype \in \oxvarctx_{1}$, where $\oxcprov$
  occurs in $\oxtype$, and $\oxnewframe{\oxvarctx_{1}}{\oxframe_1}(\oxplace) =
  \oxnewframe{\oxvarctx_{1}^\prime}{\oxframe_1}(\oxplace)$. But since $\oxcprov$
  occurs in a type in $\oxframe_1$, the place that maps to that type is such a
  place.
  
  \vspace{1em}
  \noindent \framebox[\textwidth]{ \figuresize
    \begin{mathpar}
      \inferrule[T-Drop]{
        \oxlookup{\oxvarctx}{\oxplace}{\oxsitype_\oxplace} \\
        \oxtypjudge{\oxglobalctx}{\oxtvarctx}{\oxkontctx}{
          (\oxnewframe{\oxvarctx}{\oxframe})[\oxplace \mapsto
          \oxsidtype_\oxplace]
        }{\oxexpr}{\oxsxtype}{\oxnewframe{\oxvarctx_o}{\oxframe_o}} \\
      }{
        \oxtypjudge{\oxglobalctx}{\oxtvarctx}{\oxkontctx}{\oxnewframe{\oxvarctx}{\oxframe}}{
          \oxexpr }{\oxsxtype}{\oxnewframe{\oxvarctx_o}{\oxframe_o}} }
    \end{mathpar}
  } \vspace{1em}

  In order to apply our induction hypothesis and finish the case, we only need
  to show that
  $\oxsubtypectx{\oxglobalctx}{\oxtvarctx}{(\oxnewframe{\oxvarctx}{\oxframe})[\oxplace
    \mapsto
    \oxsidtype_\oxplace]}{(\oxnewframe{\oxvarctx^\prime}{\oxframe})[\oxplace
    \mapsto \oxsidtype_\oxplace]}$, which is immediate by the definition of
  related contexts. Note that
  $(\oxnewframe{\oxvarctx^\prime}{\oxframe})[\oxplace \mapsto
  \oxsidtype_\oxplace]$ is well formed because
  $(\oxnewframe{\oxvarctx}{\oxframe})[\oxplace \mapsto \oxsidtype_\oxplace]$ is
  well formed. There cannot be any loans to $\oxplace$ because in the
  $(\oxnewframe{\oxvarctx^\prime}{\oxframe})[\oxplace \mapsto
  \oxsidtype_\oxplace]$ because those loans would be there in
  $(\oxnewframe{\oxvarctx}{\oxframe})[\oxplace \mapsto \oxsidtype_\oxplace]$.

  \vspace{1em}
  \noindent \framebox[\textwidth]{ \figuresize
    \begin{mathpar}
      \inferrule[T-App]{
        \overline{\oxenvwellformed{\oxglobalctx}{\oxtvarctx}{\oxnewframe{\oxvarctx}{\oxframe}}{\oxenv}} \\
        \overline{\oxrgnvalidity{\oxtvarctx}{\oxnewframe{\oxvarctx}{\oxframe}}{\oxprov}} \\
        \overline{\oxtypevalidity{\oxglobalctx}{\oxtvarctx}{\oxnewframe{\oxvarctx}{\oxframe}}{\oxsitype}} \\
        \delta = \oxsubstmany{
          \oxsubstmany{
            \oxsubstmany{\cdot}{\oxenv}{\oxenvvar}
          }{\oxprov}{\oxabsprov}
        }{\oxsitype}{\oxtvar} \\\\
        \oxtypjudge{\oxglobalctx}{\oxtvarctx}{\oxkontctx}{\oxnewframe{\oxvarctx}{\oxframe}}{
          \oxnoseqexpr_f }{ \oxtfunext{ \overline{\oxenvvar} \oxcomma
            \overline{\oxabsprov} \oxcomma \overline{\oxtvar} }{ \oxsitype_1
            \oxdotsc \oxsitype_n
          }{\oxsitype_f}{\oxenv_c}{\overline{\oxabsprov_1: \oxabsprov_2}}
        }{\oxnewframe{\oxvarctx_0}{\oxframe_0}} \\\\
        \forall i \in \oxset{1 \oxdots n}. \;
        \oxtypjudge{\oxglobalctx}{\oxtvarctx}{
          \oxextendctx{\oxkontctx}{
            \delta(\oxsitype_1) \oxdots \delta(\oxsitype_{i-1})
          }
        }{\oxnewframe{\oxvarctx_{i-1}}{\oxframe_{i-1}}}{
          \oxnoseqexpr_i }{\delta(\oxsitype_i)}{\oxnewframe{\oxvarctx_i}{\oxframe_i}} \\
        \forall i \in \oxset{ 1 \oxdots n }. \;
    \forall \oxcprov \in \oxfprovs{\oxsitype_i}. \;
    \oxnotreborrowed{\oxnewframe{\oxvarctx_n}{\oxframe_n}}{\oxcprov} \\
        \oxrunifymany{\oxtvarctx}{\oxkontctx}{\oxnewframe{\oxvarctx_n}{\oxframe_n}}{
          \oxsubstmany{\oxabsprov_2}{\oxprov}{\oxabsprov} }{
          \oxsubstmany{\oxabsprov_1}{\oxprov}{\oxabsprov}
        }{\oxnewframe{\oxvarctx_b}{\oxframe_b}} \\
      }{
        \oxtypjudge{\oxglobalctx}{\oxtvarctx}{\oxkontctx}{\oxnewframe{\oxvarctx}{\oxframe}}{
          \oxapp{\oxnoseqexpr_f}{ \overline{\oxenv} \oxcomma \overline{\oxprov}
            \oxcomma \overline{\oxsitype} }{ \oxnoseqexpr_1 \oxdotsc
            \oxnoseqexpr_n } }{ \oxsubstmany{ \oxsubstmany{
              \oxsubstmany{\oxsitype_f}{\oxenv}{\oxenvvar}
            }{\oxprov}{\oxabsprov} }{\oxsitype}{\oxtvar}
        }{\oxnewframe{\oxvarctx_b}{\oxframe_b}} }
    \end{mathpar}
  } \vspace{1em}

  In the case of \oxname{T-App}, we firstly must prove the well formedness
  properties:
  \begin{itemize}
  \item
    $\oxenvwellformed{\oxglobalctx}{\oxtvarctx}{\oxnewframe{\oxvarctx^\prime}{\oxframe}}{\oxenv}$.
    Since $\oxtvarctx$ is unchanged, \oxname{WF-Env} is the only interesting
    case.
    
    \vspace{1em}
    \noindent \hspace{-2.5em} \framebox[\textwidth]{ \figuresize
      \begin{mathpar}
        \inferrule[WF-Env]{
          \oxvarctxwellformed{\oxglobalctx}{\oxtvarctx}{
            \oxnewframe{\oxnewframe{\oxvarctx}{\oxframe}}{\oxframe_c} } }{
          \oxenvwellformed{\oxglobalctx}{\oxtvarctx}{\oxnewframe{\oxvarctx}{\oxframe}}{
            \oxframe_c } }
      \end{mathpar}
    } \vspace{1em}

    Let $\oxenv = \oxframe_e$. We want to show that
    $\oxctxswellformed{\oxglobalctx}{\oxtvarctx}{\oxnewframe{\oxnewframe{\oxvarctx^\prime}{\oxframe}}{\oxframe_c}}{\oxkontctx}$
    given
    $\oxctxswellformed{\oxglobalctx}{\oxtvarctx}{\oxnewframe{\oxnewframe{\oxvarctx}{\oxframe}}{\oxframe_c}}{\oxkontctx}$,
    which is immediate from \Lemma{lemma:related-contexts-well-formed}.
  \item $\oxrgnvalidity{\oxtvarctx}
    {\oxnewframe{\oxvarctx^\prime}{\oxframe}}{\oxprov}$, which is immediate from
    the premises since related loan environments have the same domains and
    $\oxtvarctx$ is the same.
  \item $\oxtypevalidity{\oxglobalctx}{\oxtvarctx}
    {\oxnewframe{\oxvarctx^\prime}{\oxframe}}{\oxsitype}$, which is immediate
    from \Lemma{lemma:types-well-formed}. We just need that for the regions that
    occur in the type, their loan sets are unchanged, but we get that from the
    premise, because the function argument is either: locally defined, in which
    case it can only use and produce types accessible in the context; an
    argument, in which case its arguments are also part of the argument type; or
    a global function, in which case these types do not contain any non abstract
    regions which are replaced with concrete regions all in $\oxframe$.
  \end{itemize}

  For the rest of the application case, we can apply our induction hypothesis on
  the function and the arguments, additionally applying the substitution lemma,
  \Lemma{lemma:substitution}, where needed. The region not reborrowed condition
  is true by the fact that at most the types between $\oxvarctx_n$ and
  $\oxvarctx_n^\prime$ are the same. The last part about outlives follows from
  \Lemma{lemma:outlives-related}, where we have the condition on the loan sets
  from the conclusion of the application of the induction hypothesis.

  \vspace{1em}
  \noindent \framebox[\textwidth]{ \figuresize
    \begin{mathpar}
      \inferrule[T-AppClosure]{
        \oxtypjudge{\oxglobalctx}{\oxtvarctx}{\oxkontctx}{\oxnewframe{\oxvarctx}{\oxframe}}{
          \oxexpr_f
        }{
          \oxtfun{}{
            \oxsitype_1 \oxdotsc \oxsitype_n
          }{\oxsitype_f}{\oxenv_c}
        }{\oxnewframe{\oxvarctx_{0s}}{\oxframe_{0s}}} \\\\
        \forall i \in \oxset{1 \oxdots n}. \;
        \oxtypjudge{\oxglobalctx}{\oxtvarctx}{
          \oxextendctx{\oxkontctx}{
            \oxsitype_1 \oxdots \oxsitype_{i-1}
          }
        }{\oxnewframe{\oxvarctx_{i-1s}}{\oxframe_{i-1s}}}{
          \oxexpr_i
        }{
          \oxsitype_{i^\prime}
        }{\oxnewframe{\oxvarctx_i}{\oxframe_i}} \\
        \oxtunify[\oxcombineunrest]{\oxtvarctx}{\oxkontctx}{\oxnewframe{\oxvarctx_i}{\oxframe_i}}
                 {\oxsitype_{i^\prime}}{\oxsitype_i}{\oxnewframe{\oxvarctx_{is}}{\oxframe_{is}}} \\\\
                 \forall i \in \oxset{ 1 \oxdots n }. \;
                 \forall \oxcprov \in \oxfprovs{\oxsitype_i}. \;
                 \oxnotreborrowed{\oxnewframe{\oxvarctx_{ns}}{\oxframe_{ns}}}{\oxcprov} \\
      }{
        \oxtypjudge{\oxglobalctx}{\oxtvarctx}{\oxkontctx}{\oxnewframe{\oxvarctx}{\oxframe}}{
          \oxapply{\oxexpr_f}{
            \oxexpr_1 \oxdotsc \oxexpr_n
          }
        }{
          \oxsitype_f
        }{\oxnewframe{\oxvarctx_{ns}}{\oxframe_{ns}}}
      }

    \end{mathpar}
  }

  In the case of \oxname{T-AppClosure}, we follow a very similar procedure to
  \oxname{T-AppFunction}, but with an empty substituion, and the addition that
  we need to apply \Lemma{lemma:unified-types-well-formed} to handle the
  rewriting.

  In the cases of \oxname{T-Move}, \oxname{T-Copy}, \oxname{T-Borrow},
  \oxname{T-BorrowIndex}, \oxname{T-BorrowSlice}, \oxname{IndexCopy}, they all
  follow from the induction hypothesis and additionally applying
  \Lemma{lemma:ownership-safety-related-envs} and
  \Lemma{lemma:tc-related-contexts}. Note we get the place having the right type
  requirement for \oxname{T-Move} from the fact that the place must be in
  $\oxframe$ since it is a free variable.

  The remaining cases of \oxname{T-Assign} and \oxname{T-AssignDeref} proceed
  similarly. Firstly, we apply the induction hypothesis on the expression, then
  \Lemma{lemma:ownership-safety-related-envs} and
  \Lemma{lemma:tc-related-contexts}, and finally we get well formedness and
  relatedness on the output environment by applying
  \Lemma{lemma:unified-types-well-formed}. Note we get the place having the same
  type requirement for type computation from the fact that the place must be in
  $\oxframe$ since it is a free variable.

\end{proof}

\begin{oxlemma}{Referent Well Formedness Preserved in Related Environments}
  {lemma:referent-related-envs} If
  $\oxsubtypectx{\oxglobalctx}{\oxtvarctx}{\oxvarctx}{\oxvarctx^\prime}$ and
  $\oxreferentvalidity{\oxglobalctx}{\oxvarctx}{\oxreferentctx[\oxplace]}{\oxxitype}$
  and $\oxvarctx(\oxplace) = \oxvarctx^\prime(\oxplace)$, then
  $\oxreferentvalidity{\oxglobalctx}{\oxvarctx^\prime}{\oxreferentctx[\oxplace]}{\oxxitype}$.
\end{oxlemma}
\begin{proof}
  Proceed by induction on the referent validity derivation. The only case that
  doesn't follow immediately from premises and the induction hypothesis in
  \oxname{WF-RefId}, which follows from the equal types premise.
\end{proof}

\begin{oxlemma}{Value Typing Preserved in Related Environments}
  {lemma:value-typing-related-envs} If
  $\oxsubtypectx{\oxglobalctx}{\oxemptyctx}{\oxvarctx} {\oxvarctx^\prime}$,
  and $\oxtypjudge{\oxglobalctx}{\oxemptyctx}{\oxemptyctx}{\oxvarctx}{\oxvalue}
  {\oxvarctx(x)}{\oxvarctx}$, then
  $\oxtypjudge{\oxglobalctx}{\oxemptyctx}{\oxemptyctx}{\oxvarctx^\prime}
  {\oxvalue}{\oxvarctx^\prime(x)}{\oxvarctx^\prime}$.
\end{oxlemma}

\begin{proof}
  Proceed by simultaneous induction on the typing derivation and the stack frame
  well formedness.
  Since we know the expression is already a value, we restrict ourselves
  only to those cases that type values: \oxname{T-Unit}, \oxname{T-u32},
  \oxname{T-True}, \oxname{T-False}, \oxname{T-Tuple}, \oxname{T-Array},
  \oxname{T-Left}, \oxname{T-Right}, \oxname{T-Dead}, \oxname{T-Pointer}, and
  \oxname{T-ClosureValue}.

  For \oxname{T-Unit}, \oxname{T-u32}, \oxname{T-Dead}, \oxname{T-True}, and
  \oxname{T-False}, this holds trivially. For \oxname{T-Tuple},
  \oxname{T-Array}, \oxname{T-Left}, and \oxname{T-Right}, this holds directly
  by repeated application of our induction hypothesis. This leaves us with
  four cases.

  \vspace{1em}
  \noindent \framebox[\textwidth]{ \figuresize
    \begin{mathpar}
      \TClosureVal
    \end{mathpar}
  } \vspace{1em}

  For the \oxname{T-ClosureValue} case, firstly we want to show
  $\oxsubstorevalidity{\oxglobalctx}{\oxvarctx^\prime}{\oxstackframe}{\oxframe_c}$.
  This follows immediatedly from applying the induction hypothesis for each value.

  Then to finish the closure case, it suffices to show\\
  $\oxtypjudge{\oxglobalctx}{\oxemptyctx}{\oxemptyctx}{\oxnewframe{\oxvarctx^\prime}{\oxframe_c}}
  {\oxexpr}{\oxsitype_r}{\oxnewframe{\oxvarctx_o^\prime}{\oxframe}}$, which
  follows immediately from \Lemma{lemma:expression-typing-related-envs}.

  \vspace{1em}
  \noindent \framebox[\textwidth]{ \figuresize
    \begin{mathpar}
      \TPointer
    \end{mathpar}
  } \vspace{1em}
  
  If $x$ was dropped, then $\oxvarctx^\prime(x) = \oxvarctx(x)^\dagger$. Then
  the proof follows immediately from \oxname{T-Dead}.
  
  If $x$ was not dropped, then $\oxvarctx(x) = \oxvarctx^\prime(x)$. All that
  is left to show is that that the referent remains well formed, and the loan
  $\oxloanpkg{\oxmuta}{\oxplace}$ is in $\oxvarctx^\prime(\oxcprov)$. The
  first condition follows from \Lemma{lemma:referent-related-envs}. The second
  condition is immediate because the only potential changes allowed in the
  related environment to loan sets is emptying the loan sets of regions if
  there's no references with the region in their type, and this particular
  reference is a reference with the region, so emptying the loan set is ruled
  out.
\end{proof}
  
\begin{oxlemma}{Value Typing Fixed on Output Environments}
  {lemma:value-typing-output} If
  $\oxtypjudge{\oxglobalctx}{\oxtvarctx}{\oxkontctx}{\oxvarctx}{\oxvalue}
  {\oxtype}{\oxvarctx^\prime}$, then
  $\oxtypjudge{\oxglobalctx}{\oxtvarctx}{\oxkontctx}{\oxvarctx^\prime}
  {\oxvalue}{\oxtype}{\oxvarctx^\prime}$.
\end{oxlemma}

\begin{proof}
  Immediate by induction on the typing derivation. The only non immediate case
  is \oxname{T-Pointer}, where we also need to apply
  \Lemma{lemma:referent-related-envs}.
\end{proof}

\begin{oxlemma}{Stack Validity is Preserved in Related Environments}
  {lemma:stack-validity-related-envs} If
  $\oxstorevalidity{\oxglobalctx}{\oxvarctx}{\oxstore}$ and
  $\oxsubtypectx{\oxglobalctx}{\oxemptyctx}{\oxvarctx}{\oxvarctx^\prime}$, then
  $\oxstorevalidity{\oxglobalctx}{\oxvarctx^\prime}{\oxstore}$.
\end{oxlemma}

\begin{proof}
  We proceed by induction over the well typedness of the store.

  \vspace{1em}
  \noindent \framebox[\textwidth]{ \figuresize
    \begin{mathpar}
      \VStack\and \VStackEmpty
    \end{mathpar}
  } \vspace{1em}

  The interesting case is when the stack is non empty. Then we have that
  $\oxstorevalidity{\oxglobalctx}{\oxvarctx^\prime}{\oxstore}$ and want to show
  that
  $\oxstorevalidity{\oxglobalctx}{\oxnewframe{\oxvarctx^\prime}{\oxframe}}{\oxnewframe{\oxstore}{\oxstackframe}}$.
  The requirement on the domain is immediate since related environments have the
  same domains. What's left to show is that the values in the store remain well
  typed under the new environment. This follows from repeated applications of
  \Lemma{lemma:value-typing-related-envs}
\end{proof}

\subsection{Preservation When Popping a Stack Frame Lemmas}
\label{sec:pop-lemmas}

\begin{oxlemma}{Stack Validity is Preserved When Popping A Stack Frame}
  {lemma:stack-envminus} If
  $\oxstorevalidity{\oxglobalctx}{\oxnewframe{\oxvarctx}{\oxframe}}{\oxnewframe{\oxstore}{\oxstackframe}}$,
  then $\oxstorevalidity{\oxglobalctx} {\oxvarctx}{\oxstore}$.
\end{oxlemma}

\begin{proof}
  Immediate by inversion on \oxname{WF-StackFrame} which gives us
  $\oxstorevalidity{\oxglobalctx}{\oxvarctx}{\oxstore}$.
\end{proof}

\subsection{Preservation under Well Typed Extension Lemmas}
\label{sec:extension-lemmas}

\begin{oxlemma}{Ownership Safety is Preserved under Well-Typed Extensions}
  {lemma:ownership-well-typed-extension} If
  \begin{enumerate}
  \item
    $\oxtypevalidity{\oxglobalctx}{\oxemptyctx}{\oxnewframe{\oxvarctx}{\oxframe}}{\oxsitype_\oxid}$
  \item
    $\oxmusafetyinner{\oxemptyctx}{\oxkontctx}{\oxnewframe{\oxvarctx}{\oxframe}}{\oxmuta}{\overline{\oxplace}}{\oxplaceexpr}{\oxset{\oxloans}}$
  \item $\oxrootof{\oxplaceexpr} \in \oxdomain{\oxframe}$
  \item $\forall \oxcprov \in \oxfprovs{\oxsitype_\oxid}. \;
    \oxnotreborrowed{\oxvarctx}{\oxcprov}$
  \item $\forall \oxcprov \in \oxfprovs{\oxsitype_\oxid}. \; \forall \oxplace_i
    \in \oxset{\overline{\oxplace}}. \; \oxvarctx(\oxplace_i) =
    \oxtref{\oxcprov^{\prime\prime}}{\oxmuta}{\oxxitype} \implies
    \oxcprov^{\prime\prime} \neq \oxcprov$
  \end{enumerate}
  then
  $\oxmusafetyinner{\oxemptyctx}{\oxkontctx}{\oxnewframe{\oxextendctx{\oxvarctx}{\oxvarctxentry{\oxid}{\oxsitype_\oxid}}}{\oxframe}}{\oxmuta}{\overline{\oxplace}}{\oxplaceexpr}{\oxset{\oxloans}}$.
\end{oxlemma}

\begin{proof}
  We proceed by induction on the ownership safety derivation.

  \vspace{1em}
  \noindent \framebox[\textwidth]{ \figuresize
    \begin{mathpar}
      \OSafePlace
    \end{mathpar}
  } \vspace{1em}

  We'd like to apply \oxname{O-SafePlace} to show
  $\oxmusafetyinner{\oxemptyctx}{\oxkontctx}{ \oxnewframe{
      \oxextendctx{\oxvarctx}{\oxvarctxentry{\oxid}{\oxsitype_\oxid}}
    }{\oxframe} }{\oxmuta}{ \overline{\oxplace_e} }{ \oxplace }{
    \oxset{\oxloanpkg{\oxmuta}{\oxplace}} }$. Let
  $\oxloanctxentry{\oxcprov^\prime}{\oxset{\oxloans}} \in \oxvarctx$. Note that
  necessarily $\oxloanctxentry{\oxcprov^\prime}{\oxset{\oxloans}} \in
  \oxextendctx{\oxvarctx}{\oxvarctxentry{\oxid}{\oxsitype_\oxid}}$.

  If $\forall \oxloanpkg{\oxmuta^\prime}{\oxplaceexprctx[\oxplace^\prime]} \in
  \oxset{\overline{\oxloan}}. (\oxmuta = \oxmut \vee \oxmuta^\prime = \oxmut)
  \implies \oxrelevant{\oxplace^\prime}{\oxplace}$ held in our original
  $\oxvarctx$, then it still holds in the extended typing $\oxnewframe{
    \oxextendctx{\oxvarctx}{\oxvarctxentry{\oxid}{\oxsitype_\oxid}} }{\oxframe}$
  since the loan sets are unchanged between the two stack typings.

  Otherwise, we must have used the second clause $\exists
  \oxvarctxentry{\oxplace^\prime}{\oxtref{\oxcprov^\prime}{\oxmuta^\prime}{\oxtype^\prime}}
  \in \oxexplode{\oxvarctx} \; \wedge \; \not \exists
  \oxtref{\oxcprov^\prime}{\oxmuta^\prime}{\oxtype^\prime} \in \oxkontctx \;
  \wedge (\forall
  \oxvarctxentry{\oxplace^\prime}{\oxtref{\oxcprov^\prime}{\oxmuta^\prime}{\oxtype^\prime}}
  \in \oxexplode{\oxvarctx}. \; \oxplace^\prime \in \oxset{
    \overline{\oxplace_e} })$ in the first place. To show that this is still
  true, we need to show that nothing in our newly-bound $\oxid$ shares a type
  with a reborrowed reference which would end up in our exclusion list. The
  reason for this is that the failure condition for this is that with such a
  reference now bound at (or reachable within) $\oxid$, we could violate the
  universally-quantified portion of this clause. Fortunately, we have from our
  premise that all the regions that appear in the type $\oxsitype_\oxid$ are
  distinct from the ones in the exclusion list ($\forall \oxcprov \in
  \oxfprovs{\oxsitype_\oxid}. \; \forall \oxplace_i \in
  \oxset{\overline{\oxplace}}. \; \oxvarctx(\oxplace_i) =
  \oxtref{\oxcprov^{\prime\prime}}{\oxmuta}{\oxxitype} \implies
  \oxcprov^{\prime\prime} \neq \oxcprov$). Thus, we know this cannot be the
  case.

  \vspace{1em}
  \noindent \framebox[\textwidth]{ \figuresize
    \begin{mathpar}
      \ODeref
    \end{mathpar}
  } \vspace{1em}

  We'd like to apply \oxname{O-Deref} to show $\oxmusafetyinner{\oxemptyctx}{\oxkontctx}{
    \oxnewframe{ \oxextendctx{\oxvarctx}{\oxvarctxentry{\oxid}{\oxsitype_\oxid}}
    }{\oxframe} }{\oxmuta}{ \overline{\oxplace_e} }{ \oxplace }{
    \oxset{\oxloanpkg{\oxmuta}{\oxplace}} }$. This requires us to show
  $\oxlookup{ (\oxnewframe{
      \oxextendctx{\oxvarctx}{\oxvarctxentry{\oxid}{\oxsitype_\oxid}}
    }{\oxframe}) }{\oxplace}{
    \oxtref{\oxcprov}{\oxmuta_\oxplace}{\oxtype_\oxplace} }$ and $\oxlookup{
    (\oxnewframe{
      \oxextendctx{\oxvarctx}{\oxvarctxentry{\oxid}{\oxsitype_\oxid}}
    }{\oxframe}) }{\oxcprov}{
    \oxset{\overline{\oxloanpkg{\oxmuta^\prime}{\oxplaceexpr}}^n} }$. The former
  follows from the disjointedness assumption for $\oxid$, i.e. that $\oxid$ is
  disjoint from all existing identifiers in $\oxvarctx$. The latter follows from
  the fact that no loan sets are changed between the two stack typings. Since
  the loan set is unchanged, we also have that the new extension for the
  exclusion list $\oxkey{excl}$ is the same. This leaves us with two pieces to
  show. First, that the recursive uses of ownership safety still succeed (for
  which we will use the induction hypothesis) and that our last obligation holds
  (which follows much as it did for \oxname{O-SafePlace}).

  For the inductive cases, we can very nearly just apply the induction
  hypothesis, but we first must show that our exclusion list invariant applies
  for the extensions to the exclusion list. That is, we have that $\forall
  \oxcprov \in \oxfprovs{\oxsitype_\oxid}. \; \forall \oxplace_i \in
  \oxset{\overline{\oxplace_e}}. \; \oxvarctx(\oxplace_i) =
  \oxtref{\oxcprov^{\prime\prime}}{\oxmuta}{\oxxitype} \implies
  \oxcprov^{\prime\prime} \neq \oxcprov$, and we need to show $\forall \oxcprov
  \in \oxfprovs{\oxsitype_\oxid}. \; \forall \oxplace_i \in \oxset{\textrm{excl}
    \oxcomma \oxplace}. \; \oxvarctx(\oxplace_i) =
  \oxtref{\oxcprov^{\prime\prime}}{\oxmuta}{\oxxitype} \implies
  \oxcprov^{\prime\prime} \neq \oxcprov$. The exclusion extension $\oxkey{excl}$
  is constructed by looking specifically at the reborrow loans associated with
  the region $\oxcprov$. Since we know that $\forall \oxcprov \in
  \oxfprovs{\oxsitype_\oxid}. \; \forall
  \oxvarctxentry{\oxplace}{\oxtref{\oxcprov}{\oxmuta}{\oxxitype}} \in
  \oxexplode{\oxvarctx}. \; \not\exists \oxcprov^\prime. \;
  \oxloanpkg{\oxmuta}{\oxderef{\! \oxplace}} \in \oxvarctx(\oxcprov^\prime)$
  (from inversion of \oxname{NRB-Region}), it
  follows directly that none of the places in $\oxkey{excl}$ can have a
  reference type with an region in $\oxsitype_\oxid$. If they did, that would
  mean syntactically that $\oxvarctx(\oxcprov)$ contains a loan for
  $\oxderef{\oxplace}$ which would give us a contradiction.

  For the last obligation, let
  $\oxloanctxentry{\oxcprov^\prime}{\oxset{\oxloans}} \in \oxvarctx$. Note that
  necessarily $\oxloanctxentry{\oxcprov^\prime}{\oxset{\oxloans}} \in
  \oxextendctx{\oxvarctx}{\oxvarctxentry{\oxid}{\oxsitype_\oxid}}$.

  If $\forall \oxloanpkg{\oxmuta^\prime}{\oxplaceexpr^{\prime\prime}} \in
  \oxset{\overline{\oxloan}}. (\oxmuta = \oxmut \vee \oxmuta^\prime = \oxmut)
  \implies \oxrelevant{\oxplaceexpr^{\prime\prime}}{
    \oxplaceexprctx[\oxderef{\oxplace}] }$ held in our original $\oxvarctx$,
  then it still holds in the extended typing $\oxnewframe{
    \oxextendctx{\oxvarctx}{\oxvarctxentry{\oxid}{\oxsitype_\oxid}} }{\oxframe}$
  since the loan sets are unchanged between the two stack typings.

  Otherwise, we must have used the second clause $\exists
  \oxvarctxentry{\oxplace^\prime}{\oxtref{\oxcprov^\prime}{\oxmuta^\prime}{\oxtype^\prime}}
  \in \oxexplode{\oxvarctx} \; \wedge \; \not \exists
  \oxtref{\oxcprov^\prime}{\oxmuta^\prime}{\oxtype^\prime} \in \oxkontctx \;
  \wedge (\forall
  \oxvarctxentry{\oxplace^\prime}{\oxtref{\oxcprov^\prime}{\oxmuta^\prime}{\oxtype^\prime}}
  \in \oxexplode{\oxvarctx}. \; \oxplace^\prime \in \oxset{
    \overline{\oxplace_e} \oxcomma \textrm{excl} \oxcomma \oxplace })$ in the
  first place. To show that this is still true, we need to show that nothing in
  our newly-bound $\oxid$ shares a type with a reborrowed reference which would
  end up in our exclusion list. The reason for this is that the failure
  condition for this is that with such a reference now bound at (or reachable
  within) $\oxid$, we could violate the universally-quantified portion of this
  clause. Fortunately, we have from our premise that all the regions that
  appear in the type $\oxsitype_\oxid$ are distinct from the ones in the
  exclusion list ($\forall \oxcprov \in \oxfprovs{\oxsitype_\oxid}. \; \forall
  \oxplace_i \in \oxset{\overline{\oxplace}}. \; \oxvarctx(\oxplace_i) =
  \oxtref{\oxcprov^{\prime\prime}}{\oxmuta}{\oxxitype} \implies
  \oxcprov^{\prime\prime} \neq \oxcprov$). Thus, we know this cannot be the
  case.

  \vspace{1em}
  \noindent \framebox[\textwidth]{ \figuresize
    \begin{mathpar}
      \ODerefAbs
    \end{mathpar}
  } \vspace{1em}

  Since $\oxtvarctx = \oxemptyctx$, there are no valid reference types that have
  an abstract region, meaning the first hypothesis is a contradiction.
\end{proof}

\begin{oxlemma}{Type Computation is Preserved under Well-Typed Extensions}
  {lemma:computety-well-typed-extension} If
  $\oxtypevalidity{\oxglobalctx}{\oxemptyctx}{\oxvarctx}{\oxsitype_\oxid}$ and
  $\oxcomputety{\oxemptyctx}{\oxvarctx}{\oxmuta}{\oxplaceexpr}{\oxtype}
  {\oxset{\overline{\oxprov}}}$ then $\oxcomputety{\oxemptyctx}{
    \oxextendctx{\oxvarctx}{\oxsitype_\oxid}
  }{\oxmuta}{\oxplaceexpr}{\oxtype}{\oxset{\overline{\oxprov}}}$.
\end{oxlemma}

\begin{proof}
  We proceed by induction on the type computation. This gives us three cases,
  \oxname{TC-Var}, \oxname{TC-Proj} and \oxname{TC-Deref}. In \oxname{TC-Var},
  the lookup yields the same type based on the assumption that our new binding
  is disjoint from our existing ones. \oxname{TC-Proj} and \oxname{TC-Deref}
  proceed directly from the induction hypothesis.
\end{proof}

\begin{oxlemma}{Outlives is Preserved under Well-Typed Extensions}
  {lemma:outlives-well-typed-extension} If
  $\oxtypevalidity{\oxglobalctx}{\oxemptyctx}{\oxvarctx}{\oxsitype_\oxid}$ and
  $\forall \oxcprov \in \oxfprovs{\oxsitype_\oxid}. \;
  \oxnotreborrowed{\oxvarctx}{\oxcprov}$ and
  $\oxrunify{\oxemptyctx}{\oxkontctx}{\oxvarctx}{\oxprov_1}{\oxprov_2}{\oxvarctx^\prime}$
  then $\oxrunify{\oxemptyctx}{\oxkontctx}{ \oxextendctx{\oxvarctx}{
      \oxvarctxentry{\oxid}{\oxsitype_\oxid} } }{\oxprov_1}{\oxprov_2}{
    \oxextendctx{\oxvarctx^\prime}{ \oxvarctxentry{\oxid}{\oxsitype_\oxid} } }$
\end{oxlemma}

\begin{proof}
  We proceed by induction on the outlives judgment. This gives us six cases,
  \oxname{OL-Refl}, \oxname{OL-Trans}, \oxname{OL-BothAbstract},
  \oxname{OL-CombineConcrete}, \oxname{OL-ConcreteAbstract} and
  \oxname{OL-AbstractConcrete}.

  \oxname{OL-Refl}, \oxname{OL-BothAbstract} and \oxname{OL-AbstractConcrete}
  are immediate.

  \oxname{OL-Trans} follows from the induction hypothesis.

  \oxname{OL-ConcreteAbstract} follows from the induction hypothesis and
  \Lemma{lemma:computety-well-typed-extension}.

  This leaves \oxname{OL-CombineConcrete} as the most interesting case. Here we
  use $\forall \oxcprov \in \oxfprovs{\oxsitype_\oxid}. \;
  \oxnotreborrowed{\oxvarctx}{\oxcprov}$ from our premise and note that since
  this holds for arbitrary regions $\oxcprov$, we know that it holds for both
  $\oxcprov_1$ and $\oxcprov_2$ in the premise of \oxname{OL-CombineConcrete} and
  thus we are done.
\end{proof}

\begin{oxlemma}{Region Rewriting is Preserved under Well-Typed Extensions}
  {lemma:subtyping-well-typed-extension} If
  $\oxtypevalidity{\oxglobalctx}{\oxemptyctx}{\oxvarctx}{\oxsitype_\oxid}$ and
  $\forall \oxcprov \in \oxfprovs{\oxsitype_\oxid}. \;
  \oxnotreborrowed{\oxvarctx}{\oxcprov}$ and
  $\oxtunify{\oxemptyctx}{\oxkontctx}{\oxvarctx}{\oxtype_1}{\oxtype_2}{\oxvarctx^\prime}$
  then $\oxtunify{\oxemptyctx}{\oxkontctx}{ \oxextendctx{\oxvarctx}{
      \oxvarctxentry{\oxid}{\oxsitype_\oxid} } }{\oxtype_1}{\oxtype_2}{
    \oxextendctx{\oxvarctx^\prime}{ \oxvarctxentry{\oxid}{\oxsitype_\oxid} } }$.
\end{oxlemma}

\begin{proof}
  We proceed by induction on the region rewriting judgment. This gives us seven
  cases, \oxname{RR-Refl}, \oxname{RR-Trans}, \oxname{RR-Array},
  \oxname{RR-Slice}, \oxname{RR-Reference}, \oxname{RR-Tuple}, and
  \oxname{RR-Dead}.

  \oxname{RR-Refl} is immediate.

  \oxname{RR-Trans}, \oxname{RR-Array}, \oxname{RR-Slice}, \oxname{RR-Tuple} and
  \oxname{RR-Dead} all follow directly from the induction hypothesis.

  This leaves \oxname{RR-Reference} which follows from
  \Lemma{lemma:outlives-well-typed-extension} and the induction hypothesis.
\end{proof}

\begin{oxlemma}{Closure Bodies are Well-Typed under Well-Typed Extensions}
  {lemma:exprs-well-typed-extension} If
  $\oxtypevalidity{\oxglobalctx}{\oxemptyctx}{\oxvarctx}{\oxsitype_\oxid}$ and
  $\forall \oxcprov \in \oxfprovs{\oxsitype_\oxid}. \;
  \oxnotreborrowed{\oxvarctx}{\oxcprov}$ and
  $\oxtypjudge{\oxglobalctx}{\oxemptyctx}{\oxkontctx}{
    \oxnewframe{\oxvarctx}{\oxframe} }{\oxexpr}{\oxsitype}{
    \oxnewframe{\oxvarctx^\prime}{\oxframe^\prime} }$, then
  $\oxtypjudge{\oxglobalctx}{\oxemptyctx}{\oxkontctx}{ \oxnewframe{
      \oxextendctx{\oxvarctx}{ \oxvarctxentry{\oxid}{\oxsitype_\oxid} }
    }{\oxframe} }{\oxexpr}{\oxsitype}{ \oxnewframe{
      \oxextendctx{\oxvarctx^\prime}{ \oxvarctxentry{\oxid}{\oxsitype_\oxid} }
    }{\oxframe^\prime} }$ and $\forall \oxcprov \in \oxfprovs{\oxsitype_\oxid}.
  \; \forall \oxvarctxentry{\oxplace}{\oxtref{\oxcprov}{\oxmuta}{\oxxitype}} \in
  \oxexplode{\oxvarctx^\prime}. \; \not\exists \oxcprov^\prime. \;
  \oxloanpkg{\oxmuta}{\oxderef{\! \oxplace}} \in
  \oxvarctx^\prime(\oxcprov^\prime)$.
\end{oxlemma}

\begin{proof}
  We proceed by induction on the typing derivation.

  \oxname{T-Function}, \oxname{T-Abort}, \oxname{T-Unit}, \oxname{T-u32},
  \oxname{T-True}, \oxname{T-False}, and \oxname{T-Dead} are all immediate.

  \oxname{T-LetRegion}, \oxname{T-While}, \oxname{T-ForArray},
  \oxname{T-ForSlice}, \oxname{T-Closure}, \oxname{T-Tuple}, \oxname{T-Slice},
  \oxname{T-Drop}, \oxname{T-Left}, \oxname{T-Right}, and \oxname{T-Shift},
  \oxname{T-Framed}, and \oxname{T-ClosureValue} all follow directly from the
  induction hypothesis.

  For \oxname{T-Move}, \oxname{T-Copy}, \oxname{T-Borrow},
  \oxname{T-BorrowIndex}, \oxname{T-BorrowSlice}, and \oxname{T-IndexCopy}, we
  rely on \Lemma{lemma:ownership-well-typed-extension} and the induction
  hypothesis for almost all of our obligations. For all of them except
  \oxname{T-Move}, we also have to show that the type computation for
  $\oxplaceexpr$ still works in the extended stack typing. This follows from
  \Lemma{lemma:computety-well-typed-extension}.

  For \oxname{T-Branch} and \oxname{T-Match}, we use the induction hypothesis in
  conjunction with \Lemma{lemma:subtyping-well-typed-extension} to get most of
  the premises. In the end, we also need to deal with the union between the
  output environments from the two region rewriting derivations. Fortunately, we
  know that by definition this operation unions corresponding loan sets for the
  same region and as such creates no new loans leaving our not-reborrowed
  property intact.

  \oxname{T-Seq} proceeds almost directly based on just the induction
  hypothesis, but with the required note that garbage collecting loans can only
  remove loans and thus leaves our not-reborrowed property intact.
  \oxname{T-Let} follows similarly to \oxname{T-Seq} but also requires the use
  of \Lemma{lemma:subtyping-well-typed-extension}.

  \oxname{T-Assign} and \oxname{T-AssignDeref} follow from the induction
  hypothesis combined with \Lemma{lemma:subtyping-well-typed-extension} and
  \Lemma{lemma:ownership-well-typed-extension}.

  \oxname{T-App} follows from the induction hypothesis and
  \Lemma{lemma:outlives-well-typed-extension}.

  \oxname{T-Pointer} follows from \Lemma{lemma:computety-well-typed-extension}.
\end{proof}

\begin{oxlemma}{Values are Well-Typed under Well-Typed Extensions}
  {lemma:values-well-typed-extension} If
  $\oxtypevalidity{\oxglobalctx}{\oxemptyctx}{\oxvarctx}{\oxsitype_\oxid}$ and
  $\forall \oxcprov \in \oxfprovs{\oxsitype_\oxid}. \;
  \oxnotreborrowed{\oxvarctx}{\oxcprov}$ and
  $\oxtypjudge{\oxglobalctx}{\oxemptyctx}{\oxkontctx}{\oxvarctx}
  {\oxvalue}{\oxsitype}{\oxvarctx}$, then
  $\oxtypjudge{\oxglobalctx}{\oxemptyctx}{\oxkontctx}{ \oxextendctx{\oxvarctx}{
      \oxvarctxentry{\oxid}{\oxsitype_\oxid} } }{\oxvalue}{\oxsitype}{
    \oxextendctx{\oxvarctx}{ \oxvarctxentry{\oxid}{\oxsitype_\oxid} } }$.
\end{oxlemma}

\begin{proof}
  We proceed by induction on the typing derivation. The only non-immediate case
  is \oxname{T-ClosureValue}.

  \vspace{1em}
  \noindent \framebox[\textwidth]{ \figuresize
    \begin{mathpar}
      \TClosureVal
    \end{mathpar}
  } \vspace{1em}

  First, we invert the stack frame typing hypothesis to get that $\forall \oxid
  \in \oxdomain{\oxstackframe}.\;
  \oxtypjudge{\oxglobalctx}{\oxemptyctx}{\oxkontctx}{\oxnewframe{\oxvarctx}{\oxframe_c}}{\oxstackframe(\oxid)}{\oxframe_c(\oxid)}{\oxnewframe{\oxvarctx}{\oxframe_c}}$.
  We can apply the induction hypothesis to each of these statements, and apply
  \oxname{WF-Frame} to get
  $\oxsubstorevalidity{\oxglobalctx}{\oxextendctx{\oxvarctx}{\oxvarctxentry{\oxid}{\oxsitype_x}}}{\oxstackframe_c}{\oxframe_c}$.

  Next, we need to show that
  $\oxtypjudge{\oxglobalctx}{\oxemptyctx}{\oxkontctx}{ \oxnewframe{
      \oxextendctx{\oxvarctx}{ \oxvarctxentry{\oxid}{\oxsitype_\oxid} } }{
      \oxframe_c \oxcomma \oxvarctxentry{\oxid_1}{\oxsitype_1} \oxdotsc
      \oxvarctxentry{\oxid_n}{\oxsitype_n} }
  }{\oxexpr}{\oxsitype_\oxcprov}{\oxnewframe{\oxvarctx^{\prime\prime}}{\oxframe^\prime}}$
  for some $\oxvarctx^{\prime\prime}$ and $\oxframe^\prime$. We get this by
  applying \Lemma{lemma:exprs-well-typed-extension} to
  $\oxtypjudge{\oxglobalctx}{\oxemptyctx}{\oxkontctx}{ \oxnewframe{\oxvarctx}{
      \oxframe_c \oxcomma \oxvarctxentry{\oxid_1}{\oxsitype_1} \oxdotsc
      \oxvarctxentry{\oxid_n}{\oxsitype_n} }
  }{\oxexpr}{\oxsitype_\oxcprov}{\oxnewframe{\oxvarctx^\prime}{\oxframe^\prime}}$
  (from the premise of \oxname{T-ClosureValue}).
\end{proof}

\begin{oxlemma}{Stack Validity is Preserved under Well-Typed Extensions}
  {lemma:store-validity-extension} If
  $\oxstorevalidity{\oxglobalctx}{\oxvarctx}{\oxstore}$ and
  $\forall \oxcprov \in \oxfprovs{\oxsitype_\oxid}. \;
  \oxnotreborrowed{\oxvarctx}{\oxcprov}$ and
  $\oxtypjudge{\oxglobalctx}{\oxtvarctx}{\oxkontctx}{\oxvarctx}{\oxvalue}
  {\oxsitype}{\oxvarctx}$, then $\oxstorevalidity{\oxglobalctx}{
    \oxextendctx{\oxvarctx}{ \oxvarctxentry{\oxid}{\oxsitype} } }{
    \oxextendctx{\oxstore}{ \oxstoreentry{\oxid}{\oxvalue} } }$.
\end{oxlemma}

\begin{proof}
  This proof follows directly from the definition of \oxname{WF-StackFrame}.
  
  \vspace{1em}
  \noindent \framebox[\textwidth]{ \figuresize
    \begin{mathpar}
      \VStack
    \end{mathpar}
  } \vspace{1em}

  In particular, inversion of \oxname{WF-StackFrame} on
  $\oxstorevalidity{\oxglobalctx}{\oxvarctx}{\oxstore}$ gives us well-formedness
  for the remainder of the stack, $\oxdomain{\oxstackframe} =
  \oxdomain{\oxframe}|_\oxid$ and $\forall \oxid \in \oxdomain{\oxstackframe}.
  \;
  \oxtypjudge{\oxglobalctx}{\oxemptyctx}{\oxemptyctx}{\oxnewframe{\oxvarctx}{\oxframe}}
  {(\oxnewframe{\oxstore}{\oxstackframe})(\oxid)}{(\oxnewframe{\oxvarctx}{\oxframe})(\oxid)}{\oxnewframe{\oxvarctx}{\oxframe}}$.
  We can then apply \Lemma{lemma:values-well-typed-extension} to each of these
  derivations to get $\forall \oxid \in \oxdomain{ \oxextendctx{\oxstackframe}{
      \oxstoreentry{\oxid}{\oxvalue} } }. \;
  \oxtypjudge{\oxglobalctx}{\oxemptyctx}{\oxemptyctx}{
    \oxextendctx{\oxnewframe{\oxvarctx}{\oxframe}}{
      \oxvarctxentry{\oxid}{\oxsitype} } }{ (\oxextendctx{\oxstackframe}{
      \oxstoreentry{\oxid}{\oxvalue} })(\oxid) }{
    (\oxextendctx{\oxnewframe{\oxvarctx}{\oxframe}}{
      \oxvarctxentry{\oxid}{\oxsitype} })(\oxid) }{
    \oxextendctx{\oxnewframe{\oxvarctx}{\oxframe}}{
      \oxvarctxentry{\oxid}{\oxsitype} } }$. We can then see that the
  well-formedness of the remainder of the stack is unaffected, and that the
  domains when extended with $\oxid$ remain equal. The last obligation is to
  show that the $\oxvalue$ is well-typed in the current stack typing, but we
  already have that from our premise. Thus, we can apply \oxname{WF-StackFrame}
  with the extended stack to get $\oxstorevalidity{\oxglobalctx}{
    \oxextendctx{\oxvarctx}{ \oxvarctxentry{\oxid}{\oxsitype} } }{
    \oxextendctx{\oxstore}{ \oxstoreentry{\oxid}{\oxvalue} } }$.
\end{proof}

\subsection{Preservation after Assignment Lemmas}
\label{sec:assign-lemmas}

\begin{oxlemma}{Ownership Safety is Preserved after Assignment}
  {lemma:ownership-safety-assignment}
  If $\oxlookup{\oxvarctx}{\oxplace_a}{\oxsxtype}$ and
  $\oxtunify[\oxnoop]{\oxtvarctx}{\oxkontctx}{
    \oxrsub{\oxvarctx}{\oxderef{\oxplace_a}}
  }{\oxsitype}{\oxsxtype}{\oxvarctx^\prime}$ and
  $\oxrgnuniqueto{\oxsxtype}{\oxplace_a}{\oxvarctx}$ and
  $\oxmusafetyinner{\oxemptyctx}{\oxkontctx}{
    \oxnewframe{\oxvarctx}{\oxframe}
  }{\oxmuta}{\overline{\oxplace_o}}{\oxplaceexpr}{\oxset{\oxloans}}$ and
  $\oxset{\overline{\oxplace_n}} = \oxset{\overline{\oxplace_o}} \setminus \oxplace_a$,
  then $\oxmusafetyinner{\oxemptyctx}{\oxkontctx}{
    \oxnewframe{
      \oxgcloans{\oxkontctx}{\oxtupdate{\oxvarctx^\prime}{\oxplace_a}{\oxsitype}}
    }{\oxframe}
  }{\oxmuta}{\overline{\oxplace_n}}{\oxplaceexpr}{\oxset{\oxloans^\prime}}$.
\end{oxlemma}

\begin{proof}
  We proceed by induction on the ownership safety derivation.

  \vspace{1em}
  \noindent \framebox[\textwidth]{ \figuresize
    \begin{mathpar}
      \OSafePlace
    \end{mathpar}
  } \vspace{1em}

  Consider an arbitrary region $\oxcprov$ from the domain of
  $\oxnewframe{\oxvarctx}{\oxframe}$. For this $\oxcprov$, we wish to show that
  either of the two clauses in \oxname{O-SafePlace} which were previously true
  are maintained after going through
  $\oxtunify[\oxnoop]{\oxtvarctx}{\oxkontctx}{
    \oxrsub{\oxvarctx}{\oxderef{\oxplace_a}}
  }{\oxsitype}{\oxsxtype}{\oxvarctx^\prime}$ and $\oxgcloans{\oxkontctx}{\cdot}$,
  which kills loans prefixed by $\oxderef{\oxplace}$, checks that the new type
  for $\oxplace_a$ is compatible with its old type, and clears out loans
  associated with its outermost region. So, we will consider each clause as a
  separate case.

  We will first consider the case where we have $\forall
  \oxloanpkg{\oxmuta^\prime}{\oxplaceexprctx[\oxplace^\prime]} \in \oxset{\overline{\oxloan}}.
  (\oxmuta = \oxmut \vee \oxmuta^\prime = \oxmut) \implies
  \oxrelevant{\oxplace^\prime}{\oxplace}$. In this case, we know by definition of
  $\rsub$ that $\forall \oxcprov \in \oxdomain{\oxvarctx}. \;
  (\oxrsub{\oxvarctx}{\oxderef{\oxplace_a}})(\oxcprov) \subseteq \oxvarctx(\oxcprov)$.
  We then know, again by definition (see \oxname{OL-CheckConcrete}), that
  $\forall \oxcprov \in \oxdomain{\oxvarctx}. \;
  (\oxrsub{\oxvarctx}{\oxderef{\oxplaceexpr}})(\oxcprov) =
  \oxvarctx^\prime(\oxcprov)$. As such, we know that the $\rsub$ and
  $\oxgcloans{\oxkontctx}{\cdot}$ has at most shrank the obligations in this case to
  having fewer disjointedness obligations, and it is otherwise unchanged.

  This leaves us to consider the second case
  $\exists \oxvarctxentry{\oxplace^\prime}{\oxtref{\oxcprov^\prime}{\oxmuta^\prime}{\oxtype^\prime}} \in \oxexplode{\oxvarctx} \; \wedge \; \not \exists
  \oxtref{\oxcprov^\prime}{\oxmuta^\prime}{\oxtype^\prime} \in \oxkontctx \;
  \wedge
  (\forall \oxvarctxentry{\oxplace^\prime}{\oxtref{\oxcprov^\prime}{\oxmuta^\prime}{\oxtype^\prime}} \in \oxexplode{\oxvarctx}. \;
  \oxplace^\prime \in \oxset{
    \overline{\oxplace_e}
  })$. Recall that, definitionally, neither $\rsub$ nor the region
  rewriting judgment change variable bindings and their associated types in the
  environment (instead both affect only the loan sets associated with regions,
  though the latter does not when run in the checking mode $\oxnoop$).
  Thus, we know that $\oxexplode{\oxvarctx} = \oxexplode{\oxvarctx^\prime}$. We then
  need to consider two distinct possibilities for how the exclusion list has
  changed. We know from the premise that $\oxset{\overline{\oxplace_n}} =
  \oxset{\overline{\oxplace_o}} \setminus \oxplace_a$ which means that either the two sets
  are exactly identical (when $\oxplace_a \not\in \oxset{\overline{\oxplace_o}}$) or
  smaller by $\oxplace_a$ in particular
  (when $\oxplace_a \in \oxset{\overline{\oxplace_o}}$). In the former case, the
  exclusion list is unchanged which means the whole clause is true for every
  $\oxcprov^\prime$ in $\oxvarctx^\prime$ for which it was true in $\oxvarctx$. In the
  latter case, the regions $\oxcprov^\prime$ is in the type of $\oxplace_a$ which
  \emph{has} been removed from the exclusion list $\oxset{\overline{\oxplace_n}}$.
  Thus, we need to show for the loans $\oxset{\overline{\oxloan}}$ associated
  with $\oxcprov^\prime$ that $\forall \oxloanpkg{\oxmuta^\prime}{\oxplaceexprctx[\oxplace^\prime]}
  \in \oxset{\overline{\oxloan}}. (\oxmuta = \oxmut \vee \oxmuta^\prime = \oxmut)
  \implies \oxrelevant{\oxplace^\prime}{\oxplace}$. By definition,
  $\oxrgnuniqueto{\oxsxtype}{\oxplace}{\oxvarctx}$ tells us that the outermost
  region $\oxcprov^\prime$ is unique to the type $\oxsxtype$ and place $\oxplace_a$,
  and thus when we replace it with $\oxsitype$, we ensure that $\oxcprov^\prime$ does
  not occur in any type in $\oxvarctx^\prime$. Thus, the surrounding call of
  $\oxgcloans{\oxkontctx}{\cdot}$ necessarily clears out the loan set meaning that
  the set associated with $\oxcprov^\prime$ is always empty in new environment,
  meaning the disjointness condition from \oxname{O-SafePlace} holds trivially.

  \vspace{1em}
  \noindent \framebox[\textwidth]{ \figuresize
    \begin{mathpar}
      \ODeref
    \end{mathpar}
  } \vspace{1em}

  We want to produce a new derivation using \oxname{O-Deref} for
  $\oxmusafetyinner{\oxemptyctx}{\oxkontctx}{
    \oxnewframe{\oxrsub{\oxvarctx}{\oxderef{\oxplace_a}}}{\oxframe}
  }{\oxmuta}{
    \overline{\oxplace_e}
  }{
    \oxplaceexprctx[\oxderef{\oxplace}]
  }{
    \oxset{\overline{\oxloan^{\prime\prime}}}
  }$. We have $\oxlookup{(\oxnewframe{\oxvarctx}{\oxframe})}{\oxplace}{
    \oxtref{\oxcprov}{\oxmuta_\oxplace}{\oxtype_\oxplace}
  }$ from the premise of \oxname{O-Deref}. We then know by the definition of
  $\rsub$ that $\oxlookup{
    (\oxnewframe{\oxrsub{\oxvarctx}{\oxderef{\oxplace_a}}}{\oxframe})
  }{\oxplace}{
    \oxtref{\oxcprov}{\oxmuta_\oxplace}{\oxtype_\oxplace}
  }$ since $\rsub$ only affects the loan set portion of the codomain of its
  input environment. We also know from the premise of \oxname{O-Deref} that
  $\oxlookup{
    (\oxnewframe{\oxvarctx}{\oxframe})
  }{\oxcprov}{
    \oxset{\overline{\oxloanpkg{\oxmuta^\prime}{\oxplaceexpr}}^n}
  }$. Again, by the definition of $\rsub$, we have that
  $\oxlookup{
    (\oxnewframe{\oxrsub{\oxvarctx}{\oxderef{\oxplace_a}}}{\oxframe})
  }{\oxcprov}{
    \oxset{\overline{\oxloanpkg{\oxmuta^k}{\oxplaceexpr^k}}^m}
  }$ where we know $m \leq n$ and
  $\oxset{\overline{\oxloanpkg{\oxmuta^k}{\oxplaceexpr^k}}^m} \subseteq
  \oxset{\overline{\oxloanpkg{\oxmuta^\prime}{\oxplaceexpr}}^n}$.

  We know by the definition of $\rsub$ also that every place expression
  $\oxplaceexpr_d$ in
  $\oxset{\overline{\oxloanpkg{\oxmuta^\prime}{\oxplaceexpr}}^n} \setminus \oxset{\overline{\oxloanpkg{\oxmuta^k}{\oxplaceexpr^k}}^m}$
  can be decomposed into $\oxplaceexprctx_d[\oxderef{\oxplace_a}]$. This means
  that for computing the set $\textrm{excl}$, it is either the same or has
  shrunk by precisely $\oxplace_a$. This lines up with our induction hypothesis
  which we apply to each of $\forall i \in \oxset{1 \oxdots n}. \;
  \oxmusafetyinner{\oxemptyctx}{\oxkontctx}{\oxvarctx}{\oxmuta}{
    \overline{\oxplace_e} \oxcomma \textrm{excl} \oxcomma \oxplace
  }{
    \oxplaceexprctx[\oxplaceexpr_i]
  }{
    \oxset{\overline{\oxloanpkg{\oxmuta}{\oxplaceexpr_i^\prime}}}
  }$ from the premise of \oxname{O-Deref}. This gives us
  $\forall i \in \oxset{1 \oxdots n}. \;
  \oxmusafetyinner{\oxemptyctx}{\oxkontctx}{
    \oxrsub{\oxvarctx}{\oxderef{\oxplace_a}}
  }{\oxmuta}{
    \overline{\oxplace_e} \oxcomma \textrm{excl} \oxcomma \oxplace
  }{
    \oxplaceexprctx[\oxplaceexpr_i]
  }{
    \oxset{\overline{\oxloanpkg{\oxmuta}{\oxplaceexpr_i^{\prime\prime}}}}
  }$.

  Finally, for the last premise of \oxname{O-Deref}, the proof precedes
  identically to the case for \oxname{O-SafePlace} since the obligation is
  exactly the same.

  \vspace{1em}
  \noindent \framebox[\textwidth]{ \figuresize
    \begin{mathpar}
      \ODerefAbs
    \end{mathpar}
  } \vspace{1em}

  Since $\oxtvarctx = \oxemptyctx$, there are no valid reference types that have
  an abstract region, meaning the first hypothesis is a contradiction.
\end{proof}

\begin{oxlemma}{Outlives is Preserved after Assignment}
  {lemma:outlives-assignment} If
  $\oxlookup{\oxvarctx}{\oxplace_a}{\oxsxtype}$
  and $\oxtunify[\oxnoop]{\oxtvarctx}{\oxkontctx}{
    \oxrsub{\oxvarctx}{\oxderef{\oxplace_a}}
  }{\oxsitype}{\oxsxtype}{\oxvarctx^\prime}$ and
  $\oxrunify{\oxemptyctx}{\oxkontctx}{\oxvarctx}{\oxprov_1}{\oxprov_2}{\oxvarctx_o}$, then
  $\oxrunify{\oxemptyctx}{\oxkontctx}{\oxvarctx^\prime}{\oxprov_1}{\oxprov_2}{
    \oxvarctx^\prime_o
  }$.
\end{oxlemma}

\begin{proof}
  We proceed by induction on the outlives judgment. This gives us seven cases,
  \oxname{OL-Refl}, \oxname{OL-Trans}, \oxname{OL-BothAbstract},
  \oxname{OL-CombineConcrete}, \oxname{OL-CombineConcrete},
  \oxname{OL-ConcreteAbstract} and \oxname{OL-AbstractConcrete}.

  \oxname{OL-Refl}, \oxname{OL-BothAbstract} and \oxname{OL-AbstractConcrete}
  are immediate.

  \oxname{OL-Trans} follows from the induction hypothesis.

  \oxname{OL-ConcreteAbstract} follows from the induction hypothesis and
  noting that the type computation does not depend on the contents of loan sets.

  This leaves \oxname{OL-CombineConcreteUnrestricted},
  \oxname{OL-CombineConcrete} and \oxname{OL-CheckConcrete} as the most
  interesting cases. We note that the checking mode $\oxnoop$ corresponds to
  making no changes to the environment, thus $\oxvarctx^\prime =
  \oxrsub{\oxvarctx}{\oxderef{\oxplace_a}}$. Then, for each, we note that the
  value of each associated loan set in the input environment only has an effect
  on the output environment and not whether or not the rule applies. Thus, since
  we know that $\oxvarctx^\prime$ (compared to $\oxvarctx$) has had some loans
  removed (those rooted at $\oxderef{\oxplace_a}$), then we can still produce a
  derivation, only with a different, potentially smaller output.
\end{proof}

\begin{oxlemma}{Region Rewriting is Preserved after Assignment}
  {lemma:subtyping-assignment} If
  $\oxlookup{\oxvarctx}{\oxplace_a}{\oxsxtype}$
  and $\oxtunify[\oxnoop]{\oxtvarctx}{\oxkontctx}{
    \oxrsub{\oxvarctx}{\oxderef{\oxplace_a}}
  }{\oxsitype}{\oxsxtype}{\oxvarctx^\prime}$ and
  $\oxtunify{\oxemptyctx}{\oxkontctx}{\oxvarctx}{\oxtype_1}{\oxtype_2}{\oxvarctx_o}$, then
  $\oxtunify{\oxemptyctx}{\oxkontctx}{\oxvarctx^\prime}{\oxtype_1}{\oxtype_2}{
    \oxvarctx^\prime_o
  }$.
\end{oxlemma}

\begin{proof}
  We proceed by induction on the region rewriting judgment. This gives us seven
  cases, \oxname{RR-Refl}, \oxname{RR-Trans}, \oxname{RR-Array},
  \oxname{RR-Slice}, \oxname{RR-Reference}, \oxname{RR-Tuple}, and
  \oxname{RR-Dead}.

  \oxname{RR-Refl} is immediate.

  \oxname{RR-Trans}, \oxname{RR-Array}, \oxname{RR-Slice}, \oxname{RR-Tuple} and
  \oxname{RR-Dead} all follow directly from the induction hypothesis.

  This leaves \oxname{RR-Reference} which follows from
  \Lemma{lemma:outlives-assignment} and the induction hypothesis.
\end{proof}

\begin{oxlemma}{Expressions are Well-Typed after Assignment}
  {lemma:exprs-well-typed-assignment}
  If $\oxlookup{\oxvarctx}{\oxplace_a}{\oxsxtype}$
  and $\oxtunify[\oxnoop]{\oxtvarctx}{\oxkontctx}{
    \oxrsub{\oxvarctx}{\oxderef{\oxplace_a}}
  }{\oxsitype}{\oxsxtype}{\oxvarctx^\prime}$ and
  $\oxmusafety{\oxemptyctx}{\oxkontctx}{\oxvarctx^\prime}{\oxmut}{\oxplace_a}{
    \oxset{\oxloanpkg{\oxmut}{\oxplace_a}}
  }$ and
  $\oxtypjudge{\oxglobalctx}{\oxemptyctx}{\oxkontctx}{
    \oxnewframe{\oxvarctx}{\oxframe}
  }{\oxexpr}{\oxsitype}{\oxnewframe{\oxvarctx_o}{\oxframe^\prime}}$, then
  $\oxtypjudge{\oxglobalctx}{\oxemptyctx}{\oxkontctx}{
    \oxnewframe{
      \oxgcloans{\oxkontctx}{\oxtupdate{\oxvarctx^\prime}{\oxplace_a}{\oxsitype}}
    }{\oxframe}
  }{\oxexpr}{\oxsitype}{
    \oxnewframe{
      \oxtupdate{\oxvarctx^\prime_o}{\oxplace_a}{\oxsitype}
    }{\oxframe^{\prime\prime}}
  }$.
\end{oxlemma}

\begin{proof}
  We proceed by induction on the typing derivation.

  \oxname{T-Function}, \oxname{T-Abort}, \oxname{T-Unit}, \oxname{T-u32},
  \oxname{T-True}, \oxname{T-False}, and \oxname{T-Dead} are all immediate.

  \oxname{T-Seq}, \oxname{T-LetRegion}, \oxname{T-While}, \oxname{T-ForArray},
  \oxname{T-ForSlice}, \oxname{T-Closure}, \oxname{T-Tuple}, \oxname{T-Slice},
  \oxname{T-Drop}, \oxname{T-Left}, \oxname{T-Right}, and \oxname{T-Shift},
  \oxname{T-Framed}, and \oxname{T-ClosureValue} all follow directly from the
  induction hypothesis.

  For \oxname{T-Move}, \oxname{T-Copy}, \oxname{T-Borrow},
  \oxname{T-BorrowIndex}, \oxname{T-BorrowSlice}, and \oxname{T-IndexCopy}, we
  rely on \Lemma{lemma:ownership-safety-assignment} and the induction
  hypothesis for almost all of our obligations. For all of them except
  \oxname{T-Move}, we also have to show that the type computation for
  $\oxplaceexpr$ still works in the updated environment. Since we know that
  $\oxplace_a$ is ownership safe from our premise, we know that $\oxplaceexpr$
  is disjoint from $\oxplace_a$ and thus the type update could not affect its
  type computation. Otherwise, the only difference is in loan sets associated
  with regions, and thus does not affect type computation.

  \oxname{T-Pointer} requires the same argument about type computation as in
  \oxname{T-Borrow}, but does not need the additional lemmas or the induction
  hypothesis.

  For \oxname{T-Branch} and \oxname{T-Match}, we use the induction hypothesis in
  conjunction with \Lemma{lemma:subtyping-assignment} to get most of

  the premises. In the end, we also need to deal with the union between the
  output environments from the two rewriting derivations. Fortunately, we know
  that by definition this operation unions corresponding loan sets for the same
  region and so commutes with $\oxgcloans{\oxkontctx}{\cdot}$ and the type update.

  \oxname{T-Let} follows similarly to \oxname{T-Branch} and \oxname{T-Match}
  using the induction hypothesis in conjunction with
  \Lemma{lemma:subtyping-assignment} without the need to address a combined
  environment.

  \oxname{T-Assign} and \oxname{T-AssignDeref} follow from the induction
  hypothesis combined with \Lemma{lemma:subtyping-assignment} and
  \Lemma{lemma:ownership-safety-assignment}.

  \oxname{T-App} follows from the induction hypothesis and
  \Lemma{lemma:outlives-assignment}.
\end{proof}

\begin{oxlemma}{Values are Well-Typed after Assignment}
  {lemma:values-well-typed-assignment}
  If $\oxlookup{\oxvarctx}{\oxplace_a}{\oxsxtype}$ and
  $\oxtunify[\oxnoop]{\oxtvarctx}{\oxkontctx}{
    \oxrsub{\oxvarctx}{\oxderef{\oxplace_a}}
  }{\oxsitype}{\oxsxtype}{\oxvarctx^\prime}$ and
  $\oxmusafety{\oxemptyctx}{\oxkontctx}{\oxvarctx^\prime}{\oxmut}{\oxplace_a}{
    \oxset{\oxloanpkg{\oxmut}{\oxplace_a}}
  }$ and
  $\oxtypjudge{\oxglobalctx}{\oxemptyctx}{\oxkontctx}{\oxvarctx}
  {\oxvalue}{\oxtype}{\oxvarctx}$, then
  $\oxtypjudge{\oxglobalctx}{\oxemptyctx}{\oxkontctx}{
    \oxgcloans{\oxkontctx}{\oxtupdate{\oxvarctx^\prime}{\oxplace_a}{\oxsitype}}
  }{
    \oxvalue
  }{\oxtype}{
    \oxgcloans{\oxkontctx}{\oxtupdate{\oxvarctx^\prime}{\oxplace_a}{\oxsitype}}
  }$.
\end{oxlemma}

\begin{proof}
  We proceed by induction on the value typing derivation. The only non-immediate
  cases are \oxname{T-Pointer} and \oxname{T-ClosureValue}.

  \vspace{1em}
  \noindent \framebox[\textwidth]{ \figuresize
    \begin{mathpar}
      \TPointer
    \end{mathpar}
  } \vspace{1em}

  In \oxname{T-Pointer}, we have a requirement that
  $\oxloanpkg{\oxmuta}{\oxplace} \in \oxvarctx(\oxcprov)$ which could
  potentially be affected by the kill rules. However, note that the definition
  of $\rsub$ is such that we only remove loans of the form
  $\oxderef{\oxplace_a}$ which necessarily cannot match this loan which has no
  dereference in it. Thus, we know that $\oxloanpkg{\oxmuta}{\oxplace} \in
  (\oxrsub{\oxvarctx}{\oxderef{\oxplace}})(\oxcprov)$ and thus,
  $\oxtypjudge{\oxglobalctx}{\oxemptyctx}{\oxkontctx}{
    \oxrsub{\oxvarctx}{\oxderef{\oxplace_a}}
  }{
    \oxptr{\oxreferentctx[\oxplace]}
  }{\oxsitype}{
    \oxrsub{\oxvarctx}{\oxderef{\oxplace_a}}
  }$. We then know that the checking mode $\oxnoop$ for rewriting does not
  change the output environment, and thus, $\oxvarctx^\prime =
  \oxrsub{\oxvarctx}{\oxderef{\oxplace_a}}$. Then, we know that
  $\oxplace \neq \oxplace_a$ (this would otherwise conflict with the ownership
  safety derivation for $\oxplace_a$ in our premise), so the type update for
  $\oxplace_a$ does not impact \oxname{T-Pointer}. Finally, the call to
  $\oxgcloans{\oxkontctx}{\cdot}$ clears out any unused regions, but the region
  $\oxcprov$ here is still in use and thus not changed.

  \vspace{1em}
  \noindent \framebox[\textwidth]{ \figuresize
    \begin{mathpar}
      \TClosureVal
    \end{mathpar}
  } \vspace{1em}

  First, we invert the stack frame typing hypothesis to get that $\forall \oxid
  \in \oxdomain{\oxstackframe}. \;
  \oxtypjudge{\oxglobalctx}{\oxemptyctx}{\oxkontctx}{
    \oxnewframe{\oxvarctx}{\oxframe_c}
  }{\oxstackframe(\oxid)}{\oxframe_c(\oxid)}
  {\oxnewframe{\oxvarctx}{\oxframe_c}}$.
  We can apply the induction hypothesis to each of these statements, and apply
  \oxname{WF-Frame} to get
  $\oxsubstorevalidity{\oxglobalctx}{
      \oxgcloans{\oxkontctx}{\oxtupdate{\oxvarctx^\prime}{\oxplace_a}{\oxsitype}}
  }{\oxstackframe_c}{\oxframe_c}$.

  Next, we need to show that
  $\oxtypjudge{\oxglobalctx}{\oxemptyctx}{\oxkontctx}{
    \oxnewframe{
      \oxgcloans{\oxkontctx}{\oxtupdate{\oxvarctx^\prime}{\oxplace_a}{\oxsitype}}
    }{
      \oxframe_c \oxcomma
      \oxvarctxentry{\oxid_1}{\oxsitype_1} \oxdotsc
      \oxvarctxentry{\oxid_n}{\oxsitype_n}
    }
  }{\oxexpr}{\oxsitype_\oxcprov}{\oxnewframe{\oxvarctx^{\prime\prime}}{\oxframe^\prime}}$
  for some $\oxvarctx^{\prime\prime}$ and $\oxframe^\prime$. We get this by
  applying \Lemma{lemma:exprs-well-typed-assignment} to
  $\oxtypjudge{\oxglobalctx}{\oxemptyctx}{\oxkontctx}{
    \oxnewframe{\oxvarctx}{
      \oxframe_c \oxcomma
      \oxvarctxentry{\oxid_1}{\oxsitype_1} \oxdotsc
      \oxvarctxentry{\oxid_n}{\oxsitype_n}
    }
  }{\oxexpr}{\oxsitype_\oxcprov}{\oxnewframe{\oxvarctx^\prime}{\oxframe^\prime}}$
  from the premise of \oxname{T-ClosureValue}.
\end{proof}

\begin{oxlemma}{Stack Validity is Preserved after Assignment}
  {lemma:stack-validity-assign}
  If $\oxstorevalidity{\oxglobalctx}{\oxvarctx}{\oxstore}$ and
  $\oxlookup{\oxvarctx}{\oxplace}{\oxsxtype}$ and
  $\oxtypjudge{\oxglobalctx}{\oxemptyctx}{\oxkontctx}{\oxvarctx}{
    \oxvalue
  }{\oxsitype}{\oxvarctx}$ and
  $\oxtunify{\oxemptyctx}{\oxkontctx}{
    \oxrsub{\oxvarctx}{\oxderef{\oxplace}}
  }{\oxsitype}{\oxsxtype}{\oxvarctx^\prime}$ and
  $\oxmusafety{\oxemptyctx}{\oxkontctx}{\oxvarctx^\prime}{\oxmut}{\oxplace}{
    \oxset{\oxloanpkg{\oxmut}{\oxplace}}
  }$ and
  $\oxnorm{\oxstore}{\oxplace}{\oxplace}{\oxvaluectx}{\_}$ and
  $\oxplace = \oxproj{\oxid}{\oxpath}$,
  then
  $\oxstorevalidity{\oxglobalctx}{
    \oxgcloans{\oxkontctx}{\oxtupdate{\oxvarctx^\prime}{\oxplace}{\oxsitype}}
  }{
    \oxvupdate{\oxstore}{\oxid}{\oxvaluectx[\oxvalue]}
  }$.
\end{oxlemma}

\begin{proof}
  The proof proceeds by induction on the stack validity judgment
  $\oxstorevalidity{\oxglobalctx}{\oxvarctx}{\oxstore}$ which has two cases,
  \oxname{WF-StackEmpty} and \oxname{WF-StackFrame}.

  \vspace{1em}
  \noindent \framebox[\textwidth]{
    \figuresize
    \begin{mathpar}
      \VStackEmpty
    \end{mathpar}
  }
  \vspace{1em}

  In this case, the stack is empty and therefore, we have a contradiction since
  our premise says that $\oxlookup{\oxvarctx}{\oxplace}{\oxsxtype}$, but
  $\oxvarctx = \oxemptyctx$ and $\oxemptyctx(\oxplace)$ necessarily fails.

  \vspace{1em}
  \noindent \framebox[\textwidth]{
    \figuresize
    \begin{mathpar}
      \VStack
    \end{mathpar}
  }
  \vspace{1em}

  In the premise of \oxname{WF-StackFrame}, we have a collection of typing
  judgments for values stored in the stack. This naturally leads us to another
  case split: either $\oxid$ (the root of $\oxplace$ from $\oxplace =
  \oxproj{\oxid}{\oxpath}$) is in the current frame or it is not.

  If $\oxid$ is not in the current frame, we apply our induction hypothesis to
  $\oxstorevalidity{\oxglobalctx}{\oxvarctx}{\oxstore}$ to get
  $\oxstorevalidity{\oxglobalctx}{
    \oxtupdate{\oxvarctx^\prime}{\oxplace}{\oxsitype}
  }{
    \oxvupdate{\oxstore}{\oxid}{\oxvaluectx[\oxvalue]}
  }$. Then, we apply \oxname{WF-StackFrame} with the same typing judgments we
  already have to reach our overall conclusion of
  $\oxstorevalidity{\oxglobalctx}{
    \oxtupdate{
      (\oxnewframe{\oxvarctx^\prime}{\oxframe})
    }{\oxplace}{\oxsitype}
  }{
    \oxvupdate{
      (\oxnewframe{\oxstore}{\oxstackframe})
    }{\oxid}{\oxvaluectx[\oxvalue]}
  }$ (noting that substituting inside or outside is definitionally equal when
  we know that $\oxid \not\in \oxdomain{\oxstore}$).

  If $\oxid$ is in the current frame, then we apply
  \Lemma{lemma:subtyping-value-typing} to
  $\oxtypjudge{\oxglobalctx}{\oxemptyctx}{\oxkontctx}{\oxvarctx}{
    \oxvalue
  }{\oxsitype}{\oxvarctx}$ and
  $\oxtunify{\oxemptyctx}{\oxkontctx}{\oxvarctx}{\oxsitype}{\oxsxtype}{\oxvarctx^\prime}$ (both
  from our premise) to get
  $\oxtypjudge{\oxglobalctx}{\oxemptyctx}{\oxkontctx}{\oxvarctx^\prime}{
    \oxvalue
  }{\oxsitype}{\oxvarctx^\prime}$. Then, we note that it would be a well-formedness
  violation for this value to depend on $\oxid$ itself (since that would mean it
  was a cyclical reference) and thus, we can get that
  $\oxtypjudge{\oxglobalctx}{\oxemptyctx}{\oxkontctx}{
    \oxtupdate{\oxvarctx^\prime}{\oxplace}{\oxsitype}
  }{
    \oxvalue
  }{\oxsitype}{
    \oxtupdate{\oxvarctx^\prime}{\oxplace}{\oxsitype}
  }$. Finally, we can garbage collect the loans from the old type to get
  $\oxtypjudge{\oxglobalctx}{\oxemptyctx}{\oxkontctx}{
    \oxgcloans{\oxkontctx}{\oxtupdate{\oxvarctx^\prime}{\oxplace}{\oxsitype}}
  }{
    \oxvalue
  }{\oxsitype}{
    \oxgcloans{\oxkontctx}{\oxtupdate{\oxvarctx^\prime}{\oxplace}{\oxsitype}}
  }$

  For the other typing judgments in this frame, we apply
  \Lemma{lemma:values-well-typed-assignment} to get
  $\forall \oxid \in \oxdomain{\oxstackframe}. \;
  \oxtypjudge{\oxglobalctx}{\oxemptyctx}{\oxemptyctx}{
    \oxgcloans{\oxkontctx}{\oxtupdate{\oxvarctx^\prime}{\oxplace}{\oxsitype}}
  }{
    (\oxnewframe{\oxstore}{\oxstackframe})(\oxid)
  }{(\oxnewframe{\oxvarctx^\prime}{\oxframe})(\oxid)}{
    \oxgcloans{\oxkontctx}{\oxtupdate{\oxvarctx^\prime}{\oxplace}{\oxsitype}}
  }$. Thus, we can apply \oxname{WF-StackFrame} to conclude
  $\oxstorevalidity{\oxglobalctx}{
    \oxgcloans{\oxkontctx}{\oxtupdate{\oxvarctx^\prime}{\oxplace}{\oxsitype}}
  }{
    \oxvupdate{\oxstore}{\oxid}{\oxvaluectx[\oxvalue]}
  }$.
\end{proof}

\subsection{Values are Well-Types at Rewritten Types Lemma}
\label{sec:values-typed-supertype-lemma}

\begin{oxlemma}{Values are Well-Typed At Rewritten Types}
  {lemma:values-typed-supertype} 
  If $\oxtypjudge{\oxglobalctx}{\oxtvarctx}{\oxkontctx}{\oxvarctx}{\oxvalue}{\oxsitype}
  {\oxvarctx_i}$ and $\oxtunify{\oxtvarctx}{\oxkontctx}{\oxvarctx_i}{\oxsitype}
  {{\oxsitype}^\prime}{\oxvarctx^\prime}$, then $\oxtypjudge{\oxglobalctx}
  {\oxtvarctx}{\oxkontctx}{\oxvarctx^\prime}{\oxvalue}{{\oxsitype}^\prime}{\oxvarctx^\prime}$.
\end{oxlemma}
\begin{proof}
  We proceed by induction on the value typing relation.

  In the case of \oxname{T-Tuple}, we need to apply the induction hypothesis for
  each entry which has a changed type, and \Lemma{lemma:subtyping-value-typing}
  for each entry which does not.

  In the case of \oxname{T-Array}, we just apply the induction hypothesis to
  each entry.

  \vspace{1em}
  \noindent \framebox[\textwidth]{
    \figuresize
    \begin{mathpar}
      \TPointer
    \end{mathpar}
  }
  \vspace{1em}

  For the \oxname{T-Pointer} case, we proceed by induction on the region
  rewriting judgement. The only interesting cases are for reference types. From
  there, we proceed by induction on the outlives relation, for which the only
  interesting case is \oxname{OL-CombineConcrete}.
 
  \vspace{1em}
  \noindent \framebox[\textwidth]{
    \figuresize
    \begin{mathpar}
      \UCombineLocalProvs
    \end{mathpar}
  }
  \vspace{1em}

  The \oxname{T-Pointer} case is immediate. We know that the referent type is
  preserved since we do not change any types in the context, and we know the
  loan is preserved since loan sets only grow.

  In all other cases, we know the types cannot change, which means $\oxvarctx =
  \oxvarctx^\prime$, so we are done.
\end{proof}

\subsection{Function Definitions are Self-Contained Lemma}
\label{sec:function-defns-lemma}

\begin{oxlemma}{Function Definitions are Self-Contained}
  {lemma:function-defns-self-contained}\hfill\\
  If $\oxctxswellformed{\oxglobalctx}{\oxemptyctx}{\oxvarctx}{\oxkontctx}$ and
  $\oxlookup{\oxglobalctx}{\oxfnname}{
    \oxfuncdef{\oxfnname}{
      \overline{\oxenvvar} \oxcomma \overline{\oxabsprov} \oxcomma \overline{\oxtvar}
    }{
      \oxascribe{\oxid_1}{\oxsitype_1} \oxdotsc
      \oxascribe{\oxid_n}{\oxsitype_n}
    }{\oxsitype_r}{\overline{\oxabsprov_1: \oxabsprov_2}}{\oxexpr}
  }$, then
  $\oxtypjudge{\oxglobalctx}{
    \overline{\oxtvarctxentry{\oxenvvar}{\oxkenv}} \oxcomma
    \overline{\oxtvarctxentry{\oxabsprov}{\oxkprov}} \oxcomma
    \overline{\oxabsprov_1 \oxoutlives \oxabsprov_2} \oxcomma
    \overline{\oxtvarctxentry{\oxtvar}{\oxktype}}
  }{\oxkontctx}{
    \oxnewframe{\oxvarctx}{
      \oxvarctxentry{\oxid_1}{\oxsitype_1} \oxdotsc
      \oxvarctxentry{\oxid_n}{\oxsitype_n}
    }
  }{
    \oxframed{\oxexpr}
  }{\oxsitype_f}{\oxvarctx}$.
\end{oxlemma}

\begin{proof}
  Begin by noting that \oxname{WF-FunctionDefinition} gives us that
  \\ $\oxtypjudge{\oxglobalctx}{\overline{\oxtvarctxentry{\oxenvvar}{\oxkenv}}
    \oxcomma \overline{\oxtvarctxentry{\oxabsprov}{\oxkprov}} \oxcomma
    \overline{\oxabsprov_1 \oxoutlives \oxabsprov_2} \oxcomma
    \overline{\oxtvarctxentry{\oxtvar}{\oxktype}}}{\oxemptyctx}{\oxnewframe{\oxemptyctx}
    {\oxvarctxentry{\oxid_1}{\oxsitype_1} \oxdotsc
      \oxvarctxentry{\oxid_n}{\oxsitype_n}}}{\oxexpr}{\oxsitype_f}{\oxvarctx^\prime}$.
  We also have by inspection of the typing rules that $\oxvarctx^\prime =
  \oxnewframe{\oxemptyctx}{\oxframe^\prime}$ for some frame $\oxframe^\prime$.
  Then by \oxname{T-Framed}, it suffices to show that
  $\oxtypjudge{\oxglobalctx}{\overline{\oxtvarctxentry{\oxenvvar}{\oxkenv}}
    \oxcomma \overline{\oxtvarctxentry{\oxabsprov}{\oxkprov}} \oxcomma
    \overline{\oxabsprov_1 \oxoutlives \oxabsprov_2} \oxcomma
    \overline{\oxtvarctxentry{\oxtvar}{\oxktype}}}{\oxkontctx}{\oxnewframe{\oxvarctx}
    {\oxvarctxentry{\oxid_1}{\oxsitype_1} \oxdotsc
      \oxvarctxentry{\oxid_n}{\oxsitype_n}}}{\oxexpr}{\oxsitype_f}{\oxnewframe{\oxvarctx}{\oxframe^\prime}}$.
  But note that this is immediate. The typing derivation with $\oxemptyctx$ and
  the current frame means that there's absolutely no reliance on context outside
  $\oxid_1, \ldots \oxid_n$, and these places are necessarily completely
  disjoint from places in $\oxvarctx$ since any regions in their types must
  be abstract.
\end{proof}

\subsection{Subset Related Environments Lemma}
\label{sec:subset-related-lemmas}

\begin{oxlemma}{Outlives Produces Subset-Related Environments}
  {lemma:outlives-subset-related-envs}
  If $\oxrunify{\oxtvarctx}{\oxkontctx}{\oxvarctx}{\oxcprov_1}{\oxcprov_2}{\oxvarctx^\prime}$, then
  $\forall \oxcprov. \; \oxvarctx(\oxcprov) \subseteq \oxvarctx^\prime(\oxcprov)$.
\end{oxlemma}

\begin{proof}
  The proof proceeds by induction on the outlives relation $\oxrunify{\oxtvarctx}{\oxkontctx}
  {\oxvarctx}{\oxcprov_1}{\oxcprov_2}{\oxvarctx^\prime}$. We will consider each case.

  \vspace{1em}
  \noindent \framebox[\textwidth]{
    \figuresize
    \begin{mathpar}
      \UReflProv \and \UAbsProvs \and \OLocalAbsProvs \and \UCheckLocalProvs
    \end{mathpar}
  }
  \vspace{1em}

  Each of \oxname{OL-Refl}, \oxname{OL-BothAbstract},
  \oxname{OL-AbstractConcrete}, and \oxname{OL-CheckConcrete} are immediate
  since $\oxvarctx^\prime = \oxvarctx$.

  \vspace{1em}
  \noindent \framebox[\textwidth]{
    \figuresize
    \begin{mathpar}
      \UTransProv \and \OAbsLocalProvs
    \end{mathpar}
  }
  \vspace{1em}

  Both \oxname{OL-Trans} and \oxname{OL-ConcreteAbstract} follow by applying the induction
  hypothesis to all instances of the outlives judgment in their premise and then relying
  on transitivity of subset.

  \vspace{1em}
  \noindent \framebox[\textwidth]{
    \figuresize
    \begin{mathpar}
      \UCombineLocalProvs \and \UCombineLocalProvsUnrest
    \end{mathpar}
  }
  \vspace{1em}

  For \oxname{OL-CombineConcrete} and \oxname{OL-CombineConcreteUnrestricted},
  the conclusion is almost immediate since $\oxvarctx^\prime$ is very nearly
  $\oxvarctx$. However, it differs in the loan set for one particular region
  $\oxcprov_2$. Fortunately, its new loan set in $\oxvarctx^\prime$ is the union
  of its loan set with the loan set for $\oxcprov_1$ and thus we immediately
  have $\oxvarctx(\oxcprov_2) \subseteq \oxvarctx^\prime(\oxcprov_2)$.
\end{proof}

\begin{oxlemma}{Region Rewriting Produces Subset-Related Environments}
  {lemma:subtyping-subset-related-envs}
  If $\oxtunify{\oxtvarctx}{\oxkontctx}{\oxvarctx}{\oxtype_1}{\oxtype_2}{\oxvarctx^\prime}$, then
  $\forall \oxcprov. \; \oxvarctx(\oxcprov) \subseteq \oxvarctx^\prime(\oxcprov)$.
\end{oxlemma}

\begin{proof}
  This proof proceeds by induction on the region rewriting relation
  $\oxtunify{\oxtvarctx}{\oxkontctx}{\oxvarctx}{\oxtype_1}{\oxtype_2}{\oxvarctx^\prime}$. We will
  consider each case. For \oxname{RR-Refl} and \oxname{RR-Uninit}, the output
  environment $\oxvarctx^\prime$ is precisely $\oxvarctx$ and thus the result is
  immediate. For \oxname{RR-Trans}, \oxname{RR-Array}, \oxname{RR-Slice} and
  \oxname{RR-Tuple}, the result follows from applying the induction hypothesis
  to every region rewriting derivation in their premise and combining the
  results by transitivity of subset. This leaves us with one more interesting
  case, \oxname{RR-Reference}. For this case, apply
  \Lemma{lemma:outlives-subset-related-envs} to the outlives derivation in the
  premise. Then, apply our induction hypothesis to the region rewriting
  derivation in their premise. Finally, combine the two by transitivity of
  subset.
\end{proof}

\begin{oxlemma}{Frame Typing Union Produces Subset-Related Environments}
  {lemma:union-subset-frames}
  If $\oxframe = \oxframe_1 \oxintersect \oxframe_2$, then
  $\forall \oxcprov. \; \oxframe_1(\oxcprov) \subseteq \oxframe(\oxcprov)$ and
  $\forall \oxcprov. \; \oxframe_2(\oxcprov) \subseteq \oxframe(\oxcprov)$.
\end{oxlemma}

\begin{proof}
  First, note that the definition of $\oxintersect$ is symmetric and thus we will only
  prove the first conclusion, the second proceeding immediately the same in all cases.

  We proceed by induction over the frame typing. For $\oxemptyctx \oxintersect
  \oxemptyctx$, the case follows immediately. For
  $\oxextendctx{\oxframe_1}{\oxvarctxentry{\oxid}{\oxtype}} \oxintersect
  \oxextendctx{\oxframe_2}{\oxvarctxentry{\oxid}{\oxtype}}$, the result follows
  directly from applying the induction hypothesis to $\oxframe_1 \oxintersect
  \oxframe_2$. The last case is the interesting one,
  $\oxextendctx{\oxframe_1}{\oxloanctxentry{\oxcprov}{\overline{\oxloan}}}
  \oxintersect
  \oxextendctx{\oxframe_2}{\oxloanctxentry{\oxcprov}{\overline{\oxloan^\prime}}}$.
  In this case, we can apply our induction hypothesis to get $\forall
  \oxcprov^\prime \in \oxdomain{\oxframe_1}$, $\oxframe_1(\oxcprov^\prime)
  \subseteq \oxframe_2(\oxcprov^\prime)$. Now we just need that
  $\oxset{\overline{\oxloan}} \subseteq \oxframe(\oxcprov)$. But this holds immediately
  since $\oxframe(\oxcprov) = \oxset{\overline{\oxloan}} \cup
  \oxset{\overline{\oxloan^\prime}}$, so we're done.
\end{proof}

\begin{oxlemma}{Stack Typing Union Produces Subset-Related Environments}
  {lemma:union-subset-related-envs}
  If $\oxvarctx = \oxvarctx_1 \oxintersect \oxvarctx_2$, then
  $\forall \oxcprov. \; \oxvarctx_1(\oxcprov) \subseteq \oxvarctx(\oxcprov)$ and
  $\forall \oxcprov. \; \oxvarctx_2(\oxcprov) \subseteq \oxvarctx(\oxcprov)$.
\end{oxlemma}

\begin{proof}
  First, note that the definition of $\oxintersect$ is symmetric and thus we will only
  prove the first conclusion, the second proceeding immediately the same in all cases.

  We proceed by induction over the Stack Typing. 

  For $\oxemptyctx \oxintersect \oxemptyctx$, the result is trivial and thus
  immediate. For $\oxnewframe{\oxvarctx_1}{\oxframe} \oxintersect
  \oxnewframe{\oxvarctx_2}{\oxframe}$, we apply the induction hypothesis and
  \Lemma{lemma:union-subset-frames}.
\end{proof}

\begin{oxlemma}{Subset-Related Frames are also Frame Typing Union Related}
  {lemma:subset-related-frame-related} If $\forall \oxcprov \in \oxframe.$
  $\oxframe^\prime(\oxcprov) \subseteq \oxframe(\oxcprov)$ and
  $\oxdomain{\oxframe} = \oxdomain{\oxframe^\prime}$, then $\exists \oxframe_o$ such
  that $\oxframe = \oxframe^\prime \oxintersect \oxframe_o$.
\end{oxlemma}

\begin{proof}
  We proceed by induction over the frame typing. For the $\oxemptyctx$ case, the
  proof follows immediately.

  For $\oxframe = \oxframe_i, \oxid : \oxtype$ and $\oxframe^\prime =
  \oxframe_i^\prime, \oxid : \oxtype$, we just apply the induction hypothesis
  on $\oxframe_i$ and $\oxframe_i^\prime$, and add $\oxid : \oxtype$ to
  $\oxframe_o$.

  For $\oxframe = \oxframe_i, \oxcprov \mapsto \oxset{\oxloans}$ and
  $\oxframe^\prime = \oxframe_i^\prime, \oxcprov \mapsto
  \oxset{\oxloans^\prime}$, we apply the induction hypothesis, and add on
  $\oxcprov \mapsto \oxset{\oxloans} \setminus \oxset{\oxloans^\prime}$, which
  is well defined because from our premise we have $\oxset{\oxloans^\prime}
  \subseteq \oxset{\oxloans}$.
\end{proof}

\begin{oxlemma}{Subset-Related Environments are also Stack Typing Union Related}
  {lemma:subset-related-stack-related} If $\forall \oxcprov \in \oxvarctx.$
  $\oxvarctx^\prime(\oxcprov) \subseteq \oxvarctx(\oxcprov)$ and
  $\oxdomain{\oxvarctx} = \oxdomain{\oxvarctx^\prime}$, then $\exists \oxvarctx_o$ such
  that $\oxvarctx = \oxvarctx^\prime \oxintersect \oxvarctx_o$.
\end{oxlemma}

\begin{proof}
  Proceed by induction on the stack typing. In the $\oxemptyctx$ case, the proof
  is immediate. In the $\oxvarctx = \oxnewframe{\oxvarctx_i}{\oxframe}$ case, we
  apply \Lemma{lemma:subset-related-frame-related} and the induction hypothesis.
\end{proof}

\subsection{Preservation in More Precise Environments Lemmas}
\label{sec:more-precise-lemmas}

\begin{oxlemma}{Type Computation Preserved in More Precise Environments}
  {lemma:computety-more-precise}\hfill\\ If $\oxdomain{\oxvarctx} =
  \oxdomain{\oxvarctx^\prime}$, and $\forall \oxid.$ $\oxvarctx(\oxid) = \oxvarctx^\prime(\oxid)$ and $\forall \oxcprov.$ $\oxvarctx^\prime(\oxcprov)
  \subseteq \oxvarctx(\oxcprov)$ and
  $\oxcomputetynoprov{\oxtvarctx}{\oxvarctx}{\oxmuta}{\oxplaceexpr}{\oxxitype}$
  then
  $\oxcomputetynoprov{\oxtvarctx}{\oxvarctx^\prime}{\oxmuta}{\oxplaceexpr}{\oxxitype}$.
\end{oxlemma}

\begin{proof}
  Proceed by induction over the type computation judgement with $\oxvarctx$. The
  only non-immediate case is \oxname{TC-Deref}.

  \vspace{1em}
  \noindent \framebox[\textwidth]{
    \figuresize
    \begin{mathpar}
      \TCDeref
    \end{mathpar}
  }
  \vspace{1em}

  This follows from the induction hypothesis and
  \Lemma{lemma:outlives-subset-related-envs}.

\end{proof}

\begin{oxlemma}{Type Well-Formedness Preserved in More Precise Environments}
  {lemma:wf-type-more-precise}\hfill\\
  If $\oxdomain{\oxvarctx} =
  \oxdomain{\oxvarctx^\prime}$, and $\forall \oxcprov.$ $\oxvarctx^\prime(\oxcprov)
  \subseteq \oxvarctx(\oxcprov)$ and
  $\oxtypevalidity{\oxglobalctx}{\oxtvarctx}{\oxvarctx}{\oxsitype}$
  then
  $\oxtypevalidity{\oxglobalctx}{\oxtvarctx}{\oxvarctx^\prime}{\oxsitype}$.
\end{oxlemma}

\begin{proof}
  Proceed by induction over the type well formedness under $\oxvarctx$. The only
  case that isn't immediate or doesn't proceed directly from the induction
  hypothesis is \oxname{WF-Ref}.

  \vspace{1em}
  \noindent \framebox[\textwidth]{
    \figuresize
    \begin{mathpar}
      \VRef
    \end{mathpar}
  }
  \vspace{1em}

  We can apply the induction hypothesis to get the premise that $\oxxitype$ is
  well formed. For our other premise, we need to show that our region is still
  well-formed. We can look at this by cases. If the region $\oxprov$ is local,
  we can apply \oxname{WF-LocalRegion} since we know $\oxdomain{\oxvarctx} =
  \oxdomain{\oxvarctx^\prime}$. If the region $\oxprov$ is abstract, we can apply
  \oxname{WF-AbstractRegion} since we know that $\oxtvarctx$ is unchanged.
\end{proof}

\begin{oxlemma}{Stack Typing Validity Preserved in More Precise Environments}
  {lemma:wf-varctx-more-precise}\hfill\\
  If $\oxdomain{\oxvarctx} =
  \oxdomain{\oxvarctx^\prime}$, and $\forall \oxcprov.$ $\oxvarctx^\prime(\oxcprov)
  \subseteq \oxvarctx(\oxcprov)$ and
  $\oxvarctxwellformed{\oxglobalctx}{\oxtvarctx}{\oxvarctx}$
  then
  $\oxvarctxwellformed{\oxglobalctx}{\oxtvarctx}{\oxvarctx^\prime}$
\end{oxlemma}

\begin{proof}
  Proceed by induction over the stack typing validity judgement for $\oxvarctx$.
  The empty case is trivial, so the interesting case is \oxname{WF-StackTyping}.

  \vspace{1em}
  \noindent \framebox[\textwidth]{
    \figuresize
    \begin{mathpar}
      \WFVarCtx
    \end{mathpar}
  }
  \vspace{1em}

  We get $\oxvarctxwellformed{\oxglobalctx}{\oxtvarctx}{\oxvarctx^\prime}$ by
  induction. We get the type well formedness from
  \Lemma{lemma:wf-type-more-precise}. We get the type computation from
  \Lemma{lemma:computety-more-precise}, and that's all we needed to show.

\end{proof}

\begin{oxlemma}{Ownership Safety Preserved in More Precise Environments}
  {lemma:ownership-safety-more-precise}\hfill\\ If $\oxdomain{\oxvarctx} =
  \oxdomain{\oxvarctx^\prime}$, and $\forall \oxcprov.$ $\oxvarctx^\prime(\oxcprov)
  \subseteq \oxvarctx(\oxcprov)$ and
  $\oxmusafety{\oxtvarctx}{\oxkontctx}{\oxvarctx}{\oxmuta}{\oxplaceexpr}{
    \oxset{\overline{\oxloan}}
  }$
  then
  $\oxmusafety{\oxtvarctx}{\oxkontctx}{\oxvarctx^\prime}{\oxmuta}{\oxplaceexpr}{
    \oxset{\overline{\oxloan^\prime}}
  }$ and
  $\oxset{\overline{\oxloan^\prime}} \subseteq \oxset{\overline{\oxloan}}$.
\end{oxlemma}

\begin{proof}
  The proof proceeds by induction on the ownership safety judgment
  $\oxmusafety{\oxtvarctx}{\oxkontctx}{\oxvarctx}{\oxmuta}{\oxplaceexpr}{
    \oxset{\overline{\oxloan}}
  }$. This gives us three cases: \oxname{O-SafePlace}, \oxname{O-Deref}, and
  \oxname{O-DerefAbs}.

  \vspace{1em}
  \noindent \framebox[\textwidth]{
    \figuresize
    \begin{mathpar}
      \OSafePlace
    \end{mathpar}
  }
  \vspace{1em}

  We need to show $\oxmusafety{\oxtvarctx}{\oxkontctx}{\oxvarctx^\prime}{\oxmuta}{\oxplaceexpr}{
    \oxset{\oxloanpkg{\oxmuta}{\oxplace}}
  }$ and that
  $\oxset{\oxloanpkg{\oxmuta}{\oxplace}} \subseteq \oxset{\oxloanpkg{\oxmuta}{\oxplace}}$.
  The latter is immediate from the definition of subset which leaves us with the former.
  For the former, we'll correspondingly wish to apply \oxname{O-SafePlace} but using
  $\oxvarctx^\prime$ as our context. This means we need to show that
  $\forall \oxloanctxentry{\oxcprov^\prime}{\oxset{\overline{\oxloan}}} \in \oxvarctx^\prime.
  \; (\forall \oxloanpkg{\oxmuta^\prime}{\oxplaceexpr} \in \oxset{\overline{\oxloan}}.
  (\oxmuta = \oxmut \vee \oxmuta^\prime = \oxmut)
  \implies
  \oxrelevant{\oxplace^\prime}{\oxplace}) \vee \;
  (\exists \oxvarctxentry{\oxplace^\prime}{\oxtref{\oxcprov^\prime}{\oxmuta^\prime}{\oxtype^\prime}} \in \oxvarctx \;
  \wedge
  (\forall \oxvarctxentry{\oxplace^\prime}{\oxtref{\oxcprov^\prime}{\oxmuta^\prime}{\oxtype^\prime}} \in \oxvarctx. \;
  \oxplace^\prime \in \oxset{
    \overline{\oxplace_e}
  }))$. Fortunately, from $\oxdomain{\oxvarctx} = \oxdomain{\oxvarctx^\prime}$, we know that
  for every $\oxcprov^\prime \in \oxdomain{\oxvarctx}$, $\oxcprov^\prime \in
  \oxdomain{\oxvarctx^\prime}$, and further that $\oxvarctx^\prime(\oxcprov^\prime)
  \subseteq \oxvarctx(\oxcprov^\prime)$. Thus, for each $\oxcprov^\prime$, we know there are only
  potentially fewer loans to show if the obligation was met using the clause of
  $\forall \oxloanpkg{\oxmuta^\prime}{\oxplaceexpr} \in \oxset{\overline{\oxloan}}.
  (\oxmuta = \oxmut \vee \oxmuta^\prime = \oxmut)
  \implies \oxrelevant{\oxplace^\prime}{\oxplace}$. If the obligation was met using the other
  clause, note that $\oxvarctx$ and $\oxvarctx^\prime$ can only differ in the loan sets they
  associate with any given region and so the exact fact must still be true for
  $\oxvarctx^\prime$.

  \vspace{1em}
  \noindent \framebox[\textwidth]{
    \figuresize
    \begin{mathpar}
      \ODeref
    \end{mathpar}
  }
  \vspace{1em}

  This case proceeds much like the \oxname{O-SafePlace} case in terms of meeting the direct
  ownership safety criterion (the last premise of \oxname{O-Deref}) for the new derivation
  using \oxname{O-Deref} with $\oxvarctx^\prime$. It differs only in that we need also apply
  our induction hypothesis to each of the $n$ derivations of ownership safety used in
  $\forall i \in \oxset{1 \oxdots n}. \;
  \oxmusafetyinner{\oxtvarctx}{\oxkontctx}{\oxvarctx}{\oxmuta}{
    \overline{\oxplace_e} \oxcomma \overline{\oxplace_i} \oxcomma \oxplace
  }{
    \oxplaceexprctx[\oxplaceexpr_i]
  }{
    \oxset{\overline{\oxloanpkg{\oxmuta}{\oxplaceexpr_i^\prime}}}
  }$. This gives us
  $\forall i \in \oxset{1 \oxdots n}. \;
  \oxmusafetyinner{\oxtvarctx}{\oxkontctx}{\oxvarctx^\prime}{\oxmuta}{
    \overline{\oxplace_e} \oxcomma \overline{\oxplace_i} \oxcomma \oxplace
  }{
    \oxplaceexprctx[\oxplaceexpr_i]
  }{
    \oxset{\overline{\oxloanpkg{\oxmuta}{\oxplaceexpr_i^{\prime\prime}}}}
  }$ and $\forall i \in \oxset{1 \oxdots n}. \;
  \oxset{\overline{\oxloanpkg{\oxmuta}{\oxplaceexpr_i^{\prime\prime}}}} \subseteq
  \oxset{\overline{\oxloanpkg{\oxmuta}{\oxplaceexpr_i^\prime}}}$. The former combined with
  the same reasoning from the \oxname{O-SafePlace} case gives us
  $\oxmusafety{\oxtvarctx}{\oxkontctx}{\oxvarctx^\prime}{\oxmuta}{
    \oxplaceexprctx[\oxderef{\oxplace}]
  }{
    \oxset{
      \overline{\oxloanpkg{\oxmuta}{\oxplaceexpr^{\prime\prime}_1}} \oxcomma
      \oxdots
      \overline{\oxloanpkg{\oxmuta}{\oxplaceexpr^{\prime\prime}_n}} \oxcomma
      \oxloanpkg{\oxmuta}{\oxplaceexprctx[\oxderef{\oxplace}]}
    }
  }$ and the latter allows us to conclude
  $\oxset{
    \overline{\oxloanpkg{\oxmuta}{\oxplaceexpr^{\prime\prime}_1}} \oxcomma
    \oxdots
    \overline{\oxloanpkg{\oxmuta}{\oxplaceexpr^{\prime\prime}_n}} \oxcomma
    \oxloanpkg{\oxmuta}{\oxplaceexprctx[\oxderef{\oxplace}]}
  } \subseteq \oxset{
    \overline{\oxloanpkg{\oxmuta}{\oxplaceexpr^\prime_1}} \oxcomma
    \oxdots
    \overline{\oxloanpkg{\oxmuta}{\oxplaceexpr^\prime_n}} \oxcomma
    \oxloanpkg{\oxmuta}{\oxplaceexprctx[\oxderef{\oxplace}]}
  }$ since we know that each individual collection of loans has the subset relation from
  above and the whole set is simply their union.

  \vspace{1em}
  \noindent \framebox[\textwidth]{
    \figuresize
    \begin{mathpar}
      \ODerefAbs
    \end{mathpar}
  }
  \vspace{1em}

  This case proceeds identically to the case for \oxname{O-SafePlace}.
  We need to show $\oxmusafety{\oxtvarctx}{\oxkontctx}{\oxvarctx^\prime}{\oxmuta}{\oxplaceexpr}{
    \oxset{\oxloanpkg{\oxmuta}{\oxplaceexprctx[\oxderef{\oxplace}]}}
  }$ and that
  $\oxset{
    \oxset{\oxloanpkg{\oxmuta}{\oxplaceexprctx[\oxderef{\oxplace}]}}
  } \subseteq \oxset{
    \oxset{\oxloanpkg{\oxmuta}{\oxplaceexprctx[\oxderef{\oxplace}]}}
  }$.
  The latter is immediate from the definition of subset which leaves us with the former.
  For the former, we'll correspondingly wish to apply \oxname{O-DerefAbs} but using
  $\oxvarctx^\prime$ as our context. This means we need to show that
  $\forall \oxloanctxentry{\oxcprov^\prime}{\oxset{\overline{\oxloan}}} \in \oxvarctx^\prime.
  \; (\forall \oxloanpkg{\oxmuta^\prime}{\oxplaceexpr} \in \oxset{\overline{\oxloan}}.
  (\oxmuta = \oxmut \vee \oxmuta^\prime = \oxmut)
  \implies
  \oxrelevant{\oxplace^\prime}{\oxplaceexprctx[\oxderef{\oxplace}]}) \vee \;
  (\exists \oxvarctxentry{\oxplace^\prime}{\oxtref{\oxcprov^\prime}{\oxmuta^\prime}{\oxtype^\prime}} \in \oxvarctx \;
  \wedge
  (\forall \oxvarctxentry{\oxplace^\prime}{\oxtref{\oxcprov^\prime}{\oxmuta^\prime}{\oxtype^\prime}} \in \oxvarctx. \;
  \oxplace^\prime \in \oxset{
    \overline{\oxplace_e}
  }))$. Fortunately, from $\oxdomain{\oxvarctx} = \oxdomain{\oxvarctx^\prime}$, we know that
  for every $\oxcprov^\prime \in \oxdomain{\oxvarctx}$, $\oxcprov^\prime \in
  \oxdomain{\oxvarctx^\prime}$, and further that $\oxvarctx^\prime(\oxcprov^\prime)
  \subseteq \oxvarctx(\oxcprov^\prime)$. Thus, for each $\oxcprov^\prime$, we know there are only
  potentially fewer loans to show if the obligation was met using the clause of
  $\forall \oxloanpkg{\oxmuta^\prime}{\oxplaceexpr} \in \oxset{\overline{\oxloan}}.
  (\oxmuta = \oxmut \vee \oxmuta^\prime = \oxmut)
  \implies \oxrelevant{\oxplace^\prime}{\oxplace}$. If the obligation was met using the other
  clause, note that $\oxvarctx$ and $\oxvarctx^\prime$ can only differ in the loan sets they
  associate with any given region and so the exact fact must still be true for
  $\oxvarctx^\prime$.
\end{proof}

\begin{oxlemma}{Expressions Remain Well-Typed in More Precise Environments}
  {lemma:expressions-typing-more-precise}\hfill\\ If $\oxdomain{\oxvarctx} =
  \oxdomain{\oxvarctx^\prime}$, and $\forall \oxcprov.$ $\oxvarctx^\prime(\oxcprov)
  \subseteq \oxvarctx(\oxcprov)$ and 
  $\oxtypjudge{\oxglobalctx}{\oxemptyctx}{\oxkontctx}{\oxvarctx}
  {\oxexpr}{\oxtype}{\oxvarctx_f}$ then
  $\oxtypjudge{\oxglobalctx}{\oxemptyctx}{\oxkontctx}{\oxvarctx^\prime}
  {\oxexpr}{\oxtype}{\oxvarctx_f^\prime}$ and
  $\oxdomain{\oxvarctx_f} = \oxdomain{\oxvarctx_f^\prime}$, and
  $\forall \oxcprov.$ $\oxvarctx_f^\prime(\oxcprov) \subseteq \oxvarctx_f(\oxcprov)$
\end{oxlemma}

\begin{proof}
  We proceed by induction over the expression typing.

  \oxname{T-Move}, \oxname{T-Copy}, and \oxname{T-Borrow} all follow from the
  \Lemma{lemma:ownership-safety-more-precise} and the \Lemma{lemma:computety-more-precise}.
  
  \oxname{T-BorrowIndex}, \oxname{T-BorrowSlice}, and \oxname{T-IndexCopy} all follow from the
  induction hypothesis, \Lemma{lemma:ownership-safety-more-precise}, and the
  \Lemma{lemma:computety-more-precise}.

  \oxname{T-Seq} follows from the observation that garbage collecting loans will
  preserve subsets and clear in exactly both or neither, and from the induction
  hypothesis.

  \oxname{T-Branch} and \oxname{T-Match} follow from applying the induction hypothesis,
  \Lemma{lemma:subtyping-subset-related-envs},
  \Lemma{lemma:union-subset-related-envs}.

  \oxname{T-Let} follows from the same observation about garbage collection in
  the \oxname{T-Seq} case, the induction hypothesis, and
  \Lemma{lemma:subtyping-subset-related-envs}.

  \oxname{T-Assign} follows from the induction hypothesis,
  \Lemma{lemma:ownership-safety-more-precise},
  \Lemma{lemma:union-subset-related-envs}.

  \oxname{T-AssignDeref} both follow from the induction hypothesis, follows from
  the induction hypothesis, \Lemma{lemma:ownership-safety-more-precise},
  \Lemma{lemma:union-subset-related-envs}, and
  \Lemma{lemma:computety-more-precise}.

  \oxname{T-AppFunction} follows from the induction hypothesis, and a few pieces
  about the well-formedness of the instantiations happening in
  \oxname{T-AppFunction} (namely, frame expressions, regions, and types). For
  the frame expression validity, we consider each case and note that
  \oxname{WF-EnvVar} depends only on $\oxtvarctx$ which is unchanged and that
  \oxname{WF-Env} appeals to stack typing validity and so it suffices to show
  that that still holds in our more precise environment which we do by appealing
  to \Lemma{lemma:wf-varctx-more-precise}. For the region validity, there are
  again two cases to consider \oxname{WF-LocalProv} which applies if the domain
  of $\oxvarctx$ is the same as the domain of $\oxvarctx^\prime$ which we have
  directly from our premise and \oxname{WF-AbstractProv} which depends only on
  $\oxtvarctx$ which is unchanged. For the type validity, we appeal to
  \Lemma{lemma:wf-type-more-precise}.

  \oxname{T-AppClosure} follows from the induction hypothesis and
  \Lemma{lemma:subtyping-subset-related-envs}.

  \oxname{T-LetRegion}, \oxname{T-While}, \oxname{T-ForArray},
  \oxname{T-ForSlice}, \oxname{T-Closure}, \oxname{T-Tuple}, \oxname{T-Array},
  \oxname{T-Slice}, \oxname{T-Drop}, \oxname{T-Left}, \oxname{T-Right} follow
  immediately from the induction hypothesis.

  \oxname{T-Function}, \oxname{T-Abort}, \oxname{T-Unit}, \oxname{T-U32},
  \oxname{T-True}, and \oxname{T-False} are immediate.

\end{proof}

\subsection{Preservation under Safe Loan Updates Lemmas}
\label{sec:safe-loan-lemmas}

\begin{oxlemma}{Ownership Safety Produces Non Conflicting Loans}
  {lemma:ownership-safety-no-conflicts}
  If
  \begin{enumerate}
  \item $\oxplaceexprctx_{\oxcprov_p}[\oxplaceexpr_{\oxcprov_p}] \in \oxset{\oxloans}$,
    where $\oxcprov_p \mapsto \oxset{\oxloans} \in \oxloanmappings{\oxvarctx}{\oxkontctx}$
  \item $\oxvarctx(\oxcprov_b) = \emptyset$
  \item $\oxnotreborrowed{\oxvarctx[\oxcprov_b \mapsto \oxset{\oxloans_b}]}{\oxcprov_p}$
  \item $\oxmusafetyinner{\oxemptyctx}{\oxkontctx}{\oxvarctx}{\oxmuta}{\overline{\oxplace_b}}{\oxplaceexpr_b}{\oxset{\oxloans_b}}$
  \end{enumerate}
  then $\forall \oxloanpkg{\oxmuta}{\oxplaceexpr^\prime} \in \oxset{\oxloans_b}$.
  $\oxrelevant{\oxplaceexpr_{\oxcprov_p}}{\oxplaceexpr^\prime}$.
\end{oxlemma}
\begin{proof}
  Proceed by induction on the ownership safety judgement for $\oxplaceexpr$ in
  the premise.
 
  \vspace{1em}
  \noindent \framebox[\textwidth]{ \figuresize
    \begin{mathpar}
      \OSafePlace
    \end{mathpar}
  } \vspace{1em}

  We need to show that $\oxrelevant{\oxplaceexpr_{\oxcprov_p}}{\oxplace}$.
  Since $\oxcprov_p \in \oxdomain{\oxloanmappings{\oxvarctx}{\oxkontctx}}$, we know from the hypothesis of
  \oxname{O-SafePlace} that either $\oxcprov_p$ is excluded, or all loans in it
  are disjoint from $\oxplace$.

  $\oxcprov_p$ cannot have been exluded because
  $\oxnotreborrowed{\oxvarctx}{\oxcprov_p}$.

  This just leaves the case where all loans in $\oxvarctx(\oxcprov_p)$ are
  disjoint from $\oxplace$. Let $\oxplace_{\oxcprov_p}$ be the inner place of
  $\oxplaceexpr_{\oxcprov_p}$. More formally, $\oxplaceexpr_{\oxcprov_p} =
  \oxplaceexprctx[\oxplace_{\oxcprov_p}]$. By this disjointness, we know
  $\oxrelevant{\oxplaceexprctx_{\oxcprov_p}[\oxplaceexprctx[\oxplace_{\oxcprov_p}]]}{\oxplace}$,
  which directly implies
  $\oxrelevant{\oxplaceexprctx[\oxplace_{\oxcprov_p}]}{\oxplace}$, which is what
  we wanted to show.
  
  \vspace{1em}
  \noindent \framebox[\textwidth]{ \figuresize
    \begin{mathpar}
      \ODeref
    \end{mathpar}
  } \vspace{1em}

  We need to show that $\oxrelevant{\oxplaceexprctx[\oxderef
      \oxplace]}{\oxplaceexpr_{\oxcprov_p}}$, and that $\forall i,
  \ \oxloanpkg{\oxmuta}{\oxplaceexpr} \in
  \oxset{\overline{\oxloanpkg{\oxmuta}{\oxplaceexpr^\prime_i}}}$. $\oxrelevant{\oxplaceexpr_{\oxcprov_p}}{\oxplaceexpr}$.

  The latter we get from applying the induction hypothesis.

  The former follows from the same reasoning as in the previous case. 

  \vspace{1em}
  \noindent \framebox[\textwidth]{ \figuresize
    \begin{mathpar}
      \ODerefAbs
    \end{mathpar}
  } \vspace{1em}

  Since $\oxtvarctx = \oxemptyctx$, there are no valid reference types that have
  an abstract region, meaning the first hypothesis is a contradiction.
\end{proof}

\begin{oxlemma}{Ownership Safety is Preserved under Safe Loan Updates}
  {lemma:ownership-safety-loan-update} If
  \begin{enumerate}
  \item $\oxmusafety{\oxemptyctx}{\oxkontctx}{\oxvarctx}{\oxmuta_b}{\oxplaceexpr_b}{\oxset{\oxloans_b}}$
  \item $\oxmusafetyinner{\oxemptyctx}{\oxkontctx}{\oxnewframe{\oxvarctx}{\oxframe}}{\oxmuta}
     {\overline{\oxplace_1}}{\oxplaceexpr}{\oxset{\oxloans}}$
   \item and $\oxlookup{\oxvarctx}{\oxcprov_b}{\{\}}$
   \item $\oxnotreborrowed{\oxvarctx[\oxcprov_b \mapsto \oxset{\oxloans_b}]}{\oxcprov_p}$
  \item $\oxplace_1 = \oxplace_2$ or $\oxplace_1 = \oxplace_2 \cup \oxset{\oxplace
    \ | \ \oxplaceexprctx[\oxderef\oxplace] \in
    \oxset{\oxloans_b}}$
  \item either $\oxrootof{\oxplaceexpr} \in \oxdomain{\oxframe}$ or
  $\exists \oxcprov_{p} \in \oxdomain{\oxnewframe{\oxvarctx}{\oxframe}}, \oxplaceexprctx, \oxplaceexpr^\prime.$
  \item $\oxplaceexpr = \oxplaceexprctx[\oxplaceexpr^\prime]$
  \item $\oxplaceexpr^\prime \in \oxnewframe{\oxvarctx}{\oxframe}(\oxcprov_p)$
  \item $\forall \oxcprov \in \oxdomain{\oxframe},
    \oxloanpkg{\oxmuta}{\oxplaceexpr} \in \oxframe(\oxcprov).$ either
    $\oxrootof{\oxplaceexpr} \in \oxdomain{\oxframe}$, or $\exists \oxcprov^\prime \mapsto \oxset{\oxloans} \in
    \oxloanmappings{\oxvarctx}{\oxkontctx}.$ $\oxloanpkg{\oxmuta}{\oxplaceexpr} \in \oxset{\oxloans}$.

      (In english, loans in the closure's frame come from the current frame or a closure's captured frame in $\oxvarctx$ or $\oxkontctx$)
  \end{enumerate}
  then $\oxmusafetyinner{\oxemptyctx}{\oxkontctx}{\oxnewframe{\oxvarctx[\oxcprov_b \mapsto
        \oxloans_b]}{\oxframe}}{\oxmuta}{\overline{\oxplace_2}}{\oxplaceexpr}{\oxset{\oxloans^\prime}}$.
\end{oxlemma}

\begin{proof}
  Proceed by induction on the ownership safety judgement for $\oxplaceexpr$ in
  the premise.
 
  \vspace{1em}
  \noindent \framebox[\textwidth]{ \figuresize
    \begin{mathpar}
      \OSafePlace
    \end{mathpar}
  } \vspace{1em}

  Let $\oxcprov^\prime$ be an arbitrary region. If the disjunction was proven
  using the right part, which talks about the exclusion list, then we can prove
  it again the same way, since none of the types are changed and the exclusion
  list $\overline{\oxplace_2}$ includes all of $\overline{\oxplace_1}$ in either case.

  If the disjunction was proven using the left part, the only interesting case
  is when $\oxcprov^\prime = \oxcprov_b$. If $\oxrootof{\oxplace} \in \oxdomain{\oxframe}$,
  then we're done, because all loans in $\oxset{\oxloans_b}$ are disjoint just
  by the well formedness of the environment $\oxvarctx$.

  Otherwise, $\oxrootof{\oxplace} \in \oxdomain{\oxvarctx}$ and $\exists
  \oxcprov_p$ such that $\oxplace \in
  \oxnewframe{\oxvarctx}{\oxframe}(\oxcprov_p)$, and we want to show that
  $\forall \oxloanpkg{\oxmuta^\prime}{\oxplaceexprctx[\oxplace^\prime]} \in
  \oxset{\oxloans_b}$, $\oxrelevant{\oxplace}{\oxplace^\prime}$.

  If $\oxcprov_p \in \oxdomain{\oxframe}$, then we're done because each loan in
  $\oxcprov_p$ either comes from $\oxdomain{\oxframe}$, in which case
  disjointness is immediate, or it comes from a loan mapping in $\oxvarctx$ or
  $\oxkontctx$, in which case we can apply
  \Lemma{lemma:ownership-safety-no-conflicts} to finish the proof.

  Otherwise, $\oxcprov_p \in \oxdomain{\oxvarctx}$. By the well formedness 
  of $\oxnewframe{\oxvarctx[\oxcprov_b \mapsto \oxset{\oxloans_b}]}{\oxframe}$,
  $\oxnotreborrowed{\oxvarctx[\oxcprov_b \mapsto
      \oxset{\oxloans_b}]}{\oxcprov^\prime}$. Given all of this we can apply
  \Lemma{lemma:ownership-safety-no-conflicts} to finish the proof.

  \vspace{1em}
  \noindent \framebox[\textwidth]{ \figuresize
    \begin{mathpar}
      \ODeref
    \end{mathpar}
  } \vspace{1em}

  Firstly, we would like to apply our induction hypothesis. In order to do so,
  we need to show that the new \texttt{excl} is either the same, or only has
  places from $\oxset{\oxloans_b}$ added. If $\oxcprov \neq \oxcprov_b$, then
  $\oxnewframe{\oxvarctx}{\oxframe}(\oxcprov) = \oxnewframe{\oxvarctx[\oxcprov_b
      \mapsto \oxset{\oxloans_b}]}{\oxframe}$, so the exclusion list is the
  same. If $\oxcprov = \oxcprov_b$, then \texttt{excl} was empty, and now
  includes the places $\oxset{\oxplace \ | \ \oxplaceexprctx[\oxderef\oxplace]
    \in \oxset{\oxloans_b}}$. We also need to show that either that either the
  place expression is in the domain of $\oxframe$ or that it has a sub place
  expression in a loan set, but this is immediate from our hypotheses. So in
  either case we satisfy the necessary hypothesis and can apply the induction
  hypothesis.

  Whats left to show is the disjointness or exclusion condition, which follows
  identically to the reasoning in the previous case.

  \vspace{1em}
  \noindent \framebox[\textwidth]{ \figuresize
    \begin{mathpar}
      \ODerefAbs
    \end{mathpar}
  } \vspace{1em}

  Since $\oxtvarctx = \oxemptyctx$, there are no valid reference types that have
  an abstract region, meaning the first hypothesis is a contradiction.
\end{proof}

\begin{oxlemma}{Ownership Safety is Preserved after Environment Union}
  {lemma:ownership-safety-environment-union} If
  $\oxmusafetyinner{\oxemptyctx}{\oxkontctx}{\oxvarctx_1}{\oxmuta}{\overline{\oxplace_1}}{\oxplaceexpr}{\oxset{\oxloans}}$
  and
  $\oxmusafetyinner{\oxemptyctx}{\oxkontctx}{\oxvarctx_2}{\oxmuta}{\overline{\oxplace_2}}{\oxplaceexpr}{\oxset{\oxloans^\prime}}$
  then $\oxmusafetyinner{\oxemptyctx}{\oxkontctx}{\oxvarctx_1 \oxintersect
    \oxvarctx_2}{\oxmuta}{\overline{\oxplace_1},\overline{\oxplace_2}}{\oxplaceexpr}{\oxset{\oxloans^{\prime\prime}}}$
  and $\oxplace_1 = \oxplace_2$ or $\oxplace_1 = \oxplace_2 \cup \oxset{\oxplace
    \ | \ \oxplaceexprctx[\oxderef\oxplace] \in
    \oxset{\overline{\oxloans_b}}}$.
\end{oxlemma}
\begin{proof}
  Proceed by induction on the ownership safety judgements in the premise. Note
  that they both have the same sequence of proof rule applications, because the
  judgement is inductive over $\oxplaceexpr$.
 
  \vspace{1em}
  \noindent \framebox[\textwidth]{ \figuresize
    \begin{mathpar}
      \OSafePlace
    \end{mathpar}
  } \vspace{1em}

  Let $\oxcprov$ be an arbitrary region.

  If either the derivation with
  $\oxvarctx_1$ or $\oxvarctx_2$ used the second part of the disjunction, we can
  proceed by the second part of the disjunction.

  Otherwise, both derivations used the first part of the disjunction. Since
  $\oxvarctx_1 \oxintersect \oxvarctx_2(\oxcprov) = \oxvarctx_1(\oxcprov) \cup
  \oxvarctx_2(\oxcprov)$, we can just combine these two facts and proceed by the
  first part of the disjunction.

  \vspace{1em}
  \noindent \framebox[\textwidth]{ \figuresize
    \begin{mathpar}
      \ODeref
    \end{mathpar}
  } \vspace{1em}

  In order to apply the induction hypothesis, we need show that our new
  \texttt{excl} is the union of the two from the derivations. This is immediate
  because $\oxvarctx_1 \oxintersect \oxvarctx_2(\oxcprov) =
  \oxvarctx_1(\oxcprov) \cup \oxvarctx_2(\oxcprov)$.

  To finish the case, follow the same reasoning from the previous case for the
  disjunction.

  \vspace{1em}
  \noindent \framebox[\textwidth]{ \figuresize
    \begin{mathpar}
      \ODerefAbs
    \end{mathpar}
  } \vspace{1em}

  Since $\oxtvarctx = \oxemptyctx$, there are no valid reference types that have
  an abstract region, meaning the first hypothesis is a contradiction.

\end{proof}

\begin{oxlemma}{Ownership Safety is Preserved after Type Checking a Closure Body}
  {lemma:ownership-safety-closure-body} If
  \begin{enumerate}
  \item $\oxmusafety{\oxemptyctx}{\oxkontctx}{\oxvarctx}{\oxmuta}{\oxplaceexpr_b}{\oxset{\oxloans_b}}$
  \item $\oxlookup{\oxvarctx}{\oxcprov}{\{\}}$ 
  \item $\forall \oxid \in \oxfvars{\oxexpr}.$ $\oxid \in \oxdomain{\oxframe}$ 
  \item $\forall \oxcprov \in \oxfprovs{\oxexpr}.$ $\oxcprov \in \oxdomain{\oxframe}$
  \item $\forall \oxcprov \in \oxdomain{\oxframe},
    \oxloanpkg{\oxmuta}{\oxplaceexpr} \in \oxframe(\oxcprov).
    \oxrootof{\oxplaceexpr} \in \oxdomain{\oxframe} \vee
    \exists \oxcprov^\prime \mapsto \oxset{\oxloans} \in
    \oxloanmappings{\oxvarctx}{\oxkontctx}.$ $\oxloanpkg{\oxmuta}{\oxplaceexpr} \in \oxset{\oxloans}$.
    
    (In English, loans in the closure's frame come from the current frame or a closure's captured
    frame in $\oxvarctx$ or $\oxkontctx$)
  \item $\oxtypjudge{\oxglobalctx}{\oxemptyctx}{\oxkontctx}{\oxnewframe{\oxvarctx}{\oxframe}}
    {\oxexpr}{\oxsitype}{\oxnewframe{\oxvarctx_o}{\oxframe_o}}$
  \end{enumerate}
  then
  \begin{enumerate}
  \item $\oxmusafety{\oxemptyctx}{\oxkontctx}{\oxvarctx_o}{\oxmuta}{\oxplaceexpr_b}{\oxset{\oxloans_b^\prime}}$
  \item $\forall \oxcprov \in \oxdomain{\oxframe_o},
    \oxloanpkg{\oxmuta}{\oxplaceexpr} \in \oxframe_o(\oxcprov). \oxrootof{\oxplaceexpr} \in \oxdomain{\oxframe_o} \vee \exists \oxcprov^\prime \mapsto \oxset{\oxloans} \in
    \oxloanmappings{\oxvarctx_o}{\oxkontctx}.$ $\oxloanpkg{\oxmuta}{\oxplaceexpr} \in \oxset{\oxloans}$.
    
    (In English, loans in the closure's frame come from the current frame or a closure's captured
    frame in $\oxvarctx_o$ or $\oxkontctx$)
  \end{enumerate}
\end{oxlemma}
\begin{proof}
  In the \oxname{T-Move} case, the context is updated, but since $\oxplace \in
  \oxdomain{\oxframe}$, $\oxvarctx_o = \oxvarctx$, so the conclusions follow
  from the premises.

  In the \oxname{T-Copy}, \oxname{T-Function}, \oxname{T-Abort},
  \oxname{T-Unit}, \oxname{T-u32}, \oxname{T-True}, and \oxname{T-False} cases,
  $\oxvarctx_o = \oxvarctx$ so the conclusions follow from the premises.

  In the \oxname{T-IndexCopy}, \oxname{T-LetRegion}, \oxname{T-While},
  \oxname{T-ForArray}, \oxname{T-ForSlice}, \oxname{T-Closure},
  \oxname{T-Tuple}, \oxname{T-Array}, \oxname{T-Slice}, \oxname{T-Drop},
  \oxname{T-Left}, and \oxname{T-Right} cases, the proof is immediate from the
  induction hypothesis.

  In the \oxname{T-Borrow}, \oxname{T-BorrowIndex}, and \oxname{T-BorrowSlice}
  cases, the proof follows from the induction hypothesis and
  \Lemma{lemma:ownership-safety-loan-update}. The last condition is the
  restriction on loans in $\oxframe$, but this is immediate by inspection of the
  ownership safety judgement. The only way to create new loans is to directly
  borrow a place, in which case we'd have that the loan is in the domain of
  $\oxframe$, and otherwise the loans originated from the loan set in
  $\oxnewframe{\oxvarctx}{\oxframe}$ of a reference being reborrowed, and we
  already know the property for $\oxnewframe{\oxvarctx}{\oxframe}$.

  In the \oxname{T-Seq} case, the proof follows from the induction hypothesis
  and \Lemma{lemma:ownership-safety-related-envs} (note that
  \texttt{gc-loans} does not change any types in the environment and produces
  related environments).

  In the \oxname{T-Branch} and \oxname{T-Match} cases, the proof follows from
  the induction hypothesis, \Lemma{lemma:subtyping-ownership-safety}, and
  \Lemma{lemma:ownership-safety-environment-union}. We also need to show that
  rewriting preserves the restriction on loans in $\oxframe$, but this is
  immediate because the most that rewriting can do is union together loan sets.
  The rest of the cases that use rewriting also use this same reasoning.

  In the \oxname{T-Let} case, the proof follows from the induction hypothesis,
  \Lemma{lemma:subtyping-ownership-safety}, and
  \Lemma{lemma:ownership-safety-related-envs} (note that
  \texttt{gc-loans} does not change any types in the environment and produces
  related environments).

  In the \oxname{T-AssignDeref} case the proof follows from the induction
  hypothesis, and \Lemma{lemma:subtyping-ownership-safety}.

  In the \oxname{T-Assign} case, the proof follows from the induction
  hypothesis, \Lemma{lemma:subtyping-ownership-safety}, and
  \Lemma{lemma:ownership-safety-assignment}.

  In the \oxname{T-App} case, the proof follows from the induction hypothesis
  and \Lemma{lemma:subtyping-ownership-safety}.
\end{proof}

\begin{oxlemma}{Outlives is Preserved under Safe Loan Updates}
  {lemma:outlives-loan-update} If
  \begin{enumerate}
  \item $\oxmusafety{\oxemptyctx}{\oxkontctx}{\oxvarctx}{\oxmuta}{\oxplaceexpr}{
    \oxset{\overline{\oxloan}} }$
  \item $\oxlookup{\oxvarctx}{\oxcprov}{\{\}}$
  \item $\oxcprov \not \in \oxdomain{\oxframe}$
  \item $\oxcprov \not = \oxcprov_1$ and $\oxcprov \not = \oxcprov_2$
  \item $\forall \oxtype^\prime \in \oxdomain{\oxvarctx}.$ $\oxcprov_1$ and
    $\oxcprov_2$ does not occur outside of a closure in $\oxtype^\prime$
  \item $\oxrunify{\oxemptyctx}{\oxkontctx}{
    \oxnewframe{\oxvarctx}{\oxframe}
  }{\oxcprov_1}{\oxcprov_2}{
    \oxnewframe{\oxvarctx^\prime}{\oxframe^\prime}
  }$
  \end{enumerate}
  then
  $\oxrunify{\oxemptyctx}{\oxkontctx}{
    \oxnewframe{\oxvarctx[\oxcprov \mapsto \oxset{\oxloans_b}]}{\oxframe}
  }{\oxcprov_1}{\oxcprov_2}{
    \oxnewframe{\oxvarctx^{\prime}[\oxcprov \mapsto \oxset{\oxloans_b}]}{\oxframe^{\prime}}
  }$
\end{oxlemma}

\begin{proof}
  Proceed by induction on the outlives relation. The only interesting cases are
  \oxname{OL-CombineConcrete}, \oxname{OL-CombineConcreteUnrestricted}, and
  \oxname{OL-CheckConcrete}. They all proceed similarly. The closure restriction
  is immediate from the premise because no types are changed. The region not
  reborrowed restriction follows from the fact that $\oxcprov_1$ and
  $\oxcprov_2$ don't occur in any types in $\oxvarctx$ outside of a closure,
  which means there's no place in the domain of $\oxvarctx$ for there to be a
  reborrow of in $\oxloans$.
\end{proof}

\begin{oxlemma}{Rewriting is Preserved under Safe Loan Updates}
  {lemma:rewriting-loan-update} If
  \begin{enumerate}
  \item $\oxmusafety{\oxemptyctx}{\oxkontctx}{\oxvarctx}{\oxmuta}{\oxplaceexpr}{
    \oxset{\overline{\oxloan}} }$
  \item $\oxlookup{\oxvarctx}{\oxcprov}{\{\}}$
  \item $\oxcprov \not \in \oxdomain{\oxframe}$
  \item $\forall \oxtype \in \oxframe$. $\oxcprov$ does not occur in $\oxtype$ or $\oxtype_1$ or $\oxtype_2$
  \item $\forall \oxcprov \in \oxfprovs{\oxtype_1} \cup \oxfprovs{\oxtype_2}$.
    $\forall \oxtype^\prime \in \oxdomain{\oxvarctx}.$ $\oxcprov$ does not occur
    outside of a closure in $\oxtype^\prime$.
  \item $\oxtunify{\oxemptyctx}{\oxkontctx}{
    \oxnewframe{\oxvarctx}{\oxframe}
  }{\oxtype_1}{\oxtype_2}{
    \oxnewframe{\oxvarctx^\prime}{\oxframe^\prime}
  }$
  \end{enumerate}
  then
  $\oxtunify{\oxemptyctx}{\oxkontctx}{
    \oxnewframe{\oxvarctx[\oxcprov \mapsto \oxset{\oxloans_b}]}{\oxframe}
  }{\oxtype_1}{\oxtype_2}{
    \oxnewframe{\oxvarctx^{\prime}[\oxcprov \mapsto \oxset{\oxloans_b}]}{\oxframe^{\prime}}
  }$.

\end{oxlemma}
\begin{proof}
  Proceed by induction on the rewriting derivation. The only interesting case is
  \oxname{RR-Reference}, in which case we apply
  \Lemma{lemma:outlives-loan-update} and the induction hypothesis. The other
  cases all follow immediately or from the induction hypothesis.
\end{proof}

\begin{oxlemma}{Closure Bodies are Well-Typed under Safe Loan Updates}
  {lemma:exprs-well-typed-loan-update}
  If
  \begin{enumerate}
  \item $\oxmusafety{\oxtvarctx}{\oxkontctx}{\oxvarctx}{\oxmuta}{\oxplaceexpr}{
    \oxset{\overline{\oxloan}}
  }$
  \item $\oxlookup{\oxvarctx}{\oxcprov}{\{\}}$
  \item $\oxcprov \not \in \oxdomain{\oxframe}$
  \item $\forall \oxtype \in \oxframe$. $\oxcprov$ does not occur in $\oxtype$
  \item $\forall \oxid \in \oxfvars{\oxexpr}.$ $\oxid \in \oxdomain{\oxframe}$
  \item $\forall \oxcprov \in \oxfprovs{\oxexpr}.$ $\oxcprov \in \oxdomain{\oxframe}$
  \item $\forall \oxcprov \in \oxdomain{\oxframe},
    \oxloanpkg{\oxmuta}{\oxplaceexpr} \in \oxframe(\oxcprov).
    \oxrootof{\oxplaceexpr} \in \oxdomain{\oxframe} \vee
    \exists \oxcprov^\prime \mapsto \oxset{\oxloans} \in
    \oxloanmappings{\oxvarctx}{\oxkontctx}.$ $\oxloanpkg{\oxmuta}{\oxplaceexpr} \in \oxset{\oxloans}$.
    
    (In English, loans in the closure's frame come from the current frame or a closure's captured
    frame in $\oxvarctx$ or $\oxkontctx$)
  \item $\forall \oxcprov \in \oxfprovs{\oxsitype}.\ \forall \oxtype^\prime \in
    \oxdomain{\oxvarctx}.\ \oxcprov\text{ does not occur in }\oxtype^\prime$
  \item $\oxtypjudge{\oxglobalctx}{\oxemptyctx}{\oxkontctx}{\oxnewframe{\oxvarctx}{\oxframe}}
    {\oxexpr}{\oxsitype}{\oxnewframe{\oxvarctx_o}{\oxframe_o}}$
  \item $\oxvarctxwellformed{\oxglobalctx}{\oxemptyctx}{\oxnewframe{
      \oxvarctx[\oxcprov \mapsto \oxset{\oxloans}]
    }{\oxframe}}$
  \end{enumerate}
  then
  \begin{enumerate}
  \item $\oxtypjudge{\oxglobalctx}{\oxemptyctx}{\oxkontctx}{
    \oxnewframe{
      \oxupdate{\oxvarctx}{
        \oxloanctxentry{\oxcprov}{ \oxset{\overline{\oxloan}} }
      }
    }{\oxframe}
  }{\oxexpr}{\oxsitype}{
    \oxnewframe{
      \oxupdate{\oxvarctx_o}{
        \oxloanctxentry{\oxcprov}{\oxset{\overline{\oxloan}}
        }
      }
    }{\oxframe_o}
  }$.
  \item $\forall \oxcprov \in \oxdomain{\oxframe_o},
    \oxloanpkg{\oxmuta}{\oxplaceexpr} \in \oxframe_o(\oxcprov).
    \oxrootof{\oxplaceexpr} \in \oxdomain{\oxframe_o} \vee
    \exists \oxcprov^\prime \mapsto \oxset{\oxloans} \in
    \oxloanmappings{\oxvarctx_o}{\oxkontctx}.$ $\oxloanpkg{\oxmuta}{\oxplaceexpr} \in \oxset{\oxloans}$.
    
    (In English, loans in the closure's frame come from the current frame or a closure's captured
    frame in $\oxvarctx_o$ or $\oxkontctx$)
  \end{enumerate}
\end{oxlemma}
\begin{proof}
  Proceed by induction over the typing derivation for $\oxexpr$.

  In \oxname{T-Move} and \oxname{T-Copy}, and \oxname{T-Borrow} cases, we can apply
  \Lemma{lemma:ownership-safety-loan-update} and note that type computation is
  unaffected by changes in loan sets.

  In the \oxname{T-Borrow} case, we can apply
  \Lemma{lemma:ownership-safety-loan-update} and note that type computation is
  unaffected by changes in loan sets. The last condition is the restriction on
  loans in $\oxframe$, but this is immediate by inspection of the ownership
  safety judgement. The only way to create new loans is to directly borrow a
  place, in which case we'd have that the loan is in the domain of $\oxframe$,
  and otherwise the loans originated from the loan set in
  $\oxnewframe{\oxvarctx}{\oxframe}$ of a reference being reborrowed, and we
  already know the loan set restriction for $\oxnewframe{\oxvarctx}{\oxframe}$.
  The rest of the cases that involve borrowing use similar reasoning.

  In the \oxname{T-BorrowIndex} and \oxname{T-IndexCopy} cases, we can apply
  \Lemma{lemma:ownership-safety-loan-update}, the induction hypothesis, and the
  note about type computation.

  In the \oxname{T-BorrowSlice} case, we can apply
  \Lemma{lemma:ownership-safety-loan-update}, the induction hypothesis, the note
  about type computation, and \Lemma{lemma:ownership-safety-closure-body}.

  In the \oxname{T-Seq} case, we can apply the induction hypothesis, for which
  we need to apply \Lemma{lemma:ownership-safety-closure-body} and use the fact
  that garbage collection produces related environments with
  \Lemma{lemma:ownership-safety-related-envs}.

  In the \oxname{T-Branch} case we can apply the induction hypothesis,
  \Lemma{lemma:ownership-safety-closure-body}, and
  \Lemma{lemma:rewriting-loan-update}. We also need to show that
  rewriting preserves the restriction on loans in $\oxframe$, but this is
  immediate because the most that rewriting can do is union together loan sets.
  The rest of the cases that use rewriting also use this same reasoning.

  In the \oxname{T-Let} case, we can apply the induction hypothesis,
  \Lemma{lemma:ownership-safety-closure-body},
  \Lemma{lemma:rewriting-loan-update}, and the fact that garbage collection
  produces related environments with
  \Lemma{lemma:ownership-safety-related-envs}. The last obligation is the region
  not reborrowed condition. Note that by environment well formedness and our
  hypothesis, any free regions in the types are not in any non closure types.
  Therefore, there are no places in $\oxvarctx$ with the region for the new
  loans in $\oxloans$ to even contain a reborrow of, meaning the regions are not
  reborrowed. The other cases which require region not reborrowed proceed by the
  same reasoning.

  In the \oxname{T-LetRegion}, \oxname{T-While}, \oxname{T-Closure},
  \oxname{T-Tuple}, \oxname{T-Array}, \oxname{T-Slice}, \oxname{T-Left}, and
  \oxname{T-Right} cases, the proofs follows from the induction hypothesis and
  \Lemma{lemma:ownership-safety-closure-body}.

  In the \oxname{T-AssignDeref} case, we can apply the induction hypothesis,
  \Lemma{lemma:ownership-safety-closure-body}, the fact that type computation is
  unaffected by loan updates, \Lemma{lemma:rewriting-loan-update}, and
  \Lemma{lemma:ownership-safety-loan-update}.

  In the \oxname{T-Assign} case, we can apply the induction hypothesis,
  \Lemma{lemma:ownership-safety-closure-body}, the fact that type computation is
  unaffected by loan updates, \Lemma{lemma:rewriting-loan-update}, and
  \Lemma{lemma:ownership-safety-loan-update}. The last obligation is the unique
  to judgement, which is unaffected by loan updates.

  In the \oxname{T-ForArray} and \oxname{T-ForSlice} cases, we can apply the
  induction hypothesis and \Lemma{lemma:ownership-safety-closure-body}. The
  remaining region not reborrowed obligation follows from the same reasoning in
  the \oxname{T-Let} case.

  In the \oxname{T-Function}, \oxname{T-Abort}, \oxname{T-Unit}, \oxname{T-u32},
  \oxname{T-True}, and \oxname{T-False} cases, the proof is immediate.

  In the \oxname{T-App} case, we can apply the induction hypothesis,
  \Lemma{lemma:ownership-safety-closure-body} and
  \Lemma{lemma:rewriting-loan-update}. Note that the well formedness judgements
  are unaffected by loan updates. The last obligation is the region not
  reborrowed judgement, follows from the same reasoning in
  the \oxname{T-Let} case.

  In the \oxname{T-Drop} case, we can apply the induction hypothesis,
  \Lemma{lemma:ownership-safety-closure-body}, and
  \Lemma{lemma:ownership-safety-related-envs}, noting that making the type of a
  place dead produces a related environment.

  In the \oxname{T-Match} case we can apply the induction hypothesis,
  \Lemma{lemma:ownership-safety-closure-body}, and
  \Lemma{lemma:rewriting-loan-update}. The last obligation is the region not
  reborrowed judgement, follows from the same reasoning in
  the \oxname{T-Let} case.
\end{proof}

\begin{oxlemma}{Values are Well-Typed under Safe Loan Updates}
  {lemma:values-well-typed-loan-update} If
  $\oxmusafety{\oxemptyctx}{\oxkontctx}{\oxvarctx}{\oxmuta}{\oxplaceexpr}{
    \oxset{\overline{\oxloan}} }$ and
  $\oxnotinclosure{\oxkontctx}{\oxvarctx}{\oxcprov}$ and
  $\oxlookup{\oxvarctx}{\oxcprov}{\emptyset}$ and
  $\oxtypjudge{\oxglobalctx}{\oxemptyctx}{\oxkontctx}{\oxvarctx}
  {\oxvalue}{\oxsitype}{\oxvarctx}$, then
  $\oxtypjudge{\oxglobalctx}{\oxemptyctx}{\oxkontctx}{ \oxupdate{\oxvarctx}{
      \oxloanctxentry{\oxcprov}{ \oxset{\overline{\oxloan}} } }
  }{\oxvalue}{\oxsitype}{ \oxupdate{\oxvarctx}{ \oxloanctxentry{\oxcprov}{
        \oxset{\overline{\oxloan}} } } }$.
\end{oxlemma}

\begin{proof}
  We proceed by induction on the value typing relation.

  For \oxname{T-u32}, \oxname{T-True}, \oxname{T-False}, the result is
  immediate.

  For \oxname{T-Tuple} and \oxname{T-Array}, we apply the induction
  hypothesis to each entry.

  \vspace{1em}
  \noindent \framebox[\textwidth]{
    \figuresize
    \begin{mathpar}
      \TPointer
    \end{mathpar}
  }
  \vspace{1em}

  In the \oxname{T-Pointer} case, both judgements in the premise are unaffected
  by taking an empty loan set and adding loans to it, so the case is immediate.

  \vspace{1em}
  \noindent \framebox[\textwidth]{
    \figuresize
    \begin{mathpar}
      \TClosureVal
    \end{mathpar}
  }
  \vspace{1em}

  In the \oxname{T-ClosureValue} case, first we invert the stack frame typing
  hypothesis to get that $\forall x \in \oxdomain{\oxstackframe}.$
  $\oxtypjudge{\oxglobalctx}{\oxemptyctx}{\oxkontctx}{
    \oxnewframe{\oxvarctx}{\oxframe_c} }{\oxstackframe(x)}{\oxframe_c(x)}{
    \oxnewframe{\oxvarctx}{\oxframe_c} }$. We can apply the induction hypothesis
  to each of these statements, and apply \oxname{WF-Frame} to get
  $\oxsubstorevalidity{\oxglobalctx}{
    \oxvarctx[\oxloanctxentry{\oxcprov}{\oxset{\overline{\oxloan}}}]
  }{\oxstackframe_c}{\oxframe_c}$.

  Next we need to show that the body remains well typed. This follows from
  \Lemma{lemma:exprs-well-typed-loan-update}.

  In all other value cases, the typing judgement holds immediately.
\end{proof}

\begin{oxlemma}{Stack Validity is Preserved under Safe Loan Updates}
  {lemma:store-validity-loan-update} If
  $\oxmusafety{\oxemptyctx}{\oxkontctx}{\oxvarctx}{\oxmuta}{\oxplaceexpr}{
    \oxset{\overline{\oxloan}} }$ and
  $\oxnotinclosure{\oxkontctx}{\oxvarctx}{\oxcprov}$ and
  and $\oxvarctx(\oxcprov) = \emptyset$ and
  $\oxstorevalidity{\oxglobalctx}{\oxvarctx}{\oxstore}$, then
  $\oxstorevalidity{\oxglobalctx}{ \oxupdate{\oxvarctx}{
      \oxloanctxentry{\oxcprov}{\oxset{\overline{\oxloan}}} } }{\oxstore}$.
\end{oxlemma}

\begin{proof}
  We proceed by induction on the stack validity. There are two cases,
  \oxname{WF-StackEmpty}, and \oxname{WF-StackFrame}. \oxname{WF-StackEmpty} is
  impossible, since we already know that $\oxcprov$ is in $\oxvarctx$.

  \vspace{1em}
  \noindent \framebox[\textwidth]{ \figuresize
    \begin{mathpar}
      \VStack\and \VStackEmpty
    \end{mathpar}
  } \vspace{1em}

  In the case of \oxname{WF-StackFrame}, we have to show that the
  values remain well-typed in the updated environment. For the remaining
  $\oxvarctx^\prime$, if $\oxcprov \in \oxvarctx^\prime$, then we apply the
  induction hypothesis, otherwise we just use the derivation from the premise.

  To show that the values in the stack are still well typed in
  $\oxvarctx[\oxcprov \mapsto \{\overline{\oxloan}\}]$, we apply
  \Lemma{lemma:values-well-typed-loan-update}.
\end{proof}

\subsection{Preservation of Rewriting under Parallel Type Checking Lemmas}
\label{sec:parallel-lemmas}

\begin{oxlemma}{Region Rewriting is Preserved by Parallel Loan Updates}
  {lemma:rewriting-parallel-ownership}
  If
  \begin{enumerate}
  \item $\oxtunify{\oxemptyctx}{\oxkontctx}{\oxvarctx^\prime}{\oxtype_1}{\oxtype_2}{\oxvarctx_s}$
  \item $\oxmusafety{\oxemptyctx}{\oxkontctx, \oxtype_1, \oxkontctx^\prime}{\oxvarctx}{\oxmuta}{\oxplaceexpr}{\oxset{\oxloans}}$
  \item $\oxmusafety{\oxemptyctx}{\oxkontctx, \oxtype_2, \oxkontctx^\prime}{\oxvarctx^\prime}{\oxmuta}{\oxplaceexpr}{\oxset{\oxloans^\prime}}$
  \item $\oxdomain{\oxvarctx} = \oxdomain{\oxvarctx^\prime}$ and $\oxdomain{\oxvarctx_o} = \oxdomain{\oxvarctx_o^\prime}$
  \item $\forall \oxcprov \in \oxdomain{\oxvarctx}.$ $\oxvarctx^\prime(\oxcprov) \subseteq \oxvarctx(\oxcprov)$
  \item $\forall \oxcprov \in \oxdomain{\oxvarctx_o}.$ $\oxvarctx^\prime_o(\oxcprov) \subseteq \oxvarctx_o(\oxcprov)$
  \end{enumerate}
  then
  $\oxtunify{\oxemptyctx}{\oxkontctx}{\oxvarctx^\prime[\oxcprov \mapsto \oxset{\oxloans^\prime}]}{\oxtype_1}{\oxtype_2}{\oxvarctx_s^\prime}$.
\end{oxlemma}

\begin{proof}
  Proceed by induction on the rewriting derivation. The only interesting case is
  \oxname{RR-Reference}, in which case we proceed by induction on the outlives
  derivation.

  The interesting cases are \oxname{OL-CombineConcrete},
  \oxname{OL-CombineConcreteUnrestricted}, and \oxname{OL-CheckConcrete} since
  the type variable environment is guaranteed to be empty and the other cases
  all involve abstract regions. The only interesting obligations are the region
  not reborrowed ones, since the types in the environments are unchanged. This
  amounts to showing that for all loans in $\oxset{\oxloans^\prime}$, none are
  reborrows of references that have either $\oxcprov_1$ or $\oxcprov_2$ as their
  region.

  Assume one such loan $\oxloanpkg{\oxmuta^\prime}{\oxderef \oxplace} \in
  \oxloans^\prime$ exists. Assume without loss of generality that $\oxplace:
  \oxtref{\oxcprov_1}{\oxmuta^{\prime\prime}}{\oxtype} \in \oxvarctx$. By the
  well formedness of $\oxkontctx, \oxtype_1, \oxkontctx^\prime$, since
  $\oxtype_1$ contains $\oxcprov_1$, it must be the case that
  $\oxvarctx(\oxcprov_1) \not = \emptyset$. When checking ownership safety for
  $\oxderef \oxplace$, we'll need to show that excluding $\oxplace$, there are
  no conflicts in $\oxcprov^\prime$ with any lons in $\oxcprov^\prime$. But
  since the exclusion clause doesn't exclude when we have a reference in theta,
  and $\oxtype_1$ contains $\oxcprov_1$, it must be the case that $\oxcprov_1$
  will not be excluded, and we will then find loan conflicts, which is a
  contradiction.
\end{proof}

\begin{oxlemma}{Region Rewriting is Preserved by Type Checking Parallel Expressions}
  {lemma:rewriting-parallel-expression} \; \\
  If
  \begin{enumerate}
  \item $\oxtunify{\oxemptyctx}{\oxkontctx}{\oxvarctx^\prime}{\oxtype_1}{\oxtype_2}{\oxvarctx_s}$
  \item $\oxtypjudge{\oxglobalctx}{\oxemptyctx}{\oxkontctx, \oxtype_1, \oxkontctx^\prime}{\oxvarctx}{\oxexpr}{\oxtype}{\oxvarctx_o}$
  \item $\oxtypjudge{\oxglobalctx}{\oxemptyctx}{\oxkontctx, \oxtype_2, \oxkontctx^\prime}{\oxvarctx^\prime}{\oxexpr}{\oxtype}{\oxvarctx_o^\prime}$
  \item $\oxdomain{\oxvarctx} = \oxdomain{\oxvarctx^\prime}$ and $\oxdomain{\oxvarctx_o} = \oxdomain{\oxvarctx_o^\prime}$
  \item $\forall \oxcprov \in \oxdomain{\oxvarctx}.$ $\oxvarctx^\prime(\oxcprov) \subseteq \oxvarctx(\oxcprov)$
  \item $\forall \oxcprov \in \oxdomain{\oxvarctx_o}.$ $\oxvarctx^\prime_o(\oxcprov) \subseteq \oxvarctx_o(\oxcprov)$
  \end{enumerate}
  then
  $\oxtunify{\oxemptyctx}{\oxkontctx}{\oxvarctx_o^\prime}{\oxtype_1}{\oxtype_2}{\oxvarctx_s^\prime}$.
\end{oxlemma}

\begin{proof}
  Proceed by induction on the typing derivation using $\oxvarctx^\prime$ (note
  that since the typing derivation is by the structure of $\oxexpr$, we can
  simultaneously induct on the typing derivation using $\oxvarctx$).

  In the \oxname{T-Copy}, \oxname{T-Function}, \oxname{T-Abort},
  \oxname{T-Unit}, \oxname{T-u32}, \oxname{T-True}, and \oxname{T-False} cases,
  the proof is immediate, with $\oxvarctx_o = \oxvarctx$ and
  $\oxvarctx^{\prime\prime} = \oxvarctx^\prime$.

  In the \oxname{T-IndexCopy}, \oxname{T-LetRegion}, \oxname{T-While},
  \oxname{T-ForArray}, \oxname{T-ForSlice}, \oxname{T-Closure},
  \oxname{T-Tuple}, \oxname{T-Array}, \oxname{T-Slice}, \oxname{T-Left}, and
  \oxname{T-Right} cases, the proof follows immediately from the induction
  hypothesis.

  In the \oxname{T-Move} case, the proof is immediate because making a type a
  dead does not add additional obligations in the rewriting judgement.

  In the \oxname{T-Borrow}, \oxname{T-BorrowIndex}, and \oxname{T-BorrowSlice}
  cases, we proceed by the induction hypothesis and apply
  \Lemma{lemma:rewriting-parallel-ownership}.

  In the \oxname{T-Seq} case, we just need to show that garbage collecting loans
  preserves rewriting. But this is immediate, because garbage collection loans
  can only clear loan sets, which makes the requirements in rewriting strictly
  easier since it could only \emph{remove} reborrows, not add them.

  In the \oxname{T-AssignDeref}, \oxname{T-AppFunction}, and
  \oxname{T-AppClosure} cases, we first apply the induction hypothesis. Note
  that the output of the region rewriting at most only combines loan sets. As
  such, the region rewriting is preserved, because the main condition, the
  region not reborrowed requirement on the regions in the types, will still
  consider the same set of loans. For all region rewriting cases below, use this
  same reasoning.

  In the \oxname{T-Assign} case, we apply the induction hypothesis and reason
  about the rewriting as above, but we additionally need to know that the type
  update maintains the region not reborrowed and closure restrictions. For both,
  it's immediate because the rewriting in the hypothesis of the typing rule will
  check the same restrictions.

  In the \oxname{T-Branch} and \oxname{T-Match} cases, we apply the induction
  hypothesis, the reasoning above about rewriting, and the fact that
  $\oxintersect$ only unions together the loan sets from $\oxvarctx_2$ and
  $\oxvarctx_3$, both of which had the rewriting restrictions true by the
  induction hypothesis.

  In the \oxname{T-Let} case, we again apply the induction hypothesis and the
  reasoning above about rewriting, but additionally use the same reasoning as
  the garbage collection case as well.

  In the \oxname{T-Drop} case, we just need to show that rewriting is preserved
  by making a place dead in order to apply the induction hypothesis. This is
  immediate though, because all of the obligations in rewriting are either the
  same difficulty or made easier by making a place dead.
\end{proof}

\begin{oxlemma}{Outlives Still Holds with Smaller Continuation Contexts}
  {lemma:outlives-smaller-kont}
  If $\oxrunify{\oxemptyctx}{\oxkontctx, \oxtype}{\oxvarctx}{\oxcprov_1}{\oxcprov_2}{\oxvarctx^\prime}$
  then $\oxrunify{\oxemptyctx}{\oxkontctx}{\oxvarctx}{\oxcprov_1}{\oxcprov_2}{\oxvarctx^\prime}$.
\end{oxlemma}
\begin{proof}
  Proceed by induction on the outlives judgement. The only interesting cases are
  the \oxname{OL-CombineConcrete}, \oxname{OL-CombineConcreteUnrestricted}, and
  \oxname{OL-CheckConcrete} cases which all proceed similarly. The main
  obligations, the region not reborrowed and closure restriction judgements, are
  immediate since they are either unaffected by or have strictly fewer
  obligations in the smaller temporary typing $\oxkontctx$.
\end{proof}

\begin{oxlemma}{Region Rewriting Still Holds with Smaller Continuation Contexts}
  {lemma:rewriting-smaller-kont}
  If $\oxtunify{\oxemptyctx}{\oxkontctx, \oxtype}{\oxvarctx}{\oxtype_1}{\oxtype_2}{\oxvarctx^\prime}$
  then $\oxtunify{\oxemptyctx}{\oxkontctx}{\oxvarctx}{\oxtype_1}{\oxtype_2}{\oxvarctx^\prime}$.
\end{oxlemma}
\begin{proof}
  Proceed by induction on the rewriting judgement. The only interesting case is
  \oxname{RR-Reference}, in which case we just apply
  \Lemma{lemma:outlives-smaller-kont}.
\end{proof}

\newpage
\subsection{Progress}
\label{sec:progress}

\begin{oxlemma}{Progress}{lemma:progress-app}
  If
  \oxtypjudge{\oxglobalctx}{\oxemptyctx}{\oxkontctx}{\oxvarctx}{\oxexpr}
  {\oxsitype}{\oxvarctx^\prime} and
  \oxstorevalidity{\oxglobalctx}{\oxvarctx}{\oxstore}, then either
  $\oxexpr$ is a value, $\oxexpr$ is an \oxabort{\,\dots\,}, or $\oxexists
  \oxstore^\prime, \oxexpr^\prime. \;
  \oxreduce{\oxglobalctx}{\oxstore}{\oxexpr}{\oxstore^\prime}{\oxexpr^\prime}$.
\end{oxlemma}

\paragraph{Proof} We proceed by induction on the derivation
\oxtypjudge{\oxglobalctx}{\oxemptyctx}{\oxkontctx}{\oxvarctx}{\oxexpr}
{\oxtype}{\oxvarctx^\prime}. \\

\oxprogresscase{T-Move}{\TMove}{\EMove}{
  Applying \Lemma{lemma:place-exprs-reduce} to
  $\oxmusafety{\oxtvarctx}{\oxvarctx}{\oxmut}{\oxplace}{\oxset{
      \oxloanpkg{\oxmut}{\oxplace}
  }}$, $\oxcomputetynoprov{\oxtvarctx}{\oxvarctx}{\oxmut}{\oxplace}{\oxsitype}$
  (from $\oxlookup{\oxvarctx}{\oxplace}{\oxsitype}$ by \oxname{TC-Place}), and
  $\oxstorevalidity{\oxglobalctx}{\oxvarctx}{\oxstore}$ to conclude that
  $\oxnorm{\oxstore}{\oxplace}{\_}{\_}{\oxvalue}$. Thus, we can step
  with \oxname{E-Move}.
}

\oxprogresscase{T-Copy}{\TCopy}{\ECopy}{
  Applying \Lemma{lemma:place-exprs-reduce} to
  $\oxmusafety{\oxtvarctx}{\oxvarctx}{\oximm}{\oxplaceexpr}{\oxset{
      \overline{\oxloan}
  }}$, $\oxcomputetynoprov{\oxtvarctx}{\oxvarctx}{\oximm}{\oxplaceexpr}
  {\oxsitype}$, and $\oxstorevalidity{\oxglobalctx}{\oxvarctx}{\oxstore}$ to
  conclude that $\oxnorm{\oxstore}{\oxplaceexpr}{\_}{\_}{\oxvalue}$. Thus, we can
  step with \oxname{E-Copy}.
}

\oxprogresscasewide{T-Borrow}{\TBorrow}{\EBorrow}{
  Applying \Lemma{lemma:place-exprs-reduce} to
  $\oxmusafety{\oxtvarctx}{\oxvarctx}{\oxmuta}{\oxplaceexpr}{\oxset{
      \overline{\oxloan}
    }}$, $\oxcomputetynoprov{\oxtvarctx}{\oxvarctx}{\oxmuta}{\oxplaceexpr}
  {\oxxitype}$, and $\oxstorevalidity{\oxglobalctx}{\oxvarctx}{\oxstore}$ to
  conclude that $\oxnorm{\oxstore}{\oxplaceexpr}{\oxreferent}{\_}{\_}$. Thus, we can
  step with \oxname{E-Borrow}.
}

\oxprogresscaseinductive{T-BorrowIndex}{\TBorrowIndex}{
  We proceed based on  whether or not $\oxexpr$ is a value. If it is not, we can
  decompose our expression into the evaluation context
  $\oxref{\oxprov}{\oxmuta}{\oxindex{\oxplaceexpr}{\oxhole}}$ and redex
  $\oxexpr$. Then, by applying our induction hypothesis to the typing derivation
  for $\oxexpr$, we know either that $\oxexpr$ is an $\oxkey{abort!}$ expression
  or it $\oxexpr$ steps to some $\oxexpr^\prime$.  In the former case, we can
  step with \oxname{E-EvalCtxAbort}.  In the latter case, we can plug
  $\oxexpr^\prime$ back into our evaluation context and step with
  \oxname{E-EvalCtx}.
}{
  If $\oxexpr$ is a value, we would like to step with one of:
}{\EBorrowIndex \and \EBorrowIndexOOB}{
  Since $\oxexpr$ is a value, we can apply \Lemma{lemma:values-dont-change} to
  get $\oxsubtypectx{\oxglobalctx}{\oxemptyctx}{\oxvarctx}{\oxvarctx^\prime}$.
  Applying \Lemma{lemma:stack-validity-related-envs}, then gives us
  $\oxstorevalidity{\oxglobalctx}{\oxvarctx^\prime}{\oxstore}$.

  Then, we can apply \Lemma{lemma:place-exprs-reduce} to
  $\oxmusafety{\oxtvarctx}{\oxvarctx^\prime}{\oxmuta}{\oxplaceexpr}{
    \oxset{\overline{\oxloan}}
  }$, $\oxcomputetynoprov{\oxtvarctx}{\oxvarctx^\prime}{\oxmuta}{\oxplaceexpr}
  {\oxxitype}$, and $\oxstorevalidity{\oxglobalctx}{\oxvarctx^\prime}{\oxstore}$
  to get $\oxnorm{\oxstore}{\oxplaceexpr}{\oxreferent}{\_}{\oxvalue}$. By
  \Lemma{lemma:canonical-forms}, we know that
  $\oxvalue =  \oxarr{\oxvalue_0 \oxdotsc \oxvalue_n}$ since the type tells us
  the shape of the resultant value.

  Since we wish to step with one of \oxname{E-BorrowIndex} and
  \oxname{E-BorrowIndexOOB}, we should observe that we now have
  their shared requirement: $\oxnorm{\oxstore}{\oxplaceexpr}{\oxreferent}{\_}{
    \oxarr{\oxvalue_0 \oxdotsc \oxvalue_n}
  }$. Their other obligations are a bounds check which together are a tautology
  (i.e. one of them must hold). Thus, we can step with the appropriate rule
  based on whether or not the bounds check succeeds.
}

\oxprogresscaseinductive{T-BorrowSlice}{\TBorrowSlice}{
  The proof proceeds along similar lines as for \oxname{T-BorrowIndex}. We
  proceed based on whether or not $\oxexpr_1$ and $\oxexpr_2$ are values.

  If $\oxexpr_1$ is not a value, then we can decompose our whole expression into
  the evaluation context
  $\oxref{\oxprov}{\oxmuta}{\oxslice{\oxplaceexpr}{\oxhole}{\oxexpr_2}}$ and
  redex $\oxexpr_1$. Then, by applying our induction hypothesis to $\oxexpr_1$,
  we know either that $\oxexpr_1$ steps to some $\oxexpr_1^\prime$ or is an
  $\oxkey{abort!}$ expression. In the former case, this satisfies our
  requirement since we can plug $\oxexpr^\prime$ back into our evaluation
  context. In the latter case, we can step with \oxname{E-EvalCtxAbort}.

  If $\oxexpr_1$ is a value and $\oxexpr_2$ is not a value, then we can
  decompose our whole expression into the evaluation context
  $\oxref{\oxprov}{\oxmuta}{\oxslice{\oxplaceexpr}{\oxvalue_1}{\oxhole}}$ and
  redex $\oxexpr_2$. Then, by applying our induction hypothesis to $\oxexpr_2$,
  we know either that $\oxexpr_2$ steps to some $\oxexpr_2^\prime$ or is an
  $\oxkey{abort!}$ expression. In the former case, this satisfies our
  requirement since we can plug $\oxexpr^\prime$ back into our evaluation
  context. In the latter case, we can step with \oxname{E-EvalCtxAbort}.
}{
  If $\oxexpr_1$ and $\oxexpr_2$ are values, we would like to step with one of:
}{\EBorrowSlice \and \EBorrowSliceOOB}{
  Since $\oxexpr_1$ is a value, we can apply \Lemma{lemma:values-dont-change} to
  get $\oxsubtypectx{\oxglobalctx}{\oxemptyctx}{\oxvarctx}{\oxvarctx_1}$.
  Then, since $\oxexpr_2$ is also a value, we can apply
  \Lemma{lemma:values-dont-change} to get
  $\oxsubtypectx{\oxglobalctx}{\oxemptyctx}{\oxvarctx_1}{\oxvarctx_2}$. Then, by
  transitivity, we get
  $\oxsubtypectx{\oxglobalctx}{\oxemptyctx}{\oxvarctx}{\oxvarctx_2}$.
  Then, applying \Lemma{lemma:stack-validity-related-envs} gives us
  $\oxstorevalidity{\oxglobalctx}{\oxvarctx_2}{\oxstore}$.

  Then, we can apply \Lemma{lemma:place-exprs-reduce} to
  $\oxmusafety{\oxtvarctx}{\oxvarctx_2}{\oxmuta}{\oxplaceexpr}{
    \oxset{\overline{\oxloan}}
  }$, $\oxcomputetynoprov{\oxtvarctx}{\oxvarctx_2}{\oxmuta}{\oxplaceexpr}
  {\oxtslice{\oxsitype}}$, and $\oxstorevalidity{\oxglobalctx}{\oxvarctx_2}{
    \oxstore
  }$ to get $\oxnorm{\oxstore}{\oxplaceexpr}{\oxreferent}{\_}{\oxvalue}$. By
  \Lemma{lemma:canonical-forms}, we know that
  $\oxvalue =  \oxarr{\oxvalue_0 \oxdotsc \oxvalue_n}$ since the type tells us
  the shape of the resultant value.

  Since we wish to step with one of \oxname{E-BorrowSlice} and
  \oxname{E-BorrowSliceOOB}, we should observe that we now have
  their shared requirement: $\oxnorm{\oxstore}{\oxplaceexpr}{\oxreferent}{\_}{
    \oxarr{\oxvalue_0 \oxdotsc \oxvalue_n}
  }$. Their other obligations are a bounds check which together are a tautology
  (i.e. one of them must hold). Thus, we can step with the appropriate rule
  based on whether or not the bounds check succeeds.
}

\oxprogresscaseinductive{T-IndexCopy}{\TIndexCopy}{
  We proceed based on whether or not $\oxexpr$ is a value. If it is not, we can
  decompose our expression into the evaluation context
  $\oxindex{\oxplaceexpr}{\oxhole}$ and redex $\oxexpr$. Then, by applying our
  induction hypothesis to $\oxexpr$, we know either that $\oxexpr$ steps to
  some $\oxexpr^\prime$ or is an $\oxkey{abort!}$ expression. In the former
  case, this satisfies our requirement since we can plug $\oxexpr^\prime$ back
  into our evaluation context. In the latter case, we can step with
  \oxname{E-EvalCtxAbort}.
}{If $\oxexpr$ is a value, we would like to step with one of:}{
  \EIndexCopy \and \EIndexCopyOOB
}{
  Since $\oxexpr$ is a value, we can apply \Lemma{lemma:values-dont-change} to
  get $\oxsubtypectx{\oxglobalctx}{\oxemptyctx}{\oxvarctx}{\oxvarctx^\prime}$.
  Applying \Lemma{lemma:stack-validity-related-envs}, then gives us
  $\oxstorevalidity{\oxglobalctx}{\oxvarctx^\prime}{\oxstore}$.

  Then, we can apply \Lemma{lemma:place-exprs-reduce} to
  $\oxmusafety{\oxtvarctx}{\oxvarctx^\prime}{\oxmuta}{\oxplaceexpr}{
    \oxset{\overline{\oxloan}}
  }$, $\oxcomputetynoprov{\oxtvarctx}{\oxvarctx^\prime}{\oxmuta}{\oxplaceexpr}
  {\oxxitype}$, and $\oxstorevalidity{\oxglobalctx}{\oxvarctx^\prime}{\oxstore}$
  to get $\oxnorm{\oxstore}{\oxplaceexpr}{\_}{\_}{\oxvalue}$. By
  \Lemma{lemma:canonical-forms}, we know that
  $\oxvalue =  \oxarr{\oxvalue_0 \oxdotsc \oxvalue_n}$ since the type tells us
  the shape of the resultant value.

  Since we wish to step with one of \oxname{E-IndexCopy} and
  \oxname{E-IndexCopyOOB}, we should observe that we now have
  their shared requirement: $\oxnorm{\oxstore}{\oxplaceexpr}{\_}{\_}{
    \oxarr{\oxvalue_0 \oxdotsc \oxvalue_n}
  }$. Their other obligations are a bounds check which together are a tautology
  (i.e. one of them must hold). Thus, we can step with the appropriate rule
  based on whether or not the bounds check succeeds.
}

\oxprogresscaseinductive{T-Seq}{\TSeq}{
  We proceed based on whether or not $\oxexpr_1$ is a value. If it is not, we can
  decompose our expression into the evaluation context
  $\oxseq{\oxhole}{\oxexpr_2}$ and redex $\oxexpr_1$. Then, by applying our
  induction hypothesis to $\oxexpr_1$, we know either that $\oxexpr_1$ steps to
  some $\oxexpr_1^\prime$ or is an $\oxkey{abort!}$ expression. In the former
  case, this satisfies our requirement since we can plug $\oxexpr_1^\prime$ back
  into our evaluation context. In the latter case, we can step with
  \oxname{E-EvalCtxAbort}.
}{If $\oxexpr_1$ is a value, we can step with:}{\ESeq}{}

\oxprogresscaseinductive{T-Branch}{\TBranch}{
  We proceed based on whether or not $\oxexpr_1$ is a value. If it is not, we can
  decompose our expression into the evaluation context
  $\oxbranch{\oxhole}{\oxexpr_2}{\oxexpr_3}$ and redex $\oxexpr_1$. Then, by
  applying our induction hypothesis to $\oxexpr_1$, we know either that
  $\oxexpr_1$ steps to some $\oxexpr_1^\prime$ or is an $\oxkey{abort!}$
  expression. In the former case, this satisfies our requirement since we can
  plug $\oxexpr_1^\prime$ back into our evaluation context. In the latter case,
  we can step with \oxname{E-EvalCtxAbort}.
}{If $\oxexpr_1$ is a value, we would like to step with one of:}{
  \EIfTrue \and \EIfFalse
}{
  Since $\oxexpr_1$ is a value, applying \Lemma{lemma:canonical-forms} tells us
  that $\oxexpr_1$ is either $\oxtrue$ or $\oxfalse$. In the former case, we can
  step with \oxname{E-IfTrue} and in the latter case, we can step with
  \oxname{E-IfFalse}
}

\oxprogresscaseinductive{T-Let}{\TLet}{
  We proceed based on whether or not $\oxexpr_1$ is a value. If it is not, we can
  decompose our expression into the evaluation context
  $\oxlet{\oxid}{\oxsitype_a}{\oxhole}{\oxexpr_2}$ and redex $\oxexpr_1$. Then,
  by applying our induction hypothesis to $\oxexpr_1$, we know either that
  $\oxexpr_1$ steps to some $\oxexpr_1^\prime$ or is an $\oxkey{abort!}$
  expression. In the former case, this satisfies our requirement since we can
  plug $\oxexpr_1^\prime$ back into our evaluation context. In the latter case,
  we can step with \oxname{E-EvalCtxAbort}.
}{If $\oxexpr_1$ is a value, we can step with:}{\ELet}{}

\oxprogresscaseinductive{T-LetRegion}{\TLetProv}{
  We proceed based on whether or not $\oxexpr$ is a value. If it is not, we can
  decompose our expression into the evaluation context
  $\oxletrgn{\oxcprov}{\oxhole}$ and redex $\oxexpr$. Then, by applying our
  induction hypothesis to $\oxexpr$, we know either that $\oxexpr$ steps to some
  $\oxexpr^\prime$ or is an $\oxkey{abort!}$ expression. In the former case,
  this satisfies our requirement since we can plug $\oxexpr^\prime$ back into
  our evaluation context. In the latter case, we can step with
  \oxname{E-EvalCtxAbort}.
}{If $\oxexpr$ is a value, we can step with:}{\ELetProv}{}

\oxprogresscaseinductive{T-Assign}{\TAssign}{
  We proceed based on whether or not $\oxexpr$ is a value. If it is not, we can
  decompose our expression into the evaluation context
  $\oxassign{\oxplaceexpr}{\oxhole}$ and redex $\oxexpr$. Then, by applying our
  induction hypothesis to $\oxexpr$, we know either that $\oxexpr$ steps to some
  $\oxexpr^\prime$ or is an $\oxkey{abort!}$ expression. In the former case,
  this satisfies our requirement since we can plug $\oxexpr^\prime$ back into
  our evaluation context. In the latter case, we can step with
  \oxname{E-EvalCtxAbort}.
}{
  If $\oxexpr$ is a value, we would like to step with:
}{\EAssign}{
  Since $\oxexpr$ is a value, we can apply \Lemma{lemma:values-dont-change} to
  get $\oxsubtypectx{\oxglobalctx}{\oxemptyctx}{\oxvarctx}{\oxvarctx^\prime}$.
  Applying \Lemma{lemma:stack-validity-related-envs}, then gives us
  $\oxstorevalidity{\oxglobalctx}{\oxvarctx^\prime}{\oxstore}$.

  Then, we can apply \Lemma{lemma:place-exprs-reduce} to
  $\oxmusafety{\oxtvarctx}{\oxvarctx^\prime}{\oxmuta}{\oxplaceexpr}{
    \oxset{\overline{\oxloan}}
  }$, $\oxcomputetynoprov{\oxtvarctx}{\oxvarctx^\prime}{\oxmuta}{\oxplaceexpr}
  {\oxxitype}$, and $\oxstorevalidity{\oxglobalctx}{\oxvarctx^\prime}{\oxstore}$
  to get $\oxnorm{\oxstore}{\oxplaceexpr}{\oxreferent}{\oxvaluectx}{\_}$.
}

\oxprogresscaseinductive{T-ForArray}{\TForArray}{
  We proceed based on whether or not $\oxexpr_1$ is a value. If it is not, we can
  decompose our expression into the evaluation context
  $\oxfor{\oxid}{\oxhole}{\oxexpr_2}$ and redex $\oxexpr_1$. Then, by applying our
  induction hypothesis to $\oxexpr_1$, we know either that $\oxexpr_1$ steps to
  some $\oxexpr_1^\prime$ or is an $\oxkey{abort!}$ expression. In the former
  case, this satisfies our requirement since we can plug $\oxexpr_1^\prime$ back
  into our evaluation context. In the latter case, we can step with
  \oxname{E-EvalCtxAbort}.
}{If $\oxexpr_1$ is a value, we would like to step with one of:}{
  \EForArray \and \EForEmptyArray
}{
  Since $\oxexpr_1$ is a value, then by \Lemma{lemma:canonical-forms}, we know
  that $\oxexpr_1$ is of the form $\oxarr{\oxvalue_1 \oxdotsc \oxvalue_n}$. If
  $n > 0$, then we can step with \oxname{E-ForArray}, and if $n = 0$, then we
  can step with \oxname{E-ForEmptyArray}.
}

\oxprogresscaseinductive{T-ForSlice}{\TForSlice}{
  We proceed based on whether or not $\oxexpr_1$ is a value. If it is not, we can
  decompose our expression into the evaluation context
  $\oxfor{\oxid}{\oxhole}{\oxexpr_2}$ and redex $\oxexpr_1$. Then, by applying our
  induction hypothesis to $\oxexpr_1$, we know either that $\oxexpr_1$ steps to
  some $\oxexpr_1^\prime$ or is an $\oxkey{abort!}$ expression. In the former
  case, this satisfies our requirement since we can plug $\oxexpr_1^\prime$ back
  into our evaluation context. In the latter case, we can step with
  \oxname{E-EvalCtxAbort}.
}{If $\oxexpr_1$ is a value, we would like to step with one of:}{
  \EForSlice \and \EForEmptySlice
}{
  If $\oxexpr_1$ is a value, then by \Lemma{lemma:canonical-forms}, we know that
  $\oxexpr_1$ is of the form $\oxsliceptr{\oxreferent}{\oxnum_1}{\oxnum_2}$.
  Further, by inversion of \oxname{T-Pointer} for the typing derivation of
  $\oxexpr_1$, we get $\oxreferentvalidity{\oxglobalctx}{\oxvarctx}{
    \oxslice{\oxreferent}{i}{j}
  }{\oxtslice{\oxsitype}}$. By inversion of \oxname{WF-RefSliceArray} or
  \oxname{WF-RefSliceSlice} (one of which must apply since the referent ends in
  a slice), we know that $i \leq j$. If $i < j$, we step with
  \oxname{E-ForSlice} and if $i = j$, we step with \oxname{E-ForEmptySlice}.
}

\oxprogresscasewide{T-Closure}{\TClosure}{\EClosure}{
  Since $\oxfvars{\cdot}$ and $\oxfncvars{\oxstore}{\cdot}$ are total, we can
  always step with \oxname{E-Closure}.
}

\oxprogresscaseinductive{T-AppClosure}{\TAppClosure}{
  We proceed based on whether or not $\oxexpr_f$ is a value. If it is not, we can
  decompose our expression into the evaluation context
  $\oxapply{\oxhole}{\oxexpr_1 \oxdotsc \oxexpr_n}$ and redex
  $\oxexpr_f$. Then, by applying our induction hypothesis to $\oxexpr_f$, we
  know either that $\oxexpr_f$ steps to some $\oxexpr_f^\prime$ or is an
  $\oxkey{abort!}$ expression. In the former case, this satisfies our
  requirement since we can plug $\oxexpr_f^\prime$ back into our evaluation
  context. In the latter case, we can step with \oxname{E-EvalCtxAbort}.

  Next, we'll proceed based on whether or not each expression $\oxexpr_i$ is a
  value. If any of them are not, we can decompose our expression into the
  evaluation context $\oxapply{\oxvalue_f}{\oxvalue_1 \oxdotsc \oxvalue_m
    \oxcomma \oxhole \oxcomma \oxexpr_1 \oxdotsc \oxexpr_{n^\prime}}$ and redex
  $\oxexpr_i$. Then, by applying our induction hypothesis to $\oxexpr_i$, we
  know either that $\oxexpr_i$ steps to some $\oxexpr_i^\prime$ or is an
  $\oxkey{abort!}$ expression. In the former case, this satisfies our
  requirement since we can plug $\oxexpr_i^\prime$ back into our evaluation
  context. In the latter case, we can step with \oxname{E-EvalCtxAbort}.
}{
  If $\oxexpr_f$ is a value and every $\oxexpr_i$ is a value, we would like to
  step with one of:
}{\EApp}{
  Since $\oxexpr_f$ is a value, then by \Lemma{lemma:canonical-forms}, we know
  that it has the form
  $\oxclosureval{\oxstore_c}{ \oxascribe{\oxid_1}{\oxsitype_1} \oxdotsc
    \oxascribe{\oxid_n}{\oxsitype_n} }{\oxsitype_r}{\oxexpr}$.
  Then, since all of the $e_i$ are values, then we can step using
  \oxname{E-AppClosure}.
}

\oxprogresscaseinductive{T-AppFunction}{\TApp}{
  We proceed based on whether or not $\oxexpr_f$ is a value. If it is not, we can
  decompose our expression into the evaluation context
  $\oxapp{\oxhole}{\overline{\oxenv} \oxcomma \overline{\oxprov} \oxcomma
    \overline{\oxsitype}}{\oxexpr_1 \oxdotsc \oxexpr_n}$ and redex
  $\oxexpr_f$. Then, by applying our induction hypothesis to $\oxexpr_f$, we
  know either that $\oxexpr_f$ steps to some $\oxexpr_f^\prime$ or is an
  $\oxkey{abort!}$ expression. In the former case, this satisfies our
  requirement since we can plug $\oxexpr_f^\prime$ back into our evaluation
  context. In the latter case, we can step with \oxname{E-EvalCtxAbort}.

  Next, we'll proceed based on whether or not each expression $\oxexpr_i$ is a
  value. If any of them are not, we can decompose our expression into the
  evaluation context
  $\oxapp{\oxvalue_f}{\overline{\oxenv} \oxcomma \overline{\oxprov} \oxcomma \overline{\oxsitype} }{\oxvalue_1 \oxdotsc \oxvalue_m \oxcomma \oxhole \oxcomma \oxexpr_1 \oxdotsc \oxexpr_{n^\prime}}$
  and redex $\oxexpr_i$. Then, by applying our induction hypothesis to
  $\oxexpr_i$, we know either that $\oxexpr_i$ steps to some $\oxexpr_i^\prime$
  or is an $\oxkey{abort!}$ expression. In the former case, this satisfies our
  requirement since we can plug $\oxexpr_i^\prime$ back into our evaluation
  context. In the latter case, we can step with \oxname{E-EvalCtxAbort}. }{ If
  $\oxexpr_f$ is a value and every $\oxexpr_i$ is a value, we would like to step
  with one of: }{\EAppFun}{ Since $\oxexpr_f$ is a value, then by
  \Lemma{lemma:canonical-forms}, we know that it has the form
  $\oxfnname$.
  Then, since all of the $e_i$ are values, then we can step using
  \oxname{E-AppFunction}.
}

\oxprogresscasevalue{T-Unit}{\TUnit}{
  By inspection of the value grammar, we know that $\oxunit$ is already a value.
}

\oxprogresscasevalue{T-u32}{\TuThreeTwo}{
  By inspection of the value grammar, we know that $\oxnum$ is already a value.
}

\oxprogresscasevalue{T-True}{\TTrue}{
  By inspection of the value grammar, we know that $\oxtrue$ is already a value.
}

\oxprogresscasevalue{T-False}{\TFalse}{
  By inspection of the value grammar, we know that $\oxfalse$ is already a
  value.
}

\oxprogresscasevalue{T-Tuple}{\TTuple}{
  We'll proceed based on whether or not each expression $\oxexpr_i$ is a
  value. If any of them are not, we can decompose our expression into the
  evaluation context $\oxprod{\oxvalue_1 \oxdotsc \oxvalue_m \oxcomma \oxhole
    \oxcomma \oxexpr_1 \oxdotsc \oxexpr_{n^\prime}}$ and redex
  $\oxexpr_i$. Then, by applying our induction hypothesis to $\oxexpr_i$, we
  know either that $\oxexpr_i$ steps to some $\oxexpr_i^\prime$ or to an
  $\oxkey{abort!}$ expression. In either case, this satisfies our requirement,
  since we can plug $\oxexpr_i^\prime$ back into our evaluation context.

  If every expression $e_i$ is a value, then the whole expression is a value by
  the definition of values.
}

\oxprogresscasevalue{T-Array}{\TArray}{
  We'll proceed based on whether or not each expression $\oxexpr_i$ is a
  value. If any of them are not, we can decompose our expression into the
  evaluation context $\oxarr{\oxvalue_1 \oxdotsc \oxvalue_m \oxcomma \oxhole
    \oxcomma \oxexpr_1 \oxdotsc \oxexpr_{n^\prime}}$ and redex
  $\oxexpr_i$. Then, by applying our induction hypothesis to $\oxexpr_i$, we
  know either that $\oxexpr_i$ steps to some $\oxexpr_i^\prime$ or to an
  $\oxkey{abort!}$ expression. In either case, this satisfies our requirement,
  since we can plug $\oxexpr_i^\prime$ back into our evaluation context.

  If every expression $e_i$ is a value, then the whole expression is a value by
  the definition of values.
}

\oxprogresscasevalue{T-Abort}{\TAbort}{
  By definition, $\oxabort{\oxdots}$ is an $\oxkey{abort!}$ expression.
}

\oxprogresscaseinductive{T-Framed}{\TFramed}{
  We proceed based on whether or not $\oxexpr$ is a value. If it is not, we can
  decompose our expression into the evaluation context
  $\oxframed{\oxhole}$ and redex $\oxexpr$. Then, by applying our
  induction hypothesis to $\oxexpr$, we know either that $\oxexpr$ steps to some
  $\oxexpr^\prime$ or to an $\oxkey{abort!}$ expression. In either case, this
  satisfies our requirement, since we can plug $\oxexpr^\prime$ back into our
  evaluation context.
}{If $\oxexpr$ is a value, then we would like to step with:}{\EFramed}{
  In order to do so, we need to know $\oxid \in \oxdomain{\oxstore}$.
  Fortunately, we know from our assumption that
  \oxstorevalidity{\oxglobalctx}{\oxvarctx}{\oxstore} (via
  \oxname{WF-Stack}). The premise of \oxname{WF-Stack} tells us that
  $\oxdomain{\oxstore} = \oxdomain{\oxvarctx}$, and thus the $\oxid \in
  \oxdomain{\oxvarctx}$ from the premise of \oxname{T-Framed} is sufficient to
  tell us that $\oxid \in \oxdomain{\oxstore}$. Thus, we can step with
  \oxname{E-Framed}.
}

\oxprogresscasevalue{T-Pointer}{\TPointer}{
  By inspection of the value grammar, we know that
  $\oxptr{\oxplace}$ is already a value.
}

\oxprogresscasevalue{T-ClosureValue}{\TClosureVal}{
  By inspection of the value grammar, we know that
  $\oxclosureval{\oxstore}{ \oxascribe{\oxid_1}{\oxsitype_1} \oxdotsc
    \oxascribe{\oxid_n}{\oxsitype_n} }{\oxsitype_r}{\oxexpr}$ is already a
  value.
}

\oxprogresscasevalue{T-Dead}{\TUninit}{
  The type $\oxsidtype$ is not in the grammar of $\oxsitype$. Thus, we have a
  contradiction.
}

\oxprogresscasevalue{T-Drop}{\TDrop}{
  By \oxname{R-Env}, we have that $\oxsubtypectx{\oxglobalctx}{\oxemptyctx}
  {\oxvarctx}{\oxtupdate{\oxvarctx}{\oxplace}{\oxsidtype_\oxplace}}$. Then,
  applying \Lemma{lemma:stack-validity-related-envs} with
  $\oxstorevalidity{\oxglobalctx}{\oxvarctx}{\oxstore}$ (from our premise) gives
  us $\oxstorevalidity{\oxglobalctx}{
    \oxtupdate{\oxvarctx}{\oxplace}{\oxsidtype_\oxplace}
  }{\oxstore}$. We can then apply our induction hypothesis to this and
  $\oxtypjudge{\oxglobalctx}{\oxemptyctx}{\oxkontctx}{
    \oxtupdate{\oxvarctx}{\oxplace}{\oxsidtype_\oxplace}
  }{\oxexpr}{\oxsxtype}{\oxvarctx_f}$ to reach our goal.
}

\oxprogresscasevalue{T-Left}{\TInl}{
  We proceed based on whether or not $\oxexpr$ is a value. If it is not, we can
  decompose our expression into the evaluation context
  $\oxinl{\oxsitype_1}{\oxsitype_2}{\oxhole}$ and redex $\oxexpr$. Then, by
  applying our induction hypothesis to $\oxexpr$, we know either that $\oxexpr$
  steps to some $\oxexpr^\prime$ or to an $\oxkey{abort!}$ expression. In either
  case, this satisfies our requirement, since we can plug $\oxexpr^\prime$ back
  into our evaluation context.

  If $\oxexpr$ is a value, then we know that the whole expression
  $\oxinl{\oxsitype_1}{\oxsitype_2}{\oxvalue}$ is a value and thus we are done.
}

\oxprogresscasevalue{T-Right}{\TInr}{
  We proceed based on whether or not $\oxexpr$ is a value. If it is not, we can
  decompose our expression into the evaluation context
  $\oxinr{\oxsitype_1}{\oxsitype_2}{\oxhole}$ and redex $\oxexpr$. Then, by
  applying our induction hypothesis to $\oxexpr$, we know either that $\oxexpr$
  steps to some $\oxexpr^\prime$ or to an $\oxkey{abort!}$ expression. In either
  case, this satisfies our requirement, since we can plug $\oxexpr^\prime$ back
  into our evaluation context.

  If $\oxexpr$ is a value, then we know that the whole expression
  $\oxinr{\oxsitype_1}{\oxsitype_2}{\oxvalue}$ is a value and thus we are done.
}

\oxprogresscaseinductive{T-Match}{\TMatch}{
  We proceed based on whether or not $\oxexpr$ is a value. If it is not, we can
  decompose our expression into the evaluation context
  $\oxmatch{\oxhole}{\oxid_1}{\oxexpr_1}{\oxid_2}{\oxexpr_2}$ and redex
  $\oxexpr$. Then, by applying our induction hypothesis to $\oxexpr$, we know
  either that $\oxexpr$ steps to some $\oxexpr^\prime$ or to an $\oxkey{abort!}$
  expression. In either case, this satisfies our requirement, since we can plug
  $\oxexpr^\prime$ back into our evaluation context.
}{If $\oxexpr$ is a value, then we would like to step with either:}{
  \EMatchLeft \and \EMatchRight
}{
  Since $\oxexpr$ is a value, applying \Lemma{lemma:canonical-forms} tells us
  that $\oxexpr$ is either of the form $\oxinl{\oxsitype_1}{\oxsitype_2}{\oxvalue}$ or
  $\oxinr{\oxsitype_1}{\oxsitype_2}{\oxvalue}$. In the former case, we can
  step with \oxname{E-MatchLeft} and in the latter case, we can step with
  \oxname{E-MatchRight}.
}


\newpage
\subsection{Preservation}
\label{sec:preservation}

\begin{oxlemma}{Preservation}{lemma:preservation-app}
  If $\oxtypjudge{\oxglobalctx}{\oxemptyctx}{\oxkontctx}{\oxvarctx}{\oxexpr}{\oxsitype_1}{\oxvarctx_f} and
  \oxstorevalidity{\oxglobalctx}{\oxvarctx}{\oxstore}$ and
  $\oxwfkontctx{\oxglobalctx}{\oxvarctx}{\overline{\oxvalue}}{\oxkontctx}$ and
  $\oxreduce{\oxglobalctx}{\oxstore}{\oxexpr}{\oxstore^\prime}{\oxexpr^\prime}$, then there exists
  $\oxvarctx_i$ such that
  $\oxstorevalidity{\oxglobalctx}{\oxvarctx_i}{\oxstore^\prime}$ and
  $\oxwfkontctx{\oxglobalctx}{\oxvarctx_i}{\overline{\oxvalue}}{\oxkontctx}$ and
  $\oxtypjudge{\oxglobalctx}{\oxemptyctx}{\oxkontctx}{\oxvarctx_i}{\oxexpr^\prime}
  {\oxsitype_2}{\oxvarctx_f^\prime}$ and
  $\oxtunify[\oxcombine]{\oxemptyctx}{\oxkontctx}{\oxvarctx_f^\prime}
  {\oxsitype_2}{\oxsitype_1}{\oxvarctx_s}$ and there exists $\oxvarctx_o$ such
  that $\oxvarctx_f = \oxvarctx_s \oxintersect \oxvarctx_o$.
\end{oxlemma}

\paragraph{Proof} We proceed by induction on the derivation
\oxtypjudge{\oxglobalctx}{\oxemptyctx}{\oxkontctx}{\oxvarctx}{\oxexpr}{\oxtype}{\oxvarctx_f} \\

\oxpreservationcase{T-Move}{\TMove}{\oxplace}{\EMove}{
  \oxtupdate{\oxvarctx}{\oxplace}{\oxsidtype}
}

\begin{preservation}
  \oxpresline{
    \oxstorevalidity{\oxglobalctx}{
      \oxtupdate{\oxvarctx}{\oxplace}{\oxsidtype}
    }{
      \oxvupdate{\oxstore}{\oxplace}{\oxuninit}
    }
  }{
    Applying \Lemma{lemma:stack-validity-related-envs} to
    $\oxsubtypectx{\oxglobalctx}{\oxemptyctx}{\oxvarctx}{
      \oxtupdate{\oxvarctx}{\oxplace}{\oxsidtype}
    }$ (immediate by \oxname{R-Env}) and
    $\oxstorevalidity{\oxglobalctx}{\oxvarctx}{\oxstore}$ (from premise) gives us
    $\oxstorevalidity{\oxglobalctx}{
      \oxtupdate{\oxvarctx}{\oxplace}{\oxsidtype}
    }{\oxstore}$. Then, since we know
    $\oxtypjudge{\oxglobalctx}{\oxemptyctx}{\oxkontctx}{
      \oxtupdate{\oxvarctx}{\oxplace}{\oxsidtype}
    }{\oxuninit}{\oxsidtype}{
      \oxtupdate{\oxvarctx}{\oxplace}{\oxsidtype}
    }$ (by \oxname{T-Dead}), we can conclude
    $\oxstorevalidity{\oxglobalctx}{
      \oxtupdate{\oxvarctx}{\oxplace}{\oxsidtype}
    }{
      \oxvupdate{\oxstore}{\oxplace}{\oxuninit}
    }$.
  }

  \oxpresline{
    \oxwfkontctx{\oxglobalctx}{
      \oxtupdate{\oxvarctx}{\oxplace}{\oxsidtype}
    }{\overline{\oxvalue}}{\oxkontctx}
  }{
    Inverting \oxname{WF-Temporaries} gives us $\forall i \in 1 \oxdots n. \;
    \oxtypjudge{\oxglobalctx}{\oxemptyctx}{\oxsitype_1 \oxdotsc \oxsitype_{i-1}}{\oxvarctx}{
      \oxvalue_i
    }{\oxsitype_i}{\oxvarctx}$. By \oxname{R-Env}, we have that
    \oxsubtypectx{\oxglobalctx}{\oxemptyctx}{\oxvarctx}{
      \oxtupdate{\oxvarctx}{\oxplace}{\oxsidtype}
    }. Thus, we can apply \Lemma{lemma:value-typing-related-envs} to each of the
    typing judgments and then apply \oxname{WF-Temporaries} to get
    $\oxwfkontctx{\oxglobalctx}{
      \oxtupdate{\oxvarctx}{\oxplace}{\oxsidtype}
    }{\overline{\oxvalue}}{\oxkontctx}$.
  }

  \oxpresline{
    \oxtypjudge{\oxglobalctx}{\oxemptyctx}{\oxkontctx}{
      \oxtupdate{\oxvarctx}{\oxplace}{\oxsidtype}
    }{\oxvalue}{\oxsitype}{
      \oxtupdate{\oxvarctx}{\oxplace}{\oxsidtype}
    }
  }{
    Applying \Lemma{lemma:place-exprs-reduce} to
    $\oxmusafety{\oxemptyctx}{\oxvarctx}{\oxmut}{\oxplace}{
      \oxset{\oxloanpkg{\oxmut}{\oxplace}}
    }$, $\oxcomputetynoprov{\oxemptyctx}{\oxvarctx}{\oxmut}{\oxplace}{\oxsitype}$
    (immediate by \oxname{TC-Place} with $\oxlookup{\oxvarctx}{\oxplace}{
      \oxsitype
    }$), and $\oxstorevalidity{\oxglobalctx}{\oxvarctx}{\oxstore}$ gives us
    $\oxtypjudge{\oxglobalctx}{\oxemptyctx}{\oxkontctx}{\oxvarctx}{\oxvalue}
    {\oxsitype}{\oxvarctx}$. Then, by applying
    \Lemma{lemma:value-typing-related-envs} with
    $\oxsubtypectx{\oxglobalctx}{\oxemptyctx}
    {\oxvarctx}{\oxtupdate{\oxvarctx}{\oxplace}{\oxsidtype}}$ (immediate by
    \oxname{R-Env}), we can conclude $\oxtypjudge{\oxglobalctx}{\oxemptyctx}{\oxkontctx}{
      \oxtupdate{\oxvarctx}{\oxplace}{\oxsidtype}
    }{\oxvalue}{\oxsitype}{
      \oxtupdate{\oxvarctx}{\oxplace}{\oxsidtype}
    }$.
  }

  \oxpresline{
    \oxtunify[\oxcombine]{\oxemptyctx}{\oxkontctx}{
      \oxtupdate{\oxvarctx}{\oxplace}{\oxsidtype}
    }{\oxsitype}{\oxsitype}{
      \oxtupdate{\oxvarctx}{\oxplace}{\oxsidtype}
    }
  }{
    Immediate by \oxname{RR-Refl}.
  }

  \oxpresline{
    \exists \oxvarctx_o. \oxtupdate{\oxvarctx}{\oxplace}{\oxsidtype} \oxintersect \oxvarctx_o
    = \oxtupdate{\oxvarctx}{\oxplace}{\oxsidtype}
  }{
    $\oxvarctx_o = \oxtupdate{\oxvarctx}{\oxplace}{\oxsidtype}$
  }
\end{preservation}

\oxpreservationcase{T-Copy}{\TCopy}{\oxplaceexpr}{\ECopy}{
  \oxvarctx
}

\begin{preservation}
  \oxpresline{
    \oxstorevalidity{\oxglobalctx}{\oxvarctx}{\oxstore}
  }{
    Immediate from our premise.
  }

  \oxpresline{
    \oxwfkontctx{\oxglobalctx}{\oxvarctx}{\overline{\oxvalue}}{\oxkontctx}
  }{
    Immediate from our premise.
  }

  \oxpresline{
    \oxtypjudge{\oxglobalctx}{\oxemptyctx}{\oxkontctx}{\oxvarctx}{\oxvalue}
    {\oxsitype}{\oxvarctx}
  }{
    Applying \Lemma{lemma:place-exprs-reduce} to
    $\oxmusafety{\oxemptyctx}{\oxvarctx}{\oximm}{\oxplaceexpr}{
      \oxset{\overline{\oxloan}}
    }$, $\oxcomputetynoprov{\oxemptyctx}{\oxvarctx}{\oximm}{\oxplaceexpr}
    {\oxsitype}$, and $\oxstorevalidity{\oxglobalctx}{\oxvarctx}{\oxstore}$
    gives us $\oxtypjudge{\oxglobalctx}{\oxemptyctx}{\oxkontctx}{\oxvarctx}{\oxvalue}
    {\oxsitype}{\oxvarctx}$.
  }

  \oxpresline{
    \oxtunify[\oxcombine]{\oxemptyctx}{\oxkontctx}{\oxvarctx}
    {\oxsitype}{\oxsitype}{\oxvarctx}
  }{
    Immediate by \oxname{RR-Refl}.
  }

  \oxpresline{
    \exists \oxvarctx_o. \oxvarctx \oxintersect \oxvarctx_o = \oxvarctx
  }{
    $\oxvarctx_o = \oxvarctx$
  }
\end{preservation}

\oxpreservationcase{T-Borrow}{\TBorrow}{
  \oxref{\oxprov}{\oxmuta}{\oxplaceexpr}
}{\EBorrow}{
  \oxupdate{\oxvarctx}{
    \oxloanctxentry{\oxcprov}{
      \oxset{\overline{\oxloan}}
    }
  }
}

\begin{preservation}
  \oxpresline{
    \oxstorevalidity{\oxglobalctx}{
      \oxupdate{\oxvarctx}{
        \oxloanctxentry{\oxcprov}{
          \oxset{\overline{\oxloan}}
        }
      }
    }{\oxstore}
  }{
    Apply \Lemma{lemma:store-validity-loan-update} to
    $\oxstorevalidity{\oxglobalctx}{\oxvarctx}{\oxstore}$ (from our premise)
    and $\oxctxswellformed{\oxglobalctx}{\oxemptyctx}{
      \oxupdate{\oxvarctx}{
        \oxloanctxentry{\oxcprov}{
          \oxset{\overline{\oxloan}}
        }
      }
    }{\oxkontctx}$ (from our premise)
    gives us $\oxstorevalidity{\oxglobalctx}{
      \oxupdate{\oxvarctx}{
        \oxloanctxentry{\oxcprov}{
          \oxset{\overline{\oxloan}}
        }
      }
    }{\oxstore}$.
  }

  \oxpresline{
    \oxwfkontctx{\oxglobalctx}{
      \oxupdate{\oxvarctx}{
        \oxloanctxentry{\oxcprov}{
          \oxset{\overline{\oxloan}}
        }
      }
    }{\overline{\oxvalue}}{\oxkontctx}
  }{
    Inverting \oxname{WF-Temporaries} gives us $\forall i \in 1 \oxdots n. \;
    \oxtypjudge{\oxglobalctx}{\oxemptyctx}{\oxsitype_1 \oxdotsc \oxsitype_{i-1}}{\oxvarctx}{
      \oxvalue_i
    }{\oxsitype_i}{\oxvarctx}$. We can then apply
    \Lemma{lemma:values-well-typed-loan-update} to each of the typing judgments
    along with
    $\oxmusafety{\oxemptyctx}{\oxkontctx}{\oxvarctx}{\oxmuta}{\oxplaceexpr}{
      \oxset{\overline{\oxloan}}
    }$ and $\oxnotinclosure{\oxkontctx}{\oxvarctx}{\oxcprov}$
    and $\oxlookup{\oxvarctx}{\oxcprov}{\emptyset}$ (all from the premise of
    \oxname{T-Borrow}) to get $\forall i \in 1 \oxdots n. \;
    \oxtypjudge{\oxglobalctx}{\oxemptyctx}{\oxsitype_1 \oxdotsc \oxsitype_{i-1}}{
      \oxupdate{\oxvarctx}{
        \oxloanctxentry{\oxcprov}{
          \oxset{\overline{\oxloan}}
        }
      }
    }{\oxvalue_i}{\oxsitype_i}{
      \oxupdate{\oxvarctx}{
        \oxloanctxentry{\oxcprov}{
          \oxset{\overline{\oxloan}}
        }
      }
    }$. We can then apply \oxname{WF-Temporaries} to get
    $\oxwfkontctx{\oxglobalctx}{
      \oxupdate{\oxvarctx}{
        \oxloanctxentry{\oxcprov}{
          \oxset{\overline{\oxloan}}
        }
      }
    }{\overline{\oxvalue}}{\oxkontctx}$.
  }

  \oxpresline{
    \oxtypjudge{\oxglobalctx}{\oxemptyctx}{\oxkontctx}{\oxvarctx_i}{
      \oxptr{\oxreferent}
    }{\oxtref{\oxcprov}{\oxmuta}{\oxxitype}}{\oxvarctx_i}
  }{
    Applying \Lemma{lemma:norm-for-valid-referents} to
    $\oxstorevalidity{\oxglobalctx}{\oxvarctx}{\oxstore}$, and
    $\oxnorm{\oxstore}{\oxplaceexpr}{\oxreferent}{\_}{\_}$ gives us
     $\oxreferentvalidity{\oxglobalctx}{\oxvarctx}{
      \oxreferentctx[\oxplace]
    }{\oxxitype}$. Then, note that referent well-formedness does not
    depend on the contents of loan sets. This means we can also conclude
    $\oxreferentvalidity{\oxglobalctx}{
      \oxupdate{\oxvarctx}{
        \oxloanctxentry{\oxcprov}{
          \oxset{\overline{\oxloan}}
        }
      }
    }{
      \oxreferentctx[\oxplace]
    }{\oxxitype}$.

    Applying \Lemma{lemma:norm-place-prefix-in-loans} to
    $\oxstorevalidity{\oxglobalctx}{\oxvarctx}{\oxstore}$,
    $\oxnorm{\oxstore}{\oxplaceexpr}{\oxreferent}{\_}{\_}$, and
    $\oxmusafety{\oxemptyctx}{\oxvarctx}{\oxmuta}{\oxplaceexpr}{
      \oxset{\overline{\oxloan}}
    }$ gives us $\oxreferent = \oxreferentctx[\oxplace]$ and
    $\oxloanpkg{\oxmuta}{\oxplace} \in \oxset{\overline{\oxloan}}$.

    Finally, we can apply \oxname{T-Pointer} to the two facts above to get
    $\oxtypjudge{\oxglobalctx}{\oxemptyctx}{\oxkontctx}{
      \oxupdate{\oxvarctx}{
        \oxloanctxentry{\oxcprov}{
          \oxset{\overline{\oxloan}}
        }
      }
    }{
      \oxptr{\oxreferent}
    }{\oxtref{\oxcprov}{\oxmuta}{\oxxitype}}{
      \oxupdate{\oxvarctx}{
        \oxloanctxentry{\oxcprov}{
          \oxset{\overline{\oxloan}}
        }
      }
    }$.
  }

  \oxpresline{
    \oxtunify[\oxcombine]{\oxemptyctx}{\oxkontctx}{\oxvarctx_i}{
      \oxtref{\oxcprov}{\oxmuta}{\oxxitype}
    }{
      \oxtref{\oxcprov}{\oxmuta}{\oxxitype}
    }{\oxvarctx_i}
  }{
    Immediate by \oxname{RR-Refl}.
  }

  \oxpresline{
    \exists \oxvarctx_o. \oxvarctx_i \oxintersect \oxvarctx_o = \oxvarctx_i
  }{
    $\oxvarctx_o = \oxvarctx_i$
  }
\end{preservation}

\oxpreservationcasemulti{T-BorrowIndex}{\TBorrowIndex}{
  \oxref{\oxprov}{\oxmuta}{\oxindex{\oxplaceexpr}{\oxexpr_i}}
}{
  \EBorrowIndex \and \EBorrowIndexOOB
}

\oxpreservationsubcaseheading{E-BorrowIndex}{
  \oxupdate{\oxvarctx^\prime}{
    \oxloanctxentry{\oxcprov}{
      \oxset{\overline{\oxloan}}
    }
  }
}

\begin{preservation}
  \oxpresline{
    \oxstorevalidity{\oxglobalctx}{
      \oxupdate{\oxvarctx^\prime}{
        \oxloanctxentry{\oxcprov}{
          \oxset{\overline{\oxloan}}
        }
      }
    }{\oxstore}
  }{
    Applying \Lemma{lemma:values-dont-change} to the typing derivation (from
    \oxname{T-BorrowIndex}) for
    $\oxexpr$ (which we know is a value from \oxname{E-BorrowIndex}) gives us
    $\oxsubtypectx{\oxglobalctx}{\oxemptyctx}{\oxvarctx}{\oxvarctx^\prime}$.

    Then, applying \Lemma{lemma:stack-validity-related-envs} to
    $\oxsubtypectx{\oxglobalctx}{\oxemptyctx}{\oxvarctx}{\oxvarctx^\prime}$ and
    $\oxstorevalidity{\oxglobalctx}{\oxvarctx}{\oxstore}$ (from premise) gives us
    $\oxstorevalidity{\oxglobalctx}{\oxvarctx^\prime}{\oxstore}$.

    Finally, applying \Lemma{lemma:store-validity-loan-update} to
    $\oxstorevalidity{\oxglobalctx}{\oxvarctx^\prime}{\oxstore}$ gives us
    $\oxstorevalidity{\oxglobalctx}{
      \oxupdate{\oxvarctx^\prime}{
        \oxloanctxentry{\oxcprov}{
          \oxset{\overline{\oxloan}}
        }
      }
    }{\oxstore}$.
  }

  \oxpresline{
    \oxwfkontctx{\oxglobalctx}{
      \oxupdate{\oxvarctx^\prime}{
        \oxloanctxentry{\oxcprov}{
          \oxset{\overline{\oxloan}}
        }
      }
    }{\overline{\oxvalue}}{\oxkontctx}
  }{
    Inverting \oxname{WF-Temporaries} gives us $\forall i \in 1 \oxdots n. \;
    \oxtypjudge{\oxglobalctx}{\oxemptyctx}{\oxsitype_1 \oxdotsc \oxsitype_{i-1}}{\oxvarctx}{
      \oxvalue_i
    }{\oxsitype_i}{\oxvarctx}$.

    Applying \Lemma{lemma:values-dont-change} to the typing derivation (from
    \oxname{T-BorrowIndex}) for
    $\oxexpr$ (which we know is a value from \oxname{E-BorrowIndex}) gives us
    $\oxsubtypectx{\oxglobalctx}{\oxemptyctx}{\oxvarctx}{\oxvarctx^\prime}$.

    Thus, we can apply \Lemma{lemma:value-typing-related-envs} to each of the
    typing judgments to get $\forall i \in 1 \oxdots n. \;
    \oxtypjudge{\oxglobalctx}{\oxemptyctx}{\oxemptyctx}{\oxvarctx^\prime}{
      \oxvalue_i
    }{\oxsitype_i}{\oxvarctx^\prime}$.

    We can then apply \Lemma{lemma:values-well-typed-loan-update} to each of the
    typing judgments along with
    $\oxmusafety{\oxemptyctx}{\oxkontctx}{\oxvarctx^\prime}{\oxmuta}{\oxplaceexpr}{
      \oxset{\overline{\oxloan}}
    }$ and $\oxnotinclosure{\oxkontctx}{\oxvarctx^\prime}{\oxcprov}$
    and $\oxlookup{\oxvarctx^\prime}{\oxcprov}{\emptyset}$ (all from the premise
    of \oxname{T-BorrowIndex}) to get $\forall i \in 1 \oxdots n. \;
    \oxtypjudge{\oxglobalctx}{\oxemptyctx}{\oxemptyctx}{
      \oxupdate{\oxvarctx^\prime}{
        \oxloanctxentry{\oxcprov}{
          \oxset{\overline{\oxloan}}
        }
      }
    }{\oxvalue_i}{\oxsitype_i}{
      \oxupdate{\oxvarctx^\prime}{
        \oxloanctxentry{\oxcprov}{
          \oxset{\overline{\oxloan}}
        }
      }
    }$. We can then apply \oxname{WF-Temporaries} to get
    $\oxwfkontctx{\oxglobalctx}{
      \oxupdate{\oxvarctx^\prime}{
        \oxloanctxentry{\oxcprov}{
          \oxset{\overline{\oxloan}}
        }
      }
    }{\overline{\oxvalue}}{\oxkontctx}$.
  }

  \oxpresline{
    \oxtypjudge{\oxglobalctx}{\oxemptyctx}{\oxkontctx}{\oxvarctx_i}{
      \oxindexptr{\oxreferent}{\oxnum_i}
    }{\oxtref{\oxcprov}{\oxmuta}{\oxsitype}}{\oxvarctx_i}
  }{
    Applying \Lemma{lemma:values-dont-change} to the typing derivation (from
    \oxname{T-BorrowIndex}) for
    $\oxexpr$ (which we know is a value from \oxname{E-BorrowIndex}) gives us
    $\oxsubtypectx{\oxglobalctx}{\oxemptyctx}{\oxvarctx}{\oxvarctx^\prime}$.

    Then, applying \Lemma{lemma:stack-validity-related-envs} to
    $\oxsubtypectx{\oxglobalctx}{\oxemptyctx}{\oxvarctx}{\oxvarctx^\prime}$ and
    $\oxstorevalidity{\oxglobalctx}{\oxvarctx}{\oxstore}$ (from premise) gives us
    $\oxstorevalidity{\oxglobalctx}{\oxvarctx^\prime}{\oxstore}$.

    Applying \Lemma{lemma:norm-for-valid-referents} to
    $\oxstorevalidity{\oxglobalctx}{\oxvarctx^\prime}{\oxstore}$, and
    $\oxnorm{\oxstore}{\oxplaceexpr}{\oxreferent}{\_}{
      \oxarr{\oxvalue_0 \oxdotsc \oxvalue_n}
    }$ gives us $\oxreferentvalidity{\oxglobalctx}{\oxvarctx^\prime}{
      \oxreferentctx[\oxplace]
    }{\oxxitype}$. Then, note that referent well-formedness does not
    depend on the contents of loan sets. This means we can also conclude
    $\oxreferentvalidity{\oxglobalctx}{
      \oxupdate{\oxvarctx^\prime}{
        \oxloanctxentry{\oxcprov}{
          \oxset{\overline{\oxloan}}
        }
      }
    }{
      \oxreferentctx[\oxplace]
    }{\oxxitype}$. We can then apply \oxname{WF-RefIndexArray} or
    \oxname{WF-RefIndexSlice} to get $\oxreferentvalidity{\oxglobalctx}{
      \oxupdate{\oxvarctx^\prime}{
        \oxloanctxentry{\oxcprov}{
          \oxset{\overline{\oxloan}}
        }
      }
    }{
      \oxindex{\oxreferentctx[\oxplace]}{\oxnum_i}
    }{\oxxitype}$.

    Applying \Lemma{lemma:norm-place-prefix-in-loans} to
    $\oxstorevalidity{\oxglobalctx}{\oxvarctx^\prime}{\oxstore}$,
    $\oxnorm{\oxstore}{\oxplaceexpr}{\oxreferent}{\_}{
      \oxarr{\oxvalue_0 \oxdotsc \oxvalue_n}
    }$, and
    $\oxmusafety{\oxemptyctx}{\oxvarctx^\prime}{\oxmuta}{\oxplaceexpr}{
      \oxset{\overline{\oxloan}}
    }$ gives us $\oxreferent = \oxreferent[\oxplace]$ and
    $\oxloanpkg{\oxmuta}{\oxplace} \in \oxset{\overline{\oxloan}}$.

    Finally, we can apply \oxname{T-Pointer} to the two facts above to get
    $\oxtypjudge{\oxglobalctx}{\oxemptyctx}{\oxkontctx}{
      \oxupdate{\oxvarctx}{
        \oxloanctxentry{\oxcprov}{
          \oxset{\overline{\oxloan}}
        }
      }
    }{
      \oxindexptr{\oxreferent}{\oxnum_i}
    }{\oxtref{\oxcprov}{\oxmuta}{\oxxitype}}{
      \oxupdate{\oxvarctx}{
        \oxloanctxentry{\oxcprov}{
          \oxset{\overline{\oxloan}}
        }
      }
    }$.
  }

  \oxpresline{
    \oxtunify[\oxcombine]{\oxemptyctx}{\oxkontctx}{\oxvarctx_i}{
      \oxtref{\oxcprov}{\oxmuta}{\oxsitype}
    }{
      \oxtref{\oxcprov}{\oxmuta}{\oxsitype}
    }{\oxvarctx_i}
  }{
    Immediate by \oxname{RR-Refl}.
  }

  \oxpresline{
    \exists \oxvarctx_o. \oxvarctx_i \oxintersect \oxvarctx_o = \oxvarctx_i
  }{
    $\oxvarctx_o = \oxvarctx_i$
  }
\end{preservation}

\oxpreservationsubcaseheading{E-BorrowIndexOOB}{
  \oxvarctx
}

\begin{preservation}
  \oxpresline{
    \oxstorevalidity{\oxglobalctx}{\oxvarctx}{\oxstore}
  }{
    Immediate from our premise.
  }

  \oxpresline{
    \oxwfkontctx{\oxglobalctx}{\oxvarctx}{\overline{\oxvalue}}{\oxkontctx}
  }{
    Immediate from our premise.
  }

  \oxpresline{
    \oxtypjudge{\oxglobalctx}{\oxemptyctx}{\oxkontctx}{\oxvarctx}
    {\oxkey{abort!}(\oxdots)}{\oxtref{\oxcprov}{\oxmuta}{\oxsitype}}{\oxvarctx}
  }{
    An \oxkey{abort!} expression is well-typed (at any type) via the rule
    \oxname{T-Abort}.
  }

  \oxpresline{
    \oxtunify[\oxcombine]{\oxemptyctx}{\oxkontctx}{\oxvarctx}{
      \oxtref{\oxcprov}{\oxmuta}{\oxsitype}
    }{
      \oxtref{\oxcprov}{\oxmuta}{\oxsitype}
    }{\oxvarctx}
  }{
    Immediate by \oxname{RR-Refl}.
  }

  \oxpresline{
    \exists \oxvarctx_o. \oxvarctx_i \oxintersect \oxvarctx_o = \oxvarctx_i
  }{
    $\oxvarctx_o = \oxvarctx_i$
  }
\end{preservation}

\oxpreservationcasemulti{T-BorrowSlice}{\TBorrowSlice}{
  \oxref{\oxcprov}{\oxmuta}{\oxslice{\oxplaceexpr}{\oxexpr_1}{\oxexpr_2}}
}{
  \EBorrowSlice \and \EBorrowSliceOOB
}

\oxpreservationsubcaseheading{E-BorrowSlice}{
  \oxupdate{\oxvarctx_2}{
    \oxloanctxentry{\oxcprov}{
      \oxset{\overline{\oxloan}}
    }
  }
}

\begin{preservation}
  \oxpresline{
    \oxstorevalidity{\oxglobalctx}{
      \oxupdate{\oxvarctx_2}{
        \oxloanctxentry{\oxcprov}{
          \oxset{\overline{\oxloan}}
        }
      }
    }{\oxstore}
  }{
    Applying \Lemma{lemma:values-dont-change} to the typing derivation (from
    \oxname{T-BorrowSlice}) for $\oxexpr_1$ (which we know is a value from
    \oxname{E-BorrowSlice}) gives us
    $\oxsubtypectx{\oxglobalctx}{\oxemptyctx}{\oxvarctx}{\oxvarctx_1}$. Then,
    applying \Lemma{lemma:values-dont-change} to the typing derivation (from
    \oxname{T-BorrowSlice}) for  $\oxexpr_2$ (which we know is a value from
    \oxname{E-BorrowSlice}) gives us
    $\oxsubtypectx{\oxglobalctx}{\oxemptyctx}{\oxvarctx_1}{\oxvarctx_2}$. Then,
    by transitivity, we have
    $\oxsubtypectx{\oxglobalctx}{\oxemptyctx}{\oxvarctx}{\oxvarctx_2}$.

    Then, applying \Lemma{lemma:stack-validity-related-envs} to
    $\oxsubtypectx{\oxglobalctx}{\oxemptyctx}{\oxvarctx}{\oxvarctx_2}$ and
    $\oxstorevalidity{\oxglobalctx}{\oxvarctx}{\oxstore}$ (from premise) gives us
    $\oxstorevalidity{\oxglobalctx}{\oxvarctx_2}{\oxstore}$. Finally,
    applying \Lemma{lemma:store-validity-loan-update} to
    $\oxstorevalidity{\oxglobalctx}{\oxvarctx_2}{\oxstore}$ gives us
    $\oxstorevalidity{\oxglobalctx}{
      \oxupdate{\oxvarctx_2}{
        \oxloanctxentry{\oxcprov}{
          \oxset{\overline{\oxloan}}
        }
      }
    }{\oxstore}$.
  }

  \oxpresline{
    \oxwfkontctx{\oxglobalctx}{
      \oxupdate{\oxvarctx_2}{
        \oxloanctxentry{\oxcprov}{
          \oxset{\overline{\oxloan}}
        }
      }
    }{\overline{\oxvalue}}{\oxkontctx}
  }{
    Inverting \oxname{WF-Temporaries} gives us $\forall i \in 1 \oxdots n. \;
    \oxtypjudge{\oxglobalctx}{\oxemptyctx}{\oxsitype_1 \oxdotsc \oxsitype_{i-1}}{\oxvarctx}{
      \oxvalue_i
    }{\oxsitype_i}{\oxvarctx}$.

    Applying \Lemma{lemma:values-dont-change} to the typing derivation (from
    \oxname{T-BorrowSlice}) for $\oxexpr_1$ (which we know is a value from
    \oxname{E-BorrowSlice}) gives us
    $\oxsubtypectx{\oxglobalctx}{\oxemptyctx}{\oxvarctx}{\oxvarctx_1}$. Then,
    applying \Lemma{lemma:values-dont-change} to the typing derivation (from
    \oxname{T-BorrowSlice}) for  $\oxexpr_2$ (which we know is a value from
    \oxname{E-BorrowSlice}) gives us
    $\oxsubtypectx{\oxglobalctx}{\oxemptyctx}{\oxvarctx_1}{\oxvarctx_2}$. Then,
    by transitivity, we have
    $\oxsubtypectx{\oxglobalctx}{\oxemptyctx}{\oxvarctx}{\oxvarctx_2}$.

    Then, we apply \Lemma{lemma:value-typing-related-envs} to each of the
    typing judgments to get $\forall i \in 1 \oxdots n. \;
    \oxtypjudge{\oxglobalctx}{\oxemptyctx}{\oxsitype_1 \oxdotsc \oxsitype_{i-1}}{\oxvarctx_2}{
      \oxvalue_i
    }{\oxsitype_i}{\oxvarctx_2}$.

    We can then apply \Lemma{lemma:values-well-typed-loan-update} to each of the
    typing judgments along with
    $\oxmusafety{\oxemptyctx}{\oxkontctx}{\oxvarctx_2}{\oxmuta}{\oxplaceexpr}{
      \oxset{\overline{\oxloan}}
    }$ and $\oxnotinclosure{\oxkontctx}{\oxvarctx_2}{\oxcprov}$
    and $\oxlookup{\oxvarctx_2}{\oxcprov}{\emptyset}$ (all from the premise
    of \oxname{T-BorrowSlice}) to get $\forall i \in 1 \oxdots n. \;
    \oxtypjudge{\oxglobalctx}{\oxemptyctx}{\oxsitype_1 \oxdotsc \oxsitype_{i-1}}{
      \oxupdate{\oxvarctx_2}{
        \oxloanctxentry{\oxcprov}{
          \oxset{\overline{\oxloan}}
        }
      }
    }{\oxvalue_i}{\oxsitype_i}{
      \oxupdate{\oxvarctx_2}{
        \oxloanctxentry{\oxcprov}{
          \oxset{\overline{\oxloan}}
        }
      }
    }$. We can then apply \oxname{WF-Temporaries} to get
    $\oxwfkontctx{\oxglobalctx}{
      \oxupdate{\oxvarctx_2}{
        \oxloanctxentry{\oxcprov}{
          \oxset{\overline{\oxloan}}
        }
      }
    }{\overline{\oxvalue}}{\oxkontctx}$.
  }

  \oxpresline{
    \oxtypjudge{\oxglobalctx}{\oxemptyctx}{\oxkontctx}{\oxvarctx_i}{
      \oxsliceptr{\oxreferent}{\oxnum_1}{\oxnum_2}
    }{
      \oxtref{\oxcprov}{\oxmuta}{\oxtslice{\oxsitype}}
    }{\oxvarctx_i}
  }{
    Applying \Lemma{lemma:values-dont-change} to the typing derivation (from
    \oxname{T-BorrowSlice}) for $\oxexpr_1$ (which we know is a value from
    \oxname{E-BorrowSlice}) gives us
    $\oxsubtypectx{\oxglobalctx}{\oxemptyctx}{\oxvarctx}{\oxvarctx_1}$. Then,
    applying \Lemma{lemma:values-dont-change} to the typing derivation (from
    \oxname{T-BorrowSlice}) for  $\oxexpr_2$ (which we know is a value from
    \oxname{E-BorrowSlice}) gives us
    $\oxsubtypectx{\oxglobalctx}{\oxemptyctx}{\oxvarctx_1}{\oxvarctx_2}$. Then,
    by transitivity, we have
    $\oxsubtypectx{\oxglobalctx}{\oxemptyctx}{\oxvarctx}{\oxvarctx_2}$.

    Then, applying \Lemma{lemma:stack-validity-related-envs} to
    $\oxsubtypectx{\oxglobalctx}{\oxemptyctx}{\oxvarctx}{\oxvarctx_2}$ and
    $\oxstorevalidity{\oxglobalctx}{\oxvarctx}{\oxstore}$ (from premise) gives us
    $\oxstorevalidity{\oxglobalctx}{\oxvarctx_2}{\oxstore}$.

    Applying \Lemma{lemma:norm-for-valid-referents} to
    $\oxstorevalidity{\oxglobalctx}{\oxvarctx_2}{\oxstore}$, and
    $\oxnorm{\oxstore}{\oxplaceexpr}{\oxreferent}{\_}{
      \oxarr{\oxvalue_0 \oxdotsc \oxvalue_n}
    }$ gives us $\oxreferentvalidity{\oxglobalctx}{\oxvarctx_2}{
      \oxreferentctx[\oxplace]
    }{\oxxitype}$. Then, note that referent well-formedness does not
    depend on the contents of loan sets. This means we can also conclude
    $\oxreferentvalidity{\oxglobalctx}{
      \oxupdate{\oxvarctx_2}{
        \oxloanctxentry{\oxcprov}{
          \oxset{\overline{\oxloan}}
        }
      }
    }{
      \oxreferentctx[\oxplace]
    }{\oxxitype}$. We can then apply \oxname{WF-RefSliceArray} or
    \oxname{WF-RefSliceSlice} to get $\oxreferentvalidity{\oxglobalctx}{
      \oxupdate{\oxvarctx^\prime}{
        \oxloanctxentry{\oxcprov}{
          \oxset{\overline{\oxloan}}
        }
      }
    }{
      \oxslice{\oxreferentctx[\oxplace]}{\oxnum_1}{\oxnum_2}
    }{\oxxitype}$.

    Applying \Lemma{lemma:norm-place-prefix-in-loans} to
    $\oxstorevalidity{\oxglobalctx}{\oxvarctx_2}{\oxstore}$,
    $\oxnorm{\oxstore}{\oxplaceexpr}{\oxreferent}{\_}{
      \oxarr{\oxvalue_0 \oxdotsc \oxvalue_n}
    }$, and
    $\oxmusafety{\oxemptyctx}{\oxvarctx_2}{\oxmuta}{\oxplaceexpr}{
      \oxset{\overline{\oxloan}}
    }$ gives us $\oxloanpkg{\oxmuta}{\oxplace} \in \oxset{\overline{\oxloan}}$.

    Finally, we can apply \oxname{T-Pointer} to the two facts above to get
    $\oxtypjudge{\oxglobalctx}{\oxemptyctx}{\oxkontctx}{
      \oxupdate{\oxvarctx}{
        \oxloanctxentry{\oxcprov}{
          \oxset{\overline{\oxloan}}
        }
      }
    }{
      \oxsliceptr{\oxreferent[\oxplace]}{\oxnum_1}{\oxnum_2}
    }{\oxtref{\oxcprov}{\oxmuta}{\oxxitype}}{
      \oxupdate{\oxvarctx}{
        \oxloanctxentry{\oxcprov}{
          \oxset{\overline{\oxloan}}
        }
      }
    }$.
  }

  \oxpresline{
    \oxtunify[\oxcombine]{\oxemptyctx}{\oxkontctx}{\oxvarctx_i}{
      \oxtref{\oxcprov}{\oxmuta}{\oxtslice{\oxsitype}}
    }{
      \oxtref{\oxcprov}{\oxmuta}{\oxtslice{\oxsitype}}
    }{\oxvarctx_i}
  }{
    Immediate by \oxname{RR-Refl}.
  }

  \oxpresline{
    \exists \oxvarctx_o. \oxvarctx_i \oxintersect \oxvarctx_o = \oxvarctx_i
  }{
    $\oxvarctx_o = \oxvarctx_i$
  }
\end{preservation}

\oxpreservationsubcaseheading{E-BorrowSliceOOB}{
  \oxvarctx
}

\begin{preservation}
  \oxpresline{
    \oxstorevalidity{\oxglobalctx}{\oxvarctx}{\oxstore}
  }{
    Immediate from our premise.
  }

  \oxpresline{
    \oxwfkontctx{\oxglobalctx}{\oxvarctx}{\overline{\oxvalue}}{\oxkontctx}
  }{
    Immediate from our premise.
  }

  \oxpresline{
    \oxtypjudge{\oxglobalctx}{\oxemptyctx}{\oxkontctx}{\oxvarctx}
    {\oxabort{\oxdots}}{\oxtref{\oxcprov}{\oxmuta}{\oxtslice{\oxsitype}}}
    {\oxvarctx^\prime}
  }{
    An \oxkey{abort!} expression is well-typed (at any type) via the rule \oxname{T-Abort}.
  }

  \oxpresline{
    \oxtunify[\oxcombine]{\oxemptyctx}{\oxkontctx}{\oxvarctx}{
      \oxtref{\oxcprov}{\oxmuta}{\oxtslice{\oxsitype}}
    }{
      \oxtref{\oxcprov}{\oxmuta}{\oxtslice{\oxsitype}}
    }{\oxvarctx}
  }{
    Immediate by \oxname{RR-Refl}.
  }

  \oxpresline{
    \exists \oxvarctx_o. \oxvarctx_i \oxintersect \oxvarctx_o = \oxvarctx_i
  }{
    $\oxvarctx_o = \oxvarctx_i$
  }
\end{preservation}

\oxpreservationcase{T-IndexCopy}{\TIndexCopy}{
  \oxindex{\oxplaceexpr}{\oxexpr_i}
}{\EIndexCopy}{
  \oxvarctx^\prime
}

\begin{preservation}
  \oxpresline{
    \oxstorevalidity{\oxglobalctx}{\oxvarctx^\prime}{\oxstore}
  }{
    Applying \Lemma{lemma:values-dont-change} to the typing derivation (from
    \oxname{T-IndexCopy}) for $\oxexpr$ (which we know is a value from
    \oxname{E-IndexCopy}) gives us
    $\oxsubtypectx{\oxglobalctx}{\oxemptyctx}{\oxvarctx}{\oxvarctx^\prime}$.
    Then, applying \Lemma{lemma:stack-validity-related-envs} to
    $\oxsubtypectx{\oxglobalctx}{\oxemptyctx}{\oxvarctx}{\oxvarctx^\prime}$ and
    $\oxstorevalidity{\oxglobalctx}{\oxvarctx}{\oxstore}$ (from premise) gives us
    $\oxstorevalidity{\oxglobalctx}{\oxvarctx^\prime}{\oxstore}$.
  }

  \oxpresline{
    \oxwfkontctx{\oxglobalctx}{\oxvarctx^\prime}{\overline{\oxvalue}}{\oxkontctx}
  }{
    Inverting \oxname{WF-Temporaries} gives us $\forall i \in 1 \oxdots n. \;
    \oxtypjudge{\oxglobalctx}{\oxemptyctx}{\oxsitype_1 \oxdotsc \oxsitype_{i-1}}{\oxvarctx}{
      \oxvalue_i
    }{\oxsitype_i}{\oxvarctx}$.

    Applying \Lemma{lemma:values-dont-change} to the typing derivation (from
    \oxname{T-IndexCopy}) for
    $\oxexpr$ (which we know is a value from \oxname{E-IndexCopy}) gives us
    $\oxsubtypectx{\oxglobalctx}{\oxemptyctx}{\oxvarctx}{\oxvarctx^\prime}$.

    Thus, we can apply \Lemma{lemma:value-typing-related-envs} to each of the
    typing judgments to get $\forall i \in 1 \oxdots n. \;
    \oxtypjudge{\oxglobalctx}{\oxemptyctx}{\oxsitype_1 \oxdotsc \oxsitype_{i-1}}{\oxvarctx^\prime}{
      \oxvalue_i
    }{\oxsitype_i}{\oxvarctx^\prime}$.

    Finally, applying \oxname{WF-Temporaries} gives us
    $\oxwfkontctx{\oxglobalctx}{\oxvarctx^\prime}{\overline{\oxvalue}}{\oxkontctx}$.
  }

  \oxpresline{
    \oxtypjudge{\oxglobalctx}{\oxemptyctx}{\oxkontctx}{\oxvarctx^\prime}{
      \oxvalue_{n_i}
    }{\oxsitype}{\oxvarctx^\prime}
  }{
    Applying \Lemma{lemma:values-dont-change} to the typing derivation (from
    \oxname{T-IndexCopy}) for $\oxexpr$ (which we know is a value from
    \oxname{E-IndexCopy}) gives us
    $\oxsubtypectx{\oxglobalctx}{\oxemptyctx}{\oxvarctx}{\oxvarctx^\prime}$.
    Then, applying \Lemma{lemma:stack-validity-related-envs} to
    $\oxsubtypectx{\oxglobalctx}{\oxemptyctx}{\oxvarctx}{\oxvarctx^\prime}$ and
    $\oxstorevalidity{\oxglobalctx}{\oxvarctx}{\oxstore}$ (from premise) gives us
    $\oxstorevalidity{\oxglobalctx}{\oxvarctx^\prime}{\oxstore}$.

    Applying \Lemma{lemma:place-exprs-reduce} to
    $\oxmusafety{\oxemptyctx}{\oxvarctx^\prime}{\oximm}{\oxplaceexpr}{
      \oxset{\overline{\oxloan}}
    }$, $\oxcomputetynoprov{\oxemptyctx}{\oxvarctx^\prime}{\oximm}{\oxplaceexpr}
    {\oxsitype}$, and $\oxstorevalidity{\oxglobalctx}{\oxvarctx^\prime}{\oxstore}$
    \oxname{T-Slice} (based on whether $\oxxitype = \oxtarr{\oxsitype}{\oxnum}$ or
    $\oxtslice{\oxsitype}$ respectively), we get
    $\forall i \in \oxset{1 \oxdots n}. \;
    \oxtypjudge{\oxglobalctx}{\oxemptyctx}{\oxkontctx}{\oxvarctx^\prime}{
      \oxvalue_i
    }{\oxsitype}{\oxvarctx^\prime}$ (after accounting for the fact that the
    constituent expressions are values and the output environment matches the
    input environment). Thus, we can pick out specifically that
    $\oxtypjudge{\oxglobalctx}{\oxemptyctx}{\oxkontctx}{\oxvarctx^\prime}{
      \oxvalue_i
    }{\oxsitype}{\oxvarctx^\prime}$.
  }

  \oxpresline{
    \oxtunify[\oxcombine]{\oxemptyctx}{\oxkontctx}{\oxvarctx^\prime}{\oxsitype}{\oxsitype}{\oxvarctx^\prime}
  }{
    Immediate by \oxname{RR-Refl}.
  }

  \oxpresline{
    \exists \oxvarctx_o. \oxvarctx^\prime \oxintersect \oxvarctx_o = \oxvarctx^\prime
  }{
    $\oxvarctx_o = \oxvarctx^\prime$
  }
\end{preservation}

\oxpreservationcase{T-Seq}{\TSeq}{\oxseq{\oxexpr_1}{\oxexpr_2}}{\ESeq}{
  \oxgcloans{\oxkontctx}{\oxvarctx_1}
}

\begin{preservation}
  \oxpresline{
    \oxstorevalidity{\oxglobalctx}{
      \oxgcloans{\oxkontctx}{\oxvarctx_1}
    }{\oxstore}
  }{
    Applying \Lemma{lemma:values-dont-change} to the typing derivation (from
    \oxname{T-Seq}) for $\oxexpr_1$ (which we know is a value from
    \oxname{E-Seq}) gives us
    $\oxsubtypectx{\oxglobalctx}{\oxemptyctx}{\oxvarctx}{\oxvarctx_1}$.
    Applying \Lemma{lemma:stack-validity-related-envs} with this and
    $\oxstorevalidity{\oxglobalctx}{\oxvarctx}{\oxstore}$ (from
    premise) gives us
    $\oxstorevalidity{\oxglobalctx}{\oxvarctx_1}{\oxstore}$. By definition of
    \oxname{R-Env}, we know that $\oxsubtypectx{\oxglobalctx}{\oxemptyctx}
    {\oxvarctx_1}{\oxgcloans{\oxkontctx}{\oxvarctx_1}}$. Then, we can apply
    \Lemma{lemma:stack-validity-related-envs} to get
    $\oxstorevalidity{\oxglobalctx}{
      \oxgcloans{\oxkontctx}{\oxvarctx_1}
    }{\oxstore}$.
  }

  \oxpresline{
    \oxwfkontctx{\oxglobalctx}{
      \oxgcloans{\oxkontctx}{\oxvarctx_1}
    }{\overline{\oxvalue}}{\oxkontctx}
  }{
    Inverting \oxname{WF-Temporaries} gives us $\forall i \in 1 \oxdots n. \;
    \oxtypjudge{\oxglobalctx}{\oxemptyctx}{\oxsitype_1 \oxdotsc \oxsitype_{i-1}}{\oxvarctx}{
      \oxvalue_i
    }{\oxsitype_i}{\oxvarctx}$.

    Applying \Lemma{lemma:values-dont-change} to the typing derivation (from
    \oxname{T-Seq}) for
    $\oxexpr_1$ (which we know is a value from \oxname{E-Seq}) gives us
    $\oxsubtypectx{\oxglobalctx}{\oxemptyctx}{\oxvarctx}{\oxvarctx_1}$. Then,
    we note that definitionally $\oxsubtypectx{\oxglobalctx}{\oxemptyctx}{\oxvarctx_1}{
      \oxgcloans{\oxkontctx}{\oxvarctx_1}
    }$ since we can only clear loans that are not present in any of the types.
    Then, by transitivity, we have
    $\oxsubtypectx{\oxglobalctx}{\oxemptyctx}{\oxvarctx}{
      \oxgcloans{\oxkontctx}{\oxvarctx_1}
    }$.

    Thus, we can apply \Lemma{lemma:value-typing-related-envs} to each of the
    typing judgments to get $\forall i \in 1 \oxdots n. \;
    \oxtypjudge{\oxglobalctx}{\oxemptyctx}{\oxsitype_1 \oxdotsc \oxsitype_{i-1}}{
      \oxgcloans{\oxkontctx}{\oxvarctx_1}
    }{
      \oxvalue_i
    }{\oxsitype_i}{
      \oxgcloans{\oxkontctx}{\oxvarctx_1}
    }$.

    Finally, applying \oxname{WF-Temporaries} gives us
    $\oxwfkontctx{\oxglobalctx}{
      \oxgcloans{\oxkontctx}{\oxvarctx_1}
    }{\overline{\oxvalue}}{\oxkontctx}$.
  }

  \oxpresline{
    \oxtypjudge{\oxglobalctx}{\oxemptyctx}{\oxkontctx}{\oxgcloans{\oxkontctx}{\oxvarctx_1}}
    {\oxexpr_2}{\oxsitype_2}{\oxvarctx_2}
  }{
    Immediate from the premise of \oxname{T-Seq}.
  }

  \oxpresline{
    \oxtunify[\oxcombine]{\oxemptyctx}{\oxkontctx}{\oxvarctx_2}
    {\oxsitype_2}{\oxsitype_2}{\oxvarctx_2}
  }{
    Immediate by \oxname{RR-Refl}.
  }

  \oxpresline{
    \exists \oxvarctx_o. \oxvarctx_2 \oxintersect \oxvarctx_o = \oxvarctx_2
  }{
    $\oxvarctx_o = \oxvarctx_2$
  }
\end{preservation}

\oxpreservationcasemulti{T-Branch}{\TBranch}{
  \oxbranch{\oxexpr_1}{\oxexpr_2}{\oxexpr_3}
}{\EIfTrue \and \EIfFalse}

\oxpreservationsubcaseheading{E-IfTrue}{
  \oxvarctx_1
}

\begin{preservation}
  \oxpresline{
    \oxstorevalidity{\oxglobalctx}{\oxvarctx_1}{\oxstore}
  }{
    Applying \Lemma{lemma:values-dont-change} to the typing derivation (from
    \oxname{T-Branch}) for $\oxexpr_1$ (which we know is a value from
    \oxname{E-IfTrue}) gives us
    $\oxsubtypectx{\oxglobalctx}{\oxemptyctx}{\oxvarctx}{\oxvarctx_1}$. Then,
    applying \Lemma{lemma:stack-validity-related-envs} with this and
    $\oxstorevalidity{\oxglobalctx}{\oxvarctx}{\oxstore}$ (from
    premise) gives us
    $\oxstorevalidity{\oxglobalctx}{\oxvarctx_1}{\oxstore}$.
  }

  \oxpresline{
    \oxwfkontctx{\oxglobalctx}{\oxvarctx_1}{\overline{\oxvalue}}{\oxkontctx}
  }{
    Inverting \oxname{WF-Temporaries} gives us $\forall i \in 1 \oxdots n. \;
    \oxtypjudge{\oxglobalctx}{\oxemptyctx}{\oxsitype_1 \oxdotsc \oxsitype_{i-1}}{\oxvarctx}{
      \oxvalue_i
    }{\oxsitype_i}{\oxvarctx}$.

    Applying \Lemma{lemma:values-dont-change} to the typing derivation (from
    \oxname{T-Branch}) for
    $\oxexpr_1$ (which we know is a value from \oxname{E-IfTrue}) gives us
    $\oxsubtypectx{\oxglobalctx}{\oxemptyctx}{\oxvarctx}{\oxvarctx_1}$.

    Thus, we can apply \Lemma{lemma:value-typing-related-envs} to each of the
    typing judgments to get $\forall i \in 1 \oxdots n. \;
    \oxtypjudge{\oxglobalctx}{\oxemptyctx}{\oxsitype_1 \oxdotsc \oxsitype_{i-1}}{\oxvarctx_1}{
      \oxvalue_i
    }{\oxsitype_i}{\oxvarctx_1}$.

    Finally, applying \oxname{WF-Temporaries} gives us
    $\oxwfkontctx{\oxglobalctx}{\oxvarctx_1}{\overline{\oxvalue}}{\oxkontctx}$.
  }

  \oxpresline{
    \oxtypjudge{\oxglobalctx}{\oxemptyctx}{\oxkontctx}{\oxvarctx_1}{\oxexpr_2}
    {\oxsitype}{\oxvarctx_2}
  }{
    Immediate from premise of \oxname{T-Branch}.
  }

  \oxpresline{
    \oxtunify[\oxcombine]{\oxemptyctx}{\oxkontctx}{\oxvarctx_2}
    {\oxsitype_2}{\oxsitype}{\oxvarctx_2^\prime}
  }{
    Immediate from premise of \oxname{T-Branch}.
  }

  \oxpresline{
    \exists \oxvarctx_o. \oxvarctx_2^\prime \oxintersect \oxvarctx_o = \oxvarctx^\prime
  }{
    $\oxvarctx_o = \oxvarctx_3^\prime$
  }
\end{preservation}

\oxpreservationsubcaseheading{E-IfFalse}{
  \oxvarctx_1
}

\begin{preservation}
  \oxpresline{
    \oxstorevalidity{\oxglobalctx}{\oxvarctx_1}{\oxstore}
  }{
    Applying \Lemma{lemma:values-dont-change} to the typing derivation (from
    \oxname{T-Branch}) for $\oxexpr_1$ (which we know is a value from
    \oxname{E-IfFalse}) gives us
    $\oxsubtypectx{\oxglobalctx}{\oxemptyctx}{\oxvarctx}{\oxvarctx_1}$. Then,
    applying \Lemma{lemma:stack-validity-related-envs} with this and
    $\oxstorevalidity{\oxglobalctx}{\oxvarctx}{\oxstore}$ (from
    premise) gives us
    $\oxstorevalidity{\oxglobalctx}{\oxvarctx_1}{\oxstore}$.
  }

  \oxpresline{
    \oxwfkontctx{\oxglobalctx}{\oxvarctx_1}{\overline{\oxvalue}}{\oxkontctx}
  }{
    Inverting \oxname{WF-Temporaries} gives us $\forall i \in 1 \oxdots n. \;
    \oxtypjudge{\oxglobalctx}{\oxemptyctx}{\oxsitype_1 \oxdotsc \oxsitype_{i-1}}{\oxvarctx}{
      \oxvalue_i
    }{\oxsitype_i}{\oxvarctx}$.

    Applying \Lemma{lemma:values-dont-change} to the typing derivation (from
    \oxname{T-Branch}) for
    $\oxexpr_1$ (which we know is a value from \oxname{E-IfFalse}) gives us
    $\oxsubtypectx{\oxglobalctx}{\oxemptyctx}{\oxvarctx}{\oxvarctx_1}$.

    Thus, we can apply \Lemma{lemma:value-typing-related-envs} to each of the
    typing judgments to get $\forall i \in 1 \oxdots n. \;
    \oxtypjudge{\oxglobalctx}{\oxemptyctx}{\oxsitype_1 \oxdotsc \oxsitype_{i-1}}{\oxvarctx_1}{
      \oxvalue_i
    }{\oxsitype_i}{\oxvarctx_1}$.

    Finally, applying \oxname{WF-Temporaries} gives us
    $\oxwfkontctx{\oxglobalctx}{\oxvarctx_1}{\overline{\oxvalue}}{\oxkontctx}$.
  }

  \oxpresline{
    \oxtypjudge{\oxglobalctx}{\oxemptyctx}{\oxkontctx}{\oxvarctx_1}
    {\oxexpr_3}{\oxsitype_3}{\oxvarctx_3}
  }{
    Immediate from premise of \oxname{T-Branch}.
  }

  \oxpresline{
    \oxtunify[\oxcombine]{\oxemptyctx}{\oxkontctx}{\oxvarctx_3}
    {\oxsitype_3}{\oxsitype}{\oxvarctx_3^\prime}
  }{
    Immediate from premise of \oxname{T-Branch}.
  }

  \oxpresline{
    \exists \oxvarctx_o. \oxvarctx_3^\prime \oxintersect \oxvarctx_o = \oxvarctx^\prime
  }{
    $\oxvarctx_o = \oxvarctx_2^\prime$ (note that $\oxintersect$ commutes)
  }
\end{preservation}

\oxpreservationcasemulti{T-Match}{\TMatch}{
  \oxmatch{\oxexpr}{\oxid_1}{\oxexpr_1}{\oxid_2}{\oxexpr_2}
}{\EMatchLeft \and \EMatchRight}

\oxpreservationsubcaseheading{E-MatchLeft}{
  \oxextendctx{\oxvarctx^\prime}{
    \oxvarctxentry{\oxid_1}{\oxsitype_l}
  }
}

\begin{preservation}
  \oxpresline{
    \oxstorevalidity{\oxglobalctx}{
      \oxextendctx{\oxvarctx^\prime}{
        \oxvarctxentry{\oxid_1}{\oxsitype_l}
      }
    }{
      \oxextendctx{\oxstore}{
        \oxstoreentry{\oxid_1}{\oxvalue}
      }
    }
  }{
    Applying \Lemma{lemma:values-dont-change} to the typing derivation (from
    \oxname{T-Match}) for $\oxexpr$ (which we know is a value from
    \oxname{E-MatchLeft}) gives us
    $\oxsubtypectx{\oxglobalctx}{\oxemptyctx}{\oxvarctx}{\oxvarctx^\prime}$. Then,
    applying \Lemma{lemma:stack-validity-related-envs} with this and
    $\oxstorevalidity{\oxglobalctx}{\oxvarctx}{\oxstore}$ (from
    premise) gives us
    $\oxstorevalidity{\oxglobalctx}{\oxvarctx^\prime}{\oxstore}$. Then, using
    $\oxtypjudge{\oxglobalctx}{\oxemptyctx}{\oxkontctx}{\oxvarctx}{\oxvalue}{\oxsitype_l}
    {\oxvarctx^\prime}$ (which we get by inversion of \oxname{T-Inl} in the
    derivation for $\oxexpr$) with \oxname{WF-StackFrame} allows us to finally
    conclude $\oxstorevalidity{\oxglobalctx}{
      \oxextendctx{\oxvarctx^\prime}{
        \oxvarctxentry{\oxid_1}{\oxsitype_l}
      }
    }{
      \oxextendctx{\oxstore}{
        \oxstoreentry{\oxid_1}{\oxvalue}
      }
    }$.
  }

  \oxpresline{
    \oxwfkontctx{\oxglobalctx}{
      \oxextendctx{\oxvarctx^\prime}{
        \oxvarctxentry{\oxid_1}{\oxsitype_l}
      }
    }{\overline{\oxvalue}}{\oxkontctx}
  }{
    Inverting \oxname{WF-Temporaries} gives us $\forall i \in 1 \oxdots n. \;
    \oxtypjudge{\oxglobalctx}{\oxemptyctx}{\oxsitype_1 \oxdotsc \oxsitype_{i-1}}{\oxvarctx}{
      \oxvalue_i
    }{\oxsitype_i}{\oxvarctx}$.

    Applying \Lemma{lemma:values-dont-change} to the typing derivation (from
    \oxname{T-Match}) for
    $\oxexpr$ (which we know is a value from \oxname{E-MatchLeft}) gives us
    $\oxsubtypectx{\oxglobalctx}{\oxemptyctx}{\oxvarctx}{\oxvarctx^\prime}$.

    Thus, we can apply \Lemma{lemma:value-typing-related-envs} to each of the
    typing judgments to get $\forall i \in 1 \oxdots n. \;
    \oxtypjudge{\oxglobalctx}{\oxemptyctx}{\oxsitype_1 \oxdotsc \oxsitype_{i-1}}{\oxvarctx^\prime}{
      \oxvalue_i
    }{\oxsitype_i}{\oxvarctx^\prime}$.

    Applying \Lemma{lemma:values-well-typed-extension} to each of the value
    typing judgments and $\forall \oxcprov \in
    \oxfprovs{\oxtsum{\oxsitype_l}{\oxsitype_r}}. \;
    \oxnotreborrowed{\oxvarctx^\prime}{\oxcprov}$ (from the premise of
    \oxname{T-Match}) gives us
    $\forall i \in 1 \oxdots n. \;
    \oxtypjudge{\oxglobalctx}{\oxemptyctx}{\oxsitype_1 \oxdotsc \oxsitype_{i-1}}{
      \oxextendctx{\oxvarctx^\prime}{
        \oxvarctxentry{\oxid_1}{\oxsitype_l}
      }
    }{
      \oxvalue_i
    }{\oxsitype_i}{
      \oxextendctx{\oxvarctx^\prime}{
        \oxvarctxentry{\oxid_1}{\oxsitype_l}
      }
    }$.

    Finally, applying \oxname{WF-Temporaries} gives us
    $\oxwfkontctx{\oxglobalctx}{
      \oxextendctx{\oxvarctx^\prime}{
        \oxvarctxentry{\oxid_1}{\oxsitype_l}
      }
    }{\overline{\oxvalue}}{\oxkontctx}$.
  }

  \oxpresline{
    \oxtypjudge{\oxglobalctx}{\oxemptyctx}{\oxkontctx}{
      \oxextendctx{\oxvarctx^\prime}{
        \oxvarctxentry{\oxid_1}{\oxsitype_l}
      }
    }{\oxexpr_1}
    {\oxsitype_1}{\oxvarctx_1}
  }{
    Immediate from applying \oxname{T-Shift} to
    $\oxtypjudge{\oxglobalctx}{\oxemptyctx}{\oxkontctx}{
      \oxextendctx{\oxvarctx^\prime}{
        \oxvarctxentry{\oxid_1}{\oxsitype_l}
      }
    }{
      \oxexpr_1
    }{\oxsitype_1}{
      \oxextendctx{\oxvarctx_1}{
        \oxvarctxentry{\oxid_1}{\oxsdtype_l}
      }
    }$ from the premise of \oxname{T-Match}.
  }

  \oxpresline{
    \oxtunify[\oxcombine]{\oxemptyctx}{\oxkontctx}{\oxvarctx_1}
    {\oxsitype_1}{\oxsitype}{\oxvarctx_1^\prime}
  }{
    Immediate from premise of \oxname{T-Match}.
  }

  \oxpresline{
    \exists \oxvarctx_o. \oxvarctx_1^\prime \oxintersect \oxvarctx_o = \oxvarctx^\prime
  }{
    $\oxvarctx_o = \oxvarctx_2^\prime$
  }
\end{preservation}

\oxpreservationsubcaseheading{E-MatchRight}{
  \oxextendctx{\oxvarctx^\prime}{
    \oxvarctxentry{\oxid_2}{\oxsitype_r}
  }
}

\begin{preservation}
  \oxpresline{
    \oxstorevalidity{\oxglobalctx}{
      \oxextendctx{\oxvarctx^\prime}{
        \oxvarctxentry{\oxid_2}{\oxsitype_r}
      }
    }{
      \oxextendctx{\oxstore}{
        \oxstoreentry{\oxid_2}{\oxvalue}
      }
    }
  }{
    Applying \Lemma{lemma:values-dont-change} to the typing derivation (from
    \oxname{T-Match}) for $\oxexpr$ (which we know is a value from
    \oxname{E-MatchRight}) gives us
    $\oxsubtypectx{\oxglobalctx}{\oxemptyctx}{\oxvarctx}{\oxvarctx^\prime}$. Then,
    applying \Lemma{lemma:stack-validity-related-envs} with this and
    $\oxstorevalidity{\oxglobalctx}{\oxvarctx}{\oxstore}$ (from
    premise) gives us
    $\oxstorevalidity{\oxglobalctx}{\oxvarctx^\prime}{\oxstore}$. Then, using
    $\oxtypjudge{\oxglobalctx}{\oxemptyctx}{\oxkontctx}{\oxvarctx}{\oxvalue}{\oxsitype_r}
    {\oxvarctx^\prime}$ (which we get by inversion of \oxname{T-Inr} in the
    derivation for $\oxexpr$) with \oxname{WF-StackFrame} allows us to finally
    conclude $\oxstorevalidity{\oxglobalctx}{
      \oxextendctx{\oxvarctx^\prime}{
        \oxvarctxentry{\oxid_2}{\oxsitype_r}
      }
    }{
      \oxextendctx{\oxstore}{
        \oxstoreentry{\oxid_2}{\oxvalue}
      }
    }$.
  }

  \oxpresline{
    \oxwfkontctx{\oxglobalctx}{
      \oxextendctx{\oxvarctx^\prime}{
        \oxvarctxentry{\oxid_2}{\oxsitype_r}
      }
    }{\overline{\oxvalue}}{\oxkontctx}
  }{
    Inverting \oxname{WF-Temporaries} gives us $\forall i \in 1 \oxdots n. \;
    \oxtypjudge{\oxglobalctx}{\oxemptyctx}{\oxsitype_1 \oxdotsc \oxsitype_{i-1}}{\oxvarctx}{
      \oxvalue_i
    }{\oxsitype_i}{\oxvarctx}$.

    Applying \Lemma{lemma:values-dont-change} to the typing derivation (from
    \oxname{T-Match}) for
    $\oxexpr$ (which we know is a value from \oxname{E-MatchLeft}) gives us
    $\oxsubtypectx{\oxglobalctx}{\oxemptyctx}{\oxvarctx}{\oxvarctx^\prime}$.

    Thus, we can apply \Lemma{lemma:value-typing-related-envs} to each of the
    typing judgments to get $\forall i \in 1 \oxdots n. \;
    \oxtypjudge{\oxglobalctx}{\oxemptyctx}{\oxsitype_1 \oxdotsc \oxsitype_{i-1}}{\oxvarctx^\prime}{
      \oxvalue_i
    }{\oxsitype_i}{\oxvarctx^\prime}$.

    Applying \Lemma{lemma:values-well-typed-extension} to each of the value
    typing judgments and $\forall \oxcprov \in
    \oxfprovs{\oxtsum{\oxsitype_l}{\oxsitype_r}}. \;
    \oxnotreborrowed{\oxvarctx^\prime}{\oxcprov}$ (from the premise of
    \oxname{T-Match}) gives us
    $\forall i \in 1 \oxdots n. \;
    \oxtypjudge{\oxglobalctx}{\oxemptyctx}{\oxsitype_1 \oxdotsc \oxsitype_{i-1}}{
      \oxextendctx{\oxvarctx^\prime}{
        \oxvarctxentry{\oxid_2}{\oxsitype_r}
      }
    }{
      \oxvalue_i
    }{\oxsitype_i}{
      \oxextendctx{\oxvarctx^\prime}{
        \oxvarctxentry{\oxid_2}{\oxsitype_r}
      }
    }$.

    Finally, applying \oxname{WF-Temporaries} gives us
    $\oxwfkontctx{\oxglobalctx}{
      \oxextendctx{\oxvarctx^\prime}{
        \oxvarctxentry{\oxid_2}{\oxsitype_r}
      }
    }{\overline{\oxvalue}}{\oxkontctx}$.
  }

  \oxpresline{
    \oxtypjudge{\oxglobalctx}{\oxemptyctx}{\oxkontctx}{
      \oxextendctx{\oxvarctx^\prime}{
        \oxvarctxentry{\oxid_2}{\oxsitype_r}
      }
    }{\oxexpr_2}
    {\oxsitype_2}{\oxvarctx_2}
  }{
    Immediate from applying \oxname{T-Shift} to
    $\oxtypjudge{\oxglobalctx}{\oxemptyctx}{\oxkontctx}{
      \oxextendctx{\oxvarctx^\prime}{
        \oxvarctxentry{\oxid_2}{\oxsitype_r}
      }
    }{
      \oxexpr_2
    }{\oxsitype_2}{
      \oxextendctx{\oxvarctx_2}{
        \oxvarctxentry{\oxid_2}{\oxsdtype_r}
      }
    }$ from the premise of \oxname{T-Match}.
  }

  \oxpresline{
    \oxtunify[\oxcombine]{\oxemptyctx}{\oxkontctx}{\oxvarctx_2}
    {\oxsitype_2}{\oxsitype}{\oxvarctx_2^\prime}
  }{
    Immediate from premise of \oxname{T-Match}.
  }

  \oxpresline{
    \exists \oxvarctx_o. \oxvarctx_2^\prime \oxintersect \oxvarctx_o = \oxvarctx^\prime
  }{
    $\oxvarctx_o = \oxvarctx_1^\prime$ (note that $\oxintersect$ commutes)
  }
\end{preservation}

\oxpreservationcase{T-Assign}{\TAssign}{
  \oxassign{\oxplace}{\oxexpr_a}
}{\EAssign}{
  \oxgcloans{\oxkontctx}{\oxtupdate{\oxvarctx^\prime}{\oxplace}{\oxsitype}}
}

\begin{preservation}
  \oxpresline{
    \oxstorevalidity{\oxglobalctx}{
      \oxgcloans{\oxkontctx}{\oxtupdate{\oxvarctx^\prime}{\oxplace}{\oxsitype}}
    }{
      \oxvupdate{\oxstore}{\oxplace}{\oxvaluectx[\oxvalue]}
    }
  }{
    From our premise, we have
    $\oxstorevalidity{\oxglobalctx}{\oxvarctx}{\oxstore}$. Then, applying
    \Lemma{lemma:values-dont-change} to the typing derivation
    $\oxtypjudge{\oxglobalctx}{\oxemptyctx}{\oxkontctx}{\oxvarctx}{
      \oxvalue
    }{\oxsitype}{\oxvarctx_1}$ (from the premise of \oxname{T-Assign} combined
    with the fact that $\oxexpr$ is a value from \oxname{E-Assign}) gives us
    $\oxsubtypectx{\oxglobalctx}{\oxemptyctx}{\oxvarctx}{\oxvarctx_1}$. Then,
    applying \Lemma{lemma:stack-validity-related-envs} to these two facts gives
    us $\oxstorevalidity{\oxglobalctx}{\oxvarctx_1}{\oxstore}$.

    Finally, we apply \Lemma{lemma:stack-validity-assign} to
    $\oxstorevalidity{\oxglobalctx}{\oxvarctx_1}{\oxstore}$ and
    $\oxlookup{{\oxvarctx_1}}{\oxplace}{\oxsxtype}$ (from premise) and
    $\oxtunify[\oxnoop]{\oxemptyctx}{\oxkontctx}{
      \oxrsub{\oxvarctx_1}{\oxplace}
    }{\oxsitype}{\oxsxtype}{\oxvarctx^\prime}$ (from the premise of
    \oxname{T-Assign}) and
    $\oxmusafety{\oxemptyctx}{\oxvarctx^\prime}{\oxmut}{\oxplace}{
      \oxset{\oxloanpkg{\oxmut}{\oxplace}}
    }$ (from premise of \oxname{T-Assign} and
    $\oxnorm{\oxstore}{\oxplace}{\oxplace}{\oxvaluectx}{\_}$ and $\oxplace =
    \oxproj{\oxid}{\oxpath}$ (both from the premise of \oxname{E-Assign}),
    $\oxtypjudge{\oxglobalctx}{\oxemptyctx}{\oxkontctx}{
      \oxrsub{\oxvarctx_1}{\oxplaceexpr} }{\oxvalue}{\oxsitype}{
      \oxrsub{\oxvarctx_1}{\oxplaceexpr} }$ (from above) and
    $\oxstorevalidity{\oxglobalctx}{
      \oxgcloans{\oxkontctx}{\oxtupdate{\oxvarctx^\prime}{\oxplace}{\oxsitype}}
    }{
      \oxvupdate{\oxstore}{\oxid}{\oxvaluectx[\oxvalue]}
    }$.
  }

  \oxpresline{
    \oxwfkontctx{\oxglobalctx}{
      \oxgcloans{\oxkontctx}{\oxtupdate{\oxvarctx^\prime}{\oxplace}{\oxsitype}}
    }{\overline{\oxvalue}}{\oxkontctx}
  }{
    Inverting \oxname{WF-Temporaries} gives us $\forall i \in 1 \oxdots n. \;
    \oxtypjudge{\oxglobalctx}{\oxemptyctx}{\oxsitype_1 \oxdotsc \oxsitype_{i-1}}{\oxvarctx}{
      \oxvalue_i
    }{\oxsitype_i}{\oxvarctx}$.

    Applying \Lemma{lemma:values-dont-change} to the typing derivation (from
    \oxname{T-Let}) for
    $\oxexpr_1$ (which we know is a value from \oxname{E-Let}) gives us
    $\oxsubtypectx{\oxglobalctx}{\oxemptyctx}{\oxvarctx}{\oxvarctx_1}$.
    Then, we can apply \Lemma{lemma:value-typing-related-envs} to each of the
    typing judgments to get $\forall i \in 1 \oxdots n. \;
    \oxtypjudge{\oxglobalctx}{\oxemptyctx}{\oxsitype_1 \oxdotsc \oxsitype_{i-1}}{\oxvarctx_1}{
      \oxvalue_i
    }{\oxsitype_i}{\oxvarctx_1}$.

    Then, we apply \Lemma{lemma:values-well-typed-assignment} to each of these
    typing judgments to get $\forall i \in 1 \oxdots n. \;
    \oxtypjudge{\oxglobalctx}{\oxemptyctx}{\oxsitype_1 \oxdotsc \oxsitype_{i-1}}{
      \oxgcloans{\oxkontctx}{\oxtupdate{\oxvarctx^\prime}{\oxplace}{\oxsitype}}
    }{
      \oxvalue_i
    }{\oxsitype_i}{
      \oxgcloans{\oxkontctx}{\oxtupdate{\oxvarctx^\prime}{\oxplace}{\oxsitype}}
    }$. After which we can apply \oxname{WF-Temporaries} to conclude
    $\oxwfkontctx{\oxglobalctx}{
      \oxgcloans{\oxkontctx}{\oxtupdate{\oxvarctx^\prime}{\oxplace}{\oxsitype}}
    }{\overline{\oxvalue}}{\oxkontctx}$.
  }

  \oxpresline{
    \oxtypjudge{\oxglobalctx}{\oxemptyctx}{\oxkontctx}{
      \oxvarctx_i
    }{\oxunit}{\oxtunit}{
      \oxvarctx_i
    }
  }{
    Immediate by \oxname{T-Unit}.
  }

  \oxpresline{
    \oxtunify[\oxcombine]{\oxemptyctx}{\oxkontctx}{
      \oxvarctx_i
    }{\oxtunit}{\oxtunit}{
      \oxvarctx_i
    }
  }{
    Immediate by \oxname{RR-Refl}.
  }

  \oxpresline{
    \exists \oxvarctx_o.
    \oxvarctx_i
    \oxintersect \oxvarctx_o =
    \oxvarctx_i
  }{
    $\oxvarctx_o = \oxvarctx_i =
    \oxgcloans{\oxkontctx}{\oxtupdate{\oxvarctx^\prime}{\oxplace}{\oxsitype}}$
  }
\end{preservation}

\oxpreservationcase{T-AssignDeref}{\TAssignDeref}{
  \oxassign{\oxplaceexpr}{\oxexpr_a}
}{\EAssign}{
  \oxvarctx^\prime
}

\begin{preservation}
  \oxpresline{
    \oxstorevalidity{\oxglobalctx}{\oxvarctx^\prime}{
      \oxvupdate{\oxstore}{\oxplace}{\oxvaluectx[\oxvalue]}
    }
  }{
    From our premise, we have
    $\oxstorevalidity{\oxglobalctx}{\oxvarctx}{\oxstore}$. Then, applying
    \Lemma{lemma:values-dont-change} to the typing derivation
    $\oxtypjudge{\oxglobalctx}{\oxemptyctx}{\oxkontctx}{\oxvarctx}{
      \oxvalue
    }{\oxsitype}{\oxvarctx_1}$ (from the premise of \oxname{T-AssignDeref} combined
    with the fact that $\oxexpr$ is a value from \oxname{E-Assign}) gives us
    $\oxsubtypectx{\oxglobalctx}{\oxemptyctx}{\oxvarctx}{\oxvarctx_1}$. Then,
    applying \Lemma{lemma:stack-validity-related-envs} to these two facts gives
    us $\oxstorevalidity{\oxglobalctx}{\oxvarctx_1}{\oxstore}$.

    Then, applying \Lemma{lemma:stack-validity-region-rewriting} to
    $\oxstorevalidity{\oxglobalctx}{\oxvarctx_1}{\oxstore}$ and
    $\oxtunify[\oxcombine]{\oxemptyctx}{\oxkontctx}{\oxvarctx_1}
    {\oxsitype_n}{\oxsitype_o}{\oxvarctx^\prime}$ (from the premise of
    \oxname{T-AssignDeref}) gives us
    $\oxstorevalidity{\oxglobalctx}{\oxvarctx^\prime}{\oxstore}$.
  }

  \oxpresline{
    \oxwfkontctx{\oxglobalctx}{
      \oxvarctx^\prime
    }{\overline{\oxvalue}}{\oxkontctx}
  }{
    Inverting \oxname{WF-Temporaries} gives us $\forall i \in 1 \oxdots n. \;
    \oxtypjudge{\oxglobalctx}{\oxemptyctx}{\oxsitype_1 \oxdotsc \oxsitype_{i-1}}{\oxvarctx}{
      \oxvalue_i
    }{\oxsitype_i}{\oxvarctx}$.

    Applying \Lemma{lemma:values-dont-change} to the typing derivation (from
    \oxname{T-AssignDeref}) for
    $\oxexpr$ (which we know is a value from \oxname{E-Assign}) gives us
    $\oxsubtypectx{\oxglobalctx}{\oxemptyctx}{\oxvarctx}{\oxvarctx_1}$.
    Then, we can apply \Lemma{lemma:value-typing-related-envs} to each of the
    typing judgments to get $\forall i \in 1 \oxdots n. \;
    \oxtypjudge{\oxglobalctx}{\oxemptyctx}{\oxsitype_1 \oxdotsc \oxsitype_{i-1}}{\oxvarctx_1}{
      \oxvalue_i
    }{\oxsitype_i}{\oxvarctx_1}$.

    Then, applying \Lemma{lemma:subtyping-value-typing} to each of these
    derivations along with
    $\oxtunify[\oxcombine]{\oxemptyctx}{\oxkontctx}{\oxvarctx_1}
    {\oxsitype_n}{\oxsitype_o}{\oxvarctx^\prime}$ (from the premise of
    \oxname{T-AssignDeref}) gives us $\forall i \in 1 \oxdots n. \;
    \oxtypjudge{\oxglobalctx}{\oxemptyctx}{\oxsitype_1 \oxdotsc \oxsitype_{i-1}}{
      \oxvarctx^\prime
    }{\oxvalue_i}{\oxsitype_i}{\oxvarctx^\prime}$. We can then apply
    \oxname{WF-Temporaries} to conclude
    $\oxwfkontctx{\oxglobalctx}{
      \oxvarctx^\prime
    }{\overline{\oxvalue}}{\oxkontctx}$.
  }

  \oxpresline{
    \oxtypjudge{\oxglobalctx}{\oxemptyctx}{\oxkontctx}{
      \oxvarctx^\prime
    }{\oxunit}{\oxtunit}{
      \oxvarctx^\prime
    }
  }{
    Immediate by \oxname{T-Unit}.
  }

  \oxpresline{
    \oxtunify[\oxcombine]{\oxemptyctx}{\oxkontctx}{
      \oxvarctx^\prime
    }{\oxtunit}{\oxtunit}{
      \oxvarctx^\prime
    }
  }{
    Immediate by \oxname{RR-Refl}.
  }

  \oxpresline{
    \exists \oxvarctx_o. \oxvarctx^\prime \oxintersect
    \oxvarctx_o = \oxvarctx^\prime
  }{
    $\oxvarctx_o = \oxvarctx^\prime$
  }
\end{preservation}

\oxpreservationcase{T-Let}{\TLet}{
  \oxlet{\oxid}{\oxsitype}{\oxexpr_1}{\oxexpr_2}
}{\ELet}{
  \oxgcloans{\oxkontctx}{
    \oxextendctx{\oxvarctx_1^\prime}{
      \oxvarctxentry{\oxid}{\oxsitype_a}
    }
  }
}

\begin{preservation}
  \oxpresline{
    \oxstorevalidity{\oxglobalctx}{
      \oxgcloans{\oxkontctx}{
        \oxextendctx{\oxvarctx_1^\prime}{
          \oxvarctxentry{\oxid}{\oxsitype_a}
        }
      }
    }{
      \oxextendctx{\oxstore}{\oxstoreentry{\oxid}{\oxvalue}}
    }
  }{
    Applying \Lemma{lemma:values-dont-change} to the typing derivation (from
    \oxname{T-Let}) for $\oxexpr_1$ (which we know is a value from
    \oxname{E-Let}) gives us
    $\oxsubtypectx{\oxglobalctx}{\oxemptyctx}{\oxvarctx}{\oxvarctx_1}$.
    Then, we can apply \Lemma{lemma:stack-validity-related-envs} to get
    $\oxstorevalidity{\oxglobalctx}{\oxvarctx_1}{\oxstore}$.
    Then,
    applying \Lemma{lemma:stack-validity-region-rewriting} to
    $\oxtunify[\oxcombine]{\oxemptyctx}{\oxkontctx}{\oxvarctx_1}{\oxsitype_1}{\oxsitype_a}
    {\oxvarctx_1^\prime}$ (from premise of \oxname{T-Let}) gives us
    $\oxstorevalidity{\oxglobalctx}{\oxvarctx_1^\prime}{\oxstore}$. We can also
    apply \Lemma{lemma:values-typed-supertype} to $\oxtypjudge{\oxglobalctx}
    {\oxemptyctx}{\oxkontctx}{\oxvarctx}{\oxvalue}{\oxsitype_1}{\oxvarctx_1}$ (from premise
    of \oxname{T-Let}) and $\oxtunify[\oxcombine]{\oxemptyctx}{\oxkontctx}{\oxvarctx_1}
    {\oxsitype_1}{\oxsitype_a}{\oxvarctx_1^\prime}$ (from premise of
    \oxname{T-Let}) gives us $\oxtypjudge{\oxglobalctx}{\oxemptyctx}{\oxkontctx}
    {\oxvarctx^\prime}{\oxexpr_1}{\oxsitype_a}{\oxvarctx^\prime}$

    Then, apply \Lemma{lemma:store-validity-extension} to
    $\oxstorevalidity{\oxglobalctx}{\oxvarctx_1^\prime}{\oxstore}$ and
    $\oxtypjudge{\oxglobalctx}{\oxemptyctx}{\oxkontctx}{\oxvarctx^\prime}
    {\oxexpr_1}{\oxsitype_a}{\oxvarctx^\prime}$ gives us
    $\oxstorevalidity{\oxglobalctx}{
      \oxextendctx{\oxvarctx_1^\prime}{
        \oxvarctxentry{\oxid}{\oxsitype_a}
      }
    }{
      \oxextendctx{\oxstore}{\oxstoreentry{\oxid}{\oxvalue}}
    }$. By definition of
    \oxname{R-Env}, we know that $\oxsubtypectx{\oxglobalctx}{\oxemptyctx}{
      \oxextendctx{\oxvarctx_1^\prime}{
        \oxvarctxentry{\oxid}{\oxsitype_a}
      }
    }{
      \oxgcloans{\oxkontctx}{
        \oxextendctx{\oxvarctx_1^\prime}{
          \oxvarctxentry{\oxid}{\oxsitype_a}
        }
      }
    }$. Then, we can apply
    \Lemma{lemma:stack-validity-related-envs} to conclude
    $\oxstorevalidity{\oxglobalctx}{
      \oxgcloans{\oxkontctx}{
        \oxextendctx{\oxvarctx_1^\prime}{
          \oxvarctxentry{\oxid}{\oxsitype_a}
        }
      }
    }{
      \oxextendctx{\oxstore}{\oxstoreentry{\oxid}{\oxvalue}}
    }$.
  }

  \oxpresline{
    \oxwfkontctx{\oxglobalctx}{
      \oxgcloans{\oxkontctx}{
        \oxextendctx{\oxvarctx_1^\prime}{
          \oxvarctxentry{\oxid}{\oxsitype_a}
        }
      }
    }{\overline{\oxvalue}}{\oxkontctx}
  }{
    Inverting \oxname{WF-Temporaries} gives us $\forall i \in 1 \oxdots n. \;
    \oxtypjudge{\oxglobalctx}{\oxemptyctx}{\oxsitype_1 \oxdotsc \oxsitype_{i-1}}{\oxvarctx}{
      \oxvalue_i
    }{\oxsitype_i}{\oxvarctx}$.

    Applying \Lemma{lemma:values-dont-change} to the typing derivation (from
    \oxname{T-Let}) for
    $\oxexpr_1$ (which we know is a value from \oxname{E-Let}) gives us
    $\oxsubtypectx{\oxglobalctx}{\oxemptyctx}{\oxvarctx}{\oxvarctx_1}$.
    Then, we can apply \Lemma{lemma:value-typing-related-envs} to each of the
    typing judgments to get $\forall i \in 1 \oxdots n. \;
    \oxtypjudge{\oxglobalctx}{\oxemptyctx}{\oxsitype_1 \oxdotsc \oxsitype_{i-1}}{\oxvarctx_1}{
      \oxvalue_i
    }{\oxsitype_i}{\oxvarctx_1}$.

    Applying \Lemma{lemma:values-well-typed-extension} to each of the value
    typing judgments and $\forall \oxcprov \in \oxfprovs{\oxsitype_a}. \;
    \oxnotreborrowed{\oxvarctx_1^\prime}{\oxcprov}$ (from the premise of
    \oxname{T-Let}) gives us
    $\forall i \in 1 \oxdots n. \;
    \oxtypjudge{\oxglobalctx}{\oxemptyctx}{\oxsitype_1 \oxdotsc \oxsitype_{i-1}}{
      \oxextendctx{\oxvarctx_1}{
        \oxvarctxentry{\oxid}{\oxsitype_a}
      }
    }{
      \oxvalue_i
    }{\oxsitype_i}{
      \oxextendctx{\oxvarctx_1}{
        \oxvarctxentry{\oxid}{\oxsitype_a}
      }
    }$.

    Then, we note that definitionally
    $\oxsubtypectx{\oxglobalctx}{\oxemptyctx}{
      \oxextendctx{\oxvarctx_1}{
        \oxvarctxentry{\oxid}{\oxsitype_a}
      }
    }{
      \oxgcloans{\oxkontctx}{
        \oxextendctx{\oxvarctx_1}{
          \oxvarctxentry{\oxid}{\oxsitype_a}
        }
      }
    }$ since we can only clear loans that are not present in any of the types.
    We can use this with each typing derivation in \Lemma{lemma:values-dont-change},
    and then finally apply \oxname{WF-Temporaries} to get
    $\oxwfkontctx{\oxglobalctx}{
      \oxgcloans{\oxkontctx}{
        \oxextendctx{\oxvarctx_1}{
          \oxvarctxentry{\oxid}{\oxsitype_a}
        }
      }
    }{\overline{\oxvalue}}{\oxkontctx}$.
  }

  \oxpresline{
    \oxtypjudge{\oxglobalctx}{\oxemptyctx}{\oxkontctx}{
      \oxgcloans{\oxkontctx}{
        \oxextendctx{\oxvarctx_1^\prime}{
          \oxvarctxentry{\oxid}{\oxsitype_a}
        }
      }
    }{\oxshift{\oxexpr}}{\oxsitype_2}{\oxvarctx_2}
  }{
    Immediate by applying \oxname{T-Shift} to the derivation
    \oxtypjudge{\oxglobalctx}{\oxemptyctx}{\oxkontctx}{
      \oxgcloans{\oxkontctx}{
        \oxextendctx{\oxvarctx_1^\prime}{
          \oxvarctxentry{\oxid}{\oxsitype_a}
        }
      }
    }{\oxexpr}{\oxsitype_2}{
      \oxextendctx{\oxvarctx_2}{
        \oxvarctxentry{\oxid}{\oxsdtype}
      }
    } (from premise of \oxname{T-Let}).
  }

  \oxpresline{
    \oxtunify[\oxcombine]{\oxemptyctx}{\oxkontctx}{\oxvarctx_2}
    {\oxsitype_2}{\oxsitype_2}{\oxvarctx_2}
  }{
    Immediate by \oxname{RR-Refl}.
  }

  \oxpresline{
    \exists \oxvarctx_o. \oxvarctx_2 \oxintersect \oxvarctx_o = \oxvarctx_2
  }{
    $\oxvarctx_o = \oxvarctx_2$
  }
\end{preservation}

\oxpreservationcase{T-LetRegion}{\TLetProv}{
  \oxletrgn{\oxcprov}{\oxexpr}
}{\ELetProv}{
  \oxvarctx^\prime
}

\begin{preservation}
  \oxpresline{
    \oxstorevalidity{\oxglobalctx}{\oxvarctx^\prime}{\oxstore}
  }{
    Applying \Lemma{lemma:values-dont-change} to the typing derivation (from
    \oxname{T-LetRegion}) for $\oxexpr$ (which we know is a value from
    \oxname{E-LetRegion}) gives us
    $\oxsubtypectx{\oxglobalctx}{\oxemptyctx}{\oxvarctx}{\oxvarctx^\prime}$. Then,
    applying \Lemma{lemma:stack-validity-related-envs} with this and
    $\oxstorevalidity{\oxglobalctx}{\oxvarctx}{\oxstore}$ (from
    premise) gives us
    $\oxstorevalidity{\oxglobalctx}{\oxvarctx^\prime}{\oxstore}$.
  }

  \oxpresline{
    \oxwfkontctx{\oxglobalctx}{\oxvarctx^\prime}{\overline{\oxvalue}}{\oxkontctx}
  }{
    Inverting \oxname{WF-Temporaries} gives us $\forall i \in 1 \oxdots n. \;
    \oxtypjudge{\oxglobalctx}{\oxemptyctx}{\oxsitype_1 \oxdotsc \oxsitype_{i-1}}{\oxvarctx}{
      \oxvalue_i
    }{\oxsitype_i}{\oxvarctx}$.

    Applying \Lemma{lemma:values-dont-change} to the typing derivation (from
    \oxname{T-LetRegion}) for
    $\oxexpr$ (which we know is a value from \oxname{E-LetRegion}) gives us
    $\oxsubtypectx{\oxglobalctx}{\oxemptyctx}{\oxvarctx}{\oxvarctx^\prime}$.

    Thus, we can apply \Lemma{lemma:value-typing-related-envs} to each of the
    typing judgments to get $\forall i \in 1 \oxdots n. \;
    \oxtypjudge{\oxglobalctx}{\oxemptyctx}{\oxsitype_1 \oxdotsc \oxsitype_{i-1}}{\oxvarctx^\prime}{
      \oxvalue_i
    }{\oxsitype_i}{\oxvarctx^\prime}$.

    Finally, applying \oxname{WF-Temporaries} gives us
    $\oxwfkontctx{\oxglobalctx}{\oxvarctx^\prime}{\overline{\oxvalue}}{\oxkontctx}$.
  }

  \oxpresline{
    \oxtypjudge{\oxglobalctx}{\oxemptyctx}{\oxkontctx}{\oxvarctx^\prime}{\oxvalue}
    {\oxsitype}{\oxvarctx^\prime}
  }{
    We know from \oxname{E-LetRegion} that $\oxexpr$ is a value $\oxvalue$. Thus,
    we can apply \Lemma{lemma:value-typing-output} to
    $\oxtypjudge{\oxglobalctx}{\oxemptyctx}{\oxkontctx}{\oxvarctx}{\oxvalue}{\oxsitype}{
      \oxextendctx{\oxvarctx^\prime}{
        \oxloanctxentry{\oxcprov}{\oxset{\overline{\oxloan}}}
      }
    }$ to get $\oxtypjudge{\oxglobalctx}{\oxemptyctx}{\oxkontctx}{
      \oxextendctx{\oxvarctx^\prime}{
        \oxloanctxentry{\oxcprov}{\oxset{\overline{\oxloan}}}
      }
    }{\oxvalue}{\oxsitype}{
      \oxextendctx{\oxvarctx^\prime}{
        \oxloanctxentry{\oxcprov}{\oxset{\overline{\oxloan}}}
      }
    }$.

    We now wish to show that
    $\oxtypjudge{\oxglobalctx}{\oxemptyctx}{\oxkontctx}{\oxvarctx^\prime}{\oxvalue}
    {\oxsitype}{\oxvarctx^\prime}$. By inspecting the grammar of values and
    their typing rules, we know that the only values who depend on the
    context are pointers and closure values. But by inversion on
    $\oxtypjudge{\oxglobalctx}{\oxemptyctx}{\oxkontctx}{\oxvarctx}{
      \oxletrgn{\oxcprov}{\oxvalue}
    }{\oxsitype}{\oxvarctx^\prime}$, we know that
    $\oxtypevalidity{\oxglobalctx}{\oxemptyctx}{\oxvarctx^\prime}{\oxsitype}$.
    Since the type is valid without the frame, we know that the values
    cannot depend on that frame. Thus, we can conclude
    $\oxtypjudge{\oxglobalctx}{\oxemptyctx}{\oxkontctx}{\oxvarctx^\prime}{\oxvalue}
    {\oxsitype}{\oxvarctx^\prime}$.
  }

  \oxpresline{
    \oxtunify[\oxcombine]{\oxemptyctx}{\oxkontctx}{\oxvarctx^\prime}
    {\oxsitype}{\oxsitype}{\oxvarctx^\prime}
  }{
    Immediate by \oxname{RR-Refl}.
  }

  \oxpresline{
    \exists \oxvarctx_o. \oxvarctx^\prime \oxintersect \oxvarctx_o = \oxvarctx^\prime
  }{
    $\oxvarctx_o = \oxvarctx^\prime$
  }
\end{preservation}

\oxpreservationcase{T-While}{\TWhile}{
  \oxwhile{\oxexpr_1}{\oxexpr_2}
}{\EWhile}{
  \oxvarctx
}

\begin{preservation}
  \oxpresline{
    \oxstorevalidity{\oxglobalctx}{\oxvarctx}{\oxstore}
  }{
    Immediate from our premise.
  }

  \oxpresline{
    \oxwfkontctx{\oxglobalctx}{\oxvarctx}{\overline{\oxvalue}}{\oxkontctx}
  }{
    Immediate from our premise.
  }

  \oxpresline{
    \oxtypjudge{\oxglobalctx}{\oxemptyctx}{\oxkontctx}{\oxvarctx}{
      \oxexpr^\prime
    }{\oxtunit}{\oxvarctx_2}
  }{
    We would like to build a derivation to show that the expression
    $\oxbranch{\oxexpr_1}{
      \oxseq{\oxexpr_2}{
        \oxwhile{\oxexpr_1}{\oxexpr_2}
      }
    }{\oxunit}$ is well-typed. We thus start by applying \oxname{T-Branch}.

    This requires us to show three things. First, $\oxtypjudge{\oxglobalctx}
    {\oxemptyctx}{\oxkontctx}{\oxvarctx}{\oxexpr_1}{\oxtbool}{\oxvarctx_1}$ which we have
    from the premise of \oxname{T-While}. Second, $\oxtypjudge{\oxglobalctx}
    {\oxemptyctx}{\oxkontctx}{\oxvarctx_1}{
      \oxseq{\oxexpr_2}{
        \oxwhile{\oxexpr_1}{\oxexpr_2}
      }
    }{\oxtunit}{\oxvarctx_2}$. We build this by applying \oxname{T-Seq} to
    $\oxtypjudge{\oxglobalctx}{\oxemptyctx}{\oxkontctx}{\oxvarctx_1}{\oxexpr_2}
    {\oxtunit}{\oxvarctx_2}$ and $\oxtypjudge{\oxglobalctx}{\oxemptyctx}{\oxkontctx}
    {\oxvarctx_2}{\oxwhile{\oxexpr_1}{\oxexpr_2}}{\oxvarctx_2}$. The former is
    directly in the premise of \oxname{T-While} and the latter can be built by
    applying \oxname{T-While} to
    $\oxtypjudge{\oxglobalctx}{\oxemptyctx}{\oxkontctx}{\oxvarctx_2}{
      \oxexpr_1
    }{\oxtbool}{\oxvarctx_2}$ and
    $\oxtypjudge{\oxglobalctx}{\oxemptyctx}{\oxkontctx}{\oxvarctx_2}{
      \oxexpr_2
    }{\oxtunit}{\oxvarctx_2}$, both from the premise of our original
    \oxname{T-While}. Finally, we need to show
    $\oxtypjudge{\oxglobalctx}{\oxemptyctx}{\oxkontctx}{\oxvarctx_2}{
      \oxunit
    }{\oxtunit}{\oxvarctx_2}$, which is immediate from \oxname{T-Unit}.
  }

  \oxpresline{
    \oxtunify[\oxcombine]{\oxemptyctx}{\oxkontctx}{\oxvarctx_2}
    {\oxtunit}{\oxtunit}{\oxvarctx_2}
  }{
    Immediate by \oxname{RR-Refl}.
  }

  \oxpresline{
    \exists \oxvarctx_o. \oxvarctx_2 \oxintersect \oxvarctx_o = \oxvarctx_2
  }{
    $\oxvarctx_o = \oxvarctx_2$
  }
\end{preservation}

\oxpreservationcasemulti{T-ForArray}{\TForArray}{
  \oxfor{\oxid}{\oxexpr_1}{\oxexpr_2}
}{
  \EForArray \and \EForEmptyArray
}

\oxpreservationsubcaseheading{E-ForArray}{
  \oxextendctx{\oxvarctx_1}{\oxvarctxentry{\oxid}{\oxsitype}}
}

\begin{preservation}
  \oxpresline{
    \oxstorevalidity{\oxglobalctx}{
      \oxextendctx{\oxvarctx_1}{\oxvarctxentry{\oxid}{\oxsitype}}
    }{
      \oxextendctx{\oxstore}{
        \oxstoreentry{\oxid}{\oxvalue_0}
      }
    }
  }{
    Applying \Lemma{lemma:values-dont-change} to the typing derivation (from
    \oxname{T-ForArray}) for $\oxexpr_1$ (which we know is a value from
    \oxname{E-ForArray}) gives us
    $\oxsubtypectx{\oxglobalctx}{\oxemptyctx}{\oxvarctx}{\oxvarctx_1}$. Then,
    we can apply \Lemma{lemma:stack-validity-related-envs} to get
    $\oxstorevalidity{\oxglobalctx}{\oxvarctx_1}{\oxstore}$. Applying
    \Lemma{lemma:value-typing-output} to the derivation
    $\oxtypjudge{\oxglobalctx}{\oxemptyctx}{\oxkontctx}{\oxvarctx}
    {\oxarr{\oxvalue_0 \oxdotsc \oxvalue_n}}{\oxtarr{\oxsitype}{\oxnum}}{\oxvarctx_1}$
    gives us $\oxtypjudge{\oxglobalctx}{\oxemptyctx}{\oxkontctx}{\oxvarctx_1}
    {\oxarr{\oxvalue_0 \oxdotsc \oxvalue_n}}{\oxtarr{\oxsitype}{\oxnum}}{\oxvarctx_1}$.
    Then, using inversion (of \oxname{T-Array}), we get
    $\oxtypjudge{\oxglobalctx}{\oxemptyctx}{\oxkontctx}{\oxvarctx_1}
    {\oxvalue_0}{\oxsitype}{\oxvarctx_1}$. Finally, applying
    \Lemma{lemma:store-validity-extension} to
    $\oxstorevalidity{\oxglobalctx}{\oxvarctx_1}{\oxstore}$ and
    $\oxtypjudge{\oxglobalctx}{\oxemptyctx}{\oxkontctx}{\oxvarctx_1}
    {\oxvalue_0}{\oxsitype}{\oxvarctx_1}$ gives us
    $\oxstorevalidity{\oxglobalctx}{
      \oxextendctx{\oxvarctx_1}{
        \oxvarctxentry{\oxid}{\oxsitype}
      }
    }{
      \oxextendctx{\oxstore}{\oxstoreentry{\oxid}{\oxvalue_0}}
    }$.
  }

  \oxpresline{
    \oxwfkontctx{\oxglobalctx}{
      \oxextendctx{\oxvarctx_1}{
        \oxvarctxentry{\oxid}{\oxsitype}
      }
    }{\overline{\oxvalue}}{\oxkontctx}
  }{
    Inverting \oxname{WF-Temporaries} gives us $\forall i \in 1 \oxdots n. \;
    \oxtypjudge{\oxglobalctx}{\oxemptyctx}{\oxsitype_1 \oxdotsc \oxsitype_{i-1}}{\oxvarctx}{
      \oxvalue_i
    }{\oxsitype_i}{\oxvarctx}$.

    Applying \Lemma{lemma:values-dont-change} to the typing derivation (from
    \oxname{T-ForArray}) for
    $\oxexpr_1$ (which we know is a value from \oxname{E-ForArray}) gives us
    $\oxsubtypectx{\oxglobalctx}{\oxemptyctx}{\oxvarctx}{\oxvarctx_1}$.

    Thus, we can apply \Lemma{lemma:value-typing-related-envs} to each of the
    typing judgments to get $\forall i \in 1 \oxdots n. \;
    \oxtypjudge{\oxglobalctx}{\oxemptyctx}{\oxemptyctx}{\oxvarctx_1}{
      \oxvalue_i
    }{\oxsitype_i}{\oxvarctx_1}$.

    Applying \Lemma{lemma:values-well-typed-extension} to each of the value
    typing judgments and $\forall \oxcprov \in \oxfprovs{\oxsitype}. \;
    \oxnotreborrowed{\oxvarctx_1}{\oxcprov}$ (from the premise of
    \oxname{T-ForArray}) gives us
    $\forall i \in 1 \oxdots n. \;
    \oxtypjudge{\oxglobalctx}{\oxemptyctx}{\oxemptyctx}{
      \oxextendctx{\oxvarctx_1}{
        \oxvarctxentry{\oxid}{\oxsitype}
      }
    }{
      \oxvalue_i
    }{\oxsitype_i}{
      \oxextendctx{\oxvarctx_1}{
        \oxvarctxentry{\oxid}{\oxsitype}
      }
    }$.
  }

  \oxpresline{
    \oxtypjudge{\oxglobalctx}{\oxemptyctx}{\oxkontctx}{
      \oxextendctx{\oxvarctx_1}{\oxvarctxentry{\oxid}{\oxsitype}}
    }{\oxexpr^\prime}{\oxtunit}{\oxvarctx_1}
  }{
    We need to build a derivation for the expression $\oxseq{\oxshift{\oxexpr}}{
      \oxfor{\oxid}{\oxarr{\oxvalue_1 \oxdotsc \oxvalue_n}}{\oxexpr}
    }$. The bottom of this derivation will be \oxname{T-Seq} which requires us
    to show that $\oxtypjudge{\oxglobalctx}{\oxemptyctx}{\oxkontctx}{
      \oxextendctx{\oxvarctx_1}{
        \oxvarctxentry{\oxid}{\oxsitype}
      }
    }{\oxshift{\oxexpr}}{\oxtunit}{\oxvarctx_1}$ and that
    $\oxtypjudge{\oxglobalctx}{\oxemptyctx}{\oxkontctx}{\oxvarctx_1}{
      \oxfor{\oxid}{\oxarr{\oxvalue_1 \oxdotsc \oxvalue_n}}{\oxexpr}
    }{\oxtunit}{\oxvarctx_1}$.

    To show $\oxtypjudge{\oxglobalctx}{\oxemptyctx}{\oxkontctx}{
      \oxextendctx{\oxvarctx_1}{
        \oxvarctxentry{\oxid}{\oxsitype}
      }
    }{\oxshift{\oxexpr}}{\oxtunit}{\oxvarctx_1}$, we apply \oxname{T-Shift} to
    $\oxtypjudge{\oxglobalctx}{\oxemptyctx}{\oxkontctx}{
      \oxextendctx{\oxvarctx_1}{
        \oxvarctxentry{\oxid}{\oxsitype}
      }
    }{\oxexpr}{\oxtunit}{
      \oxextendctx{\oxvarctx_1}{
        \oxvarctxentry{\oxid}{\oxsdtype}
      }
    }$ (from the premise of \oxname{T-ForArray}).

    To show $\oxtypjudge{\oxglobalctx}{\oxemptyctx}{\oxkontctx}{\oxvarctx_1}{
      \oxfor{\oxid}{\oxarr{\oxvalue_1 \oxdotsc \oxvalue_n}}{\oxexpr}
    }{\oxtunit}{\oxvarctx_1}$, we apply \Lemma{lemma:value-typing-output} to the
    derivation $\oxtypjudge{\oxglobalctx}{\oxemptyctx}{\oxkontctx}{\oxvarctx}
    {\oxarr{\oxvalue_0 \oxdotsc
        \oxvalue_n}}{\oxtarr{\oxsitype}{\oxnum}}{\oxvarctx_1}$ to get
    $\oxtypjudge{\oxglobalctx}{\oxemptyctx}{\oxkontctx}{\oxvarctx_1} {\oxarr{\oxvalue_0
        \oxdotsc \oxvalue_n}}{\oxtarr{\oxsitype}{\oxnum}}{\oxvarctx_1}$. Then,
    we rewrite the derivation (inverting and then reapply \oxname{T-Array}) to
    exclude $\oxvalue_0$ giving us
    $\oxtypjudge{\oxglobalctx}{\oxemptyctx}{\oxkontctx}{\oxvarctx_1} {\oxarr{\oxvalue_1
        \oxdotsc \oxvalue_n}}{\oxtarr{\oxsitype}{\oxnum - 1}}{\oxvarctx_1}$.
    Finally, we apply \oxname{T-ForArray} using this combined with
    $\oxtypjudge{\oxglobalctx}{\oxemptyctx}{\oxkontctx}{\oxvarctx_1}
    {\oxexpr}{\oxtunit}{\oxvarctx_1}$ (from the premise of \oxname{T-ForArray}).
  }

  \oxpresline{
    \oxtunify[\oxcombine]{\oxemptyctx}{\oxkontctx}{\oxvarctx_1}
    {\oxtunit}{\oxtunit}{\oxvarctx_1}
  }{
    Immediate by \oxname{RR-Refl}.
  }

  \oxpresline{
    \exists \oxvarctx_o. \oxvarctx_1 \oxintersect \oxvarctx_o = \oxvarctx_1
  }{
    $\oxvarctx_o = \oxvarctx_1$
  }
\end{preservation}

\oxpreservationsubcaseheading{E-ForEmptyArray}{
  \oxvarctx
}

\begin{preservation}
  \oxpresline{
    \oxstorevalidity{\oxglobalctx}{\oxvarctx}{\oxstore}
  }{
    Immediate from our premise.
  }

  \oxpresline{
    \oxwfkontctx{\oxglobalctx}{\oxvarctx}{\overline{\oxvalue}}{\oxkontctx}
  }{
    Immediate from our premise.
  }

  \oxpresline{
    \oxtypjudge{\oxglobalctx}{\oxemptyctx}{\oxkontctx}{\oxvarctx}
    {\oxunit}{\oxtunit}{\oxvarctx}
  }{
    Immediate by \oxname{T-Unit}.
  }

  \oxpresline{
    \oxtunify[\oxcombine]{\oxemptyctx}{\oxkontctx}{\oxvarctx}
    {\oxtunit}{\oxtunit}{\oxvarctx}
  }{
    Immediate by \oxname{RR-Refl}.
  }

  \oxpresline{
    \exists \oxvarctx_o. \oxvarctx \oxintersect \oxvarctx_o = \oxvarctx
  }{
    $\oxvarctx_o = \oxvarctx$
  }
\end{preservation}

\oxpreservationcasemulti{T-ForSlice}{\TForSlice}{
  \oxfor{\oxid}{\oxexpr_1}{\oxexpr_2}
}{
  \EForSlice \and \EForEmptySlice
}

\oxpreservationsubcaseheading{E-ForSlice}{
  \oxextendctx{\oxvarctx_1}{
    \oxvarctxentry{\oxid}{
      \oxtref{\oxcprov}{\oxmuta}{\oxsitype}
    }
  }
}

\begin{preservation}
  \oxpresline{
    \oxstorevalidity{\oxglobalctx}{
      \oxextendctx{\oxvarctx_1}{
        \oxvarctxentry{\oxid}{
          \oxtref{\oxcprov}{\oxmuta}{\oxsitype}
        }
      }
    }{
      \oxextendctx{\oxstore}{
        \oxstoreentry{\oxid}{
          \oxindexptr{\oxreferent}{\oxnum_1}
        }
      }
    }
  }{
    Applying \Lemma{lemma:values-dont-change} to the typing derivation (from
    \oxname{T-ForSlice}) for $\oxexpr_1$ (which we know is a value from
    \oxname{E-ForSlice}) gives us
    $\oxsubtypectx{\oxglobalctx}{\oxemptyctx}{\oxvarctx}{\oxvarctx_1}$. Then,
    we can apply \Lemma{lemma:stack-validity-related-envs} to get
    $\oxstorevalidity{\oxglobalctx}{\oxvarctx_1}{\oxstore}$. Applying
    \Lemma{lemma:value-typing-output} to the derivation
    $\oxtypjudge{\oxglobalctx}{\oxemptyctx}{\oxkontctx}{\oxvarctx}
    {\oxsliceptr{\oxreferent}{i}{j}}{
      \oxtref{\oxcprov}{\oxmuta}{\oxtslice{\oxsitype}}
    }{\oxvarctx_1}$
    gives us $\oxtypjudge{\oxglobalctx}{\oxemptyctx}{\oxkontctx}{\oxvarctx_1}
    {\oxsliceptr{\oxreferent}{i}{j}}{
      \oxtref{\oxcprov}{\oxmuta}{\oxtslice{\oxsitype}}
    }{\oxvarctx_1}$.
    Then, using inversion (on \oxname{T-Pointer}), we get
    $\oxreferentvalidity{\oxglobalctx}{\oxvarctx_1}{
      \oxslice{\oxreferent}{i}{j}
    }{\oxxitype}$ (where $\oxxitype = \oxtarr{\oxsitype}{\oxnum}$ or
    $\oxtslice{\oxsitype}$) and
    $\oxloanpkg{\oxmuta}{\oxplace} \in \oxvarctx_1(\oxcprov)$ (where
    $\oxreferent = \oxreferentctx[\oxplace]$). We can invert
    \oxname{WF-RefSliceArray} or \oxname{WF-RefSliceSlice} (based on
    $\oxxitype$) for $\oxreferentvalidity{\oxglobalctx}{\oxvarctx_1}{
      \oxslice{\oxreferent}{i}{j}
    }{\oxxitype}$ and then apply \oxname{WF-RefIndexArray} or
    \oxname{WF-RefIndexSlice} appropriately to get
    $\oxreferentvalidity{\oxglobalctx}{\oxvarctx_1}{
      \oxindex{\oxreferent}{i}
    }{\oxxitype}$. We can then use
    \oxname{T-Pointer} to get
    $\oxtypjudge{\oxglobalctx}{\oxemptyctx}{\oxkontctx}{\oxvarctx_1}
    {\oxindexptr{\oxreferent}{\oxnum_1}}{
      \oxtref{\oxcprov}{\oxmuta}{\oxsitype}
    }{\oxvarctx_1}$. Finally, applying
    \Lemma{lemma:store-validity-extension} to
    $\oxstorevalidity{\oxglobalctx}{\oxvarctx_1}{\oxstore}$ and
    $\oxtypjudge{\oxglobalctx}{\oxemptyctx}{\oxkontctx}{\oxvarctx_1}
    {\oxindexptr{\oxreferent}{\oxnum_1}}{
      \oxtref{\oxcprov}{\oxmuta}{\oxsitype}
    }{\oxvarctx_1}$ gives us
    $\oxstorevalidity{\oxglobalctx}{
      \oxextendctx{\oxvarctx_1}{
        \oxvarctxentry{\oxid}{\oxtref{\oxcprov}{\oxmuta}{\oxsitype}}
      }
    }{
      \oxextendctx{\oxstore}{
        \oxstoreentry{\oxid}{\oxindexptr{\oxreferent}{\oxnum_1}}
      }
    }$.
  }

  \oxpresline{
    \oxwfkontctx{\oxglobalctx}{
      \oxextendctx{\oxvarctx_1}{
        \oxvarctxentry{\oxid}{
          \oxtref{\oxcprov}{\oxmuta}{\oxsitype}
        }
      }
    }{\overline{\oxvalue}}{\oxkontctx}
  }{
    Inverting \oxname{WF-Temporaries} gives us $\forall i \in 1 \oxdots n. \;
    \oxtypjudge{\oxglobalctx}{\oxemptyctx}{\oxsitype_1 \oxdotsc \oxsitype_{i-1}}{\oxvarctx}{
      \oxvalue_i
    }{\oxsitype_i}{\oxvarctx}$.

    Applying \Lemma{lemma:values-dont-change} to the typing derivation (from
    \oxname{T-ForSlice}) for
    $\oxexpr_1$ (which we know is a value from \oxname{E-ForSlice}) gives us
    $\oxsubtypectx{\oxglobalctx}{\oxemptyctx}{\oxvarctx}{\oxvarctx_1}$.

    Thus, we can apply \Lemma{lemma:value-typing-related-envs} to each of the
    typing judgments to get $\forall i \in 1 \oxdots n. \;
    \oxtypjudge{\oxglobalctx}{\oxemptyctx}{\oxemptyctx}{\oxvarctx_1}{
      \oxvalue_i
    }{\oxsitype_i}{\oxvarctx_1}$.

    Applying \Lemma{lemma:values-well-typed-extension} to each of the value
    typing judgments and $\forall \oxcprov \in
    \oxfprovs{\oxtref{\oxprov}{\oxmuta}{\oxsitype}}. \;
    \oxnotreborrowed{\oxvarctx_1}{\oxcprov}$ (from the premise of
    \oxname{T-ForSlice}) gives us
    $\forall i \in 1 \oxdots n. \;
    \oxtypjudge{\oxglobalctx}{\oxemptyctx}{\oxemptyctx}{
      \oxextendctx{\oxvarctx_1}{
        \oxvarctxentry{\oxid}{
          \oxtref{\oxcprov}{\oxmuta}{\oxsitype}
        }
      }
    }{
      \oxvalue_i
    }{\oxsitype_i}{
      \oxextendctx{\oxvarctx_1}{
        \oxvarctxentry{\oxid}{
          \oxtref{\oxcprov}{\oxmuta}{\oxsitype}
        }
      }
    }$.
  }

  \oxpresline{
    \oxtypjudge{\oxglobalctx}{\oxemptyctx}{\oxkontctx}{
      \oxextendctx{\oxvarctx_1}{
        \oxvarctxentry{\oxid}{
          \oxtref{\oxcprov}{\oxmuta}{\oxsitype}
        }
      }
    }{\oxexpr^\prime}{\oxtunit}{\oxvarctx_1}
  }{
    We need to build a derivation for the expression $\oxseq{\oxshift{\oxexpr}}{
      \oxfor{\oxid}{
        \oxsliceptr{\oxreferent}{\oxnum_1^\prime}{\oxnum_2}
      }{\oxexpr}
    }$. The bottom of this derivation will be \oxname{T-Seq} which requires us
    to show that $\oxtypjudge{\oxglobalctx}{\oxemptyctx}{\oxkontctx}{
      \oxextendctx{\oxvarctx_1}{
        \oxvarctxentry{\oxid}{\oxsitype}
      }
    }{\oxshift{\oxexpr}}{\oxtunit}{\oxvarctx_1}$ and that
    $\oxtypjudge{\oxglobalctx}{\oxemptyctx}{\oxkontctx}{\oxvarctx_1}{
      \oxfor{\oxid}{
        \oxsliceptr{\oxreferent}{\oxnum_1^\prime}{\oxnum_2}
      }{\oxexpr}
    }{\oxtunit}{\oxvarctx_1}$.

    To show $\oxtypjudge{\oxglobalctx}{\oxemptyctx}{\oxkontctx}{
      \oxextendctx{\oxvarctx_1}{
        \oxvarctxentry{\oxid}{\oxsitype}
      }
    }{\oxshift{\oxexpr}}{\oxtunit}{\oxvarctx_1}$, we apply \oxname{T-Shift} to
    $\oxtypjudge{\oxglobalctx}{\oxemptyctx}{\oxkontctx}{
      \oxextendctx{\oxvarctx_1}{
        \oxvarctxentry{\oxid}{\oxsitype}
      }
    }{\oxexpr}{\oxtunit}{
      \oxextendctx{\oxvarctx_1}{
        \oxvarctxentry{\oxid}{\oxsdtype}
      }
    }$ (from the premise of \oxname{T-ForSlice}).

    To show $\oxtypjudge{\oxglobalctx}{\oxemptyctx}{\oxkontctx}{\oxvarctx_1}{
      \oxfor{\oxid}{\oxsliceptr{\oxreferent}{\oxnum_1^\prime}{\oxnum_2}}{\oxexpr}
    }{\oxtunit}{\oxvarctx_1}$, we apply
    \Lemma{lemma:value-typing-output} to the derivation
    $\oxtypjudge{\oxglobalctx}{\oxemptyctx}{\oxkontctx}{\oxvarctx}
    {\oxsliceptr{\oxreferent}{\oxnum_1}{\oxnum_2}}{
      \oxtref{\oxcprov}{\oxmuta}{\oxtslice{\oxsitype}}
    }{\oxvarctx_1}$
    to get $\oxtypjudge{\oxglobalctx}{\oxemptyctx}{\oxkontctx}{\oxvarctx_1}
    {\oxsliceptr{\oxreferent}{\oxnum_1}{\oxnum_2}}{
      \oxtref{\oxcprov}{\oxmuta}{\oxtslice{\oxsitype}}
    }{\oxvarctx_1}$. Then, we rewrite the derivation (inverting and then reapply
    \oxname{T-Pointer}) to increment $\oxnum_1$ to $\oxnum_1^\prime$ giving us
    $\oxtypjudge{\oxglobalctx}{\oxemptyctx}{\oxkontctx}{\oxvarctx_1}
    {\oxsliceptr{\oxreferent}{\oxnum_1^\prime}{\oxnum_2}}{
      \oxtref{\oxcprov}{\oxmuta}{\oxtslice{\oxsitype}}
    }{\oxvarctx_1}$.
    Finally, we apply \oxname{T-ForSlice} using this combined with
    $\oxtypjudge{\oxglobalctx}{\oxemptyctx}{\oxkontctx}{\oxvarctx_1}
    {\oxexpr}{\oxtunit}{\oxvarctx_1}$ (from the premise of \oxname{T-ForSlice}).
  }

  \oxpresline{
    \oxtunify[\oxcombine]{\oxemptyctx}{\oxkontctx}{\oxvarctx_1}
    {\oxtunit}{\oxtunit}{\oxvarctx_1}
  }{
    Immediate by \oxname{RR-Refl}.
  }

  \oxpresline{
    \exists \oxvarctx_o. \oxvarctx_1 \oxintersect \oxvarctx_o = \oxvarctx_1
  }{
    $\oxvarctx_o = \oxvarctx_1$
  }
\end{preservation}

\oxpreservationsubcaseheading{E-ForEmptySlice}{
  \oxvarctx
}

\begin{preservation}
  \oxpresline{
    \oxstorevalidity{\oxglobalctx}{\oxvarctx}{\oxstore}
  }{
    Immediate from our premise.
  }

  \oxpresline{
    \oxwfkontctx{\oxglobalctx}{\oxvarctx}{\overline{\oxvalue}}{\oxkontctx}
  }{
    Immediate from our premise.
  }

  \oxpresline{
    \oxtypjudge{\oxglobalctx}{\oxemptyctx}{\oxkontctx}{\oxvarctx}
    {\oxunit}{\oxtunit}{\oxvarctx}
  }{
    Immediate by \oxname{T-Unit}.
  }

  \oxpresline{
    \oxtunify[\oxcombine]{\oxemptyctx}{\oxkontctx}{\oxvarctx}
    {\oxtunit}{\oxtunit}{\oxvarctx}
  }{
    Immediate by \oxname{RR-Refl}.
  }

  \oxpresline{
    \exists \oxvarctx_o. \oxvarctx \oxintersect \oxvarctx_o = \oxvarctx
  }{
    $\oxvarctx_o = \oxvarctx$
  }
\end{preservation}

\oxpreservationcase[
]{T-AppFunction}{\TApp}{
  \oxapp{\oxexpr_f}{
    \overline{\oxenv} \oxcomma
    \overline{\oxprov^\prime} \oxcomma
    \overline{\oxsitype}
  }{
    \oxexpr_1 \oxdotsc \oxexpr_n
  }
}{\EAppFun}{
  \oxnewframe{\oxvarctx_b}{
    \oxvarctxentry{\oxid_1}{\delta(\oxsitype_1)} \oxdotsc
    \oxvarctxentry{\oxid_n}{\delta(\oxsitype_n)}
  }
}

\begin{preservation}
  \oxpresline{
    \oxstorevalidity{\oxglobalctx}{
      \oxvarctx_i
    }{
      \oxstore^\prime
    }
  }{
    Applying \Lemma{lemma:values-dont-change} to the derivation for $\oxvalue_f$
    gives us $\oxsubtypectx{\oxglobalctx}{\oxemptyctx}{\oxvarctx}{\oxvarctx_0}$.
    Then, applying \Lemma{lemma:values-dont-change} to the derivations for
    every $\oxvalue_i$ gives us $\forall i \in \oxset{1 \oxdots n}. \;
    \oxsubtypectx{\oxglobalctx}{\oxemptyctx}{\oxvarctx_{i-1}}{\oxvarctx_i}$.

    Inversion on $\oxrunifymany{\oxemptyctx}{\oxkontctx}{\oxvarctx_n}{
      \oxsubstmany{\oxabsprov_2}{\oxprov}{\oxabsprov}
    }{
      \oxsubstmany{\oxabsprov_1}{\oxprov}{\oxabsprov}
    }{\oxvarctx_b}$ gives us a sequence of outlives relations with
    intermediate contexts. Applying  \Lemma{lemma:outlives-related} to each of
    them and then combining the result by transitivity gives us
    $\oxsubtypectx{\oxglobalctx}{\oxemptyctx}{\oxvarctx_n}{\oxvarctx_b}$.
    Combining both by transitivity, we have
    $\oxsubtypectx{\oxglobalctx}{\oxemptyctx}{\oxvarctx}{\oxvarctx_b}$.

    We can then apply \Lemma{lemma:stack-validity-related-envs} with
    $\oxstorevalidity{\oxglobalctx}{\oxvarctx}{\oxstore}$ to get
    $\oxstorevalidity{\oxglobalctx}{\oxvarctx_b}{\oxstore}$.

    We can apply \oxname{WF-StackFrame} to get
    $\oxstorevalidity{\oxglobalctx}{
      \oxnewframe{\oxvarctx_b}{\oxemptyctx}
    }{
      \oxnewframe{\oxstore}{\oxemptyctx}
    }$. Then, we repeatedly apply \Lemma{lemma:store-validity-extension} to
    the derivations for the arguments ($\oxvalue_1 \oxdots \oxvalue_n$) to get
    $\oxstorevalidity{\oxglobalctx}{
      \oxnewframe{\oxvarctx}{
        \oxvarctxentry{\oxid_1}{\delta(\oxsitype_1)} \oxdotsc
        \oxvarctxentry{\oxid_n}{\delta(\oxsitype_n)}
      }
    }{
      \oxnewframe{\oxstore}{
        \oxstoreentry{\oxid_1}{\oxvalue_1} \oxdotsc
        \oxstoreentry{\oxid_n}{\oxvalue_n}
      }
    }$.
  }

  \oxpresline{
    \oxwfkontctx{\oxglobalctx}{\oxvarctx_i}{\overline{\oxvalue}}{\oxkontctx}
  }{
    Inverting \oxname{WF-Temporaries} gives us $\forall i \in 1 \oxdots n. \;
    \oxtypjudge{\oxglobalctx}{\oxemptyctx}{\oxsitype_1 \oxdotsc \oxsitype_{i-1}}{\oxvarctx}{
      \oxvalue_i
    }{\oxsitype_i}{\oxvarctx}$.

    Applying \Lemma{lemma:values-dont-change} to the derivation for $\oxvalue_f$
    gives us $\oxsubtypectx{\oxglobalctx}{\oxemptyctx}{\oxvarctx}{\oxvarctx_0}$.
    Then, applying \Lemma{lemma:values-dont-change} to the derivations for
    every $\oxvalue_i$ gives us $\forall i \in \oxset{1 \oxdots n}. \;
    \oxsubtypectx{\oxglobalctx}{\oxemptyctx}{\oxvarctx_{i-1}}{\oxvarctx_i}$.

    Inversion on $\oxrunifymany{\oxemptyctx}{\oxkontctx}{\oxvarctx_n}{
      \oxsubstmany{\oxabsprov_2}{\oxprov}{\oxabsprov}
    }{
      \oxsubstmany{\oxabsprov_1}{\oxprov}{\oxabsprov}
    }{\oxvarctx_b}$ gives us a sequence of outlives relations with
    intermediate contexts. Applying  \Lemma{lemma:outlives-related} to each of
    them and then combining the result by transitivity gives us
    $\oxsubtypectx{\oxglobalctx}{\oxemptyctx}{\oxvarctx_n}{\oxvarctx_b}$.
    Combining both by transitivity, we have
    $\oxsubtypectx{\oxglobalctx}{\oxemptyctx}{\oxvarctx}{\oxvarctx_b}$.

    Thus, we can apply \Lemma{lemma:value-typing-related-envs} to each of the
    typing judgments to get $\forall i \in 1 \oxdots n. \;
    \oxtypjudge{\oxglobalctx}{\oxemptyctx}{\oxemptyctx}{\oxvarctx_b}{
      \oxvalue_i
    }{\oxsitype_i}{\oxvarctx_b}$.

    We also immediately have that adding a new empty frame leaves all the typing
    derivations good since it doesn't actually bind any new names on its own.
    Thus, we have $\forall i \in 1 \oxdots n. \;
    \oxtypjudge{\oxglobalctx}{\oxemptyctx}{\oxemptyctx}{
      \oxnewframe{\oxvarctx_b}{\oxemptyctx}
    }{
      \oxvalue_i
    }{\oxsitype_i}{
      \oxnewframe{\oxvarctx_b}{\oxemptyctx}
    }$.

    Repeatedly applying \Lemma{lemma:values-well-typed-extension} to each of the value
    typing judgments gives us
    $\forall i \in 1 \oxdots n. \;
    \oxtypjudge{\oxglobalctx}{\oxemptyctx}{\oxemptyctx}{
      \oxnewframe{\oxvarctx_b}{
        \oxvarctxentry{\oxid_1}{\delta(\oxsitype_1)} \oxdotsc
        \oxvarctxentry{\oxid_n}{\delta(\oxsitype_n)}
      }
    }{
      \oxvalue_i
    }{\oxsitype_i}{
      \oxnewframe{\oxvarctx_b}{
        \oxvarctxentry{\oxid_1}{\delta(\oxsitype_1)} \oxdotsc
        \oxvarctxentry{\oxid_n}{\delta(\oxsitype_n)}
      }
    }$.
  }

  \oxpresline{
    \oxtypjudge{\oxglobalctx}{\oxemptyctx}{\oxkontctx}{\oxvarctx_i}{
      \oxexpr^\prime
    }{\delta(\oxsitype_f)}{\oxvarctx_b}
  }{
    Applying \Lemma{lemma:function-defns-self-contained} with
    $\oxctxswellformed{\oxglobalctx}{\oxemptyctx}{\oxvarctx_b}{\oxkontctx}$ and
    $\oxlookup{\oxglobalctx}{\oxfnname}{
      \oxfuncdef{\oxfnname}{
        \overline{\oxenvvar} \oxcomma \overline{\oxabsprov} \oxcomma \overline{\oxtvar}
      }{
        \oxascribe{\oxid_1}{\oxsitype_1} \oxdotsc
        \oxascribe{\oxid_n}{\oxsitype_n}
      }{\oxsitype_r}{\overline{\oxabsprov_1: \oxabsprov_2}}{\oxexpr}
    }$ (from the premise of on \oxname{E-AppFunction}) gives us
    $\oxtypjudge{\oxglobalctx}{
      \overline{\oxtvarctxentry{\oxenvvar}{\oxkenv}} \oxcomma
      \overline{\oxtvarctxentry{\oxabsprov}{\oxkprov}} \oxcomma
      \overline{\oxabsprov_1 \oxoutlives \oxabsprov_2} \oxcomma
      \overline{\oxtvarctxentry{\oxtvar}{\oxktype}}
    }{\oxkontctx}{
      \oxnewframe{\oxvarctx_b}{
        \oxvarctxentry{\oxid_1}{\oxsitype_1} \oxdotsc
        \oxvarctxentry{\oxid_n}{\oxsitype_n}
      }
    }{
      \oxframed{\oxexpr}
    }{
      \oxsitype_f
    }{\oxvarctx_b}$.

    We then repeatedly apply \Lemma{lemma:substitution} for all of our type,
    region, and environment variables to get
    $\oxtypjudge{\oxglobalctx}{\oxemptyctx}{\oxkontctx}{
      \oxnewframe{\oxvarctx_b}{
        \oxvarctxentry{\oxid_1}{\delta(\oxsitype_1)} \oxdotsc
        \oxvarctxentry{\oxid_n}{\delta(\oxsitype_n)}
      }
    }{
      \oxframed{\oxexpr}
    }{\delta(\oxsitype_f)}{\oxvarctx_b}$.
  }

  \oxpresline{
    \oxtunify[\oxcombine]{\oxemptyctx}{\oxkontctx}{\oxvarctx_b}{
      \oxtype^\prime
    }{
      \oxtype^\prime
    }{\oxvarctx_b}
  }{
    Immediate by \oxname{RR-Refl}.
  }

  \oxpresline{
    \exists \oxvarctx_o. \oxvarctx_b \oxintersect \oxvarctx_o = \oxvarctx_b
  }{
    $\oxvarctx_o = \oxvarctx_b$
  }
\end{preservation}

\oxpreservationcase{T-AppClosure}{\TAppClosure}{
  \oxapply{\oxexpr_f}{
    \oxexpr_1 \oxdotsc \oxexpr_n
  }
}{\EApp}{
  \oxnewframe{\oxvarctx^\prime_n}{
    \oxextendctx{\oxframe_c}{
      \oxvarctxentry{\oxid_1}{\oxsitype_1} \oxdotsc
      \oxvarctxentry{\oxid_n}{\oxsitype_n}
    }
  }
}

\begin{preservation}
  \oxpresline{
    \oxstorevalidity{\oxglobalctx}{
      \oxvarctx_i
    }{
      \oxstore^\prime
    }
  }{
    Applying \Lemma{lemma:values-dont-change} to the derivation for $\oxvalue_f$
    gives us $\oxsubtypectx{\oxglobalctx}{\oxemptyctx}{\oxvarctx}{\oxvarctx_0}$.
    We can apply \Lemma{lemma:stack-validity-related-envs} with
    $\oxstorevalidity{\oxglobalctx}{\oxvarctx}{\oxstore}$ to get
    $\oxstorevalidity{\oxglobalctx}{\oxvarctx_0}{\oxstore}$.

    Then, for each pair of value typing judgment and region rewriting, we follow
    the same pattern we will describe in terms of an arbitrary index $i$. First,
    we apply \Lemma{lemma:values-dont-change} to the derivation for $\oxvalue_i$
    to get $\oxsubtypectx{\oxglobalctx}{\oxemptyctx}
    {\oxvarctx_{i-1}}{\oxvarctx_i}$. Then, applying
    \Lemma{lemma:stack-validity-related-envs} gives us
    $\oxstorevalidity{\oxglobalctx}{\oxvarctx_i}{\oxstore}$. Then, we can apply
    \Lemma{lemma:stack-validity-region-rewriting} to
    $\oxtunify[\oxcombineunrest]{\oxemptyctx}{\oxkontctx}{\oxvarctx_i}
    {\oxsitype_{i^\prime}}{\oxsitype_i}{\oxvarctx_i^\prime}$ to get
    $\oxstorevalidity{\oxglobalctx}{\oxvarctx_i^\prime}{\oxstore}$. Going through
    this for all the values in sequence gives us
    $\oxstorevalidity{\oxglobalctx}{\oxvarctx_n^\prime}{\oxstore}$

    Then, inversion of \oxname{T-ClosureValue} for the typing derivation for
    $\oxvalue_f$ gives us
    $\oxsubstorevalidity{\oxglobalctx}{\oxvarctx}{\oxstackframe_c}{\oxframe_c}$.
    We can then invert \oxname{WF-Frame} here to get
    $\oxdomain{\oxstackframe} = \oxdomain{\oxframe_c}|_\oxid$
    $\forall \oxid \in \oxdomain{\oxstackframe}. \;
    \oxtypjudge{\oxglobalctx}{\oxemptyctx}{\oxkontctx}{\oxnewframe{\oxvarctx}{\oxframe_c}}
    {\oxstackframe(\oxid)}{\oxframe_c(\oxid)}{\oxnewframe{\oxvarctx}{\oxframe_c}}$
    which we can then use with $\oxstorevalidity{\oxglobalctx}{\oxvarctx_n}{\oxstore}$
    in \oxname{WF-StackFrame} to get
    $\oxstorevalidity{\oxglobalctx}{
      \oxnewframe{\oxvarctx_n^{\prime}}{\oxframe_c}
    }{
      \oxnewframe{\oxstore}{\oxstackframe_c}
    }$.

    Finally, we repeatedly apply \Lemma{lemma:store-validity-extension} to
    the derivations for the arguments ($\oxvalue_1 \oxdots \oxvalue_n$) to get
    $\oxstorevalidity{\oxglobalctx}{
      \oxvarctx_i
    }{
      \oxstore^\prime
    }$.
  }

  \oxpresline{
    \oxwfkontctx{\oxglobalctx}{\oxvarctx_i}{\overline{\oxvalue^\prime}}{\oxkontctx}
  }{
    Inverting \oxname{WF-Temporaries} gives us $\forall j \in 1 \oxdots n. \;
    \oxtypjudge{\oxglobalctx}{\oxemptyctx}{\oxsitype_1 \oxdotsc \oxsitype_{i-1}}{\oxvarctx}{
      \oxvalue^\prime_j
    }{\oxsitype_j}{\oxvarctx}$.

    Applying \Lemma{lemma:values-dont-change} to the derivation for $\oxvalue_f$
    gives us $\oxsubtypectx{\oxglobalctx}{\oxemptyctx}{\oxvarctx}{\oxvarctx_0}$.
    For each of $\forall j \in 1 \oxdots n. \;
    \oxtypjudge{\oxglobalctx}{\oxemptyctx}{\oxemptyctx}{\oxvarctx}{
      \oxvalue^\prime_j
    }{\oxsitype_i}{\oxvarctx}$, we apply \Lemma{lemma:value-typing-related-envs}
    to get $\forall j \in 1 \oxdots n. \;
    \oxtypjudge{\oxglobalctx}{\oxemptyctx}{\oxemptyctx}{\oxvarctx_0}{
      \oxvalue^\prime_j
    }{\oxsitype_j}{\oxvarctx_0}$

    From here, we have an alternating pattern of lemmas to apply to deal with
    the interleaved value typing and then region rewriting in the premise of
    \oxname{T-AppClosure}. For each value $\oxvalue_i$ in \oxname{T-AppClosure},
    we'll first apply \Lemma{lemma:values-dont-change} to get
    $\oxsubtypectx{\oxglobalctx}{\oxemptyctx}{\oxvarctx_{i-1}}{\oxvarctx_i}$.
    Then, we'll apply \Lemma{lemma:value-typing-related-envs} to
    $\forall j \in 1 \oxdots n. \;
    \oxtypjudge{\oxglobalctx}{\oxemptyctx}{\oxemptyctx}{\oxvarctx_{i-1}}{
      \oxvalue^\prime_j
    }{\oxsitype_j}{\oxvarctx_{i-1}}$ to get $\forall j \in 1 \oxdots n. \;
    \oxtypjudge{\oxglobalctx}{\oxemptyctx}{\oxemptyctx}{\oxvarctx_i}{
      \oxvalue^\prime_j
    }{\oxsitype_j}{\oxvarctx_i}$. Then, we apply
    \Lemma{lemma:subtyping-value-typing} to each of these along with the
    rewriting $\oxtunify[\oxcombineunrest]{\oxemptyctx}{\oxkontctx}{\oxvarctx_i}
    {\oxsitype_{i^\prime}}{\oxsitype_i}{\oxvarctx_i^\prime}$ to get
    $\forall j \in 1 \oxdots n. \;
    \oxtypjudge{\oxglobalctx}{\oxemptyctx}{\oxemptyctx}{\oxvarctx_i^\prime}{
      \oxvalue^\prime_j
    }{\oxsitype_j}{\oxvarctx_i^\prime}$. After going through this for each value
    $\oxvalue_i$, we have $\forall j \in 1 \oxdots n. \;
    \oxtypjudge{\oxglobalctx}{\oxemptyctx}{\oxemptyctx}{\oxvarctx_n^\prime}{
      \oxvalue^\prime_j
    }{\oxsitype_j}{\oxvarctx_n^\prime}$.

    We also immediately have that adding a new empty frame leaves all the typing
    derivations good since it doesn't actually bind any new names on its own.
    Thus, we have $\forall i \in 1 \oxdots n. \;
    \oxtypjudge{\oxglobalctx}{\oxemptyctx}{\oxemptyctx}{
      \oxnewframe{\oxvarctx_n^\prime}{\oxemptyctx}
    }{
      \oxvalue_i
    }{\oxsitype_i}{
      \oxnewframe{\oxvarctx_n^\prime}{\oxemptyctx}
    }$.

    Repeatedly applying \Lemma{lemma:values-well-typed-extension} to each of the
    value typing judgments and each of $\forall i \in \oxset{ 1 \oxdots n }. \;
    \forall \oxcprov \in \oxfprovs{\oxsitype_i}. \;
    \oxnotreborrowed{\oxvarctx_1}{\oxcprov}$ (from the premise of
    \oxname{T-App}) gives us
    $\forall i \in 1 \oxdots n. \;
    \oxtypjudge{\oxglobalctx}{\oxemptyctx}{\oxemptyctx}{
      \oxnewframe{\oxvarctx_n^\prime}{
        \oxextendctx{\oxframe_c}{
          \oxvarctxentry{\oxid_1}{\oxsitype_1} \oxdotsc
          \oxvarctxentry{\oxid_n}{\oxsitype_n}
        }
      }
    }{
      \oxvalue_i
    }{\oxsitype_i}{
      \oxnewframe{\oxvarctx_n^\prime}{
        \oxextendctx{\oxframe_c}{
          \oxvarctxentry{\oxid_1}{\oxsitype_1} \oxdotsc
          \oxvarctxentry{\oxid_n}{\oxsitype_n}
        }
      }
    }$.
  }
\end{preservation} 
\begin{preservation}
  \oxpresline{
    \oxtypjudge{\oxglobalctx}{\oxemptyctx}{\oxkontctx}{\oxvarctx_i}{
      \oxexpr^\prime
    }{\oxsitype_f}{\oxvarctx^\prime_n}
  }{
    Consider first the typing derivation for $\oxvalue_f$ of
    $\oxtypjudge{\oxglobalctx}{\oxemptyctx}{\oxkontctx}{\oxvarctx}{
      \oxvalue_f
    }{
      \oxtfun{}{
        \oxsitype_1 \oxdotsc \oxsitype_n
      }{\oxsitype_f}{\oxenv_c}
    }{\oxvarctx_0^\prime}$.

    Applying \Lemma{lemma:value-typing-output} gives us
    $\oxtypjudge{\oxglobalctx}{\oxemptyctx}{\oxkontctx}{\oxvarctx_0^\prime}{
      \oxvalue_f
    }{
      \oxtfun{}{
        \oxsitype_1 \oxdotsc \oxsitype_n
      }{\oxsitype_f}{\oxenv_c}
    }{\oxvarctx_0^\prime}$.

    Then, as in the previous cases, we can apply
    \Lemma{lemma:values-dont-change}, \Lemma{lemma:value-typing-related-envs},
    and \Lemma{lemma:subtyping-value-typing} in sequence starting with the above
    typing derivation with each of the typing derivations for the arguments
    $\oxvalue_i$ and their subsequent corresponding region rewriting derivation.
    At the end, this gives us
    $\oxtypjudge{\oxglobalctx}{\oxemptyctx}{\oxkontctx}{\oxvarctx_n^\prime}{
      \oxvalue_f
    }{
      \oxtfun{}{
        \oxsitype_1 \oxdotsc \oxsitype_n
      }{\oxsitype_f}{\oxenv_c}
    }{\oxvarctx_n^\prime}$

    Then, inversion on \oxname{T-ClosureValue} on this typing derivation for
    $\oxvalue_f$ gives us $\oxfvars{\oxexpr} \setminus \overline{\oxid} =
    \overline{\oxid_f} = \oxdomain{\oxframe_c}|_\oxid$,
    $\overline{\oxcprov} = \overline{\oxfprovs{\oxvarctx(\oxid_f)}} \oxcomma
    \oxfprovs{\oxexpr} = \oxdomain{\oxframe_c}|_\oxcprov$, and
    $\oxtypjudge{\oxglobalctx}{\oxemptyctx}{\oxkontctx}{
      \oxnewframe{\oxvarctx_n^\prime}{
        \oxframe_c \oxcomma
        \oxvarctxentry{\oxid_1}{\oxsitype_1} \oxdotsc
        \oxvarctxentry{\oxid_n}{\oxsitype_n}
      }
    }{\oxexpr}{\oxsitype_r}{\oxnewframe{\oxvarctx_n^\prime}{\oxframe}}$. We can
    then apply \oxname{T-Framed} to get
    $\oxtypjudge{\oxglobalctx}{\oxemptyctx}{\oxkontctx}{
      \oxnewframe{\oxvarctx_n}{
        \oxextendctx{\oxframe_c}{
          \oxvarctxentry{\oxid_1}{\oxsitype_1} \oxdotsc
          \oxvarctxentry{\oxid_n}{\oxsitype_n}
        }
      }
    }{
      \oxframed{\oxexpr}
    }{\oxsitype_f}{\oxvarctx_n}$.
  }

  \oxpresline{
    \oxtunify[\oxcombine]{\oxemptyctx}{\oxkontctx}{\oxvarctx^\prime_n}{
      \oxsitype_f
    }{
      \oxsitype_f
    }{\oxvarctx^\prime_n}
  }{
    Immediate by \oxname{RR-Refl}.
  }

  \oxpresline{
    \exists \oxvarctx_o. \oxvarctx^\prime_n \oxintersect \oxvarctx_o =
    \oxvarctx^\prime_n
  }{
    $\oxvarctx_o = \oxvarctx^\prime_n$
  }
\end{preservation}

\oxpreservationcase{T-Function}{\TFunction}{
  \oxtfun{\overline{\oxprov} \oxcomma \overline{\oxtvar}}{
    \oxsitype_1 \oxdotsc \oxsitype_n
  }{\oxsitype_r}{}
}{\EFunction}{
  \oxvarctx
}

\begin{preservation}
  \oxpresline{\oxstorevalidity{\oxglobalctx}{\oxvarctx}{\oxstore}}{
    Immediate from our premise.
  }

  \oxpresline{
    \oxwfkontctx{\oxglobalctx}{\oxvarctx}{\overline{\oxvalue}}{\oxkontctx}
  }{
    Immediate from our premise.
  }

  \oxpresline{
    \oxtypjudge{\oxglobalctx}{\oxemptyctx}{\oxkontctx}{\oxvarctx}
    {\oxexpr^\prime}{\oxtype^\prime}{\oxvarctx}
  }{
    By \oxname{T-ClosureValue} since $\oxstore_c$ is empty, and we know that the
    body itself is well-typed as a consequence of inversion on
    \oxname{WF-FunctionDefinition} for \oxfnname.
  }

  \oxpresline{
    \oxtunify[\oxcombine]{\oxemptyctx}{\oxkontctx}{\oxvarctx}
    {\oxtype^\prime}{\oxtype}{\oxvarctx}
  }{
    Immediate by \oxname{RR-Refl}.
  }

  \oxpresline{
    \exists \oxvarctx_o. \oxvarctx \oxintersect \oxvarctx_o = \oxvarctx
  }{
    $\oxvarctx_o = \oxvarctx$
  }
\end{preservation}

\oxpreservationcase{T-Closure}{\TClosure}{
  \oxfunction{
    \overline{\oxenvvar} \oxcomma \overline{\oxprov} \oxcomma \overline{\oxtvar}
  }{
    \oxascribe{\oxid_1}{\oxsitype_1} \oxdotsc
    \oxascribe{\oxid_n}{\oxsitype_n}
  }{\oxsitype_r}{
    \oxexpr
  }
}{\EClosure}{
  \oxtupdatemany{\oxvarctx}{\oxid_{nc}}{\oxtuninit{\oxvarctx(\oxid_{nc})}}
}

\begin{preservation}
  \oxpresline{
    \oxstorevalidity{\oxglobalctx}{
      \oxtupdatemany{\oxvarctx}{\oxid_{nc}}{\oxtuninit{\oxvarctx(\oxid_{nc})}}
    }{
      \oxupdate{\oxstore}{
        \overline{\oxstoreentry{\oxid_{nc}}{\oxuninit}}
      }
    }
  }{
    Compared to $\oxvarctx$, we know that
    $\oxtupdatemany{\oxvarctx}{\oxid_{nc}}{\oxtuninit{\oxvarctx(\oxid_{nc})}}$
    has more things marked dead and no other changes. Thus, we can apply
    \oxname{R-Env} to get $\oxsubtypectx{\oxglobalctx}{\oxemptyctx}{\oxvarctx}{
      \oxtupdatemany{\oxvarctx}{\oxid_{nc}}{\oxtuninit{\oxvarctx(\oxid_{nc})}}
    }$. Then, we can apply \Lemma{lemma:stack-validity-related-envs} to get
    $\oxstorevalidity{\oxglobalctx}{
      \oxtupdatemany{\oxvarctx}{\oxid_{nc}}{\oxtuninit{\oxvarctx(\oxid_{nc})}}
    }{\oxstore}$. Since $\oxuninit$ is good at every dead type $\oxsdtype$ by
    \oxname{T-Dead}, we can then build a new derivation using that rule instead
    for every $\oxid_{nc}$ that is now at a dead type. This gives us
    $\oxstorevalidity{\oxglobalctx}{
      \oxtupdatemany{\oxvarctx}{\oxid_{nc}}{\oxtuninit{\oxvarctx(\oxid_{nc})}}
    }{
      \oxupdate{\oxstore}{
        \overline{\oxstoreentry{\oxid_{nc}}{\oxuninit}}
      }
    }$.
  }

  \oxpresline{
    \oxwfkontctx{\oxglobalctx}{
      \oxtupdatemany{\oxvarctx}{\oxid_{nc}}{\oxtuninit{\oxvarctx(\oxid_{nc})}}
    }{\overline{\oxvalue}}{\oxkontctx}
  }{
    Inverting \oxname{WF-Temporaries} gives us $\forall i \in 1 \oxdots n. \;
    \oxtypjudge{\oxglobalctx}{\oxemptyctx}{\oxsitype_1 \oxdotsc \oxsitype_{i-1}}{\oxvarctx}{
      \oxvalue_i
    }{\oxsitype_i}{\oxvarctx}$. By \oxname{R-Env}, we have that
    \oxsubtypectx{\oxglobalctx}{\oxemptyctx}{\oxvarctx}{
      \oxtupdatemany{\oxvarctx}{\oxid_{nc}}{\oxtuninit{\oxvarctx(\oxid_{nc})}}
    }. Thus, we can apply \Lemma{lemma:value-typing-related-envs} to each of the
    typing judgments and then apply \oxname{WF-Temporaries} to get
    $\oxwfkontctx{\oxglobalctx}{
      \oxtupdatemany{\oxvarctx}{\oxid_{nc}}{\oxtuninit{\oxvarctx(\oxid_{nc})}}
    }{\overline{\oxvalue}}{\oxkontctx}$.
  }

  \oxpresline{
    \oxtypjudge{\oxglobalctx}{\oxemptyctx}{\oxkontctx}{\oxvarctx_i}
    {\oxexpr^\prime}{\oxtype^\prime}{\oxvarctx_i}
  }{
    Immediate by inversion of \oxname{T-Closure} and application of
    \oxname{T-ClosureValue}. Note that they have identical premises.
  }

  \oxpresline{
    \oxtunify[\oxcombine]{\oxemptyctx}{\oxkontctx}{\oxvarctx_i}
    {\oxtype^\prime}{\oxtype^\prime}{\oxvarctx_i}
  }{
    Immediate by \oxname{RR-Refl}.
  }

  \oxpresline{
    \exists \oxvarctx_o. \oxvarctx_i \oxintersect \oxvarctx_o = \oxvarctx_i
  }{
    $\oxvarctx_o = \oxvarctx_i$
  }
\end{preservation}

\oxpreservationcase{T-Shift}{\TShift}{
  \oxshift{\oxexpr_i}
}{\EShift}{\oxvarctx^\prime}

\begin{preservation}
  \oxpresline{
    \oxstorevalidity{\oxglobalctx}{\oxvarctx^\prime}{\oxstore}
  }{
    By inversion of \oxname{WF-StackFrame} on $\oxstorevalidity{\oxglobalctx}{
      \oxextendctx{\oxvarctx^\prime}{\oxvarctxentry{\oxid}{\oxsdtype}}
    }{\oxextendctx{\oxstore}{\oxstoreentry{\oxid}{\oxvalue^\prime}}}$, we get
    $\oxstorevalidity{\oxglobalctx}{\oxvarctx_i}{\oxstore}$,
    $\oxdomain{
      \oxextendctx{\oxstackframe}{
        \oxstoreentry{\oxid}{\oxvalue^\prime}
      }
    } = \oxdomain{
      \oxextendctx{\oxframe}{
        \oxvarctxentry{\oxid}{\oxsdtype}
      }
    }|_\oxid$, and
    $\forall \oxid \in \oxdomain{
      \oxnewframe{\oxstore}{
        \oxextendctx{\oxstackframe}{
          \oxstoreentry{\oxid}{\oxvalue^\prime}
        }
      }
    }. \;
    \oxtypjudge{\oxglobalctx}{\oxemptyctx}{\oxkontctx}{
      \oxnewframe{\oxvarctx_i}{
        \oxextendctx{\oxframe}{
          \oxvarctxentry{\oxid}{\oxsdtype}
        }
      }
    }{
      (\oxnewframe{\oxstore}{
        \oxextendctx{\oxstackframe}{
          \oxstoreentry{\oxid}{\oxvalue^\prime}
        }
      })(\oxid)
    }{(\oxnewframe{\oxvarctx_i}{
        \oxextendctx{\oxframe}{
          \oxvarctxentry{\oxid}{\oxsdtype}
        }
      })(\oxid)}{
      \oxnewframe{\oxvarctx_i}{
        \oxextendctx{\oxframe}{
          \oxvarctxentry{\oxid}{\oxsdtype}
        }
      }
    }$. Note that $\oxnewframe{\oxvarctx_i}{\oxframe} = \oxvarctx^\prime$.
    We can then immediately see that the above implies
    $\oxdomain{\oxstackframe} = \oxdomain{\oxframe}|_\oxid$ and
    $\forall \oxid \in \oxdomain{
      \oxnewframe{\oxstore}{
        \oxextendctx{\oxstackframe}{
          \oxstoreentry{\oxid}{\oxvalue^\prime}
        }
      }
    }. \;
    \oxtypjudge{\oxglobalctx}{\oxemptyctx}{\oxkontctx}{
      \oxnewframe{\oxvarctx_i}{\oxframe}
    }{
      (\oxnewframe{\oxstore}{\oxstackframe})(\oxid)
    }{(\oxnewframe{\oxvarctx_i}{\oxframe})(\oxid)}{
      \oxnewframe{\oxvarctx_i}{\oxframe}
    }$. Thus, we can apply \oxname{WF-StackFrame} to get
    $\oxstorevalidity{\oxglobalctx}{
      \oxnewframe{\oxvarctx_i}{\oxstackframe}
    }{\oxstore}$ which can be rewritten as
    $\oxstorevalidity{\oxglobalctx}{\oxvarctx^\prime}{\oxstore}$.
  }

  \oxpresline{
    \oxwfkontctx{\oxglobalctx}{\oxvarctx^\prime}{\overline{\oxvalue}}{\oxkontctx}
  }{
    Inverting \oxname{WF-Temporaries} gives us $\forall i \in 1 \oxdots n. \;
    \oxtypjudge{\oxglobalctx}{\oxemptyctx}{\oxsitype_1 \oxdotsc \oxsitype_{i-1}}{\oxvarctx}{
      \oxvalue_i
    }{\oxsitype_i}{\oxvarctx}$.

    Applying \Lemma{lemma:values-dont-change} to the typing derivation (from
    \oxname{T-Shift}) for
    $\oxexpr$ (which we know is a value from \oxname{E-Shift}) gives us
    $\oxsubtypectx{\oxglobalctx}{\oxemptyctx}{\oxvarctx}{\oxvarctx^\prime}$.

    Thus, we can apply \Lemma{lemma:value-typing-related-envs} to each of the
    typing judgments to get $\forall i \in 1 \oxdots n. \;
    \oxtypjudge{\oxglobalctx}{\oxemptyctx}{\oxsitype_1 \oxdotsc \oxsitype_{i-1}}{\oxvarctx^\prime}{
      \oxvalue_i
    }{\oxsitype_i}{\oxvarctx^\prime}$.

    Finally, applying \oxname{WF-Temporaries} gives us
    $\oxwfkontctx{\oxglobalctx}{\oxvarctx^\prime}{\overline{\oxvalue}}{\oxkontctx}$.
  }

  \oxpresline{
    \oxtypjudge{\oxglobalctx}{\oxemptyctx}{\oxkontctx}{\oxvarctx^\prime}{\oxvalue}
    {\oxsitype}{\oxvarctx^\prime}
  }{
    We know from \oxname{E-Shift} that $\oxexpr$ is a value $\oxvalue$. Thus,
    we can apply \Lemma{lemma:value-typing-output} to
    $\oxtypjudge{\oxglobalctx}{\oxemptyctx}{\oxkontctx}{\oxvarctx}{\oxvalue}{\oxsitype}{
      \oxextendctx{\oxvarctx^\prime}{
        \oxvarctxentry{\oxid}{\oxsdtype}
      }
    }$ to get $\oxtypjudge{\oxglobalctx}{\oxemptyctx}{\oxkontctx}{
      \oxextendctx{\oxvarctx^\prime}{
        \oxvarctxentry{\oxid}{\oxsdtype}
      }
    }{\oxvalue}{\oxsitype}{
      \oxextendctx{\oxvarctx^\prime}{
        \oxvarctxentry{\oxid}{\oxsdtype}
      }
    }$.

    We now wish to show that
    $\oxtypjudge{\oxglobalctx}{\oxemptyctx}{\oxkontctx}{\oxvarctx^\prime}{\oxvalue}
    {\oxsitype}{\oxvarctx^\prime}$. By inspecting the grammar of values and
    their typing rules, we know that the only values who depend on the
    context are pointers and closure values. But by inversion on
    $\oxtypjudge{\oxglobalctx}{\oxemptyctx}{\oxkontctx}{\oxvarctx}{
      \oxshift{\oxvalue}
    }{\oxsitype}{\oxvarctx^\prime}$, we know that
    $\oxtypevalidity{\oxglobalctx}{\oxemptyctx}{\oxvarctx^\prime}{\oxsitype}$.
    Since the type is valid without the frame, we know that the values
    cannot depend on that frame. Thus, we can conclude
    $\oxtypjudge{\oxglobalctx}{\oxemptyctx}{\oxkontctx}{\oxvarctx^\prime}{\oxvalue}
    {\oxsitype}{\oxvarctx^\prime}$.
  }

  \oxpresline{
    \oxtunify[\oxcombine]{\oxemptyctx}{\oxkontctx}{\oxvarctx^\prime}
    {\oxsitype}{\oxsitype}{\oxvarctx^\prime}
  }{
    Immediate by \oxname{RR-Refl}.
  }

  \oxpresline{
    \exists \oxvarctx_o. \oxvarctx^\prime \oxintersect \oxvarctx_o = \oxvarctx^\prime
  }{
    $\oxvarctx_o = \oxvarctx^\prime$
  }
\end{preservation}

\oxpreservationcase{T-Framed}{\TFramed}{
  \oxframed{\oxexpr_i}
}{\EFramed}{\oxvarctx^\prime}

\begin{preservation}
  \oxpresline{
    \oxstorevalidity{\oxglobalctx}{\oxvarctx^\prime}{\oxstore}
  }{
    Applying \Lemma{lemma:stack-envminus} to $\oxstorevalidity{\oxglobalctx}{
      \oxnewframe{\oxvarctx^\prime}{\oxframe}
    }{\oxnewframe{\oxstore}{\oxstackframe}}$ gives us
    $\oxstorevalidity{\oxglobalctx}{\oxvarctx^\prime}{\oxstore}$.
  }

  \oxpresline{
    \oxwfkontctx{\oxglobalctx}{\oxvarctx^\prime}{\overline{\oxvalue}}{\oxkontctx}
  }{
    Inverting \oxname{WF-Temporaries} gives us $\forall i \in 1 \oxdots n. \;
    \oxtypjudge{\oxglobalctx}{\oxemptyctx}{\oxsitype_1 \oxdotsc \oxsitype_{i-1}}{\oxvarctx}{
      \oxvalue_i
    }{\oxsitype_i}{\oxvarctx}$.

    Applying \Lemma{lemma:values-dont-change} to the typing derivation (from
    \oxname{T-Framed}) for
    $\oxexpr$ (which we know is a value from \oxname{E-Framed}) gives us
    $\oxsubtypectx{\oxglobalctx}{\oxemptyctx}{\oxvarctx}{\oxvarctx^\prime}$.

    Thus, we can apply \Lemma{lemma:value-typing-related-envs} to each of the
    typing judgments to get $\forall i \in 1 \oxdots n. \;
    \oxtypjudge{\oxglobalctx}{\oxemptyctx}{\oxsitype_1 \oxdotsc \oxsitype_{i-1}}{\oxvarctx^\prime}{
      \oxvalue_i
    }{\oxsitype_i}{\oxvarctx^\prime}$.

    Finally, applying \oxname{WF-Temporaries} gives us
    $\oxwfkontctx{\oxglobalctx}{\oxvarctx^\prime}{\overline{\oxvalue}}{\oxkontctx}$.
  }

  \oxpresline{
    \oxtypjudge{\oxglobalctx}{\oxemptyctx}{\oxkontctx}{\oxvarctx^\prime}{\oxvalue}
    {\oxsitype}{\oxvarctx^\prime}
  }{
    We know from \oxname{E-Framed} that $\oxexpr$ is a value $\oxvalue$. Thus,
    we can apply \Lemma{lemma:value-typing-output} to
    $\oxtypjudge{\oxglobalctx}{\oxemptyctx}{\oxkontctx}{
      \oxnewframe{\oxvarctx}{\oxframe}
    }{\oxvalue}{\oxsitype}{
      \oxnewframe{\oxvarctx^\prime}{\oxframe^\prime}
    }$ to get $\oxtypjudge{\oxglobalctx}{\oxemptyctx}{\oxkontctx}{
      \oxnewframe{\oxvarctx^\prime}{\oxframe^\prime}
    }{\oxvalue}{\oxsitype}{
      \oxnewframe{\oxvarctx^\prime}{\oxframe^\prime}
    }$.

    We now wish to show that
    $\oxtypjudge{\oxglobalctx}{\oxemptyctx}{\oxkontctx}{\oxvarctx^\prime}{\oxvalue}
    {\oxsitype}{\oxvarctx^\prime}$. By inspecting the grammar of values and
    their typing rules, we know that the only values who depend on the
    context are pointers and closure values. But by inversion on
    $\oxtypjudge{\oxglobalctx}{\oxemptyctx}{\oxkontctx}{\oxvarctx}{
      \oxframed{\oxvalue}
    }{\oxsitype}{\oxvarctx^\prime}$, we know that
    $\oxtypevalidity{\oxglobalctx}{\oxemptyctx}{\oxvarctx^\prime}{\oxsitype}$.
    Since the type is valid without the frame, we know that the values
    cannot depend on that frame. Thus, we can conclude
    $\oxtypjudge{\oxglobalctx}{\oxemptyctx}{\oxkontctx}{\oxvarctx^\prime}{\oxvalue}
    {\oxsitype}{\oxvarctx^\prime}$.
  }

  \oxpresline{
    \oxtunify[\oxcombine]{\oxemptyctx}{\oxkontctx}{\oxvarctx^\prime}
    {\oxsitype}{\oxsitype}{\oxvarctx^\prime}
  }{
    Immediate by \oxname{RR-Refl}.
  }

  \oxpresline{
    \exists \oxvarctx_o. \oxvarctx^\prime \oxintersect \oxvarctx_o = \oxvarctx^\prime
  }{
    $\oxvarctx_o = \oxvarctx^\prime$
  }
\end{preservation}

\noindent Case \oxname{T-BorrowIndex}: \\[0.5em]
\begin{tabular}{l}
  \pbox{\textwidth}{\vspace*{0.5em} From premise: \vspace*{0.5em}} \\

  \framebox[\textwidth]{
    \figuresize
    \begin{mathpar}
      \TBorrowIndex
    \end{mathpar}
  } \\

  \pbox{\textwidth}{
    \vspace*{0.5em}
    Since $\oxexpr =
    \oxref{\oxcprov}{\oxmuta}{\oxindex{\oxplaceexpr}{\oxexpr}}$, by inspection of
    the reduction rules, we know that $\oxexpr$ steps with the following rule:
    \vspace*{0.5em}
  } \\

  \framebox[\textwidth]{
    \figuresize
    \begin{mathpar}
      \inferrule[E-EvalCtx]{
        \oxreduce{\oxglobalctx}{\oxstore}{\oxexpr}
                 {\oxstore^\prime}{\oxexpr^\prime}
      }{
        \oxreduce{\oxglobalctx}{\oxstore}{\oxref{\oxcprov}{\oxmuta}{\oxindex{\oxplaceexpr}{\oxexpr}}}
                 {\oxstore^\prime}{\oxref{\oxcprov}{\oxmuta}{\oxindex{\oxplaceexpr}{\oxexpr^\prime}}}
      }
    \end{mathpar}
  } \\[1.5em]

\end{tabular}

We begin by applying the induction hypothesis on $\oxexpr$. We get that $\exists
\oxvarctx_i$ such that
$\oxstorevalidity{\oxglobalctx}{\oxvarctx_i}{\oxstore^\prime}$ and
$\oxwfkontctx{\oxglobalctx}{\oxvarctx_i}{\overline{\oxvalue}}{\oxkontctx}$ and
$\oxtypjudge{\oxglobalctx}{\oxemptyctx}{\oxkontctx}{\oxvarctx_i}{\oxexpr^\prime}{\oxtnum}{\oxvarctx_f^\prime}$
and
$\oxtunify[\oxcombine]{\oxemptyctx}{\oxkontctx}{\oxvarctx_f^\prime}{\oxtnum}{\oxtnum}{\oxvarctx_s}$
and there exists $\oxvarctx_o$ such that $\oxvarctx^\prime = \oxvarctx_s
\oxintersect \oxvarctx_o$. Note that since rewriting does nothing with
$\oxtnum$, $\oxvarctx_f^\prime = \oxvarctx_s$. We chose $\oxvarctx_i$ in the
lemma statement to be \framebox{$\oxvarctx_i$} and need to show:

\begin{preservation}
  \oxpresline{
    \oxstorevalidity{\oxglobalctx}{\oxvarctx_i}{\oxstore^\prime}
  }{
    Immediate.
  }
  \oxpresline{
    \oxwfkontctx{\oxglobalctx}{\oxvarctx_i}{\overline{\oxvalue}}{\oxkontctx}
  }{
    Immediate
  }

  \oxpresline{
    \oxtypjudge{\oxglobalctx}{\oxemptyctx}{\oxkontctx}{\oxvarctx_i}{\oxref{\oxcprov}{\oxmuta}
      {\oxplaceexpr[\oxexpr^\prime]}}{\oxtref{\oxcprov}{\oxmuta}{\oxsitype}}
               {\oxvarctx_f^\prime[\oxcprov \mapsto \oxset{\oxloans^\prime}]}
  }{
    We'd like to apply \oxname{T-BorrowIndex}. In order to do so, we still need
    to show $\oxvarctx^\prime_f(\oxcprov) = \emptyset$,
    $\oxnotinclosure{\oxkontctx}{\oxvarctx^\prime_f}{\oxcprov}$,
    $\oxmusafety{\oxglobalctx}{\oxemptyctx}{\oxvarctx_f^\prime}{\oxmuta}{\oxplaceexpr}{\oxset{\oxloans^\prime}}$,
    and
    $\oxcomputetynoprov{\oxemptyctx}{\oxvarctx^\prime}{\oxmuta}{\oxplaceexpr}{\oxxitype}$.

    $\oxvarctx^\prime_f(\oxcprov) = \emptyset$ is immediate because $\oxvarctx^\prime_f(\oxcprov) \subseteq \oxvarctx^\prime(\oxcprov)$.

    Both $\oxnotinclosure{\oxkontctx}{\oxvarctx^\prime_f}{\oxcprov}$ and
    $\oxcomputetynoprov{\oxemptyctx}{\oxvarctx^\prime}{\oxmuta}{\oxplaceexpr}{\oxxitype}$
    follow from the fact that types are unchanged between $\oxvarctx^\prime$ and
    $\oxvarctx^\prime_f$.

    The last obligation is
    $\oxmusafety{\oxglobalctx}{\oxemptyctx}{\oxvarctx_f^\prime}{\oxmuta}{\oxplaceexpr}{\oxset{\oxloans^\prime}}$,
    which follows from \Lemma{lemma:ownership-safety-more-precise}
  }

  \oxpresline{
    \oxtunify[\oxcombine]{\oxemptyctx}{\oxkontctx}{\oxvarctx_f^{\prime}[\oxcprov \mapsto \oxset{\oxloans^\prime}]}
    {\oxtnum}{\oxtnum}{\oxvarctx_s}
  }{
    Immediate, with $\oxvarctx_s = \oxvarctx_f^\prime[\oxcprov \mapsto \oxset{\oxloans^\prime}]$.
  }

  \oxpresline{
    \exists \oxvarctx_o. \oxvarctx_s \oxintersect \oxvarctx_o =
    \oxvarctx^\prime[\oxcprov \mapsto \oxset{\oxloans}]
  }{
    From our application of the induction hypothesis above, we have that
    $\oxvarctx^\prime = \oxvarctx_f^\prime \oxintersect \oxvarctx_o$. As we
    found in the typing judgement section above, $\oxset{\oxloans^\prime}
    \subseteq \oxset{\oxloans}$. Therefore, for this judgement, we can use
    $\oxvarctx_o[\oxcprov \mapsto \oxset{\oxloans} \setminus
      \oxset{\oxloans^\prime}]$.
  }
\end{preservation}

\noindent Case \oxname{T-Tuple}: \\[0.5em]
\begin{tabular}{l}
  \pbox{\textwidth}{\vspace*{0.5em} From premise: \vspace*{0.5em}} \\

  \framebox[\textwidth]{
    \figuresize
    \begin{mathpar}
      \TTuple
    \end{mathpar}
  } \\
  \pbox{\textwidth}{
    \vspace*{0.5em}
    Since $\oxexpr = \oxprod{\oxvalue_1 \oxdotsc \oxvalue_{i-1}, \oxexpr_i \oxdotsc \oxexpr_n}$, by inspection of the
    reduction rules, we know that $\oxexpr$ steps with the following rule:
    \vspace*{0.5em}
  } \\

  \framebox[\textwidth]{
    \figuresize
    \begin{mathpar}
      \inferrule[E-EvalCtx]{
        \oxreduce{\oxglobalctx}{\oxstore}{\oxexpr_i}
                 {\oxstore^\prime}{\oxexpr_i^\prime}
      }{
        \oxreduce{\oxglobalctx}{\oxstore}{\oxprod{\oxvalue_1 \oxdotsc \oxvalue_{i-1}, \oxexpr_i \oxdotsc \oxexpr_n}}
                 {\oxstore^\prime}{\oxprod{\oxvalue_1 \oxdotsc \oxvalue_{i-1}, \oxexpr_i^\prime, \oxexpr_{i+1} \oxdotsc \oxexpr_n}}
      }
    \end{mathpar}
  } \\[1.5em]
\end{tabular} \\[1em]

We want to apply our induction hypothesis to the premise of \oxname{E-EvalCtx}
with the typing judgement $\oxtypjudge{\oxglobalctx}{\oxemptyctx}{\oxkontctx,
  \oxsitype_1 \oxdotsc \oxsitype_{i-1}}{\oxvarctx_{i-1}}
{\oxexpr_i}{\oxsitype_i}{\oxvarctx_i}$. In order to do so we need to show that
$\forall \ \overline{\oxvalue}$.
$\oxwfkontctx{\oxglobalctx}{\oxvarctx_{i-1}}{\overline{\oxvalue}, \oxvalue_1
  \oxdotsc \oxvalue_{i-1}}{\oxkontctx, \oxsitype_1 \oxdotsc \oxsitype_{i-1}}$
given that
$\oxwfkontctx{\oxglobalctx}{\oxvarctx_0}{\overline{\oxvalue}}{\oxkontctx}$. So
let $\oxvalue \in \overline{\oxvalue}$. For each $\oxvalue_{j}$, where $0 < j <
i$, we can repeatedly apply \Lemma{lemma:values-dont-change} to get
$\oxsubtypectx{\oxglobalctx}{\oxemptyctx}{\oxvarctx_{j-1}}{\oxvarctx_{i-1}}$, and
then apply \Lemma{lemma:value-typing-related-envs} to get that $\forall \oxvalue_k \in \overline{\oxvalue}$. 
$\oxtypjudge{\oxglobalctx}{\oxemptyctx}{\oxkontctx, \oxsitype_1 \oxdotsc
  \oxsitype_{k-1}}{\oxvarctx_{i-1}}{\oxvalue_k}{\oxkontctx[k]}{\oxvarctx_{i-1}}$.

Now we can apply the induction hypothesis to get there is some
$\oxvarctx_{i-1}^\prime$ such that $\forall\, \overline{\oxvalue}$.
$\oxwfkontctx{\oxglobalctx}{\oxvarctx_{i-1}^\prime}{\overline{\oxvalue}}
{\oxkontctx, \oxsitype_1 \oxdotsc \oxsitype_{i-1}}$ and
$\oxtypjudge{\oxglobalctx}{\oxemptyctx}{\oxkontctx, \oxsitype_1 \oxdotsc \oxsitype_{i-1}}{\oxvarctx_{i-1}^{\prime}}
{\oxexpr_i^{\prime}}{\oxtype_i^{\textsc{si} \prime}}{\oxvarctx_i^{\prime}}$,
with $\oxstorevalidity{\oxglobalctx}{\oxvarctx_{i-1}^\prime}{\oxstore^\prime}$
and $\oxtunify[\oxcombine]{\oxemptyctx}{\oxkontctx, \oxsitype_1 \oxdotsc \oxsitype_{i-1}}{\oxvarctx_i^\prime}
{\oxtype_i^{\textsc{si}\prime}}{\oxsitype_i}{\oxvarctx_{si}}$ and there is some
$\oxvarctx_{oi}$ such that $\oxvarctx_i = \oxvarctx_{si} \oxintersect \oxvarctx_{oi}$.

We then pick $\oxvarctx_i$ in the lemma statement to be
\framebox{$\oxvarctx_{i-1}^{\prime}$} and need to show:

\begin{preservation}
  \oxpresline{
    \oxstorevalidity{\oxglobalctx}{\oxvarctx_{i-1}^{\prime}}{\oxstore^\prime}
  }{
    Immediate.
  }
  \oxpresline{
    \oxwfkontctx{\oxglobalctx}{\oxvarctx_{i-1}^\prime}{\overline{\oxvalue}}{\oxkontctx}
  }{
    Immediate
  }

  \oxpresline{
    \begin{array}{r}
    \oxglobalctx; \, \oxemptyctx; \, \oxkontctx; \, \oxvarctx_{i-1}^{ \prime} \vdash
    \framebox{$\oxprod{\oxvalue_1 \oxdotsc \oxvalue_{i-1}, \oxexpr_i^\prime,
        \oxexpr_{i+1} \oxdotsc \oxexpr_n}$}\\
    : \oxprod{\oxsitype_1 \oxdotsc
        \oxsitype_{i-1}, \oxtype_i^{\textsc{si} \prime}, \oxsitype_{i+1}
        \oxdotsc \oxsitype_{n}} \Rightarrow \oxvarctx_n^{\prime}
    \end{array}
  }{ We want to apply \oxname{T-Tuple} in order to typecheck this tuple
    expression. We firstly want to show that we can typecheck $\oxvalue_1$
    through $\oxvalue_{i-1}$ with initial input context
    $\oxvarctx_{i-1}^{\prime}$. For each $\oxvalue_{j}$, where $0 < j < i$, we
    can repeatedly apply \Lemma{lemma:values-dont-change} to get
    $\oxsubtypectx{\oxglobalctx}{\oxemptyctx}{\oxvarctx_{j-1}}{\oxvarctx_{i-1}}$,
    and then apply \Lemma{lemma:value-typing-related-envs} to get that
    $\oxtypjudge{\oxglobalctx}{\oxemptyctx}{\oxkontctx, \oxsitype_1 \oxdotsc \oxsitype_{j-1}}{\oxvarctx_{i-1}}{\oxvalue_j}{\oxtype_j}{\oxvarctx_{i-1}}$,
    and finally applying our continuation typing hypothesis gives us
    $\oxtypjudge{\oxglobalctx}{\oxemptyctx}{\oxkontctx, \oxsitype_1 \oxdotsc \oxsitype_{j-1}}{\oxvarctx_{i-1}^\prime}{\oxvalue_j}{\oxtype_j}{\oxvarctx_{i-1}^\prime}$.

    Then we can use the result of our induction hypothesis,
    $\oxtypjudge{\oxglobalctx}{\oxemptyctx}{\oxkontctx, \oxsitype_1 \oxdotsc
      \oxsitype_{i-1}}{\oxvarctx_{i-1}^{\prime}}{\oxexpr_i^{\prime}}{\oxtype_i^{\textsc{si}
        \prime}}{\oxvarctx_i^{\prime}}$.

    And finally for $\oxexpr_{i+1} \oxdotsc \oxexpr_n$, we can repeatedly apply
    \Lemma{lemma:expressions-typing-more-precise}. }

  \oxpresline{
    \oxtunify[\oxcombine]{\oxemptyctx}{\oxkontctx}{\oxvarctx_n^{\prime}}
             {\oxprod{\oxsitype_1 \oxdotsc \oxsitype_{i-1},
                 \oxtype_i^{\textsc{si} \prime}, \oxsitype_{i+1} \oxdotsc
                 \oxsitype_{n}}}{\oxprod{\oxsitype_1 \oxdotsc
                 \oxsitype_n}}{\oxvarctx_s}
  }{
    Apply \oxname{RR-Tuple} to
    \oxname{RR-Refl} for every pair of types that are identical, leaving just
    $\oxtype_i^{\textsc{si}\prime}$ and $\oxsitype_i$ for which we can use
    $\oxtunify[\oxcombine]{\oxemptyctx}
    {\oxkontctx, \oxsitype_1 \oxdotsc \oxsitype_{i-1}}{\oxvarctx_i^\prime}
    {\oxtype_i^{\textsc{si}\prime}}{\oxsitype_i}{\oxvarctx_{si}}$ from
    our induction hypothesis with
    \Lemma{lemma:rewriting-parallel-expression} and
    \Lemma{lemma:rewriting-smaller-kont} to get that
    $\oxtunify[\oxcombine]{\oxemptyctx}{\oxkontctx}{\oxvarctx_n^\prime}
    {\oxtype_i^{\textsc{si}\prime}}{\oxsitype_i}{\oxvarctx_s}$.
  }

  \oxpresline{
    \exists \oxvarctx_o. \oxvarctx_s \oxintersect \oxvarctx_o = \oxvarctx_n
  }{
    Note from the induction hypothesis above, we have $\oxvarctx_{i} =
    \oxvarctx_{si} \oxintersect \oxvarctx_{oi}$. This means that for whatever
    loan sets were unioned between $\oxvarctx_i^\prime$ and $\oxvarctx_{si}$,
    those loan sets will be unioned between $\oxvarctx^\prime_n$ and
    $\oxvarctx_s$. And since $\oxvarctx_i$ contained $\oxvarctx_{si}$, it will
    also be true that $\oxvarctx_n$ will contain $\oxvarctx_s$, which means we
    can use $\oxvarctx_o = \oxvarctx_n \setminus \oxvarctx_s$.
   }
\end{preservation}

\oxpreservationcaseheader{T-Drop}{\TDrop}

Since \oxname{T-Drop} applies to \emph{any} expression $\oxexpr$, we cannot
determine anything about what rule we stepped with. So, we will instead try to
apply our induction hypothesis to the typing derivation in the premise of
\oxname{T-Drop}
\oxtypjudge{(\oxglobalctx}{\oxemptyctx}{\oxkontctx}{
  \oxtupdate{\oxvarctx}{\oxplace}{\oxsidtype_\oxplace}
}{\oxexpr}{\oxsxtype}{\oxvarctx_f)}. To do this, we need
to establish the premises of our inductive hypothesis.

Namely, we need to show:
\begin{enumerate}
\item \oxtypjudge{\oxglobalctx}{\oxemptyctx}{\oxkontctx}{
    \oxtupdate{\oxvarctx}{\oxplace}{\oxsidtype_\oxplace}
  }{\oxexpr}{\oxsxtype}{\oxvarctx_f},
\item $\oxstorevalidity{\oxglobalctx}{
    \oxtupdate{\oxvarctx}{\oxplace}{\oxsidtype_\oxplace}
  }{\oxstore}$,
\item $\oxreduce{\oxglobalctx}{\oxstore}{\oxexpr}
  {\oxstore^\prime}{\oxexpr^\prime}$.
\end{enumerate}

\vspace{0.5em}

\begin{enumerate}
\item follows immediately from our premise.
\item follows almost directly from our premise, which tells us that
  $\oxstorevalidity{\oxglobalctx}{\oxvarctx}{\oxstore}$.
  We just need to show that the value at $\oxid$ (where $\oxid$ is the root of
  $\oxplace$) is still valid at its new type. Fortunately, it's old derivation
  works almost perfectly except for typing the part that corresponds directly to
  $\oxplace$. In this case, we can use \oxname{T-Dead} on the value to get the
  new derivation with that part of the aggregate structure at the uninitialized
  type $\oxsidtype_\oxplace$.
\item follows immediately from our premise.
\end{enumerate}

\vspace{0.5em}

This allows us to use our induction hypothesis to conclude that there exists
$\oxvarctx^\prime_i$ such that:
\begin{enumerate}
  \setcounter{enumi}{4}
\item $\oxstorevalidity{\oxglobalctx}{\oxvarctx^\prime_i}{\oxstore^\prime}$,
\item $\oxwfkontctx{\oxglobalctx}{\oxvarctx^\prime_i}{\overline{\oxvalue}}{\oxkontctx}$,
\item $\oxtypjudge{\oxglobalctx}{\oxemptyctx}{\oxkontctx}{\oxvarctx^\prime_i}
  {\oxexpr^\prime}{\oxtype^\prime}{\oxvarctx_f^\prime}$,
\item $\oxtunify[\oxcombine]{\oxemptyctx}{\oxkontctx}{\oxvarctx_f^\prime}
  {\oxtype^\prime}{\oxsxtype}{\oxvarctx_s}$, and
\item $\exists \oxvarctx_o. \; \oxvarctx_s \oxintersect \oxvarctx_o = \oxvarctx_f$.
\end{enumerate}
\hfill$\square$


\subsection{Type Safety}
\label{sec:type-safety}

\begin{oxtheorem}{Type Safety}{theorem:type-safety-app}
  If $\oxtypjudge{\oxglobalctx}{\oxemptyctx}{\oxemptyctx}{\oxemptyctx}{\oxexpr}
  {\oxsitype}{\oxvarctx}$ and $\oxglobalctxvalidity{\oxglobalctx}$, then
  $\oxreducemany{\oxglobalctx}{\oxemptyctx}{\oxexpr}{\oxstore^\prime}{\oxvalue}$ or
  evaluation of $\oxexpr$ aborts or diverges.
\end{oxtheorem}

\begin{proof}
  By the interleaved use of \textbf{Progress} and \textbf{Preservation}.
\end{proof}


\end{document}
